\documentclass[12pt,expanded,copyright]{fsuthesis}
\usepackage{amsmath}
\usepackage{graphicx} 
\usepackage[american, shorthands=off]{babel}
\graphicspath{ {./images/Cu/BronzeOnNb}{./images/Cu/CuSnThickness}{./images/Cu/Diffusion}{./images/intro}{./images/Nb}{./images/Metallurgy}{./images/Sapphire}{./images}{./images/Cu}{./images/Methodology}{./images/Future} }
\usepackage[section]{placeins} 
\usepackage{booktabs}
\usepackage{array}
\usepackage[utf8]{inputenc}
\usepackage{textcomp} 
\usepackage[
    backend=biber,
    sorting=none,
    doi=true,
    url=false,
    eprint=true
]{biblatex}

\usepackage{blindtext}
\usepackage{subcaption}
\usepackage{siunitx}
\usepackage{pifont}
\usepackage{tcolorbox}
\usepackage{dirtytalk}
\usepackage{bm}
\usepackage{csquotes}
\usepackage{amssymb}
\usepackage{float}
\makeatletter
\newenvironment{chapquote}[2][2em]
  {\setlength{\@tempdima}{#1}%
   \def\chapquote@author{#2}%
   \parshape 1 \@tempdima \dimexpr\textwidth-2\@tempdima\relax%
   \itshape}
  {\par\normalfont\hfill--\ \chapquote@author\hspace*{\@tempdima}\par\bigskip}
\makeatother

\appto{\bibsetup}{\sloppy}

\usepackage{nomencl}
\makenomenclature
\usepackage{etoolbox}
\usepackage{xcolor}
\usepackage{needspace}

\renewcommand\nomgroup[1]{%
  \needspace{5\baselineskip}
  \vspace{2em}%
  \item[\Large\bfseries%
  \ifstrequal{#1}{P}{%
    \textcolor{blue!70!black}{Physical Quantities}%
  }{%
  \ifstrequal{#1}{G}{%
    \textcolor{red!70!black}{Greek Symbols}%
  }{%
  \ifstrequal{#1}{A}{%
    \textcolor{green!50!black}{Abbreviations}%
  }{}}}%
  ]%
  \nopagebreak[4]
  \vspace{0.3em}%
  \nopagebreak[4]%
  \hrule height 2pt%
  \nopagebreak[4]%
  \vspace{0.8em}%
  \nopagebreak[4]%
}

\nomenclature[P01]{$\mathbf{B}$}{Magnetic field vector (T)}
\nomenclature[P02]{$B$}{Magnetic field magnitude (T)}
\nomenclature[P03]{$B_0$}{External DC magnetic field (T)}
\nomenclature[P04]{$\mathbf{B}_{AC}$}{AC magnetic field component (T)}
\nomenclature[P05]{$\mathbf{B}_{DC}$}{DC magnetic field component (T)}
\nomenclature[P06]{$C_\alpha$}{Form factor (cavity mode overlap)}
\nomenclature[P07]{$c$}{Speed of light in vacuum ($2.998 \times 10^8$ m/s)}
\nomenclature[P08]{$c_{11}$}{Elastic shear modulus for vortices}
\nomenclature[P09]{$d$}{Grain size (nm)}
\nomenclature[P10]{$e$}{Elementary charge ($1.602 \times 10^{-19}$ C)}
\nomenclature[P11]{$\mathbf{E}$}{Electric field vector (V/m)}
\nomenclature[P12]{$E$}{Electric field magnitude (V/m)}
\nomenclature[P13]{$\mathbf{E}_{AC}$}{AC electric field component (V/m)}
\nomenclature[P14]{$E_{acc}$}{Accelerating electric field (MV/m)}
\nomenclature[P15]{$f$}{Frequency (Hz or GHz)}
\nomenclature[P16]{$f_0$}{Resonant frequency (Hz)}
\nomenclature[P17]{$f_a$}{Axion decay constant (GeV)}
\nomenclature[P18]{$\mathbf{F}_L$}{Lorentz force vector (N)}
\nomenclature[P19]{$F_P$}{Pinning force per unit length (N/m)}
\nomenclature[P20]{$f_p$}{Pinning force density}
\nomenclature[P21]{$G$}{Geometric factor ($\Omega$)}
\nomenclature[P22]{$g_{a\gamma\gamma}$}{Axion-photon coupling constant (GeV$^{-1}$)}
\nomenclature[P23]{$g_\gamma$}{Model-dependent coupling constant}
\nomenclature[P24]{$h$}{Planck constant ($6.626 \times 10^{-34}$ J$\cdot$s)}
\nomenclature[P25]{$\hbar$}{Reduced Planck constant ($1.055 \times 10^{-34}$ J$\cdot$s)}
\nomenclature[P26]{$\mathbf{H}$}{Magnetic field intensity vector (A/m)}
\nomenclature[P27]{$H$}{Magnetic field intensity magnitude (A/m)}
\nomenclature[P28]{$\mu_0H_{c}$}{Thermodynamic critical field (T)}
\nomenclature[P29]{$\mu_0H_{c1}$}{Lower critical field (T)}
\nomenclature[P30]{$\mu_0H_{c2}$}{Upper critical field (T)}
\nomenclature[P31]{$\mathbf{J}$}{Current density vector (A/m²)}
\nomenclature[P32]{$J_c$}{Critical current density (A/cm²)}
\nomenclature[P33]{$\mathbf{J}_{AC}$}{AC current density (A/m²)}
\nomenclature[P34]{$\mathbf{j}_s$}{Supercurrent density (A/m²)}
\nomenclature[P35]{$k$}{Wave number (m$^{-1}$)}
\nomenclature[P36]{$k_B$}{Boltzmann constant ($1.381 \times 10^{-23}$ J/K)}
\nomenclature[P37]{$\ell$}{Mean free path (nm)}
\nomenclature[P38]{$m$}{Electron mass (kg)}
\nomenclature[P39]{$m_e$}{Electron rest mass ($9.109 \times 10^{-31}$ kg)}
\nomenclature[P40]{$m_a$}{Axion mass (eV)}
\nomenclature[P41]{$m^*$}{Effective mass (kg)}
\nomenclature[P42]{$n$}{Electron density (m$^{-3}$)}
\nomenclature[P43]{$n_s$}{Superconducting electron density (m$^{-3}$)}
\nomenclature[P44]{$n_n$}{Normal electron density (m$^{-3}$)}
\nomenclature[P45]{$n_v$}{Vortex density (m$^{-2}$)}
\nomenclature[P46]{$P_{a\gamma\gamma}$}{Axion-to-photon conversion power (W)}
\nomenclature[P47]{$Q$}{Quality factor (dimensionless)}
\nomenclature[P48]{$Q_0$}{Unloaded quality factor}
\nomenclature[P49]{$Q_L$}{Loaded quality factor}
\nomenclature[P50]{$Q_a$}{Axion quality factor}
\nomenclature[P51]{$Q_1, Q_2$}{Quality factors from antenna 1 and 2}
\nomenclature[P52]{$R_s$}{Surface resistance ($\Omega$)}
\nomenclature[P53]{$R_{BCS}$}{BCS surface resistance ($\Omega$)}
\nomenclature[P54]{$R_0$}{Residual resistance ($\Omega$)}
\nomenclature[P55]{$R_c$}{Campbell resistance ($\Omega$)}
\nomenclature[P56]{$T$}{Temperature (K)}
\nomenclature[P57]{$T_c$}{Critical temperature (K)}
\nomenclature[P58]{$T_{sys}$}{System noise temperature (K)}
\nomenclature[P59]{$t$}{Time (s)}
\nomenclature[P60]{$U$}{Stored energy (J)}
\nomenclature[P61]{$V$}{Cavity volume (m³)}
\nomenclature[P62]{$\mathbf{v}$}{Velocity vector (m/s)}
\nomenclature[P63]{$v_F$}{Fermi velocity (m/s)}
\nomenclature[P64]{$v_d$}{Drift velocity (m/s)}

\nomenclature[G01]{$\alpha$}{Fine structure constant ($\approx 1/137$)}
\nomenclature[G02]{$\alpha_L$}{Labusch parameter (restoring force per unit displacement)}
\nomenclature[G03]{$\beta$}{Antenna coupling coefficient}
\nomenclature[G04]{$\beta_1, \beta_2$}{Coupling coefficients for antennas 1 and 2}
\nomenclature[G05]{$\Gamma$}{Reflection coefficient}
\nomenclature[G06]{$\Delta$}{Superconducting energy gap (eV)}
\nomenclature[G07]{$\Delta T_c$}{Critical temperature transition width (K)}
\nomenclature[G08]{$\Delta f_{\text{signal}}$}{Axion bandwidth (Hz)}
\nomenclature[G09]{$\Delta f_c$}{Cavity bandwidth (Hz)}
\nomenclature[G10]{$\delta$}{Skin depth (m)}
\nomenclature[G11]{$\delta_{anom}$}{Anomalous skin depth (m)}
\nomenclature[G12]{$\epsilon$}{Strain (dimensionless)}
\nomenclature[G13]{$\epsilon_0$}{Permittivity of free space ($8.854 \times 10^{-12}$ F/m)}
\nomenclature[G14]{$\xi$}{Coherence length (nm)}
\nomenclature[G15]{$\xi_0$}{Clean-limit coherence length (nm)}
\nomenclature[G16]{$\kappa$}{Ginzburg-Landau parameter ($\lambda/\xi$)}
\nomenclature[G17]{$\lambda$}{London penetration depth (nm)}
\nomenclature[G18]{$\lambda_c$}{Campbell penetration depth (nm)}
\nomenclature[G19]{$\mu_0$}{Permeability of free space ($4\pi \times 10^{-7}$ H/m)}
\nomenclature[G20]{$\rho$}{Electrical resistivity ($\Omega\cdot$m)}
\nomenclature[G21]{$\rho_n$}{Normal state resistivity ($\Omega\cdot$m)}
\nomenclature[G22]{$\rho_a$}{Axion density (GeV/cm³)}
\nomenclature[G23]{$\rho_{eff}$}{Effective resistivity from flux flow ($\Omega\cdot$m)}
\nomenclature[G24]{$\sigma$}{Electrical conductivity (S/m)}
\nomenclature[G25]{$\sigma_n$}{Normal state conductivity (S/m)}
\nomenclature[G26]{$\sigma_{eff}$}{Effective conductivity in anomalous skin regime (S/m)}
\nomenclature[G27]{$\tau$}{Electron scattering time (s)}
\nomenclature[G28]{$\Phi_0$}{Magnetic flux quantum ($h/2e = 2.07 \times 10^{-15}$ Wb)}
\nomenclature[G29]{$\omega$}{Angular frequency (rad/s)}
\nomenclature[G30]{$\omega_a$}{Axion angular frequency (rad/s)}
\nomenclature[G31]{$\psi$}{Superconducting order parameter}

\nomenclature[A01]{AC}{Alternating Current}
\nomenclature[A02]{ADMX}{Axion Dark Matter eXperiment}
\nomenclature[A03]{AFM}{Atomic Force Microscopy}
\nomenclature[A04]{APT}{Atom Probe Tomography}
\nomenclature[A05]{ASC}{Applied Superconductivity Center}
\nomenclature[A06]{A15}{Crystal structure type (e.g., Nb$_3$Sn)}
\nomenclature[A07]{BCC}{Body-Centered Cubic}
\nomenclature[A08]{BCS}{Bardeen-Cooper-Schrieffer (theory)}
\nomenclature[A09]{BSCCO}{Bismuth Strontium Calcium Copper Oxide}
\nomenclature[A10]{CAST}{CERN Axion Solar Telescope}
\nomenclature[A11]{CERN}{European Organization for Nuclear Research}
\nomenclature[A12]{CTE}{Coefficient of Thermal Expansion}
\nomenclature[A13]{Cu}{Copper}
\nomenclature[A14]{Cu-Sn}{Copper-Tin (bronze)}
\nomenclature[A15]{CVD}{Chemical Vapor Deposition}
\nomenclature[A16]{DC}{Direct Current}
\nomenclature[A17]{DFSZ}{Dine-Fischler-Srednicki-Zhitnitsky (axion model)}
\nomenclature[A18]{DOE}{Department of Energy}
\nomenclature[A19]{DUNE}{Deep Underground Neutrino Experiment}
\nomenclature[A20]{EDS}{Energy Dispersive X-ray Spectroscopy}
\nomenclature[A21]{FIB}{Focused Ion Beam}
\nomenclature[A22]{FSU}{Florida State University}
\nomenclature[A23]{HAYSTAC}{Haloscope At Yale Sensitive To Axion Cold dark matter}
\nomenclature[A24]{HiPIMS}{High Power Impulse Magnetron Sputtering}
\nomenclature[A25]{HTS}{High-Temperature Superconductor}
\nomenclature[A26]{KEK}{High Energy Accelerator Research Organization (Japan)}
\nomenclature[A27]{KSVZ}{Kim-Shifman-Vainshtein-Zakharov (axion model)}
\nomenclature[A28]{LIGO}{Laser Interferometer Gravitational-Wave Observatory}
\nomenclature[A29]{LTS}{Low-Temperature Superconductor}
\nomenclature[A30]{MKID}{Microwave Kinetic Inductance Detector}
\nomenclature[A31]{Mo}{Molybdenum}
\nomenclature[A32]{MRI}{Magnetic Resonance Imaging}
\nomenclature[A33]{Nb}{Niobium}
\nomenclature[A34]{Nb$_3$Sn}{Niobium-3-Tin}
\nomenclature[A35]{Nb$_6$Sn$_5$}{Niobium-6-Tin-5 (undesired phase)}
\nomenclature[A36]{NbSn$_2$}{Niobium-Tin-2 (undesired phase)}
\nomenclature[A37]{NbTi}{Niobium-Titanium}
\nomenclature[A38]{OFHC}{Oxygen-Free High Conductivity (copper)}
\nomenclature[A39]{ORGAN}{Oscillating Resonant Group AxioN}
\nomenclature[A40]{PPMS}{Physical Property Measurement System}
\nomenclature[A41]{PQ}{Peccei-Quinn (symmetry)}
\nomenclature[A42]{QCD}{Quantum Chromodynamics}
\nomenclature[A43]{QCM}{Quartz Crystal Microbalance}
\nomenclature[A44]{QPR}{Quadrupole Resonator}
\nomenclature[A45]{RBS}{Rutherford Backscattering Spectrometry}
\nomenclature[A46]{REBCO}{Rare-Earth Barium Copper Oxide}
\nomenclature[A47]{RF}{Radio Frequency}
\nomenclature[A48]{RRR}{Residual Resistivity Ratio}
\nomenclature[A49]{RRP}{Restacked Rod Process (Nb$_3$Sn wire architecture)}
\nomenclature[A50]{SE}{Secondary Electrons}
\nomenclature[A51]{SEM}{Scanning Electron Microscopy}
\nomenclature[A52]{SIMS}{Secondary Ion Mass Spectrometry}
\nomenclature[A53]{SLAC}{Stanford Linear Accelerator Center}
\nomenclature[A54]{Sn}{Tin}
\nomenclature[A55]{SNR}{Signal-to-Noise Ratio}
\nomenclature[A56]{SQUID}{Superconducting Quantum Interference Device}
\nomenclature[A57]{SRF}{Superconducting Radio Frequency}
\nomenclature[A58]{Ta}{Tantalum}
\nomenclature[A59]{TE}{Transverse Electric (mode)}
\nomenclature[A60]{TEM}{Transmission Electron Microscopy}
\nomenclature[A61]{TES}{Transition Edge Sensor}
\nomenclature[A62]{Ti}{Titanium}
\nomenclature[A63]{TM}{Transverse Magnetic (mode)}
\nomenclature[A64]{UHV}{Ultra-High Vacuum}
\nomenclature[A65]{VNA}{Vector Network Analyzer}
\nomenclature[A66]{XPS}{X-ray Photoelectron Spectroscopy}
\nomenclature[A67]{XRD}{X-Ray Diffraction}

\addto\captionsamerican{\renewcommand*{\contentsname}{Table of Contents}}

\usepackage[%
    colorlinks=true,
    pdfborder={0 0 0},
    citecolor=black,
    urlcolor=black,
    hypertexnames=false   
]{hyperref}

\usepackage{bookmark}     
\usepackage{breakurl}
\usepackage{cleveref}

\usepackage{bookmark}

\usepackage{breakurl}  
\usepackage{cleveref}

\title{Thesis 5_22_25}
\author{Andre Robert Juliao}
\date{May 2025}

\title{\texorpdfstring{Nb$_3$Sn}{Nb3Sn} Thin Films Using a Cu-Sn Route \protect\\ for Dark Matter Detection}
\author{Andre Robert Juliao}  
\college{College of Arts and Sciences}     
\department{Department of Physics}  
\manuscripttype{Dissertation}               
\degree{Doctor of Philosophy}            
\degreeyear{2025}                     
\defensedate{November 7, 2025}             

\subject{Dissertation Formatting}     

\committeeperson{Lance Cooley}{Professor Co-Directing Dissertation}
\committeeperson{Irinel Chiorescu}{Professor Co-Directing Dissertation}
\committeeperson{Ingo Wiedenhover}{Committee Member}
\committeeperson{Takemichi Okui}{Committee Member}
\committeeperson{David Larbalestier}{Committee Member}

\begin{document}

\frontmatter          
\maketitle            
\makecommitteepage    

\chapter*{}
\vspace*{\fill}
\begin{center}
\textit{To myself at 13,\\
who was amazed by the universe,\\
decided to venture out into the unknown,\\
and never looked back.}
\end{center}
\vspace*{\fill}
\chapter*{Acknowledgments}

I'd like to thank my parents Federico and Cynthia Juliao for inspiring and supporting me with their love. Thank you to all my friends especially Andy, Alex, Patrick, Chippy, Ruth, Caleb, Jerry, Andre Q, and Nick.

I would like to express my deepest appreciation to my advisor, Dr. Lance Cooley, whose ideas have helped shape both this PhD and the scientist I have become. His willingness to remain available throughout my time, despite his busy schedule, was invaluable.\textit{}

I want to thank Wenura Withanage for setting up the lab and much of the groundwork for the experiments performed in this dissertation. Without him, there would not have been a lab to work in or experiments to follow.

I am grateful to my committee members, Dr. David Larbalestier, Dr. Irinel Chiorescu, Dr. Takemichi Okui, and Dr. Ingo Wiedenhoever, for their time, insight, and constructive feedback on this work.

The Applied Superconductivity Center (ASC) has been an extraordinary environment where I could experiment freely and delve deeper into any question of relevance to the scientific frontier. I thank the faculty at ASC, Dr. Peter Lee, Dr. Shreyas Balachandran, Dr. Eric Hellstrom, Dr. Chiara Tarantini, and Dr. Fumitake Kametani, for fostering a collaborative and rigorous scientific culture. I am indebted to the staff, particularly Connie Linville, Van Griffin, and Bill Starch, whose support kept the lab running smoothly. I also thank my fellow researchers: postdocs Santosh Chetri and Rastislav Ries; graduate students Manish Mandal, Griffin Bradford, Jonathan Lee, and Emma Martin; and undergraduates Nikolya Cadavid and Liora Louis. Special thanks to former ASC members Nawaraj Paudel and John Tietsworth for their camaraderie and collaboration. I also appreciate the skilled work of Emilio Morillo and James Gillman in the machine shop. 

I thank our collaborators in the ADMX Collaboration, particularly at Lawrence Livermore National Laboratory, Gianpaolo Carosi, Nick Du, and Sean Durham, and at Pacific Northwest National Laboratory, Thomas Braine, for their partnership on axion dark matter detection research.

I am grateful to Bill Brey, Malathy Elumalai, Thierry Dubroca, and Ilya Litvak at the National High Magnetic Field Laboratory for their expertise and assistance.

I also acknowledge the international thin film SRF community, including colleagues at INFN, CEA, CERN, UKRI, KEK, and HZB, for valuable discussions and shared knowledge that advanced this work.

\vspace{1em}

This work was supported by the U.S. Department of Energy, Office of Science, Office of High Energy Physics, under the program High-Field Cavities for Axion Dark Matter DE-SC0023656. Additional support was provided by the Office of Science Graduate Student Research (SCGSR) Program and the Accelerator Research and Development Program (ARDAP) under Award No. DE-SC0009960. A portion of this work was performed at the National High Magnetic Field Laboratory, which is supported by National Science Foundation Cooperative Agreement Nos. DMR-1644779 (2018--2022) and DMR-2128556 (2023--2027), and the State of Florida.

\bgroup 
\hypersetup{linkcolor = black}
\tableofcontents
\listoftables
\listoffigures

\cleardoublepage
\phantomsection
\addcontentsline{toc}{frontmatter}{Nomenclature}
\printnomenclature


\egroup

\begin{abstract}

Axion dark matter searches require superconducting radio-frequency (SRF) cavities on copper (Cu) substrates capable of maintaining quality factors $Q > 10^{5}$ in multi-tesla magnetic fields. Copper is an ideal substrate as it reduces thermal noise and offers excellent formability for complex cavity geometries across a broad frequency range. Among the superconductors available for these applications, Nb$_3$Sn is a compelling candidate due to its superior superconducting properties and well-established fabrication history. However, achieving compositional uniformity and high critical temperature ($T_c$) in Nb$_3$Sn thin films on Cu is complicated by Sn loss and substrate-induced strain.

This work addresses Sn loss by investigating solid-state diffusion of Sn from high-Sn Cu-Sn alloys into Nb layers to form Nb$_3$Sn at temperatures compatible with Cu substrates ($650-750^\circ$C), in contrast to traditional Sn vapor diffusion methods requiring $\sim1100^\circ$C. By systematically varying the Cu-Sn alloy composition, an optimal formulation was identified that maintained high Sn activity throughout the Nb$_3$Sn formation reaction. 

Compositional analysis of the Nb$_3$Sn phase, combined with thermal expansion mismatch calculations, revealed that $T_c$ was suppressed below the optimal value of 18~K due to strain effects from the Cu substrate. To isolate strain contributions, complementary studies on Nb and sapphire substrates, which have thermal expansion coefficients closer to Nb$_3$Sn, helped elucidate the impact of strain from the Cu-Sn layer and the Cu substrate. Two viable routes emerged: (1) Cu-Sn deposited on Ta-coated Cu, followed by hot Nb sputtering to form Nb$_3$Sn during deposition, and (2) Nb deposited on Ta/Cu, followed by Cu-Sn evaporation and ex-situ reaction to form Nb$_3$Sn. Route 1 achieved $T_c = 16$~K with a narrow transition width ($\Delta T_c \sim 1$~K), while Route 2 yielded exceptionally uniform Nb$_3$Sn layers and was selected for cavity coating.

To demonstrate RF performance, a hexagonal cavity design was developed by merging ideas from the University of Washington~\cite{Braine:2024nzi} and the Center for Axion and Precision Physics~\cite{Ahn:2021fgb}. This cavity accommodated flat coating surfaces and fit within the bore of a high-field magnet system at the Applied Superconductivity Center and a dilution fridge at Lawrence Livermore National Laboratory. The cavity with coating from Route 2 was characterized at temperatures down to 50~mK and magnetic fields up to 9~T. At zero field, the Nb$_3$Sn-coated cavity achieved $Q = 77{,}000$, a 40\% improvement over bare Cu ($Q = 55{,}000$). However, Q degraded rapidly with increasing magnetic field, dropping to $Q \sim 4{,}000$ at 9~T.

This dissertation demonstrates that Nb$_3$Sn coatings on Cu cavities can outperform bare Cu at zero field, with clear pathways for further optimization. Since axion detection rates scale with cavity quality factor, improvements in Nb$_3$Sn processing could significantly accelerate the search for axion dark matter. The processing routes developed here provide a foundation for future work toward Nb$_3$Sn processing routes for Cu cavity bodies.

\end{abstract}

\mainmatter

\chapter{Introduction}

\begin{chapquote}{Diogenes Laërtius, Democritus, -400 BC}
Nothing exists except atoms and empty space; everything else is opinion.
\end{chapquote}

\section{Background and Motivation}

Science has profoundly transformed our understanding of our place in the universe, from early Greek philosophers such as Democritus, who first speculated about indivisible atoms, to our modern understanding of particle physics within the Standard Model. New experiments promise to extend this view further. The Standard Model of particle physics~\cite{Donoghue:1992dd, chengGaugeTheoryElementary2000} has withstood many experimental tests to $5\sigma$ accuracy. However, cracks in its framework, such as its inability to explain dark matter (85\% of all matter in the universe), hint at the need for improvements to this fundamental model~\cite{Witten:1997sc}.

Among the most promising candidates for dark matter is the theoretical particle called the axion. This hypothetical particle was initially proposed to solve the strong charge-parity (CP) problem, a long-standing problem in quantum chromodynamics (QCD)~\cite{pecceiMathrmCPConservation1977}. The Axion Dark Matter eXperiment (ADMX) is one of the leading efforts to detect axions, using tunable resonant microwave cavities immersed in strong magnetic fields~\cite{duffyHighResolutionSearch2006}. In this experiment, the axion couples to an external magnetic field and decays into detectable photons via the Primakoff effect~\cite{primakoffDoubleBetaDecay1959, Sikivie:1983ip} within the cavity volume. A key challenge for this experiment is distinguishing the weak narrow-band axion signal from the noise. In addition, the ADMX experiment requires scanning over a wide axion mass range, so the scan rate must be maximized to probe the parameter space efficiently. The detector scan rate depends on the magnetic field ($B^4$), the volume ($V^2$), and the quality factor ($Q$). However, axion frequencies of interest correspond to cavity diameters ranging from a few to a hundred centimeters, and trade-offs between magnet bore size, field strength, and cost constrain both $B$ and $V$. For instance, plans for the next-generation ADMX experiment will use a large-bore 9~T magnet previously used only for medical imaging research~\cite{Braine:2024nzi}. Therefore, improving cavity $Q$ with superconducting thin films is the most accessible route for increasing the scan rate.

Currently, only normal-conducting resonant cavities are used in high-magnetic-field applications because they degrade minimally under such conditions. Cu has been the primary conductor used for alternating current (AC), radio-frequency (RF), high-field applications because it can easily be refined to a very high purity (RRR $>$ 300), resulting in low surface resistance of $\sim 5$~\unit{m\ohm} and a high quality factor of $Q \sim 3 \times 10^5$. However, as will be discussed in more detail later, in high-frequency low-temperature operation, the anomalous skin effect limits how low the surface resistance of normal metals can drop~\cite{pippardf.r.s.SurfaceImpedanceSuperconductors1947}. Superconductors are not limited by this mechanism~\cite{tinkhamIntroductionSuperconductivity2004}, and can achieve surface resistances orders of magnitude lower than that of Cu, e.g., a few \unit{n\ohm}'s for superconducting Nb~\cite{ciovatiSurfaceResistanceHigh}. Unfortunately, conventional superconductors such as niobium, which have been used previously to make resonant cavities, have a low upper critical field ($H_{c2} \approx 0.2$ T for Nb) and, therefore, break down at the magnetic fields of interest for axion detection.

Niobium-tin (Nb$_3$Sn) represents the most promising near-term solution for high-field superconducting cavities. Recent demonstrations using Nb$_3$Sn thin films~\cite{Posen:2022tbs} and other superconductors such as NbTi bulk~\cite{FirstQUAXGalactic2020} and REBCO tape~\cite{ahnSuperconductingCavityHigh2020, Ahn:2021fgb} have achieved order-of-magnitude improvements in quality factor under high magnetic fields, sufficient to reduce axion experiment runtime from 4 years to 2 years if brought to scale. Nb$_3$Sn offers particular advantages with a higher upper critical field ($H_{c2} \approx 30$ T) than NbTi (the material used for large medical imaging magnets), lower surface resistance than REBCO~\cite{gurevichUseNotUse2011}, and compatibility with Cu substrates~\cite{padamseeRFSuperconductivity2009, antoineHowAchieveBest2015} for superior thermal management.

Achieving low surface resistance with high-field superconductors requires careful material optimization. Superconductors with high $H_{c2}$ typically have high normal-state resistivity ($\rho_n$) thousands of times higher than high-purity Cu. In high magnetic fields, normal electrons exist within vortex cores that cause dissipation that scales as $R_s \propto f^2\sqrt{\rho_n}$~\cite{Mattis:1958yid}. Normal electrons also exist in a bulk superconductor due to impurities and thermal excitations. The impact of vortices can be minimized by designing the cavity shape and positioning it in the magnetic field to avoid strong interactions with the vortex cores. Thermal excitations can be limited by operating axion detectors at sub-Kelvin temperatures. To achieve a net surface resistance below that of Cu, all contributions from highly resistive normal electrons must be minimized through careful experiment, material, and recipe selection.

This dissertation employs a ternary Nb–Cu–Sn process to fabricate Nb$_3$Sn. Tin is supplied via solid-state diffusion reaction from a Cu-Sn alloy to Nb at $\sim700~^\circ$~C to form the Nb$_3$Sn compound. These reaction parameters are compatible with Cu substrates, and have been built upon established techniques for optimizing Nb$_3$Sn wires for magnet technology. Therefore, the wire knowledge base may be extended to thin-film geometries. Literature has yet to explore how these Nb$_3$Sn processes and the microstructures they produce affect superconducting properties in high-frequency, high-field regimes. This dissertation is a step towards a better understanding of these properties.

\section{Dissertation Contribution to Broader Knowledgebase}

The dissertation study resulted in three main achievements in Nb$_3$Sn thin films for the broader field, all of which were produced on Cu substrates and Ta diffusion barriers using common deposition equipment and reaction temperatures between 650 and 750$~^\circ$~C common to the wire fabrication industry:

\begin{enumerate}
    \item Depositing Nb via magnetron sputtering on Cu-Sn alloy layers held at high temperature, thin ($\sim300$~nm) Nb$_3$Sn films with maximum Sn content, high $T_c$ (16~K) and sharp superconducting transitions ($\Delta T_c \sim 1$~K) have been demonstrated. Strain due to thermal expansion mismatch reduces $T_c$ from its optimal value of $\sim18$~K.
    \item Depositing a thick ($\sim3$~\unit{\um}) Nb layer on a Cu-Sn alloy layer held at high temperature resulted in a $T_c$ onset of 17.7~K. This suggests that mitigation of strain effects using thick layers is possible. 
    \item Sputtering Nb, coating with a Cu-Sn alloy, and then reacting, a Nb$_3$Sn layer with very uniform grain morphology was produced. Films exhibited $T_c = 15$~K and $\Delta T_c \sim 2$~K . 
\end{enumerate}

Additionally, characterizing Nb$_3$Sn films under RF conditions in high magnetic fields required the design and testing of a small cavity resonator. The starting point was Braine's cylindrical cavity~\cite{Braine:2024nzi}, sized to fit the 26~mm bore of a high-field laboratory magnet with the cylinder axis aligned to the field. The resonant frequency of approximately 10~GHz permits quality factor measurements using standard RF equipment and a vector network analyzer. However, Braine's cylindrical geometry was incompatible with the thin-film deposition methods explored in this dissertation. Ahn et al.~\cite{ahnSuperconductingCavityHigh2020, Ahn:2021fgb} addressed a similar constraint using wedge segments with seams aligned parallel to the cavity axis. Combining these approaches, this work developed and tested a hexagonal cavity in which flat coating surfaces, compatible with planar deposition, were assembled into a prism with its long axis aligned to the magnetic field.

This dissertation developed an electromagnetic model of the cavity to extract parameters such as quality factor and surface resistance. The electromagnetic model was also built for a cylinder to permit cross-referencing to other cavity measurements. Seams between the prism walls were included to account for the real conditions in the actual cavity. The cavity was tested under various conditions, including a high magnetic field and a temperature of $\sim50$mK separately. Actual axion detectors would operate under both of these conditions simultaneously. Data was obtained under conditions rarely reported in the literature. It was demonstrated that hexagonal cavities with Nb$_3$Sn-coated pieces achieved a higher quality factor than those with copper alone.

This dissertation provides an assessment of how the thermally evaporated Cu-Sn route can improve Nb$_3$Sn thin film performance and discusses limitations of this work:

\textbf{Outcomes that support further work:}
\begin{itemize}
    \item Low-temperature ternary reactions (650--750°C enables Cu substrate compatibility)
    \item Sharp superconducting transitions ($\Delta T_c \sim 1$~K through post-reaction optimization)
    \item High-purity, high-Sn concentration source
\end{itemize}

\textbf{Practical limitations encountered:}
\begin{itemize}
    \item Line-of-sight deposition (hard to coat complex geometries)
    \item Surface preparation and a lack of substrate rotation in the thermal evaporator currently limits Q to 77,000 (vs Cu at 55,000), indicating a need for improved cleaning protocols
    \item 2~K depression in $T_c$ onset relative to Nb substrates (16~K vs 17.9~K) is well understood as a strain effect, with clear paths forward (thicker diffusion barriers)
    \item Cu in grain boundaries (beneficial for grain refinement, but higher impurities = higher residual resistance)
\end{itemize}

By documenting achieved performance (sharp $\Delta T_c$, uniform thickness, cavity validation) alongside practical constraints (geometry limitations, strain effects, surface preparation challenges), this dissertation provides data that enables the SRF community to compare films in this work with other recipes. The route offers an alternative to the limited number of fabrication approaches that have demonstrated high-Q, high-field cavities. This work demonstrates that a thermally evaporated Cu-Sn process can produce Nb$_3$Sn films with properties comparable to those of the best-reported films. Future work can advance proof-of-concept demonstrations by optimizing cavity design and coating more cavities with the recipes outlined in this dissertation.

\section{Dissertation Outline}

This dissertation is organized as follows. Chapter 2 introduces axions and axion detectors. Chapter 3 covers the properties of conductors and superconductors relevant to RF cavities. Chapter 4 describes the experimental methods used to characterize superconductors. Chapter 5 presents the properties and standard fabrication techniques for Nb$_3$Sn. Chapter 6 details the thin-film deposition method developed in this work and its results. Chapter 7 reviews prototype superconducting axion detector cavities from recent literature. Chapter 8 reports the RF cavity designs and measurements from this dissertation. Chapter 9 summarizes the findings and suggests future directions. Appendix A provides detailed step-by-step information for future researchers in this area of study to continue the work.

\chapter{Axion Physics and Detection with Resonant Cavities}

This chapter begins by introducing dark matter and how a theoretical particle, the axion, may resolve this long-standing enigma. A brief historical section on particle detectors is presented to provide context for current-generation detector experiments. A discussion of current experiments to detect axions is also given, including various strategies to probe different mass ranges and types of coupling. Finally, a deep dive into the resonant haloscope detector design is provided.

    \section{The Axion} \label{sec:axion}
    
\begin{figure}
        \centering
        \includegraphics[width=.8\textwidth]{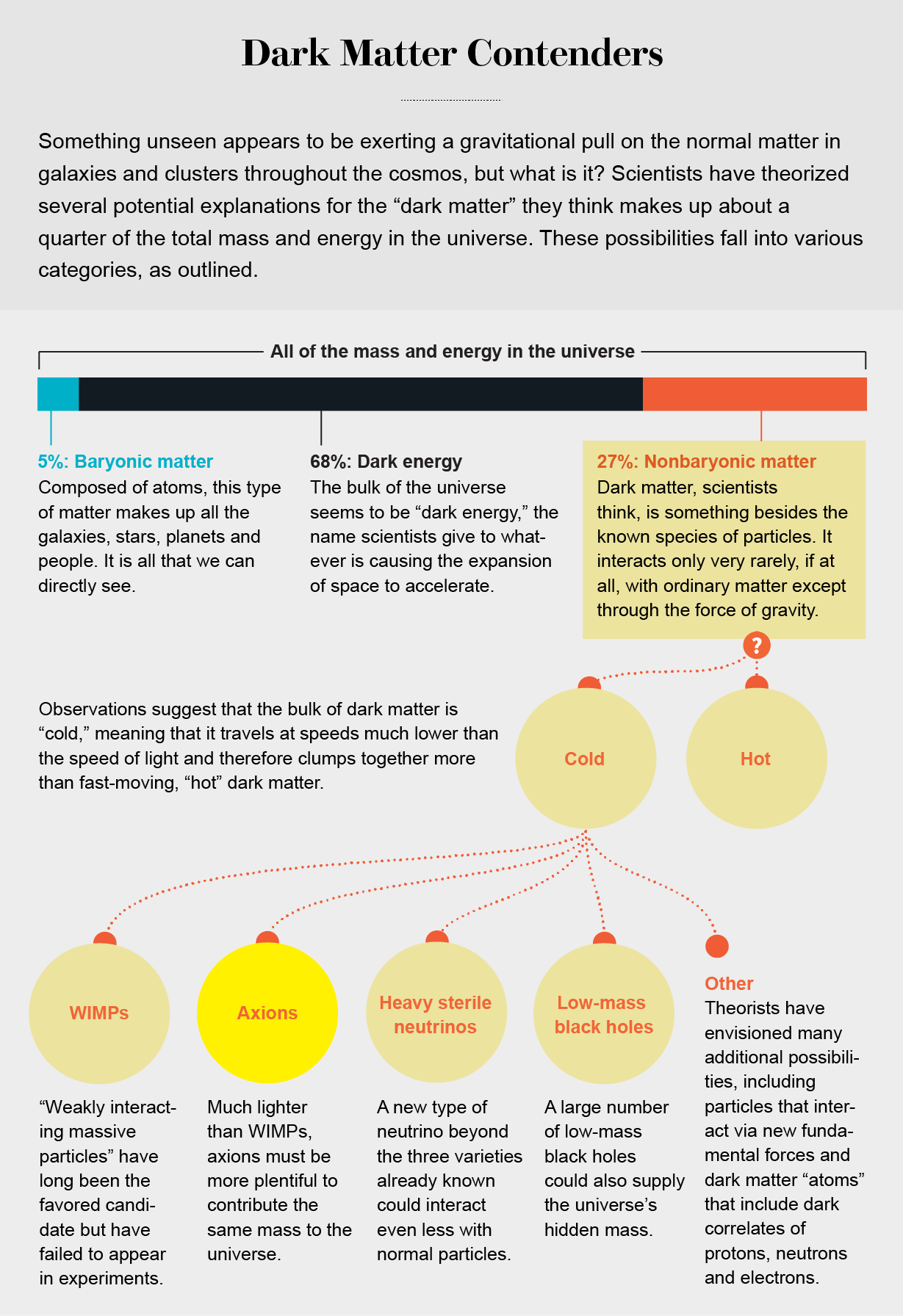}
        \caption[Proposed Dark Matter candidates.]{Proposed Dark Matter candidates. By Jen Christiansen; Article by Leslie Rosenberg, Scientific American, January 2018~\cite{rosenbergSearchingDarkHunt2018}.}
        \label{fig:DarkMatterContenders}
    \end{figure}

    The $\Lambda$CDM cosmological model~\cite{popoloSmallScaleProblems2017} indicates that approximately 85\% of the universe’s matter consists of dark matter, which may account for deviations observed in galactic rotation curves~\cite{Begeman:1991iy}, gravitational lensing~\cite{Clowe:2006eq}, cosmic microwave background anisotropies~\cite{WMAP:2010qai}, and galaxy cluster collisions~\cite{zwickyRotverschiebungExtragalaktischenNebeln1933}. The matter is called dark because it interacts weakly with light. Particle theorists have proposed many solutions to the dark matter problem. Weakly Interacting Massive Particles (WIMPs) were one of the first dark matter candidates, but these particles have not yet been found at the energies predicted. This suggests that WIMPs might not account for the observed dark matter~\cite{Malik:2012sa, lowetteSupersymmetrySearchesATLAS2012}, so alternative theories have been proposed (Fig.~\ref{fig:DarkMatterContenders})~\cite{rosenbergSearchingDarkHunt2018}.


    A new particle called the axion is gaining traction as the best possible candidate for dark matter, first proposed by Peccei and Quinn (PQ symmetry breaking)~\cite{pecceiMathrmCPConservation1977} in 1977 and Weinberg in 1978~\cite{Weinberg:1978ma}. There may be many axion-like particles (termed the axiverse), some of which naturally arise in fundamental models such as string theory~\cite{Witten:1997sc}. This leads to optimism in the community, since the axion may naturally emerge from a deeper theory, but also pessimism that the axion signal may be further suppressed, given many different axion-like particles.


    Properties of the axion mainly depend on the value of the axion decay constant $f_a$, which is inversely proportional to the axion mass $m_a$~\cite{Weinberg:1978ma}:
    
    \begin{equation}
        m_a \approx  6 \times10^{-5}~\text{eV} \left( \frac{10^{12}~\text{GeV}}{f_a} \right) \,.
        \label{eq:massscaling}
    \end{equation}
    
    \noindent This theory claims that in the early universe, the axion field underwent ``vacuum realignment", which produced cold dark matter with density $\rho_a \propto f_a^2 \theta_{in}^2$, where $\theta_{in}$ is the initial misalignment angle. In the simplest axion model, when $\theta_{in} \sim \mathcal{O}(1)$ and the axion density equals the dark matter density $\rho_a \simeq \rho_{\text{DM}} \approx 0.4$ GeV/cm$^3$, the predicted axion mass is on the order of $10^{-5}$ eV (corresponding to $f_a \sim 10^{12}$ GeV) \cite{Braine:2024nzi}. 
    
    The axion mass has large uncertainties due to the homogenization of the axion field during inflation, topological defects, and the unknown composition of dark matter. Literature has bounded the axion mass to a range of $10^{-13}$ to $10^{-2}$ eV. The lower bound comes from the age of the universe, and the upper bound from accelerator and stellar evolution constraints~\cite{Sikivie:1983ip}. Early experiments searching for the axion yielded null results, as its coupling to ordinary matter was speculated to be too weak for detection~\cite{Sikivie:2020zpn}. As will be discussed in the next section, particle detector sensitivity has improved significantly over time, and the new generation of detectors may be able to detect the weak axion signal.
        
\section{Particle Detectors}

    \subsection{Antiquity to Modern}
        
    \begin{figure}
        \centering
        \includegraphics[width=.2\textwidth]{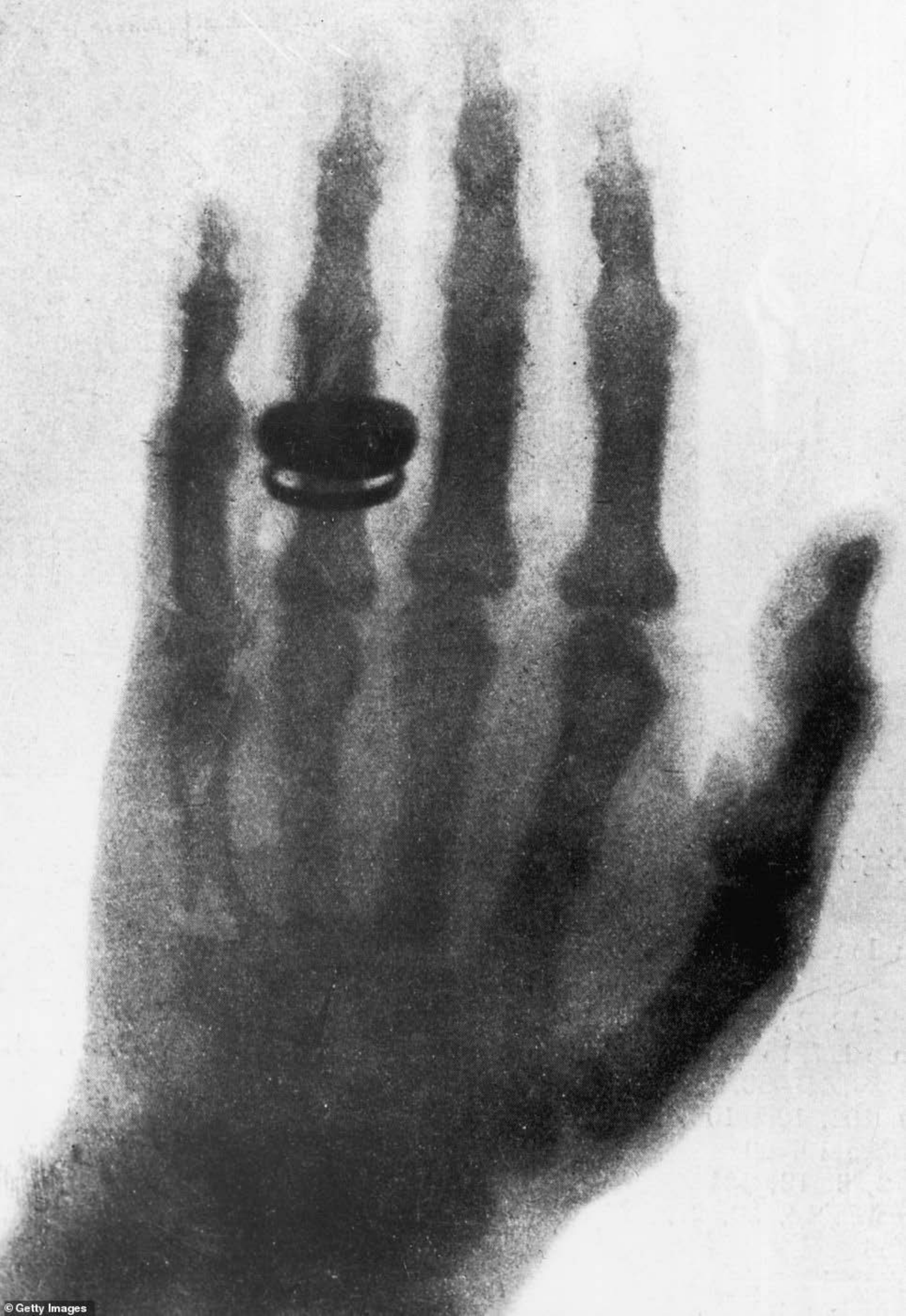}
        \caption{Anna Bertha's hand, imaged using Wilhelm Röntgen's X-ray detector in 1895.}
        \label{fig:RontgenWifeHand}
    \end{figure}
    
    The historical context in this section gives perspective on how greatly detectors have improved over the past 130 years. In 1895, Wilhelm Röntgen painted a fluorescent material on a piece of cardboard, creating the world's first particle detector~\cite{glasserWilhelmConradRontgen1993}. This fluorescent material glowed brightly when Wilhelm accelerated electrons into a metal target, producing X-rays that then hit the screen. He asked his wife to put her hand in front of the X-rays, and the first X-ray image was produced (Figure~\ref{fig:RontgenWifeHand}). 

\begin{figure}
            \centering
            \includegraphics[width=.5\textwidth]{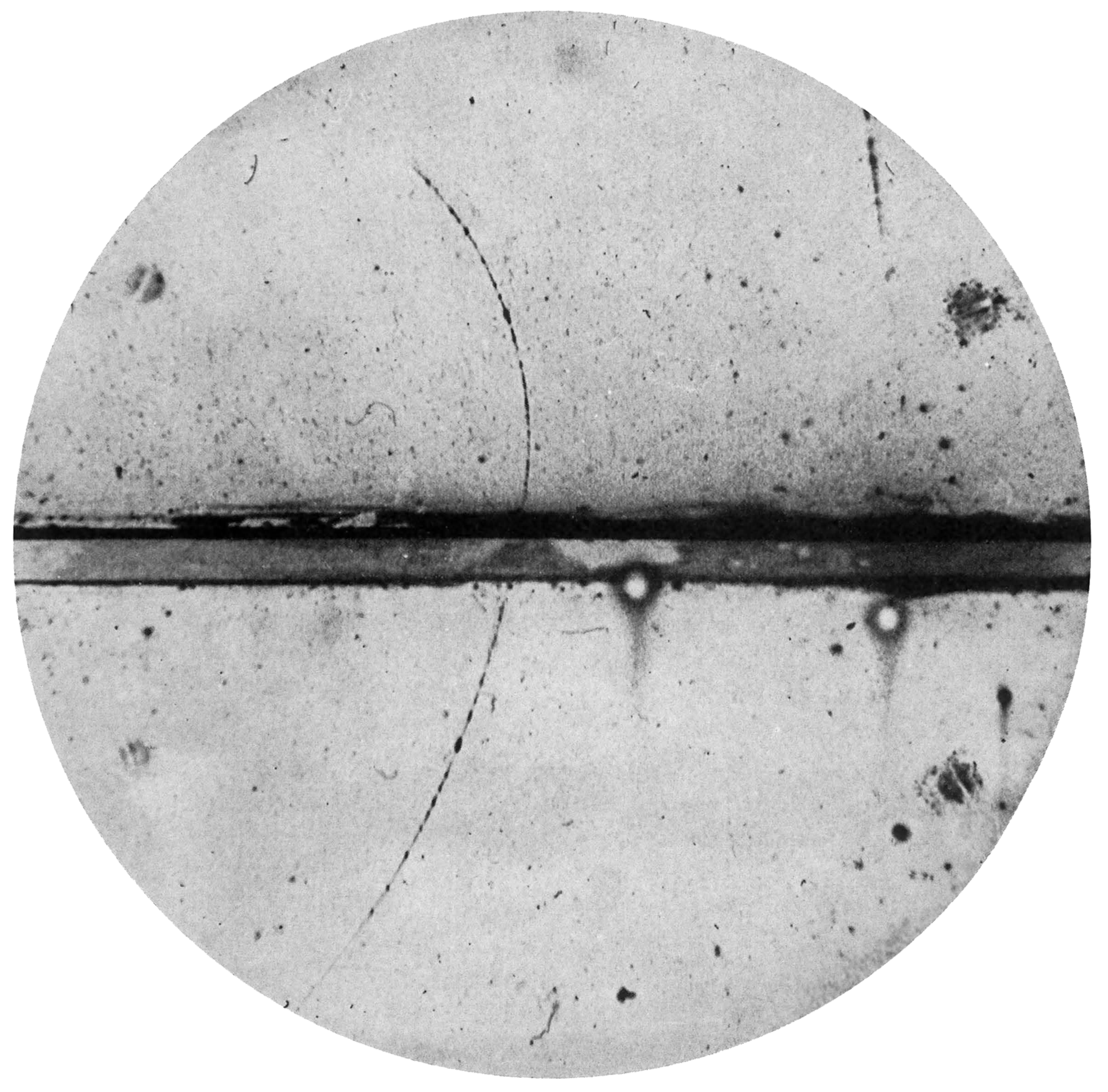}
            \caption{Cloud chamber track used to prove the existence of the positron in 1932 by Carl Anderson.}
            \label{fig:positron}
    \end{figure}

    Radioactivity, the electron, and the structure of the atom were all analyzed using fluorescent detectors. However, a fluorescent screen can only detect where a particle has ended up and provides no information about the particle’s momentum. Shortly after, in the early 1900s, the cloud chamber was invented. This enabled the visualization of particle tracks and momentum estimation, leading to the discovery of new particles such as the positron (Fig.~\ref{fig:positron}), muon, and kaon~\cite{charlesthomsonreeswilsonMethodMakingVisible1911, andersonPositiveElectron1933, andersonCloudChamberObservations1936, rochesterdr.EvidenceExistenceNew1947}.

    The scientific community has continued to innovate particle detectors, using knowledge from the semiconductor industry to create modern-day scintillators~\cite{greskovichCERAMICSCINTILLATORS1997}. These near-perfect doped-silicon crystals demonstrate significant improvement in sensitivity and detection time. New superconducting thin film technology such as transition edge sensors (TES)~\cite{irwinTransitionEdgeSensors2005} and microwave kinetic induction devices (MKIDs)~\cite{mazinSuperconductingMaterialsMicrowave2021}, have shown orders of magnitude improvement in sensitivity due to their ability to detect small changes in temperature and frequency by taking advantage of the very rapid change of resistance during the transition of a superconductor to the normal state. As will be described later, superconductivity plays a key role in axion detectors. With these new advancements developed through collaborations across many institutions, the community has an unprecedented capability to analyze more data and detect weaker interactions.
    
    A recent example of success with modern detector technology and large-scale collaboration is the Laser Interferometer Gravitational-Wave Observatory (LIGO), which directly detected gravitational waves using interferometers at two sites separated by 3000 km across the United States~\cite{LIGOScientific:2007fwp}. The Deep Underground Neutrino Experiment (DUNE) is another large-scale effort, using a 700-ton container of liquid argon connected to $10^{6}$ scintillators to study neutrino oscillations~\cite{DUNE:2021tad}. Of direct relevance to this dissertation, the Axion Dark Matter eXperiment (ADMX) employs state-of-the-art superconducting quantum sensors (TES and MKIDs) and high-field, large-bore magnets, with plans to integrate superconducting thin-film resonators~\cite{duffyHighResolutionSearch2006, ADMX:2018gho}.
    
    \subsection{Summary of Axion Detection Experiments}
    
    Axions can couple with photons, spins, or gluons, leading to many different types of experiments that probe for axion dark matter. A non-exhaustive list of these experiments include: ADMX~\cite{duffyHighResolutionSearch2006}, investigates axion-photon coupling with a large resonant cavity; QUAX~\cite{FirstQUAXGalactic2020}, exploits spin-precession effects; HAYSTAC~\cite{HAYSTAC:2018rwy}, uses squeezed-state receivers to lower the quantum noise limit; DMRadio~\cite{Silva-Feaver:2016qhh}, optimized for low-mass axions uses lumped-element resonators; MADMAX~\cite{garciaFirstSearchAxion2025}, uses a dielectric haloscopes to scan higher mass ranges; ORGAN~\cite{McAllister:2017lkb}, uses many small cavities; and BREAD~\cite{BREAD:2021tpx}, uses a broadband dish antenna for THz probing. 
    
\begin{figure}[htb]
        \centering
        \includegraphics[width=.8\textwidth]{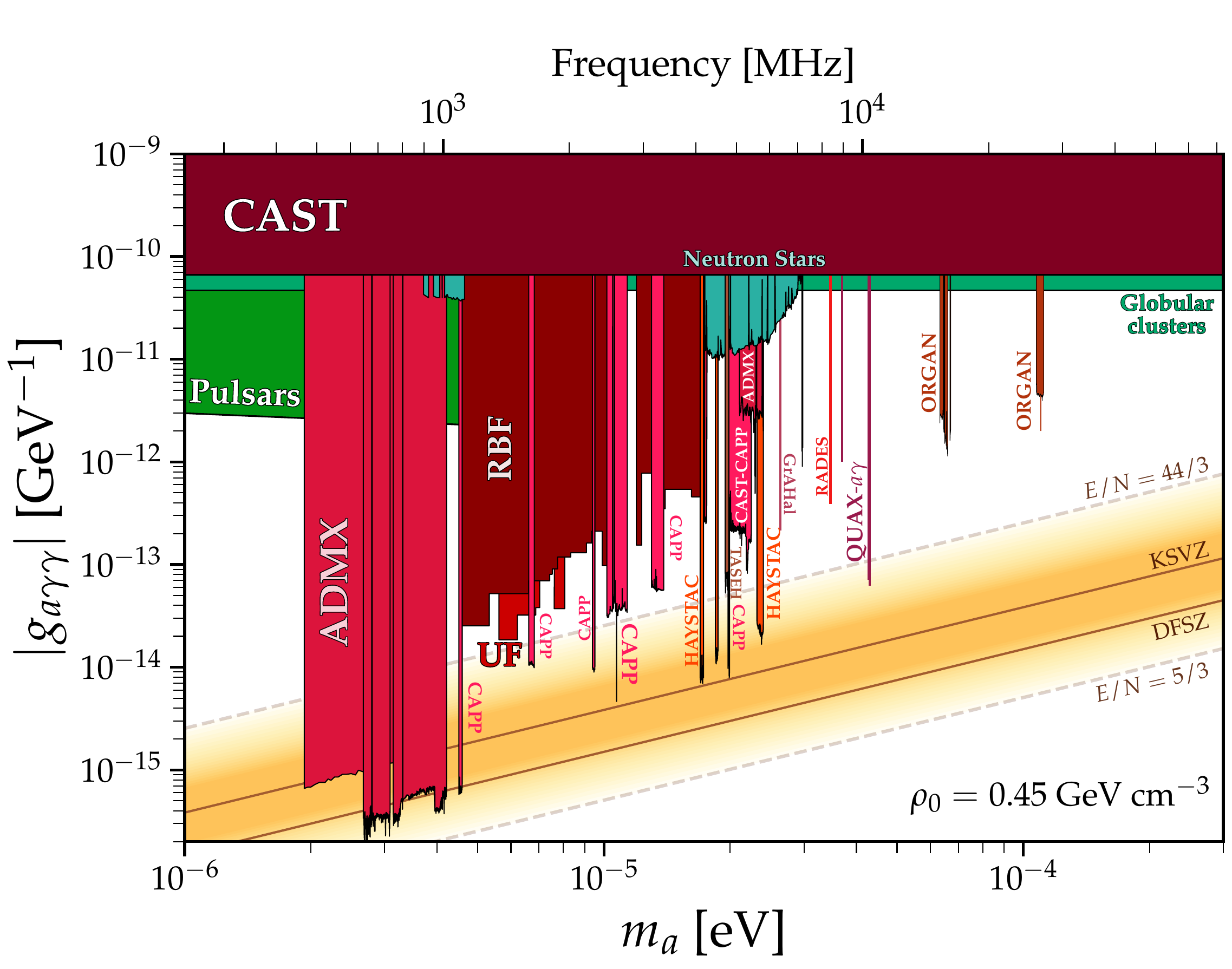}
        \caption{Axion-photon coupling strength vs axion mass for cavity haloscopes~\cite{ciarana.j.ohareAxionLimits2020}.}
        \label{fig:AxionPhoton_RadioFreqCloseup}
    \end{figure}

    This dissertation is closely associated with the ADMX collaboration. The experiments that investigate the photon-axion coupling near GHz frequencies, including ADMX, are given in Fig.~\ref{fig:AxionPhoton_RadioFreqCloseup}.

\begin{figure}[htbp]
        \centering
        \includegraphics[width=.8\textwidth]{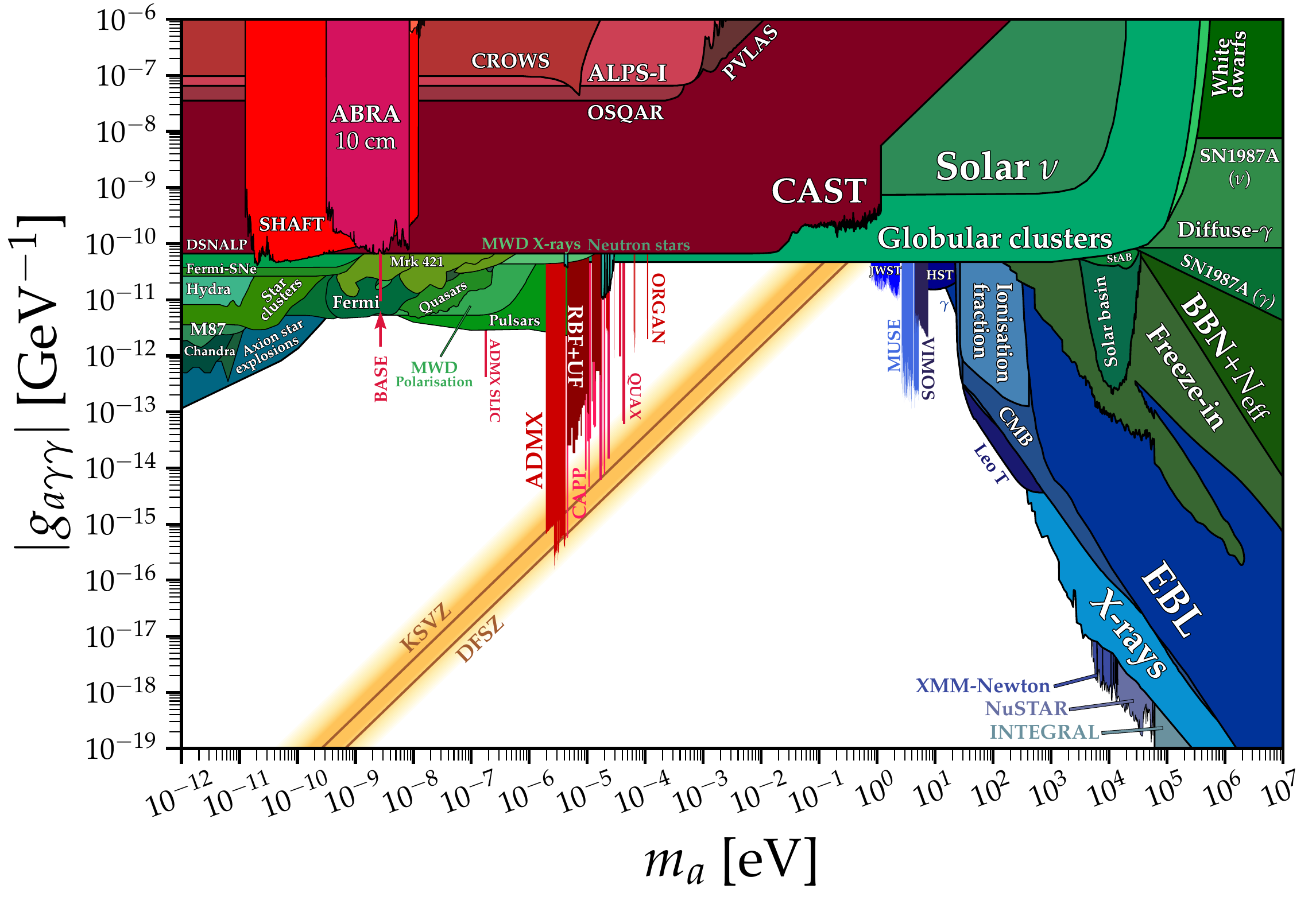}
        \caption{Axion-photon coupling strength vs axion mass for full axion-photon coupling mass range~\cite{ciarana.j.ohareAxionLimits2020}.}
        \label{fig:AxionPhoton}
    \end{figure}

    As outlined in an extensive review by Sikivie~\cite{Sikivie:2020zpn}, the excluded areas on the plot, denoted by different shaded colored regions, come from a variety of experiments including high-energy astrophysical models, helioscope experiments (axions generated from the sun), heavy axion experiments, and ultralight axion experiments. Sikivie also reviews other coupling schemes, although new and improved modes and corresponding experiments are being proposed every year.  The full theoretical mass range of the QCD-axion experiments is shown in Fig.~\ref{fig:AxionPhoton}.

\section{Resonant Cavity Haloscopes}

    Eight years after the axion was theorized, Pierre Sikivie proposed an experiment to search for the hypothetical axion particles using a resonant cavity haloscope~\cite{Sikivie:1985yu}. This device enables axion detection by combining a strong external magnetic field with a resonant cavity. The Primakoff effect~\cite{primakoffDoubleBetaDecay1959}) converts axions to photons in the magnetic field, and resonant cavities were suggested by Sikivie as the best means to acquire and amplify the photon signal.
    
\begin{figure}[htbp]
        \centering
        \includegraphics[width=.5\textwidth]{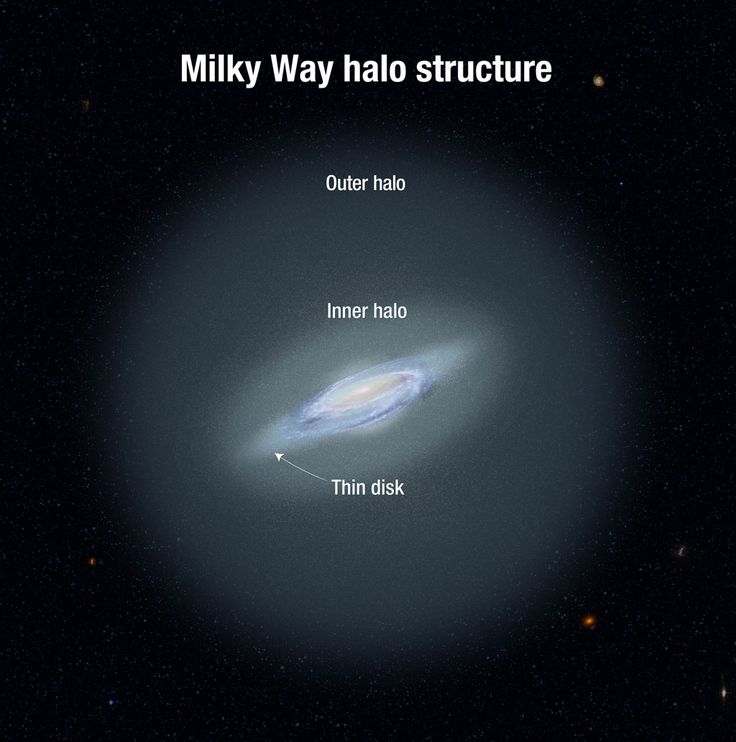}
        \caption[Spherical halo containing dark matter surrounding the visible Milky Way disk.]{Spherical halo containing dark matter surrounding the visible Milky Way disk. Credit: NASA, ESA, and A. Feild}
        \label{fig:milkwayhalo}
    \end{figure}

    The haloscope assumes a specific variety of axions are moving at low velocity, sometimes called ``cold dark matter". This axion is detectable on Earth, since we are immersed in the dark matter halo surrounding our Milky Way galaxy (Fig~\ref{fig:milkwayhalo})~\cite{Battaglia:2005rj, Kafle:2014xfa}. The first axion-probing device was a Cu haloscope cavity made by the ADMX collaboration at the University of Washington~\cite{ADMX:2009iij}. The axion search using a resonant cavity haloscope proceeds through four steps:

\begin{samepage}
   
\begin{enumerate}
    \item The cavity is tuned to the target frequency corresponding to a theoretical axion mass.
    \item The resonance response ($S_{21}$) is measured to verify tuning and extract the loaded quality factor ($Q_L$).
    \item The output signal from the cavity is recorded and integrated to detect possible excess power.
    \item Step to the next frequency and repeat to scan the desired mass range. The desired scan time per frequency is of order 100~s, as a trade-off between detector noise limit and ability to effectively scan across mass ranges over the duration of the experiment.
\end{enumerate}
 
\end{samepage}

    \subsection{Axion-Photon Conversion Physics}

    When the axion is present within the volume of the resonant cavity, it will couple with the externally applied magnetic field $\textbf{B}_0$~\cite{Sikivie:2020zpn}. This axion-photon coupling is described by the interaction Lagrangian:

    \begin{equation}
         \mathcal{L}_{a\gamma \gamma} = -g_{a\gamma\gamma} \frac{\alpha}{\pi} \frac{1}{f_a} a \textbf{E} \cdot \textbf{B}
         \label{eq:Lagammgamma}
    \end{equation}
    
    \noindent where $\alpha$ is the fine structure constant, $a$ is the axion field operator, $\textbf{E}$ and $\textbf{B}$ are the electromagnetic field operators, and $g_{a\gamma\gamma}$ is a model dependent coupling parameter. In a resonant cavity with an applied magnetic field, this coupling manifests as an induced current density ($\mathbf{J}_{\text{axion}}$) given by:
    
    \begin{equation}
          \mathbf{J}_{\text{axion}} = g_{a\gamma\gamma} \mathbf{E}_{\text{cavity}} \cdot \mathbf{B}_{\text{0}} 
          \label{eq:Jaxion}
    \end{equation}
    
    \noindent where $\mathbf{E}_{\text{cavity}}$ is the resonant cavity's AC electric field and $\mathbf{B}_{\text{0}}$ is the applied DC magnetic field. The axion current density inputs power into the cavity that creates a signal that experiments aim to detect~\cite{Sikivie:2020zpn}.

    \subsubsection{Parallel Field Requirement}

    The axion-photon coupling current (Eq.~\ref{eq:Jaxion}) is maximized when $\mathbf{E}_{\text{cavity}} $ is parallel to $ \mathbf{B}_{\text{0}}$. However, given a single electromagnetic wave, $\mathbf{E}$ and $\mathbf{B}$ are always perpendicular to each other. Therefore, it is not possible to generate parallel electric and magnetic fields from a single source. Instead, two independent field sources are required:

    \noindent ADMX use a cylindrical cavity with its axis aligned with an external magnetic field:
    
    \begin{enumerate}
        \item DC magnetic field ($\mathbf{B}_{\text{0}}$): Generated by an external superconducting solenoid oriented along $z$.
        \item AC electromagnetic field ($\mathbf{E}_{\text{cavity}}$): Generated by a resonant cavity mode where the AC electric field is oriented in the axial-direction, parallel to $\mathbf{B_0}$ (e.g., TM modes).
    \end{enumerate} 

    The coupling between the external magnetic field and the cavity mode is captured in a form factor $C_\alpha$:
        
    \begin{equation}
        C_\alpha = \frac{\left| \int_V \mathbf{E}_{\text{cavity}} \cdot \mathbf{B}_0 \, dV \right|^2}{V \mathbf{B}_0^2 \int_V \epsilon |\mathbf{E}_{\text{cavity}}|^2 \, dV} \, .
        \label{eq:formfactor}
    \end{equation}
    
    \noindent This dimensionless quantity ranges from 0 to 1, with 1 corresponding to maximum dot-product between cavity electric field and external magnetic field ($\mathbf{E}_{\text{cavity}} \parallel \mathbf{B}_0$). For the TM$_{010}$ mode in a cylindrical cavity aligned along the solenoid axis, typical $C_\alpha$ values are $0.4-0.5$ depending on actual cavity  geometry~\cite{Braine:2024nzi}. Shapes optimized for detection are not upright cylinders, because losses at the endcaps reduce the quality factor (discussed in \ref{sec:campbell}). Thus, there is a trade-off between optimizing the cavity shape for high quality factor and optimizing the form factor.

    \subsubsection{Axion Conversion Power}

    The power developed in the cavity when the resonant mode overlaps with the frequency of photons from axion conversion is given by~\cite{ADMX:2018gho}:

    \begin{equation}
        P_{a\gamma\gamma} = g_{a\gamma\gamma}^2 \,  B_{0}^2 \, \frac{\rho_a}{m_a} \, V\,  Q_{L}\, C_{\alpha} \,.\label{eq:FullPower1}  
    \end{equation}

    \noindent where $V$ is the cavity volume, $\rho_a$ is the local axion dark matter density, $m_a$ is the axion mass, and $Q_L$ is the loaded quality factor (measured quality factor). Using typical experimental parameters~\cite{Braine:2024nzi}:

    \begin{equation}
    \begin{split}
        P_{axion} = 1.9 \times 10^{-22} \ \si{W} \left( \frac{V}{136\ \si{l}} \right) \left( \frac{B}{6.8\  \si{T}} \right)^2 \left( \frac{C_{\alpha}}{0.4} \right) \left( \frac{g_{a\gamma\gamma}}{ 0.97}\right)^2 \\ 
        \times \left( \frac{\rho_a}{ 0.45\ \si{GeV} \ \si{cm}^{-3}} \right) \left( \frac{f}{ 650\ \si{MHz}} \right) \left( \frac{Q_{L}}{ 50{,}000} \right).
        \label{eq:FullPower2}  
    \end{split}
    \end{equation}

    \noindent where $g_{a\gamma\gamma}$ is the dimensionless model-dependent coupling constant with values of --0.97 and 0.36 for KSVZ and DFSZ models respectively~\cite{ADMX:2018gho}, and $f$ is the frequency of the cavity mode. The unknown variables are the axion density $\rho_a$, the axion-photon conversion frequency $f$, and the coupling constant $g_{a\gamma\gamma}$.

    \subsection{Resonance Enhancement Mechanism}
    
    \subsubsection{Frequency Matching Requirement}

    The conversion factor between the axion mass and the corresponding photon frequency during the axion-photon conversion in the haloscope cavity is given by the energy-mass equivalence~\cite{alberteinsteinElectrodynmaicsMovingBodies1905, griffithsIntroductionElectrodynamics2023}:
    
    \begin{equation}
        E = m_ac^2 = hf_{\gamma} \,
        \label{eq:masstofrequency}
    \end{equation}
    
    \noindent where $f_{\gamma}$ is the photon frequency when the axion is at rest. With the simplest model where $m_a \sim 10^{-5}$~eV~\cite{Sikivie:2020zpn}, the matching cavity frequency is $f_{\gamma} \approx 2.4$~GHz, when inserting values for fundamental constants $c$ and $h$ in \ref{eq:masstofrequency}. 
    
    When matching photon frequency ($f_{\gamma}$) to cavity resonant frequency $f_0$, dimensions scale inversely with frequency, as shown in Table~\ref{tab:axion_cavity_sizes}, which presents approximate cavity diameters for cylindrical resonators operating in the TM$_{010}$ mode at various axion masses.

    \begin{table}[h]
    \centering
    \caption{Axion mass, corresponding resonant frequency, and cylindrical cavity diameter for axion detection}
    \begin{tabular}{ccc}
    \hline
    Axion Mass [eV] & Frequency [GHz] & Cavity Diameter [mm] \\
    \hline
    $10^{-6}$  & 0.24  & 1300  \\
    $10^{-5}$  & 2.4  & 130  \\
    $10^{-4}$  & 24  & 13  \\
    \hline
    \end{tabular}
    \label{tab:axion_cavity_sizes}
    \end{table}

    At 2.4 GHz, the cavity diameter is approximately 130~mm, which conveniently fits inside the bore of a high-field superconducting MRI magnet~\cite{thulbornChallengesIntegrating94T2006}. At lower masses $m_a<10^{-6}$~eV, the cavity is much too large to fit inside the bore of a high-field magnet, and at higher axion masses $m_a<10^{-4}$~eV, cavity diameters shrink to $13$~mm, presenting challenges for maintaining a large axion-photon cross-section with a shrinking volume. Higher frequencies can be covered by introducing multiple smaller cavities into the magnetic field, which is an active area of exploration~\cite{McAllister:2017lkb, knirckADMXExtendedFrequency2023}. A plot of the QCD-axion parameter space for axion detector experiments in the range of $m_a<10^{-6}$~eV to $m_a<10^{-4}$~eV is given in Fig.~\ref{fig:AxionPhoton_RadioFreqCloseup}. Much of the unexplored space in this figure corresponds to cavities with frequencies between 2 and 20~GHz.

    \subsubsection{Energy Buildup at Resonance}

    The optimal strategy for increasing $\mathbf{E}_{\text{cavity}}$ is to improve the stored energy in the cavity. At resonance, a high-Q cavity accumulates energy over many oscillation cycles before dissipating it, allowing large $\textbf{E}$ field amplitudes to build up even with modest input power. The stored energy in the cavity is given by~\cite{padamseeRFSuperconductivity2009}:

    \begin{equation}
        U = \frac{1}{4} \int_V \left( \epsilon_0 |\mathbf{E}_{\text{AC}}|^2 + \frac{1}{\mu_0} |\mathbf{B}_{\text{AC}}|^2 \right) dV
        \label{eq:stored_energy}
    \end{equation}
    
    \noindent where $\mathbf{E}_{\text{AC}}$ and $\mathbf{B}_{\text{AC}}$ are the oscillating electromagnetic fields of the cavity's resonant mode. Higher stored energy produces larger $\mathbf{E}_{\text{AC}}$ amplitude which, when combined with strong external $\mathbf{B}_0$ for a solenoid magnet, maximizes the axion coupling $\mathbf{E}_{\text{AC}} \cdot \mathbf{B}_0$.

    \subsubsection{Cavity and Axion Bandwidth}

    A cavity at resonant frequency $f_0$ amplifies signals within the bandwidth:

    \begin{equation}
        \Delta f_c = \frac{f_0}{Q_L}\,.
        \label{eq:bandwidth}
    \end{equation}
    
    \noindent For a cavity with $f_0 = 2.4$~GHz and $Q_L = 50{,}000$, the bandwidth is $\Delta f_c \sim 50$~kHz, amplifying signals only within $\pm25$~kHz of resonance.

    The axion signal itself has a finite bandwidth due to the dark matter particles' velocity distribution. Including kinetic energy in the total energy:
        
    \begin{equation}
       E =  m_ac^2 +  \frac{1}{2}m_av^2 = hf_{\text{signal}} \,
        \label{eq:totalEnergy}
    \end{equation}
        
    \noindent Combining with Eq.~\ref{eq:masstofrequency} ($f_{\gamma} = m_ac^2/h$):

    \begin{equation}
        f_{\gamma} \left( 1+ \frac{v^2}{2c^2}\right) = f_{\text{signal}} \,.
        \label{eq:TotalE}
    \end{equation}

    \noindent The velocity spread $\Delta v$ induces a frequency spread. Differentiating Eq.~\ref{eq:TotalE}:

    \begin{equation}
        \frac{\Delta f_{\text{signal}}}{f_{\gamma}} \approx \frac{v \cdot \Delta v}{c^2} \,.
        \label{eq:FSpreadVSpread}
    \end{equation}

    \noindent For QCD axions constituting the galactic dark matter halo, $v \sim 270$~km/s $\approx 10^{-3}c$~\cite{lewinReviewMathematicsNumerical1996, evansRefinementStandardHalo2019}. Virial equilibrium implies $\Delta v \sim v$, giving a signal bandwidth:
    
    \begin{equation}
        \Delta f_{\text{signal}} \sim 10^{-6} \cdot f_{\gamma} \,,
    \end{equation}
    
    \noindent corresponding to an axion quality factor $Q_a \equiv f_{\gamma}/\Delta f_{\text{signal}} \sim 10^6$. This sets the intrinsic width of the axion signal in frequency space. 

    It was previously thought that once the cavity quality factor exceeded the axion quality factor ($Q_L > Q_a$), further increases in cavity quality factor would not improve scan rate. However, \citeauthor{Cervantes:2022gtv}~\cite{Cervantes:2022gtv} found that scan rate remains proportional to $Q_L$ even when $Q_L \gg Q_a$, motivating the pursuit of ultra-high-Q superconducting cavities for axion detection.
    
    \subsection{Quality Factor}\label{sec:QualityCh2}

    The loaded quality factor ($Q_L$) of the cavity is determined by the losses from the cavity walls ($Q_0$) and all other loss mechanisms ($Q_S$):

    \begin{equation}
        \frac{1}{Q_L}= \frac{1}{Q_0} + \frac{1}{Q_s} .  \label{eq:QualityHW} 
    \end{equation}

    \noindent With highly optimized materials (high $Q_0$) and minimizing the impact of other losses (ensuring minimal RF leakage and weak antenna coupling $Q_s>>Q_0$ see App.~\ref{sec:RFCharacterization}), the loaded quality factor $Q_L$ can be maximized. This explains how a better understanding of the materials science of cavity materials can lead to improved quality factor, which is the central thrust of this dissertation. Quality factor can also be calculated using:
        
    \begin{align}
        Q = \frac{G}{R_s} \label{eq:Q0_geometric} \,,
    \end{align}
    
    \noindent where $G$ is the geometric factor (units: $\Omega$) and $R_s$ is the AC surface resistance (units: $\Omega$). 

    \subsubsection{Geometric Factor}
    
    The geometric factor is defined as~\cite{padamseeRFSuperconductivity2009}:
    
    \begin{align}
       G = \frac{\omega_0 \text{~(stored magnetic energy)}}{\text{(power loss per unit } R_s \text{)}} = \frac{2\pi f_0 \mu_0 \int_V |\vec{H}|^2 \, dV}{\int_S |\vec{H}_t|^2 \, dA}  \label{eq:geometric_factor} ,
    \end{align}
   
    \noindent where $f_0$ is the resonant frequency ($2\pi f_0 = \omega_0$), $\mu_0 = 4\pi \times 10^{-7}$~H/m is the permeability of free space, $\textbf{H}$ is the AC magnetic field of the resonant cavity mode in volume $V$, $\textbf{H}_t$ is the tangential AC magnetic field at the cavity surface $S$, and $dV$, $dA$ are internal volume and surface elements of the cavity respectively. 
    
    Note that $G$ is a purely geometric property of the cavity and its resonating mode. $G$ can be analytically solved for simple geometries, but is usually numerically solved using finite element (FEM) eigenfrequency simulation. 

    \subsubsection{Loss Decomposition}

    One can assign geometric factors to different cavity surfaces:

    \begin{equation}
        \frac{1}{G_{\text{total}}} = \frac{1}{G_1} + \frac{1}{G_2} + \cdots + \frac{1}{G_N}
        \label{eq:G_parallel}
    \end{equation}
        
    The total cavity quality factor (Eq.~\ref{eq:QualityHW}) can be expanded using different surfaces:

    \begin{equation}
        \frac{1}{Q_L} = \underbrace{\frac{R_{s,\text{walls}}}{G_{\text{walls}}}}_{\frac{1}{Q_0}} + \underbrace{\left(\frac{R_{s,\text{seams}}}{G_{\text{seams}}} + \frac{R_{s,\text{antenna}}}{G_{\text{antenna}}} + \cdots \right)}_{\frac{1}{Q_S}}
        \label{eq:Q0_decomposed_grouped}
    \end{equation}

    This surface decomposition reveals that $Q_L$ can still be maximized even when the resistance contributions from the seams and antennas are high. This is provided that their geometric factors are large, thereby minimizing their contribution to total losses.
        
    Surface resistance $R_s$ is a material property that depends on the conductor type, temperature, and frequency. For normal conductors, $R_s$ increases with frequency and decreases with temperature as phonon scattering reduces~\cite{ashcroftSolidStatePhysics1976}. For superconductors, $R_s$ depends on the superconducting transition temperature, normal-state resistivity, and applied magnetic field through mechanisms including the Bardeen-Cooper-Schrieffer (BCS) surface resistance~\cite{bardeenTheoryMeissnerEffect1955} and dissipation due to flux motion~\cite{campbellResponsePinnedFlux1969}.
    
    The detailed physics of surface resistance for both normal conductors and superconductors, including behavior in high magnetic fields, will be described in Chapter~\ref{ch:MaterialScience}.

    \subsection{Experimental Scan Strategy}

\begin{figure}[H]
        \centering
        \includegraphics[width=.6\textwidth]{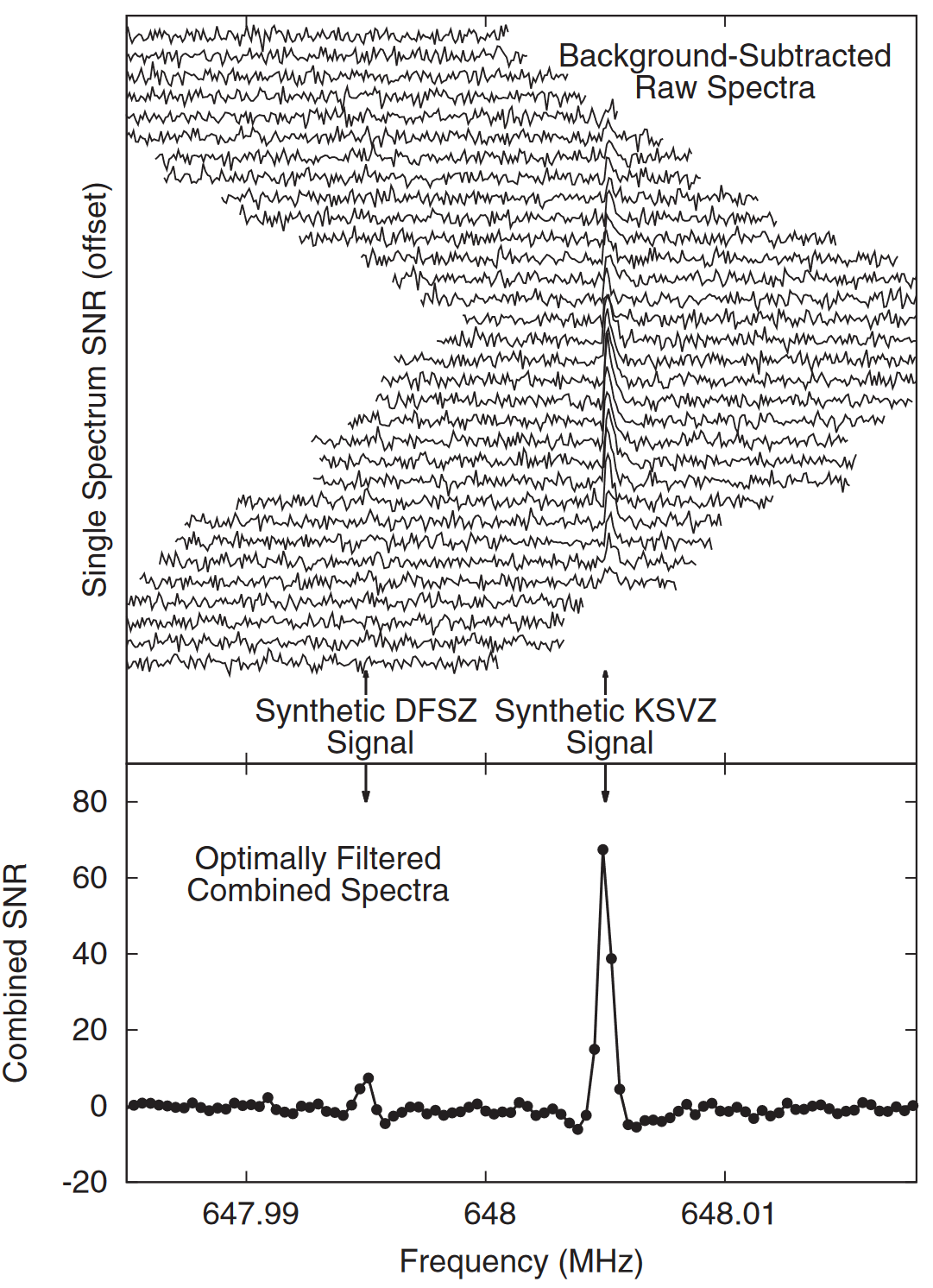}
        \caption[(Upper) Background-subtracted single scans with synthetic axion signals at DFSZ and KSVZ coupling strengths~\cite{Lentz:2017aay}.]{(Upper) Background-subtracted single scans with synthetic axion signals at DFSZ and KSVZ coupling strengths~\cite{Lentz:2017aay}. (Lower) Same data after filtering; both signals become visible~\cite{ADMX:2018gho}.}
        \label{fig:axionsignal}
    \end{figure}

    The measurement strategy relies on placing an antenna inside the cavity that couples to inner resonant surface. During normal operation, without contributions from decaying axions, a phase-locked loop recycles power leaving the cavity back into the cavity. Excess power from an axion will shift the output phase, which can be extracted as a signal. The phase-shifted power coupled out of the cavity is amplified and spectrum-analyzed, resulting in a measurable signal $\mathcal{O} (10^{-23}$ W)~\cite{ADMX:2018gho}. Figure~\ref{fig:axionsignal} shows a simulated axion signal imposed on real data. If a peak persists after signal averaging, it can be distinguished from noise or environmental interference by four characteristic properties of the axion signal~\cite{Sikivie:2020zpn}:
    
    \begin{enumerate}
        \item Signal must be independent of microwave shielding around the cavity.
        \item Signal cannot be detected by an antenna outside the cavity.
        \item Its central frequency $\omega_\alpha$ exhibits a Lorentzian lineshape with width set by the axion velocity distribution.
        \item The signal power scales as $\mathbf{B}_0^2$ (proportional to squared magnetic field).
    \end{enumerate}

    \FloatBarrier

    \subsubsection{Signal Detection and Discrimination}

    The signal-to-noise ratio is defined as~\cite{Braine:2024nzi}:
    
    \begin{equation}
        SNR = \frac{P_{a\gamma\gamma} }{k_B T_{sys}} \sqrt{\frac{t}{B}}
        \label{eq:signalt:noise}
    \end{equation}
       
    \noindent where $P_{a\gamma\gamma}$ is the axion power deposited into the cavity (Eq.~\ref{eq:FullPower1}), $t$ is the integration time over which the Fourier transform is taken, and $\Delta f$ is the integrated bandwidth of the Fourier transform~\cite{Braine:2024nzi}. ADMX typically uses an SNR of about 3.5, providing a good trade-off between staying on resonance to allow the axion signal to appear above the noise and maintaining an adequate scan rate to cover the frequency space efficiently without prolonging the experiment.

    \subsubsection{Frequency Tuning}

    Scanning to different axion masses with the same cavity requires a movable metal or dielectric rod inserted into the cavity volume~\cite{ADMX:2023rgo}. This tuning rod changes the effective volume seen by the electric field, shifting the resonant frequency by at most $\pm 40\%$ from the base value. Fine-tuning achieves frequency steps smaller than the cavity bandwidth ($\Delta f_c = f_0/Q_L$). For a cavity with $f_0 = 2.4$ GHz and $Q_L = 50{,}000$, the bandwidth is $\Delta f_c \sim 50$ kHz as discussed previously. To cover 2~GHz in one year requires a dwell time per frequency point of approximately $(1/3 \text{ year}) \times 1/Q_L \sim 100$ seconds.

    \subsubsection{Scan Rate Optimization}

    The scan rate, or how quickly the experiment can cover axion parameter space, depends on five experimental parameters~\cite{Braine:2024nzi}:

    \begin{equation}
        \frac{df}{dt} \propto T_{\text{sys}}^{-2} \, \text{SNR}^{-2}  \,   V^{2} \, Q_{L} \, \text{B}^{4} \,.
        \label{eq:scanrate}
    \end{equation}
        
    \noindent Each parameter faces different optimization constraints:
    
    \noindent \text{System Temperature ($T_{\text{sys}}$):} Already minimized using dilution refrigerators (millikelvin operation) combined with quantum-limited amplifiers (Josephson parametric amplifiers and traveling-wave parametric amplifiers).
   
    \noindent \text{Signal-to-Noise Ratio (SNR):} Fixed by statistical detection confidence requirements, cannot be arbitrarily reduced without sacrificing discovery potential.
    
    \noindent \text{Cavity Volume ($V$):} Directly increases signal power (Eq.~\ref{eq:FullPower1}), but diameter determines frequency ($f \propto 1/D$ for TM$_{010}$ mode), so it can not be arbitrarily increased. The cavity can be elongated axially to increase its volume while maintaining a constant resonant frequency, but its length is constrained by the homogeneous field zone in the magnet bore. A multi-cavity array is proposed to increase volume while operating at a higher frequency, $2-4$~GHz (ADMX extended frequency range - EFR~\cite{Braine:2024nzi}).  
    
    \noindent \text{Magnetic Field ($B$):} Despite strong $B^4$ scaling, practical improvements face severe limits. Achieving the same scan-rate improvement when increasing $Q$ from 50,000 to $10^6$ would require increasing the field from 9 T to 19 T, which is beyond the capabilities of any commercially available large-bore superconducting magnet.
    
    \noindent \text{Quality Factor ($Q_L$):} Although scan rate scales only linearly with $Q$, practical improvements span 1--2 orders of magnitude: from $Q \sim 50{,}000$ for high-purity copper cavities to $Q > 10^6$ for optimized superconducting films. This represents the most accessible path to order-of-magnitude improvements in scan rate.
        
    In this context, replacing copper cavities with superconducting thin films offers greater gains than increasing other variables, owing to lower complexity and cost. The development of high-$Q$ superconducting coatings compatible with high magnetic fields is, therefore, critical for accelerating the axion search, which motivated the research presented in this dissertation.

\chapter{Conductors and Superconductors}\label{ch:MaterialScience}

This chapter starts by considering how normal conductors and superconductors respond to AC and DC fields. Surface resistance is defined for conductors and superconductors under the operating conditions of axion detectors. With a fundamental understanding of superconductors, thin films of superconducting materials can be coated on the inner surfaces of cavities to achieve high-$Q$ at high fields. 

\section{Normal Conductors}
    
    \subsection{Static Field}
    
    This discussion of conductors follows from Ashcroft and Mermin~\cite{ashcroftSolidStatePhysics1976}. The behavior of metals can be predicted and described using the free electron model. This simple model assumes that an atom's outer electrons are free to wander in the metal ion lattice. When these electrons sense an electric field, they move due to the force $\mathbf{F}=e\mathbf{E}$. The electrons may scatter off impurities, phonons (lattice vibrations), or other electrons, leading to an average time between collisions $\tau$. The mean free path $\ell = v_F\tau$ (where $v_F$ is the Fermi velocity) characterizes the average distance an electron travels between collisions. The equation of motion for the electron is
    
    \begin{align}
        m\frac{dv_d}{dt} = -e\mathbf{E}(t) - \frac{m}{\tau}v_d\label{eq:EOMcondcutor}
    \end{align}   

    \noindent where $m$ is the electron mass, $e$ is the electron charge, and $v_d$ is the drift velocity. The first term $-e\mathbf{E}(t)$ represents the driving force, while the second term $- \frac{m}{\tau}v_d$ represents dissipation from scattering. With the constant DC electric field, the steady-state drift velocity is
    
    \begin{align}
        v_d = \frac{eE\tau}{m}\,. \label{eq:driftvelocity}
    \end{align}
    
    \noindent These electrons are charged particles, so moving through the material creates a measurable current in the material:
    
    \begin{align}
        \mathbf{j} = -ne\mathbf{v}_d\,, \label{eq:normalcurrent}
    \end{align}
    
    \noindent where $n$ is the number density of conduction electrons. Combining equations~\ref{eq:driftvelocity},~\ref{eq:normalcurrent}, and $J=\sigma \textbf{E}$ yields the DC electrical conductivity $\sigma$ or resistivity $\rho$: 
    
    \begin{align}
        \sigma = \frac{1}{\rho} = \frac{ne^2\tau}{m} \,. \label{eq:normalconductivity}
    \end{align}

\begin{figure}
        \centering
        \includegraphics[width=.75\textwidth]{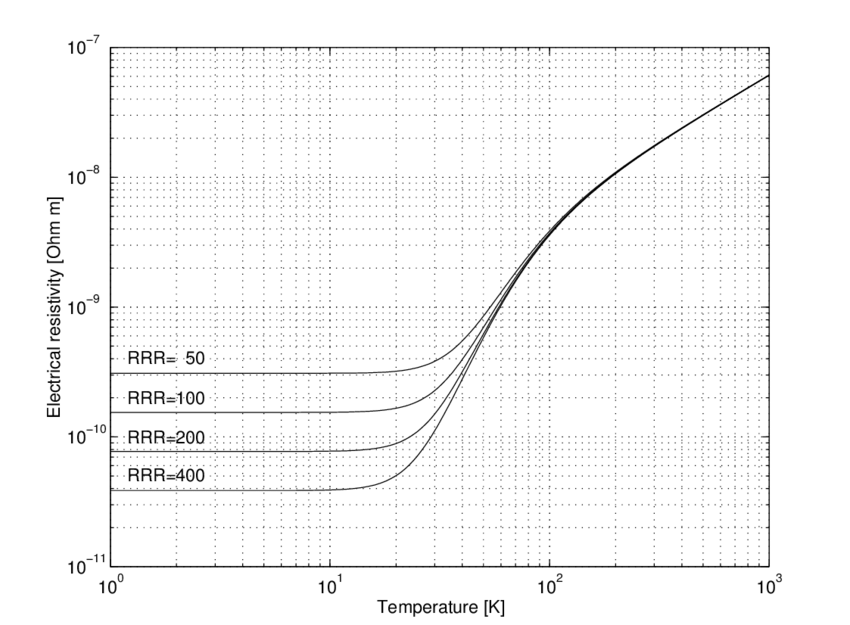}
        \caption{Resistivity of Cu with varying levels of purity (Residual resistivity ratio - RRR) as a function of temperature~\cite{krainzQuenchProtectionPowering1997}.}
        \label{fig:RRRResistivity}
    \end{figure}

    As the material temperature decreases, thermal oscillations (phonons) decrease, leading to reduced electron scattering and lower resistivity. At low temperature $T < 20$~K, phonons freeze out, and $\rho$ saturates to a residual resistance related to the purity of the material (Fig.~\ref{fig:RRRResistivity}). The purity is quantified with the residual resistivity ratio (RRR), defined as the ratio between room temperature and residual resistivity (resistivity value saturating at $T = 0$~K):
    
    \begin{align}
        RRR = \frac{\rho(300~\text{K})}{\rho_0}. \label{eq:RRR}
    \end{align}

    \noindent The electrical conductivity of high purity Cu (RRR = 300) is $\sigma \approx 2 \times10^{10}$ S/m.
    
    \FloatBarrier
    
    \subsection{Classical Skin Effect in AC field}

    Conductors resist the penetration of time-varying magnetic fields (AC-currents) by inducing eddy currents in the conductor. These currents concentrate near the surface over a characteristic depth called the skin depth~\cite{johndavidjacksonClassicalElectrodynamicsThird1999}:

    \begin{align}
        \delta = \frac{1}{\sqrt{\pi f \mu_0 \sigma}} \label{eq:skindepth}
    \end{align}
    
    \noindent where $f$ is the AC frequency and $\mu_0=4 \pi\times10^{-7}$~H/m is the permeability of free space. For copper, the skin depth at 300~K and 10~GHz is $\approx 0.6$~\unit{\um}. With these currents restricted to the conductor surface, the usable cross-sectional area of the conductor is reduced, leading to higher resistance. This is modeled as a current flowing through a thin resistive square sheet of area $L\times L$ and thickness $\delta$. The surface resistance is

    \begin{align}
        R_s = \rho \cdot \frac{\text{length}}{\text{Cross-sectional area}} = \frac{1}{\sigma} \cdot \frac{L}{L \cdot \delta}\,,
       \label{eq:SurfaceResistance1}
    \end{align}

    \noindent which, substituting Eq.~\ref{eq:skindepth}, simplifies to:

    \begin{align}
        R_s = \sqrt{\frac{\pi f \mu_0}{\sigma}} \,.
       \label{eq:SurfaceResistance2}
    \end{align}

    \subsection{Anomalous Skin Depth}

    At low-temperature $T<20$~K and high-frequency $f>100$~MHz, the classical skin depth equation~\ref{eq:SurfaceResistance2} doesn't hold, and the surface resistance no longer depends on conductivity~\cite{pippardf.r.s.SurfaceImpedanceSuperconductors1947}. This anomalous skin-depth regime can be attributed to the skin depth becoming smaller than the scattering length, i.e., $\ell > \delta$. As the electrons travel through the skin depth, they no longer scatter at the same rate. This reduces the conductor's effectiveness at AC-field screening, leading to increased surface resistance. The effective anomalous skin depth is defined as~\cite{Braine:2024nzi}:

    \begin{align}
        \delta_{\text{anom}} = \left(\frac{\sqrt3c^2 m v_F}{8\pi^2 \omega_0 n e^2}\right)^{1/3} , \label{eq:anomalousskin}
    \end{align}

    \noindent For Cu, $v_F = 1.57 \times 10^8$ cm/s and $n = 8.50 \times 10^{22}$ cm$^{-3}$, which simplifies the above equation to:
    
    \begin{align}
        \delta_{\text{anom,Cu}} = 2.84 \times 10^{-4} \text{ m} \left(\frac{1}{f}\right)^{1/3} \, \label{eq:anomalouscopperskin}
    \end{align}

    \noindent with frequency in Hz. For copper at 10~GHz and 20~K, the effective skin depth is $\approx0.2$~\unit{\um}.

    Most electromagnetic simulation software (COMSOL, ANSYS) assumes the classical skin effect. To use these modeling tools in the anomalous regime, an effective conductivity $\sigma_{\text{eff}}$ can be defined that reproduces the correct surface resistance when inserted into the classical formula $R_s = \sqrt{\pi f \mu_0 / \sigma_{\text{eff}}}$.

    Starting from the requirement that $R_s = 1/(\sigma_{\text{eff}} \,\delta_{\text{anom}})$ must equal the classical form Eq.~\ref{eq:SurfaceResistance2}, the effective conductivity for Cu at cryogenic temperatures becomes 
    
    \begin{align}
        \sigma_{\text{eff,anom,Cu}} = \frac{1}{2\mu_0 c \pi f \delta_{\text{anom}}^2} = 3.14 \times 10^{12} \text{ S/m} \cdot f^{-1/3} \,.\label{eq:effectiveconductivity}
    \end{align}

    \noindent While DC conductivity improves 340$\times$ from room temperature to  4~K, the anomalous skin effect reduces this benefit to $5-20\times$ improvement (Table~\ref{tab:sigma_eff_anomalous}). This frequency-dependent degradation becomes more severe at higher frequencies~\cite{padamseeRFSuperconductivity2009}.


    The anomalous skin effect leads to a stronger frequency dependence on surface resistance, where $R_s \propto f^{2/3}$, compared to the classical $R_s \propto f^{1/2}$ scaling. Calculating surface resistance and quality factor in this anomalous regime with equations~\ref{eq:SurfaceResistance2} and~\ref{eq:effectiveconductivity}, we find that $Q$ drops by $\sim 75\%$ from 1-10 GHz (seen in table~\ref{tab:Rs_Q_anomalous}). The strong degradation in $Q$ with frequency becomes problematic as axion searches push beyond $10$~GHz ($m_a \sim 10^{-5}$~eV). Conductors have remained the material of choice for high-field RF applications due to their surface resistance degrading very weakly in large magnetic fields.

    \begin{table}[ht]
        \centering
        \caption{Conductivity of Cu (high-purity, RRR=300) at 4~K showing degradation from anomalous skin effect}
        \label{tab:sigma_eff_anomalous}
        \begin{tabular}{ccc}
        \toprule
        \textbf{Frequency (GHz)} & \textbf{$\bm{\sigma}$ or $\bm{\sigma_{\text{eff}}}$ (S/m)} & \textbf{Improvement from 300~K} \\
        \midrule
        DC & $2.0 \times 10^{10}$ & 340$\times$ \\
        \midrule
        1  & $3.14 \times 10^9$ & 54$\times$ \\
        5  & $1.84 \times 10^9$ & 32$\times$ \\
        10 & $1.46 \times 10^9$ & 25$\times$ \\
        20 & $1.16 \times 10^9$ & 20$\times$ \\
        \bottomrule
        \end{tabular}
    \end{table}

  \begin{table}[ht]
        \centering
        \caption{Theoretical best surface resistance and quality factor for Cu (high-purity, RRR=300) at 4~K in the anomalous regime ($G = 350~\Omega$)}
        \label{tab:Rs_Q_anomalous}
        \begin{tabular}{ccc}
        \toprule
        \textbf{Frequency (GHz)} & \textbf{$\bm{R_s}$ (m$\bm{\Omega}$)} & \textbf{$\bm{Q}$} \\
        \midrule
        1  & 1 & 300{,}000 \\
        5  & 3 & 100{,}000 \\
        10 & 5 & 70{,}000 \\
        20 & 8 & 40{,}000 \\
        \bottomrule
    \end{tabular}
    \end{table}

\section{Superconductors}\label{sec:superconductor}

    Superconductivity is a phenomenon that occurs in many metals, alloys, and compounds where DC electrical resistance falls to zero below a characteristic temperature, called the critical temperature $T_c$, and a limiting magnetic field, usually the upper critical field $\mu_0 H_{c2}$~T. For RF cavity applications, superconductors are not limited by the anomalous skin depth and exhibit fundamentally different loss mechanisms in AC fields. This leads to much higher quality factors, $Q\sim10^{11}$~\cite{bafiaSignaturesEnhancedSuperconducting2025}, orders of magnitude higher than those of the best normal conductors. 
    
    However, for superconducting cavities made from pure metals (e.g., Pb, Nb), superconductivity breaks down well below the multi-tesla fields of interest for axion searches. Niobium possesses the highest upper critical field of any elemental superconductor, with $\mu_0 H_{c2} \approx 0.4$~T at 0~K. While this enables excellent performance in zero-field RF applications with ductile pure metals~\cite{padamseeRFSuperconductivity2009}, it is insufficient for the 9~T operating conditions of an axion detector. To maintain superconductivity in these high-field regimes, it is necessary to move beyond elements and explore superconducting compounds. These materials exhibit upper critical fields that are orders of magnitude higher than those of Niobium, though they introduce significant fabrication challenges.

    \subsection{Fundamentals of Superconductivity}

    Superconductivity was first discovered in 1911 by Kamerlingh Onnes, who cooled down mercury to cryogenic temperatures and observed zero resistance~\cite{onnesLiquefactionHelium1991}. Since this discovery, scientists have sought to harness this zero-resistance state for practical applications. The primary commercial application of superconductors has been in high-field magnets for Magnetic Resonance Imaging (MRI)~\cite{cooleyBusinessModelsAssure2023}. The number of superconducting applications is too large to be exhaustively reviewed in this dissertation. However, the emergence of quantum sensors and detectors based on superconducting thin films~\cite{irwinTransitionEdgeSensors2005, mazinSuperconductingMaterialsMicrowave2021} is a topical area relevant to dark-matter detectors.
    
    The first theory of superconductivity was introduced in the 1930s by the London brothers~\cite{londonProblemMolecularTheory1948}. Their equations were derived from classical electrodynamics, assuming that in the superconducting state, electrons move with zero resistance and minimize their free energy. These equations explained the Meissner effect, an experimentally observed phenomenon in which the magnetic field is expelled from the bulk of the superconductor. The screening currents that expel this magnetic field are arranged on the surface of the superconductor to a depth called the London penetration depth $\lambda$, given by~\cite{tinkhamIntroductionSuperconductivity2004}:
    
    \begin{equation}
        \lambda = \sqrt{\frac{m}{\mu_0 ne^2}} \,,
        \label{eq:lambda}
    \end{equation}

    \noindent where $m$, $e$, and $n$ are the electron mass, elementary charge, and number density of superconducting carriers, respectively. Quantitatively, this length scale defines where the magnetic field amplitude drops to $1/\text{e}$ intensity relative to a parallel magnetic field at the surface (surface along z-direction, and x-direction going into the bulk of the superconductor):

    \begin{equation}
            B_{z}(x)=B_{0}\text{e}^{-x/\lambda} \,.
        \label{eq:MagneticFieldPen}
    \end{equation}  
    
    \noindent The screening currents at this depth are described by classical Ampere's law:
    
    \begin{equation}
        {\displaystyle \nabla \times \mathbf {B} =\mu _{0}\mathbf {j_s} }
        \label{eq:ampereslaw}
    \end{equation}

    \noindent where $j_s$ is the surface current. 
    
    To help explain the transition from the normal state to the superconducting state, a more complex theory of phase transitions was needed. The subsequent significant theoretical development came in 1950, when Ginzburg and Landau introduced a phenomenological description of superconductivity~\cite{ginzburgTheorySuperconductivity1950}. In this framework, the superconducting state is characterized by a complex order parameter $\psi(\mathbf{r}, t)$, whose magnitude $|\psi|^2$ represents the local density of superconducting carriers (superconducting electrons). When $\psi = 0$, the electrons behave normally as in common metals (the ``normal state"), when $\psi$ is finite, superconductivity exists. This model defines critical variables associated with the breakdown of the superconducting order parameter beyond a critical magnetic field ($H_c$) and critical temperature ($T_c$). This theory introduces the Ginzburg-Landau coherence length $\xi_{GL}$, which governs the spatial variation of the order parameter. Near surfaces, grain boundaries, and impurities at this length scale ($\sim3$~nm for Nb$_3$Sn), the order parameter can drop to zero, locally breaking superconductivity.
    
    In 1957, the Bardeen–Cooper–Schrieffer (BCS) theory~\cite{Bardeen:1957mv} was proposed; it was the first microscopic theory of superconductivity and successfully predicted superconducting phenomena down to low temperatures ($T \ll T_c$). This theory describes how a small attraction between electrons can arise from interactions between conduction electrons and phonons, creating a 2-electron bound state called a Cooper pair. BCS theory accurately predicts the superconducting gap $\Delta \approx 1.76 k_B Tc$, equivalent to the binding energy of the Cooper pair~\cite{bardeenTheoryMeissnerEffect1955}. 
    
    At $T > 0$, thermal excitations break Cooper pairs, leading to a two-fluid model, where the total electron density $n_{\text{total}}$ is divided into superconducting carriers ($n_s$) and normal electrons ($n_n$), where $n_s + n_n = n_{\text{total}}$. Normal electrons that arise from thermally excited Cooper pairs follow a Boltzmann distribution~\cite{landauTheorySuperfluidity1949, tinkhamIntroductionSuperconductivity2004}:

    \begin{align}
        n_{n} \propto \exp \left( -\frac{\Delta}{k_B T}   \right )\,, \label{eq:normalelectrons}
    \end{align}

    \noindent where $k_B$ is the Boltzmann constant. The number of normal electrons increases exponentially with an increase in temperature until $T=T_c$, where superconductivity breaks down, and all electrons move to the normal state. This motivates operating an RF cavity at $T<<T_c$. BCS theory defines the clean coherence length ($\xi_0$), related to the expected separation of the bound electrons in a Cooper pair in the limit of no impurity scattering~\cite{tinkhamIntroductionSuperconductivity2004}:


    \begin{align}
        \xi_0 \equiv \frac{\hbar v_F}{\pi \Delta(0)} \propto \frac{\hbar v_F}{k_B T_c}  .\label{eq:coherencelengthpure}
    \end{align}

    \noindent $\xi_0 \rightarrow\infty$ at $T = T_c$. 

    Before BCS theory, Pippard~\cite{pippardCoherenceConceptSuperconductivity1953} introduced the concept of the effective coherence length ($\xi$) to explain how the London equations could be generalized in a non-local way, in particular under microwave irradiation. The expression:
    
    \begin{equation}
        \frac{1}{\xi} = \frac{1}{\xi_0} + \frac{1}{\ell} .\label{eq:coherencelength}
    \end{equation} 
    
    \noindent is Pippard's equation to explain what was observed when Bi was added to Pb to make an alloy, while follows from the solution of Chambers' non-local derivation of Ohm's law~\cite{chambersAnomalousSkinEffect1952}. This Eq.~\ref{eq:coherencelength} allows for a description in the dirty limit of superconductivity where $\ell << \xi_0$ (BCS is naturally in the clean limit $\ell >> \xi_0$). Increased impurity scattering shortens the mean free path $\ell$, which reduces the coherence length $\xi$ according to Eq.~\ref{eq:coherencelength}~\cite{tinkhamIntroductionSuperconductivity2004}. 
    

    The intrinsic material properties set upper bounds on $v_F$, $n$, $T_c$, and $\ell$, while processing quality determines how closely these optimal values are achieved. Poor processing can dramatically degrade all parameters, while optimized processing approaches the material's intrinsic limits.

    \subsection{Type~II Superconductors in DC Fields}

    Superconductors can be classified into type~I or type~II, depending on the ratio between the magnetic penetration depth $\lambda$ and the coherence length $\xi$, where the Ginzburg-Landau parameter $\kappa = \lambda / \xi$ can be used to define the separation between types. When $\kappa < 1/\sqrt{2}$, the superconductor is type~I, and when $\kappa > 1/\sqrt{2}$, the superconductor is type~II~\cite{tinkhamIntroductionSuperconductivity2004}. Type~I superconductors expel all magnetic field below the critical field $H_c$, and lose superconductivity above $H_c$. As discussed briefly earlier, most pure metals are type-I superconductors and are not suitable for use as axion detector cavities. Type~II superconductors have a lower ($H_{c1}$) and an upper ($H_{c2}$) critical field, where $H_{c1} \sim H_c/ \sqrt{2} \kappa$ and $H_{c2} = \sqrt{2} \kappa H_c$. While bulk superconductivity is lost for fields above $H_{c2}$, a mixed state of normal and superconducting electron phases allows superconductivity to persist at high magnetic fields. All useful superconductors for high-field applications are strongly type~II.
    
    Below the lower critical field $H_{c1}$ ($38$~mT for Nb$_3$Sn~\cite{godekeNb3SnRadioFrequency2006}), a type~II superconductor is in the Meissner state. It behaves like a type-I superconductor, where all magnetic field is expelled from the bulk of the material. Above the lower critical field:

    \begin{equation}
        \mu_0 H_{c1} = \frac{\Phi_0}{4\pi \lambda^2}\ln\kappa \,,
        ~\label{eq:FirstCriticalField}
    \end{equation} 

    \noindent vortices penetrate the superconductor in the form of magnetic flux quanta $\Phi_0 = h/2e \approx2.07\times 10^{-15}$~T\,m$^2$. These vortices enter at the edge of the sample and move into the material's interior. 

    The structure of the vortex core follows from incorporating Abrikosov's solutions~\cite{abrikosovMagneticPropertiesSuperconducting1957} for the Ginzburg-Landau equations and later studies from Brandt~\cite{brandtFluxlineLatticeSuperconductors1995}. The order parameter ($\psi$) falls to 0 at the vortex core, implying that all electrons are in the normal state within the core. Outside the core, the superconducting order parameter rises to its equilibrium value. A thread of magnetic flux (a flux quantum $\phi_0$) is contained within the vortex core, and screening currents circulate the core producing a magnetic field described by a modified Bessel function (similar to Eq~\ref{eq:MagneticFieldPen}), decaying on the characteristic length scale $\lambda$.
    
    The number density of vortices is dependent on the magnetic field:

    \begin{equation}
        n_\text{v} = \frac{\text{B}}{\Phi_0}
        \,.\label{eq:numberofvortices}
    \end{equation}

    \noindent As the number density of vortices increases with magnetic field, the screening current around vortices overlaps with currents from nearby vortices. This causes the vortex lattice to stiffen at a magnetic field around~\cite{larkinPinningTypeII1979, blatterVorticesHightemperatureSuperconductors1994}:

    \begin{equation}
        \mu_0 H \approx \frac{\Phi_0}{\lambda^2}
        \,.\label{eq:Hstiff}
    \end{equation} 

    \noindent Once the magnetic field exceeds this value ($\approx200$~mT for Nb$_3$Sn), vortices begin to arrange in a triangular lattice with regular spacing throughout the material~\cite{traubleFluxLineArrangementSuperconductors1968, abrikosovMagneticPropertiesSuperconducting1957, brandtFluxlineLatticeSuperconductors1995}. The vortex lattice can be distorted due to vortex-vortex repulsion, interactions with impurities, vortex pinning~\cite{campbellPinningFluxVortices1968}.
    
    Since axion cavities operate at 9~T, many vortices exist within the cavity body, with separation $a_\text{vortex} \approx \sqrt{\frac{\phi_0}{\text{B}}} \approx 15$~nm~\cite{blatterVorticesHightemperatureSuperconductors1994}. More vortices enter the material as the magnetic field increases, until the cores are separated by a coherence length. The whole superconducting state breaks down beyond this field, and is termed the upper critical field:


     \begin{equation}
        \mu_0 H_{c2} = \frac{\phi_0}{2\pi\xi^2} \,.
        \label{eq:Hc2andCoherence}
    \end{equation}
    
    \noindent From the above equation, it is found that there is a relation between a small coherence length and a high $H_c2$. However, a small coherence length implies a short electron mean free path (Eq.~\ref{eq:coherencelength}), therefore there is a correlation between materials that have many impurities and therefore a high normal state resistivity, and materials with a high $H_c2$. Therefore, optimizing superconducting materials for high-field applications requires balancing disorder-enhanced $H_c2$ against disorder-induced increases in surface resistance.
    
    When these superconductors are used to generate large magnetic fields, as in superconducting solenoid magnets, a large transport current is injected into a superconducting wire. There is zero resistance as long as the vortices in the bulk remain stationary. These vortices remain pinned if the pinning force $F_P$ (pinning force per unit length of vortex) is stronger than the Lorentz force $F_L$ induced by the transport current:

    \begin{equation}
        \mathbf{F}_L = \mathbf{J} \times \mathbf{B}\,.
        \label{eq:lorentzforce}
    \end{equation}
    
    \noindent If the Lorentz force exceeds the pinning force (transport current exceeds critical current $J>J_c$), the normal electrons in the vortex cores flow through the bulk, causing ohmic loss ($V=IR$) described by Bardeen-Stephen Theory~\cite{Bardeen:1965zz}:

    \begin{equation}
        \rho_{\text{eff}} = \rho_n \frac{\text{B}}{B_{c2}}\,.
        \label{eq:rhoeffective}
    \end{equation}

     \noindent This free flux flow limit describes an equivalence between moving vortices and electrical resistance analogous to an ohmic wire. Flux-flow losses constitute a significant concern in RF cavities used for particle acceleration~\cite{padamseeRFSuperconductivity2009}. In axion detector applications, however, AC currents should be small, so vortices remain pinned by vortex-vortex interactions and mild interactions with material defects.

    \subsection{Type~II Superconductivity in Weak RF Fields}
    
    
    
    When an AC magnetic field is present, screening currents arise within the penetration depth, just as in the DC case. However, the dynamic magnetic field generates an electric field within the penetration depth. The E-field interacts with the Boltzmann distribution of normal electrons in the bulk (Eq.~\ref{eq:normalelectrons}), leading to surface resistance that scales with normal state resistivity and temperature~\cite{padamseeRFSuperconductivity2009}: 

    \begin{align}
        R_{\text{BCS}}(T) = \frac{2 \pi^2 \mu_0^2 f^2 \lambda^3 \Delta}{ k_B T \sigma_n} \exp\left(-\frac{\Delta}{k_B T}\right)\,,
        \label{eq:RBCS}
    \end{align}  

    \noindent where $\sigma_n$ is the normal state conductivity. This equation is valid for type~II superconductors when $f << \Delta$ and $T<<T_c$. However, at very low temperatures, when $T \rightarrow 0$~K (mK temperatures), the Boltzmann population of electrons $n_\text{n} \rightarrow 0$ which leads to $R_{\text{BCS}}\rightarrow 0$, the total surface resistance still does not go to zero, but saturates to a residual resistance $R_0$. 
    
    The residual resistance component combines dissipative phenomena from multiple sources, including flux trapping~\cite{ciovatiHighFieldSlope2010}, surface roughness~\cite{padamseeRFSuperconductivity2009}, and non-superconducting impurities~\cite{alexgurevich33ThermalRF2006}. For clean and uniform films under optimal experimental conditions, $R_0 \approx 1~\text{n}\Omega$ has been reached for pure Nb cavities in accelerators~\cite{romanenkoUltrahighQualityFactors2014}. For compound superconductors like Nb$_3$Sn, achieving similarly low residual resistance is much more challenging; $R_0 \sim 10~\text{n}\Omega$ represents excellent performance~\cite{Posen:2020kei}.
    
    \subsection{Type~II Superconductivity in Weak RF Fields and High DC Fields}\label{sec:campbell}
    
    For axion detector applications, cavity haloscopes are immersed in a strong DC magnetic field and excited with a weak AC signal. The external DC field pushes vortices into the cavity, while an AC field can oscillate these vortices. The AC field exerts a Lorentz force (Eq.~\ref{eq:lorentzforce}) on the vortices within a few penetration depths ($\lambda$) from the surface~\cite{campbellResponsePinnedFlux1969, brandtFluxlineLatticeSuperconductors1995}. For a cylindrical cavity aligned with the solenoid axis (z-direction) operating in TM$_{010}$ mode:
    
    \noindent Field orientation relative to the surface determines vortex response:
    
    \begin{enumerate}
        \item \textbf{Endcaps ($\mathbf{B}_{\text{DC}} \perp$ surface):} 
        Surface currents flow radially $\mathbf{J}_{\text{endcap}} = J_{\text{AC}} \hat{r}$, producing azimuthal Lorentz force:
        \begin{equation}
            \mathbf{F}_L = (J_{\text{AC}} \hat{r}) \times (B_0 \hat{z}) = -J_{\text{AC}} B_0 \hat{\phi}
        \end{equation}
        This perpendicular force maximizes vortex oscillation amplitude.
        
        \item \textbf{Walls ($\mathbf{B}_{\text{DC}} \parallel$ surface):} 
        Surface currents flow axially $\mathbf{J}_{\text{wall}} = J_{\text{AC}} \hat{z}$, yielding zero Lorentz force:
        \begin{equation}
            \mathbf{F}_L = (J_{\text{AC}} \hat{z}) \times (B_0 \hat{z}) = 0
        \end{equation}
        Parallel geometry eliminates direct vortex driving.
    \end{enumerate}
    
    The parallel configuration produces zero Lorentz force, minimizing vortex motion and dissipation. Additional dissipative mechanisms on the walls include vortex entry at the edges and field inhomogeneities. These contributions to surface resistance have been observed experimentally, where bulk Nb-Ti cavities exhibit higher dissipation on the endcaps than on the walls (Fig.~\ref{fig:NbTiendcapsl}). For thin films, surface roughness can tilt vortices slightly out of the field direction, creating non-zero Lorentz forces even on nominally parallel surfaces.
    
    \subsubsection{Surface Resistance in DC Magnetic Fields}
    
    Vortices that undergo small-amplitude elastic oscillations cause dissipation to depths greater than the London penetration depth, characterized by the Campbell penetration depth $\lambda_c$~\cite{campbellResponsePinnedFlux1969}:
    
    \begin{align}
       \lambda_c^2 = \lambda^2 \frac{c_{11}}{\alpha_L} \label{eq:lambdaprime}
    \end{align}  
    
    \noindent where $\alpha_L$ is the Labusch parameter (restoring force per unit displacement) and $c_{11} = B^2/(\mu_0\lambda^2)$ is the elastic compression modulus of the triangular vortex lattice. The Labusch parameter \cite{labuschElasticConstantsFluxoid1969, panLabuschParameterDriven2000} quantifies the restoring force for vortex displacements from pinning centers, even for small displacements where a vortex is not liberated from its pinning energy well. It is affected by the material's defect characteristics and is usually measured across different microstructures. Oscillating vortex cores dissipate energy through motion of the normal electrons in and around vortex cores, producing the Campbell resistance~\cite{markcoffeyUnifiedTheoryEffects1991}:
    
    \begin{align}
        R_c(B) = \frac{\mu_0^2 \omega^2 \lambda^3 \sigma_n}{2}\left(\frac{c_{11}}{\alpha_L}\right)^{3/2}
        \label{eq:Rcampbell}
    \end{align}
  
    \noindent Further theoretical discussion of collective pinning in AC fields can be found in Clem and Coffey~\cite{markcoffeyUnifiedTheoryEffects1991} and recent work by Pompeo et al.~\cite{nicolapompeoReliableDeterminationVortex2008}.

    \subsection{Total Resistance} 
        
    The total surface resistance $R_s$ is therefore a sum of three components when a superconductor operates in the axion detector parameter space: BCS resistance (Eq.~\ref{eq:RBCS}), residual resistance, and Campbell resistance (Eq.~\ref{eq:Rcampbell}):
    
    \begin{align}
        R_{s,sc} = R_{\text{BCS}}(T) + R_0 + R_C(B)  \label{eq:Rs} .
    \end{align}
    
    These components can be isolated experimentally through systematic measurements. At millikelvin temperatures in zero field, $R_{\text{BCS}}$ becomes negligible, and $R_0$ dominates, representing residual losses from non-superconducting impurities, defects, and surface quality~\cite{alexgurevich33ThermalRF2006}. When high DC fields are applied at low temperature, the field-dependent Campbell resistance $R_C$ also has a strong effect. A goal for axion detector development is to reduce the total surface resistance below that of high-purity Cu (RRR~=~300) at 1~GHz and 4~K, for which values from table \ref{tab:Rs_Q_anomalous} give an estimation of $R_{s,nc} \sim 5~\text{m}\Omega$.
    
    Different superconductors exhibit distinct field-dependent behavior in measurements. NbTi cavities show significant resistance $R_s>~1 \text{m}\Omega$ with increasing magnetic field, indicating Campbell resistance becomes the limiting factor~\cite{Braine:2024nzi}. In contrast, REBCO films maintain low surface resistance $R_s<~1 \text{m}\Omega$ even in strong fields (Fig.~\ref{fig:axionprototypes})~\cite{posenMeasurementHighQuality2022, ahnSuperconductingCavityHigh2020}, suggesting instead that residual resistance is dominant.

\chapter{Methodology}\label{sec:chapter3}

The framework established in the previous chapters has prioritized the need for cavities with high-quality factors in magnetic fields for axion detector experiments. This chapter describes:

\begin{itemize}
    \item Fabrication methods to make Nb$_3$Sn films 
    \item Methods to analyze the structure and composition of Nb$_3$Sn film samples 
    \item Methods to characterize the normal-state and superconducting properties of Nb$_3$Sn film samples
    \item Methods to characterize the RF properties of conductors and superconductors
\end{itemize}

After fabrication, the film quality of small samples can be characterized using critical temperature ($T_c$) measurements, visible light microscopy, and scanning electron microscopy (SEM) with energy-dispersive spectroscopy (EDS). These techniques require little setup time and provide a wealth of information about the films.  It should be noted that building an entire cavity is often necessary to assess the RF behavior of a film or material, which will be covered in later chapters, because this is the end product of this dissertation.


\section{Making \texorpdfstring{Nb$_3$Sn}{Nb3Sn} Thin Films}

Standard fabrication techniques are described in the following thin-film textbook~\cite{seshanHandbookThinFilm2002}. The specific recipes used in this work to fabricate Nb$_3$Sn are provided in Sec.~\ref{sec:nb3snresults}.

    \subsection{Substrate Preparation}
    
    Surface preparation is critical for thin film adhesion and quality. Substrates underwent mechanical polishing from carbide paper through diamond slurry to colloidal silica, progressively reducing surface roughness to $<50$~nm RMS (detailed protocol in Appendix~\ref{appsec:substrateprep}). Substrates were cut differently depending on the use case: $<1$~cm square (small samples), 2-inch disk (mushroom cavity), and custom geometries (RF measurements).

    \subsection{Thermal Evaporation}
    
\begin{figure}[tb]
        \centering
        \includegraphics[width=.8\textwidth]{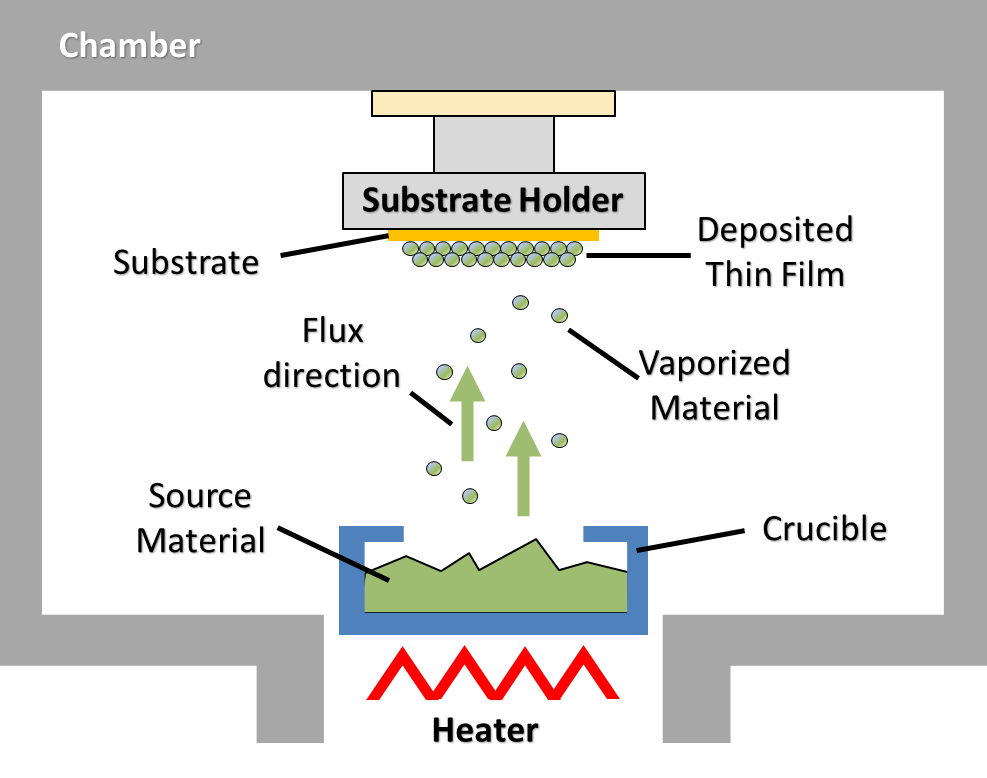}
        \caption[Source material heated and evaporated onto a target.]{Source material heated and evaporated onto a target. Material is moving from a hot location (crucible) to a cold location (substrate).}
        \label{fig:ThermalEvap}
    \end{figure}

    Thermal evaporation is a physical vapor deposition technique where source material is heated to high temperatures in vacuum until it vaporizes and condenses on a substrate (Fig.~\ref{fig:ThermalEvap}). Evaporation occurs when atoms gain sufficient thermal energy to overcome surface binding forces. An electric current passes through a refractory-metal container (boat, basket, or crucible) containing the source material. The container's electrical resistance causes Joule heating ($P = I^2R$), and as current increases, the temperature rises until the source material reaches its vaporization temperature. A standard heating element is tungsten (melting point 3422~°C), chosen for its high melting point and low reactivity. 

    The atomic-scale deposition process involves several sequential steps. At the source, thermal energy drives atomic diffusion towards the surface, followed by desorption when atoms gain energy exceeding the surface work function. The desorbed atoms undergo ballistic motion through the vacuum chamber (enabled by mean free paths exceeding chamber dimensions at typical pressures), then adsorb onto the substrate surface. Surface dynamics include ad-atom lateral mobility, re-desorption events, atomic displacement by incoming flux (``atomic peening"), and defect formation when mobility is insufficient for ideal crystal growth. Bulk diffusion is generally negligible at typical substrate temperatures.
    
\begin{figure}[tb]
        \centering
        \includegraphics[width=.8\textwidth]{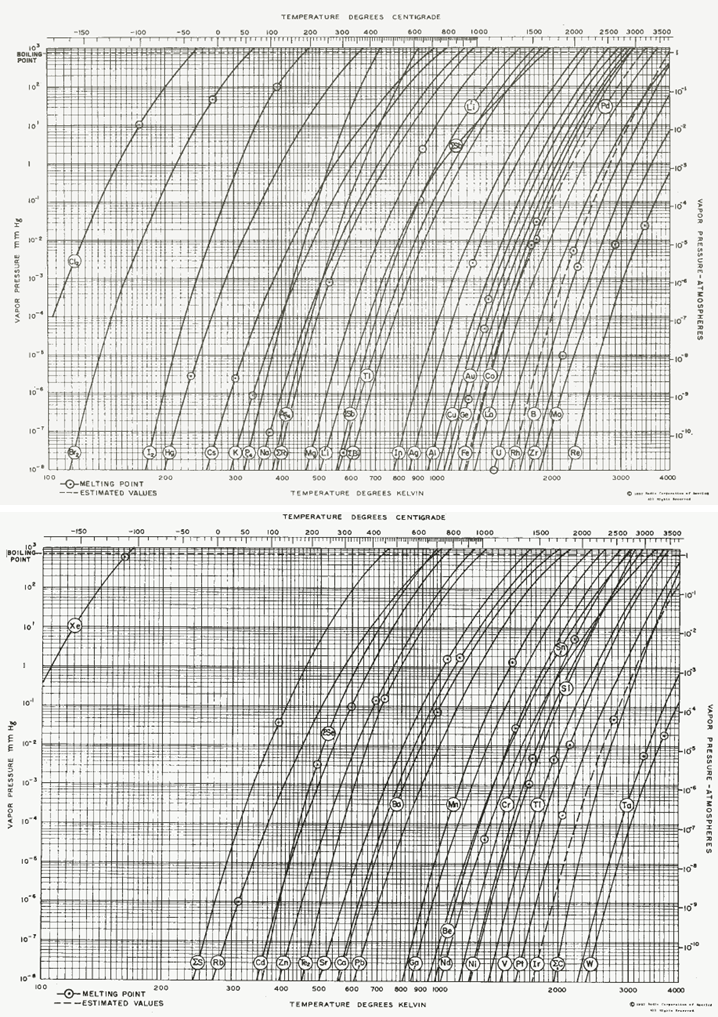}
        \caption{Vapor pressure curves for common elements \cite{leonmaisselHandbookThinFilm1995}.}
        \label{fig:VaporPressure}
    \end{figure}

    The vapor pressure of a material varies with temperature (Fig.~\ref{fig:VaporPressure}). For materials relevant to Nb$_3$Sn synthesis: tin has a relatively low melting point of 232~°C and evaporates readily, while copper melts at 1085~°C and niobium at 2477~°C. The chamber pressure must be significantly lower than the material's vapor pressure to allow free-molecular flow. At typical chamber pressures ($\sim10^{-6}$ Torr), the mean free path exceeds the chamber dimensions, enabling line-of-sight deposition.

    \FloatBarrier
    
    When evaporating alloys or co-evaporating multiple elements, components evaporate at different rates according to their individual vapor pressures, thereby altering the film's elemental composition relative to the source elemental composition. This is particularly important for materials such as Cu-Sn alloys, where tin's higher vapor pressure leads to preferential evaporation. The deposition rate follows a cosine distribution from the source, with maximum thickness directly above the source and decreasing with increasing angle. Thickness monitoring is typically achieved using a quartz crystal microbalance (QCM) or through post-deposition techniques such as profilometry or cross-sectional imaging.
    
    Thermal evaporation offers several advantages, including high-purity depositions, reasonable rate control, and the ability to deposit multiple materials sequentially without breaking vacuum. However, it is limited to line-of-sight deposition, resulting in poor step coverage, non-uniform thicknesses over large areas, and difficulty in depositing very high-melting-point materials~\cite{leonmaisselHandbookThinFilm1995}. Pre-outgassing the source material at a moderate temperature before evaporation helps reduce contamination from adsorbed gases. While substrate heating promotes crystallinity and substrate rotation could improve thickness uniformity, the evaporation chamber was not equipped with these tools, which constrained the work in this dissertation. Small samples were used to mitigate the adverse effects of these constraints. However, in the large RF samples used, defects were found that may have resulted from these constraints. 

    \subsection{Magnetron Sputtering}

    In a weak vacuum, a high electric field can ionize gas and produce a plasma for deposition. The Paschen condition specifies the required field or voltage for plasma discharge~\cite{paschenUeberFunkenubergangLuft1889}. The gas type and pressure are selected based on voltage and chamber characteristics. Inert gas ions are typically used to avoid unwanted reactions with the target material. Ions from the resulting discharge can have high energy, more than sufficient to liberate atoms from the surface of a metal target, which is called sputtering.

    The inert gas plasma (typically Ar) accelerates ions toward a target material, where momentum transfer ejects target atoms. These atoms traverse the vacuum chamber and nucleate on a substrate, building a thin film atom by atom. Control over plasma conditions, chamber pressure, and substrate temperature allows tuning of film composition, microstructure, and deposition rate.

\begin{figure}[tb]
        \centering
        \includegraphics[width=\textwidth]{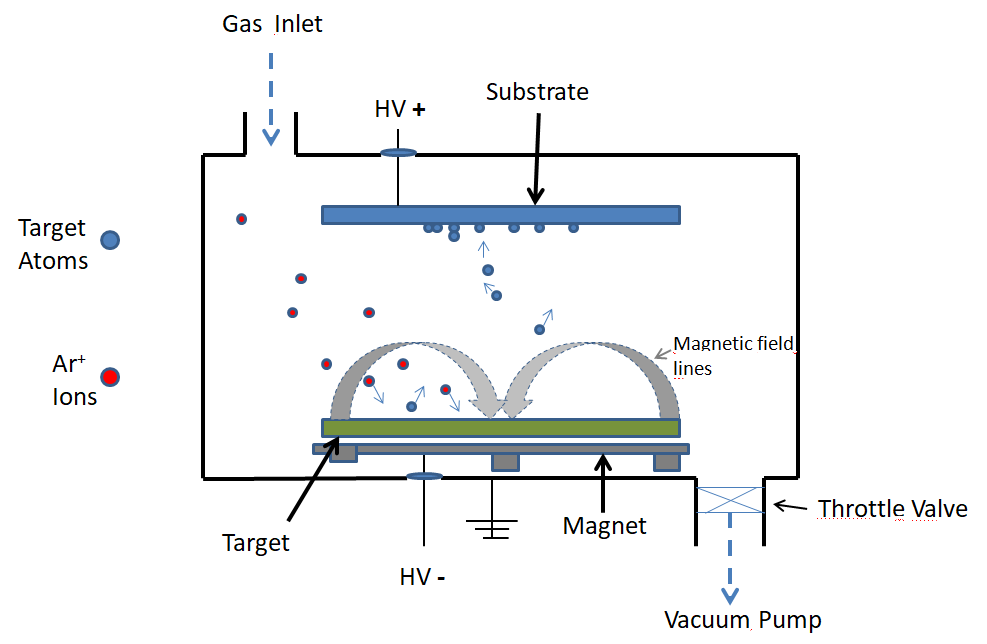}
        \caption[An ultra-high vacuum chamber with a working gas (Argon) that hits the target material and liberates target atoms.]{An ultra-high vacuum chamber with a working gas (Argon) that hits the target material and liberates target atoms. The argon atoms hit the target many times over as the magnet underneath the target accelerates them back to the target \cite{mullerIntroductionVacuumCoating}.}
        \label{fig:MagnetronSputtering}
    \end{figure}

    Magnetic fields increase the efficiency of harvesting gas ions for sputtering. The magnetic field contours trap ions near the target (Fig.~\ref{fig:MagnetronSputtering}), confining the plasma to a dense region above the target and increasing deposition rates by $10-100\times$ while reducing substrate heating~\cite{mattoxHandbookPhysicalVapor2010}. DC sputtering was used for all metal depositions in this work. RF sputtering at 13.56~MHz was used only for in-situ substrate cleaning before deposition.

    \FloatBarrier
        
    Sputtering can deposit any material regardless of melting point, unlike thermal evaporation. Sputtered films are generally denser and more adherent due to the higher kinetic energy of arriving atoms~\cite{mattoxHandbookPhysicalVapor2010}. When sputtered atoms arrive at the substrate surface, they become adatoms with limited mobility. The substrate temperature and kinetic energy of arriving atoms determine how far adatoms can diffuse before they nucleate at favorable sites or join existing islands. The competition between arrival rate and surface diffusion rate determines the resulting film morphology~\cite{ohringMaterialsScienceThin2002}.
    
\begin{figure}[tb]
        \centering
        \includegraphics[width=\textwidth]{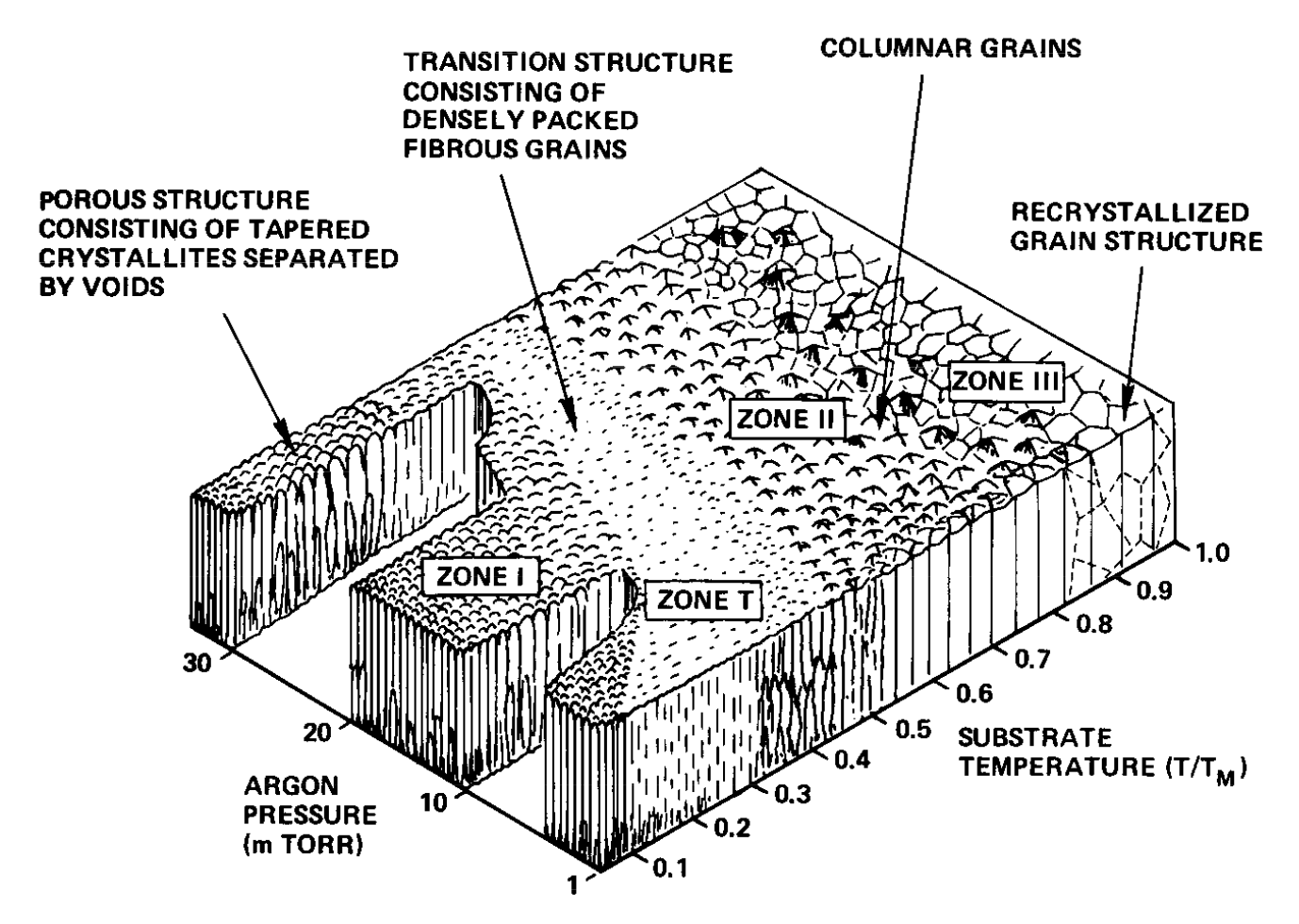}
        \caption{Thornton structure zone model diagram from~\cite{thorntonInfluenceSubstrateTemperature1975}.}
        \label{fig:ThorntonDiagram}
    \end{figure}

    This relationship is captured by the Thornton structure zone model (Fig.~\ref{fig:ThorntonDiagram}), which maps film microstructure as a function of homologous temperature $T/T_m$ (substrate temperature normalized to the melting point) and working gas pressure. At low $T/T_m$ and high pressure (Zone 1), adatom mobility is limited, and porous columnar films with voided grain boundaries form. As $T/T_m$ increases or pressure decreases (Zone T and Zone 2), surface diffusion increases and adatoms can fill voids, producing denser columnar grains. At the highest $T/T_m$ (Zone 3), bulk diffusion dominates and large equiaxed grains form.
    
    For Nb deposition at 8~mTorr Ar, a substrate temperature of 200$^\circ$C gives $T/T_m \approx 0.17$ ($T_m = 2477^\circ$C for Nb), placing the film in Zone 1 or the Zone T boundary. Columnar growth with voided boundaries is expected. At 700$^\circ$C, $T/T_m \approx 0.35$, which falls in Zone 2 where surface diffusion produces denser columnar films with fewer voids. The moderate Ar pressure of 8~mTorr allows sputtered atoms to retain kinetic energy upon arrival, promoting adatom mobility compared to higher pressures where gas-phase scattering thermalizes the arriving flux.
    
    \FloatBarrier

    \subsection{Post-Deposition Processing}
    
    \subsubsection{Heat-Treatment}
        
    As discussed in more detail later, this dissertation relies upon a diffusion reaction between Cu-Sn and Nb to form Nb$_3$Sn. A necessary step in Nb$_3$Sn film fabrication is a heat treatment in which the film materials are held at high temperatures to enable diffusion reactions. Ultra-high vacuum (UHV) furnaces enable high-temperature reactions in clean environments. Oxygen, nitrogen, and other impurities are problematic in many types of reactions~\cite{Sun:2023ejj}, so UHV furnaces reduce air impurities so the resulting reaction has low impurity content. Ambient pressure is at 1 atm = 760 Torr, high-vacuum pressure is considered $10^{-3}$~to~$10^{-7}$~Torr, and ultra-high vacuum is $10^{-7}$~to~$10^{-12}$~Torr~\cite{rothVacuumTechnology2012}. 
    
    Impurities collect on the surface of newly reacted or deposited material. The rate at which these impurity monolayers form is proportional to the background pressure. The time to form one monolayer of residual gas on a surface is approximately $t_\text{{monolayer}} \approx 3 \times 10^{-8}/P$~seconds, where $P$ is pressure in Torr \cite{ohringMaterialsScienceThin2002}. For example, at atmospheric pressure, a monolayer forms in about three nanoseconds, at $10^{-6}$~Torr in three milliseconds, and at $10^{-9}$~Torr in 3 seconds. This means that even during high-vacuum heat treatments lasting several hours, surfaces are continuously exposed to impurity adsorption. 
    
    Positive-pressure argon furnaces allow further impurity reduction by first pumping down to vacuum, then adding a non-reactive gas until a positive pressure is reached. Continual argon flows through the chamber actively flushing out outgassing impurities. This can allow vacuum chambers with low pumping power to remain clean enough for high-quality films (discussed in Sec.~\ref{sec:Nbsub}).
    
    \subsubsection{Chemical Etching}
    
    Chemical etching removes material through controlled dissolution in acidic or oxidizing solutions, used for surface cleaning, thickness reduction, or complete film removal. Common etchants for our application include ammonium persulfate ((NH$_4$)$_2$S$_2$O$_8$) and nitric acid (HNO$_3$) for Cu and Cu-Sn alloys, and hydrofluoric acid (HF) for niobium. The etch rate depends on solution concentration, temperature, and agitation, with heating typically accelerating the reaction significantly. 
    
    Etching proceeds through oxidation of the metal surface, followed by dissolution of the oxidized species into solution. The reaction is self-limiting in that fresh etchant must continuously reach the surface, making agitation or stirring important for uniform etching. Most etchants leave a thin native oxide layer on the exposed surface after rinsing and drying, which may need removal via plasma cleaning or in-situ ion bombardment before subsequent processing steps. Chemical etching offers simplicity and low cost, but provides limited control over thickness compared to physical methods such as ion milling. Corrosive chemistry requires appropriate safety precautions, including fume hoods and acid-resistant equipment.

\section{Analyzing Film Structure and Composition}

The uniformity, composition, and microstructure of a film directly influence its quality factor ($Q$) through surface resistance (see Sec.~\ref{sec:superconductor}). Consequently, $Q$ can be estimated by characterizing small-coupon samples before full cavity implementation. 

While complex or unknown film features may necessitate further probing via X-ray Diffraction (XRD), X-ray Photoelectron Spectroscopy (XPS), Transmission Electron Microscopy (TEM), Secondary Ion Mass Spectrometry (SIMS), or Rutherford Backscattering Spectrometry (RBS), these are not always required. These techniques offer insights into phase identification, surface chemistry, atomic-scale microstructure, and depth-resolved elemental profiles. In this work, however, Scanning Electron Microscopy (SEM) and Energy-Dispersive X-ray Spectroscopy (EDS) proved sufficient to optimize deposition recipes and identify the most promising fabrication routes for cavity-scale production.

\subsection{Preparation of Cross-Sections and Surfaces for Analysis}

    
    Surface analysis using SEM/EDS provides compositional information averaged through the entire film thickness at a specific lateral position. However, superconducting thin films synthesized by diffusion-driven reactions (such as Nb$_3$Sn formed by Sn diffusion into Nb) often exhibit compositional gradients perpendicular to the substrate. Resolving these depth-dependent variations requires cross-sectional imaging.
    
\begin{figure}[tb]
        \centering
        \includegraphics[width=.7\textwidth]{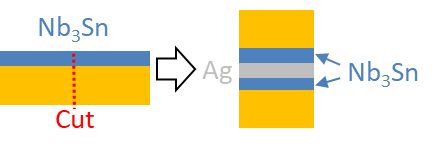}
        \caption{Schematic of film that is prepared for cross-section by cutting and gluing together using a sandwich structure.}
        \label{fig:CuttingCrossSection}
    \end{figure}

    Two primary approaches enable cross-sectional analysis, each with distinct trade-offs:
    
    \noindent \textbf{Mechanical cross-sectioning:} First, a thin silver film was thermally evaporated onto the sample surface to protect the film during subsequent processing. The sample was cut in half using a diamond saw. Silver paste was applied to the coated surface, and the two halves were pressed together face-to-face, creating a symmetric "sandwich" structure (seen in Fig.~\ref{fig:CuttingCrossSection}). This configuration positions the films of interest at the center of mass of the mounted sample, where polishing pressure is most uniform. The assembly was embedded in a conductive mounting medium and polished using a sequential abrasive protocol: 600- and 1200-grit SiC paper, followed by 5, 3, and 1~\unit{\um} diamond suspensions, and finished with colloidal silica on a vibratory polisher. After vibratory polishing, samples were cleaned with soap and water.

    \FloatBarrier

    When imaged with the polished edge perpendicular to the electron beam, this geometry enables accurate quantitative EDS measurements without angular corrections. However, mechanical stress during cutting can damage brittle films, introduce artifacts such as smearing or pull-out of softer phases, or cause delamination at the film-substrate interface.

    \begin{figure}[H]
        \centering
        \includegraphics[width=.9\textwidth]{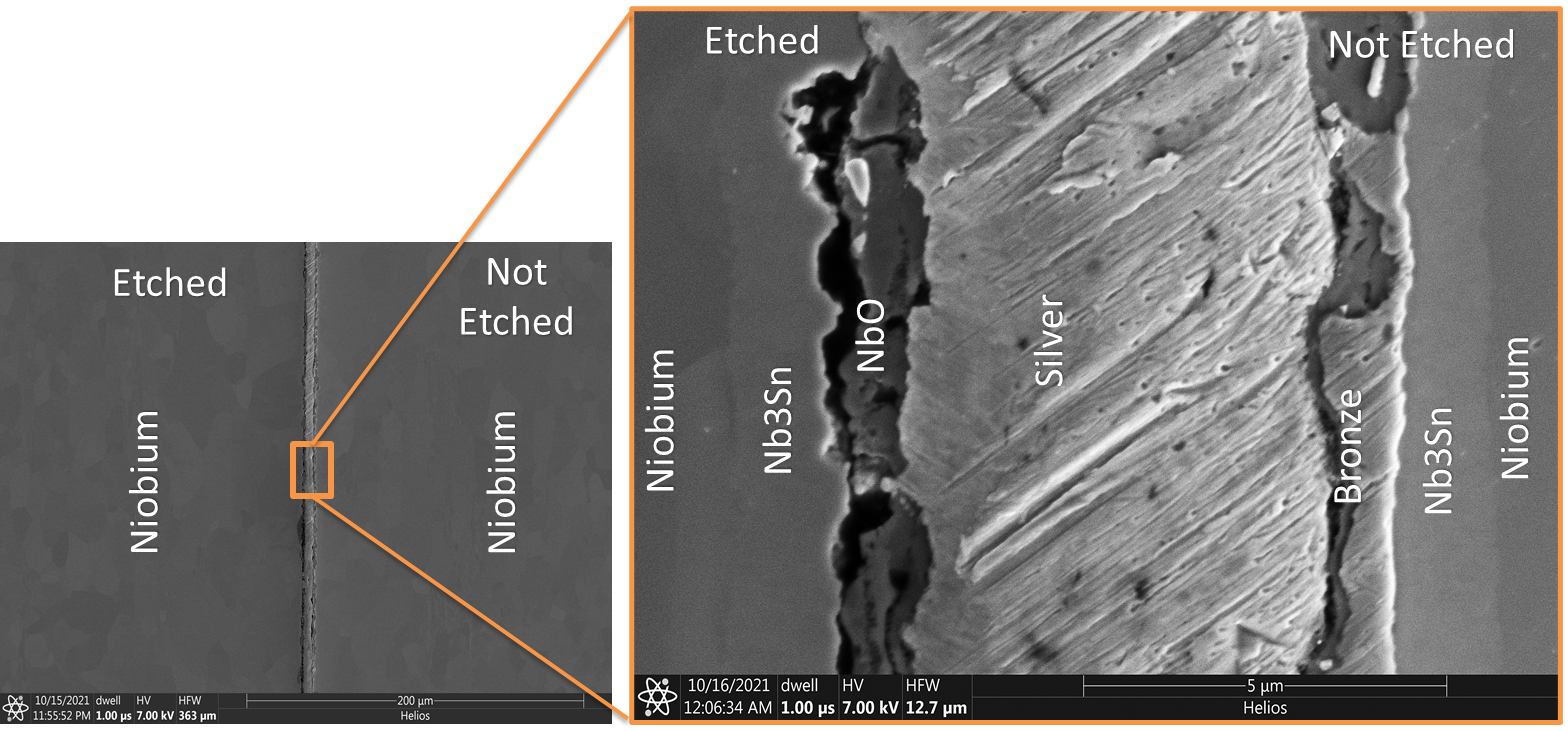}
        \caption{SEM images of a Nb$_3$Sn film on a Nb substrate that was cut in half, and one half was etched and the other half not etched, and sandwiched together.}
        \label{fig:NbsubetchSEM}
    \end{figure}  
    \FloatBarrier
    
    \noindent \textbf{Focused Ion Beam (FIB) milling:} An ion beam mills a trench into the sample, exposing the cross-section in situ with minimal mechanical contact or stress. FIB minimizes damage and provides precise site-specific sectioning, allowing investigation of localized features (defects, grain boundaries). The primary limitation is geometric: standard FIB systems image the milled cross-section at a 52~° stage tilt. For thin films ($\sim500$~nm), this oblique viewing angle causes the electron beam interaction volume to span multiple layers, complicating compositional analysis of the sharp interface.
    
    To overcome the 52~° limitation, a thin lamella ($\sim1-2$~\unit{\um} thick) can be extracted from the cross-section and remounted perpendicular to the electron beam. This lift-out technique positions the film thickness edge-on for direct imaging and perpendicular EDS analysis, eliminating angular artifacts. The thin geometry also reduces electron scattering from material behind the region of interest, improving spatial resolution. If atomic-scale investigation is required, the lamella can be further thinned to electron transparency ($<50$~nm) for transmission electron microscopy (TEM). However, lamella preparation is time-intensive ($2-4$ hours per sample) and was reserved for films exhibiting anomalous features requiring detailed investigation. FIB protocols are detailed in Appendix~\ref{appsec:FIBProtocol}.

    \begin{figure}[H]
        \centering
        \includegraphics[width=.7\textwidth]{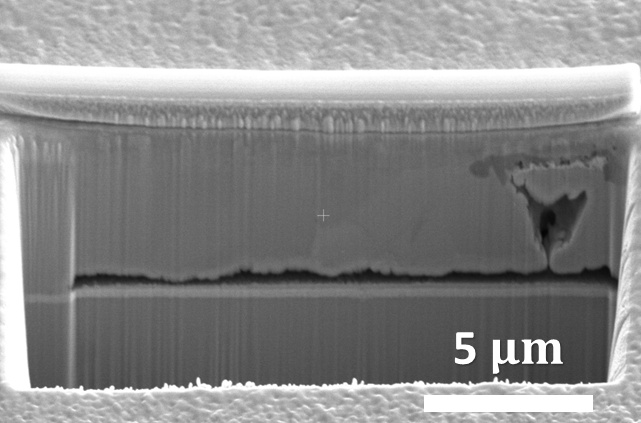}
        \caption{Example FIB cross section SEM image showing voids above the diffusion layer.}
        \label{fig:ExampleFIBCrosssection}
    \end{figure}  
    \FloatBarrier

    \subsection{Microscopy and Micro-chemical Analysis}
        
    Film morphology and composition were characterized using scanning electron microscopy (SEM) with energy dispersive spectroscopy (EDS). Visible light microscopy provided initial screening, and transmission electron microscopy (TEM) was used selectively for atomic-scale analysis. Both surface and cross-sectional imaging were performed to correlate microstructural features with RF performance. 
    
\begin{figure}[H]
        \centering
        \includegraphics[width=.9\textwidth]{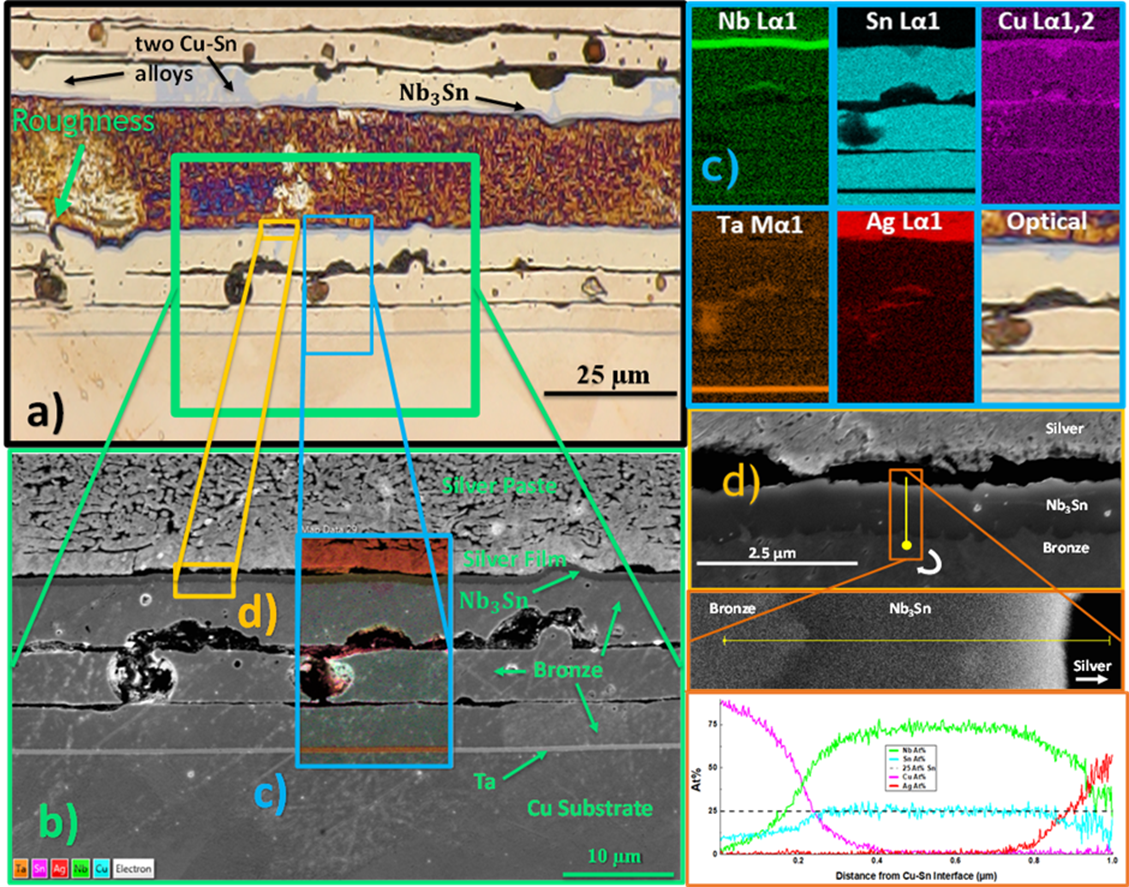}
        \caption[Cross-sectional characterization of a Cu-Sn first film: (a) visible light micrograph and (b) SEM image showing the layered structure from Cu substrate through Ta barrier, bronze, Nb$_3$Sn, and residual Cu-Sn.]{Cross-sectional characterization of a Cu-Sn first film: (a) visible light micrograph and (b) SEM image showing the layered structure from Cu substrate through Ta barrier, bronze, Nb$_3$Sn, and residual Cu-Sn. Colored boxes indicate regions examined in (c) and (d). (c) EDS elemental maps of the cyan-boxed region. (d) High-magnification SEM and EDS line scan across the Nb$_3$Sn layer (orange-boxed region).}
        \label{fig:FullNb3Sn10micron2}
    \end{figure}

    Figure~\ref{fig:FullNb3Sn10micron2} illustrates a typical cross-sectional characterization workflow on a Cu-Sn first sample. The visible light micrograph and SEM image reveal non-uniform surface morphology and thickness variations in the Nb$_3$Sn film, resulting from the underlying bronze layer's porosity and roughness. Multiple Cu-Sn alloy phases near the Nb$_3$Sn surface appear as distinct regions with different contrast. The EDS elemental maps show that Sn penetrated through the Ta diffusion barrier, with comparable Sn concentration on both sides of the Ta layer, indicating barrier failure in this sample. The EDS line scan confirms approximately 25~at.\% Sn uniformly distributed across the film thickness, consistent with stoichiometric Nb$_3$Sn.
        
    \FloatBarrier

  \subsubsection{Visible Light Microscopy}
    
    Surface topography and cross-sections were analyzed using an Olympus DSX1000 digital microscope. The instrument constructs three-dimensional surface maps via focus variation, acquiring images at sequential focal planes and computationally stitching in-focus regions. Lateral resolution is diffraction-limited to the wavelength of visible light ($\sim$400--700~nm), but using a 50$\times$ objective, roughness parameters such as $R_a$ can be quantified down to approximately 5~nm. This is sufficient for characterizing substrate roughness before and after deposition.

    \subsubsection{Scanning Electron Microscopy (SEM)}
    
    Scanning electron microscopy uses a focused electron beam to image surface morphology and composition with resolution down to $\sim$5~nm. When the electron beam strikes the sample, it generates multiple signals that provide complementary information:
    
    \begin{itemize}
        \item \textbf{Secondary electrons (SE)}: Emitted from the near-surface region (<10~nm depth), providing high-resolution topographic contrast ideal for imaging grain boundaries, surface roughness, and morphological features
        \item \textbf{Backscattered electrons (BSE)}: Higher-energy electrons scattered from deeper regions ($\sim$100~nm--1~\unit{\um}), providing compositional contrast where heavier elements (higher atomic number Z) appear brighter
        \item \textbf{Characteristic X-rays}: Enable elemental identification and quantification through energy dispersive spectroscopy (described below)
    \end{itemize}
    
    The electron beam is rastered across the sample surface in a grid pattern, and the detected signal intensity at each point creates an image pixel-by-pixel. Unlike visible light microscopy, which is limited by light wavelength to $\sim$200~nm resolution, SEM achieves much finer resolution due to the short de Broglie wavelength of accelerated electrons ($\lambda \sim 0.01$~nm at 10~keV). 
    
    Acceleration voltage selection involves trade-offs: lower voltages (1--5~kV) reduce beam penetration depth, improving surface sensitivity and minimizing charging on insulating samples, while higher voltages (15--30~kV) increase signal generation from deeper regions, excite higher-energy characteristic X-rays from heavy elements, and improve image contrast for thick films. Detailed imaging protocols are provided in Appendix~\ref{appsec:SEMProtocol}.

    \subsubsection{Energy Dispersive Spectroscopy (EDS)}
        
    Energy dispersive spectroscopy (EDS) analyzes characteristic X-rays emitted when the electron beam excites inner-shell electrons in the sample. When an inner-shell electron is ejected, an outer-shell electron fills the vacancy, releasing an X-ray with element-specific energy.
        
\begin{figure}[H]
        \centering
        \includegraphics[width=.9\textwidth]{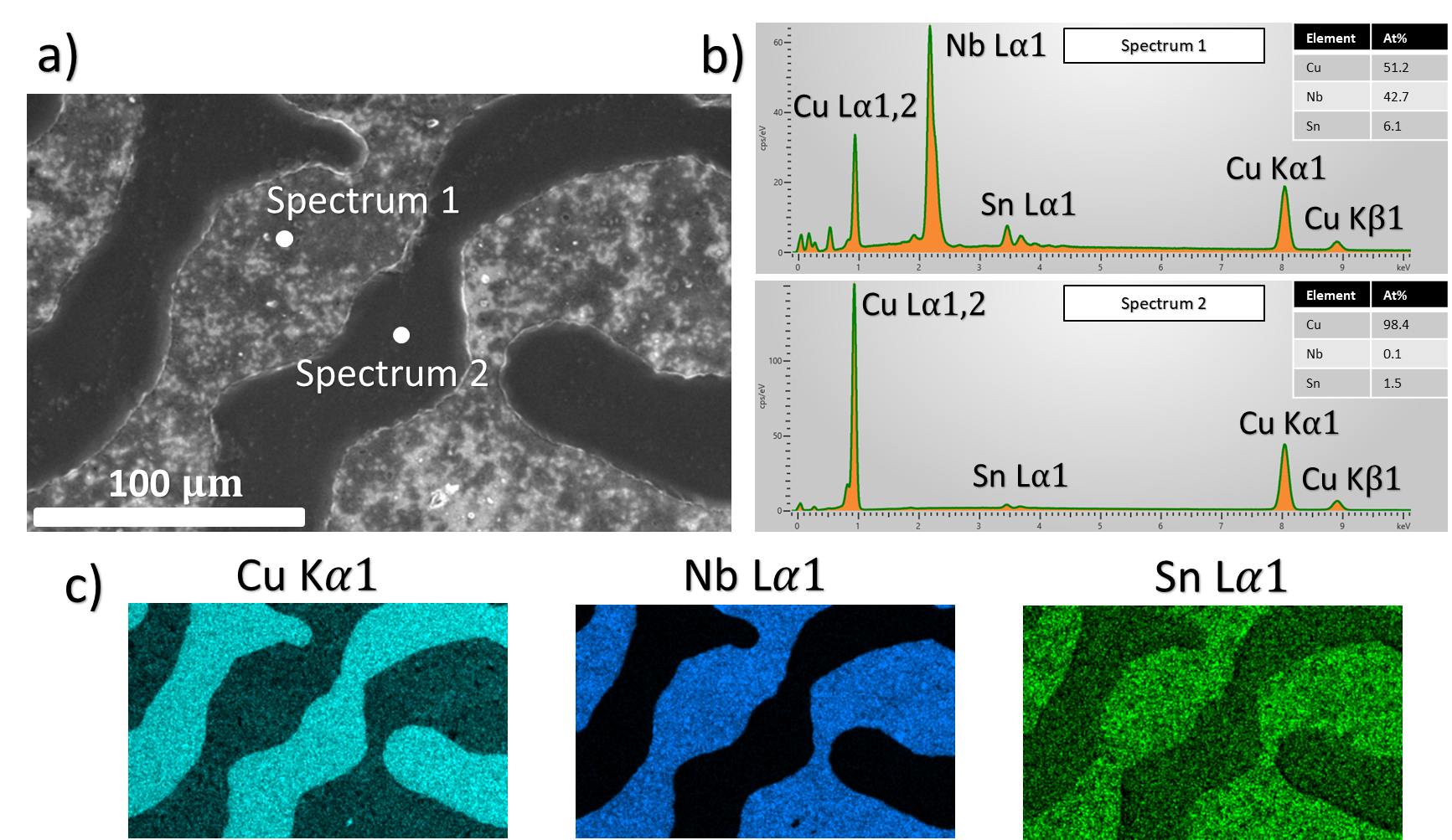}
        \caption{SEM image (a) with (b) EDS point scan and (c) EDS mapping showing elemental content in different regions.}
        \label{fig:EDSSpectrum}
    \end{figure}

    Figure~\ref{fig:EDSSpectrum} shows EDS spectra from two locations on a Nb$_3$Sn film. The detector sorts X-rays by energy, producing a histogram of counts versus keV. Each peak corresponds to a characteristic transition: Nb~L$\alpha$1 at 2.17~keV, Sn~L$\alpha$1 at 3.44~keV, and Cu~K$\alpha$1 at 8.04~keV. Spectrum~1, from a Nb-rich region, shows strong Nb, Sn, and Cu peaks; Spectrum~2 from the surrounding matrix is predominantly Cu.
        
    EDS provides three measurement modes~\cite{AZtecUserManual2023}:
        
    \textbf{Point analysis} acquires a spectrum from a stationary beam position, typically dwelling 30--120~seconds. The analyzed volume is approximately 1~\unit{\um}$^3$ at 15--20~kV.
        
    \textbf{Line scans} acquire spectra sequentially along a path, stepping point-to-point with 0.1--1~s dwell times. This produces compositional profiles useful for measuring gradients across interfaces or layered structures.
        
    \textbf{Elemental mapping} rasters the beam across a region, collecting X-rays at each pixel. The system defines energy ``windows" around peaks of interest. Figure~\ref{fig:EDSSpectrum} shows Cu, Nb, and Sn maps from the same region as the point spectra. Map acquisition typically requires 30 minutes to several hours.
    
    Quantitative accuracy is $\sim$1--2~at.\% for elements with Z $\geq$ 11. Light elements (Z $<$ 11) produce weak signals due to low fluorescence yield.

    \FloatBarrier
    
\section{Critical Temperature Measurements}\label{sec:TcDetail}
    
    \subsection{SQUID Magnetometry}
    
    \subsubsection{Operating Principle}
    
    Traditional AC susceptometry measurements measure a change in magnetic flux:
    
    \begin{equation}
    V = \frac{d\Phi}{dt}
    \end{equation}
    
    \noindent where $\Phi$ is the flux contained in the coil. However, in a SQUID system, the output voltage is proportional to the magnetic flux in the pick-up coil instead of its time derivative. The measurement involves quantization in superconducting loops. As a sample is moved within the area enclosed by a superconducting loop, a persistent screening current is induced and coupled through a transformer to a Josephson junction that detects flux changes with extreme sensitivity.
    
    Multiple loops comprising four superconducting coils, wound in a counterwound configuration (two clockwise, two counterclockwise), form a second-order gradiometer that rejects uniform field flux while detecting local moments. Therefore, if a homogeneous sample is longer than the distance between the outer loops, the magnetic moment is counted as a uniform external field and is not measured. For this reason, samples must be small ($<10$~mm).
    
    \begin{figure}[tb]
        \includegraphics[width=.8\textwidth]{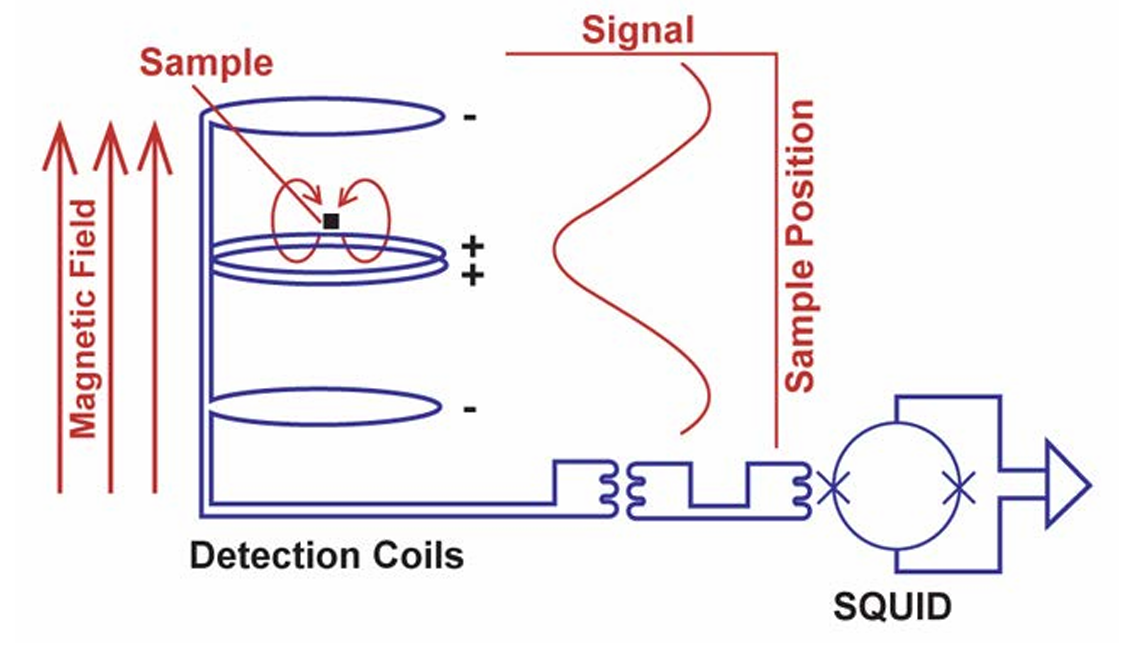}
        \centering
        \caption{Superconducting Quantum Interference Device (SQUID) schematic showing sample translation through pickup coils and Josephson junction detection \cite{designquantumMagneticPropertyMeasurement2009}.}
        \label{fig:SQUID}
    \end{figure}

    The sample is mounted on a long plastic straw that cannot be magnetized. The straw is continuous along the entire sample path, ensuring the coils experience no change in flux due to the straw. As the magnetized sample moves through the detection coils (Fig.~\ref{fig:SQUID}), the magnetic flux $\Phi = BA$ is felt by the coils, where $B = \mu_0(H + M)$. The pick-up coils are wound such that a material that increases the enclosed $B$ (i.e., $M > 0$ for $H > 0$) produces a positive voltage, while a diamagnetic response ($M < 0$) produces a negative voltage. Since the voltage amplitude is proportional to the magnetic moment, small-volume samples require high magnetization to produce a measurable signal.

\begin{figure}[tb]
        \includegraphics[width=\textwidth]{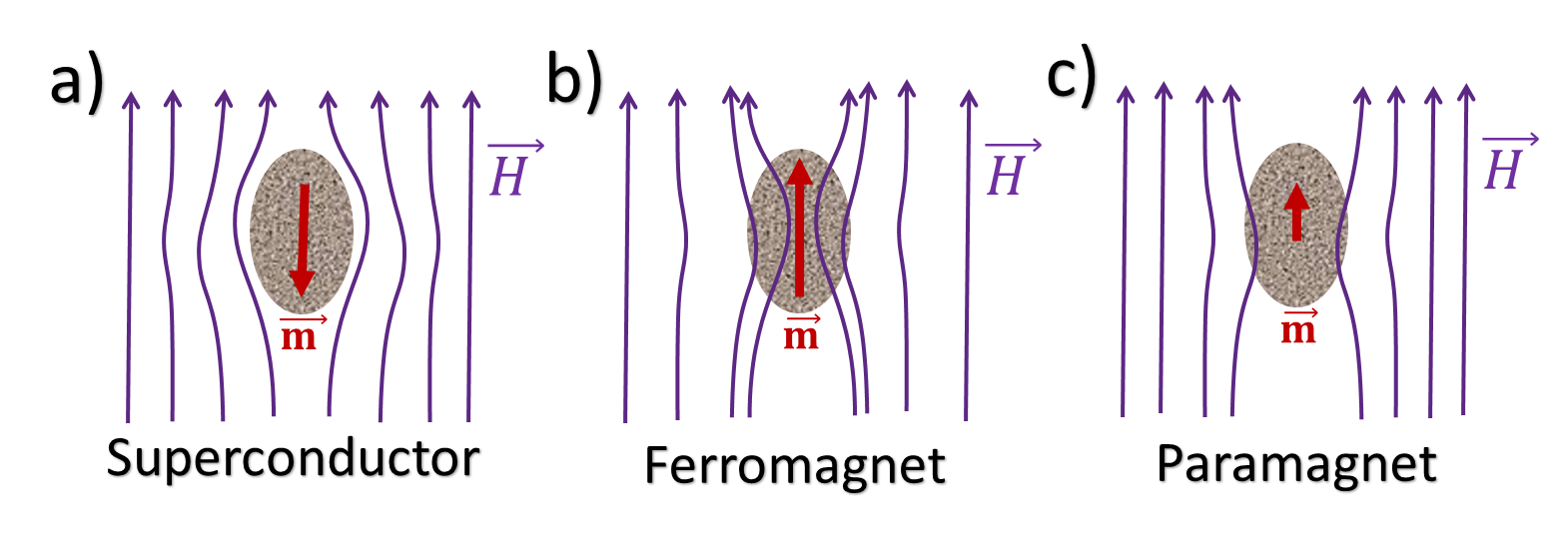}
        \caption{Ideal flux response to a small magnetic field for different samples: a) superconductor, b) ferromagnet, and c) paramagnet.}
        \label{fig:FluxResponseInMaterials}
    \end{figure}

    Magnetized samples can be characterized into three groups based on their response to an applied magnetic field (Fig.~\ref{fig:FluxResponseInMaterials}). A superconductor exhibits a diamagnetic response where $M = -H_0$, expelling the applied field from its interior so that $B = 0$ inside the bulk. A ferromagnet exhibits a strong positive magnetization $M \gg H_0$, greatly increasing the flux density inside the material. An insulator or weakly magnetic material exhibits a paramagnetic response where $M \approx 0$, producing a small positive change in the local flux density.

\begin{figure}[tb]
        \includegraphics[width=\textwidth]{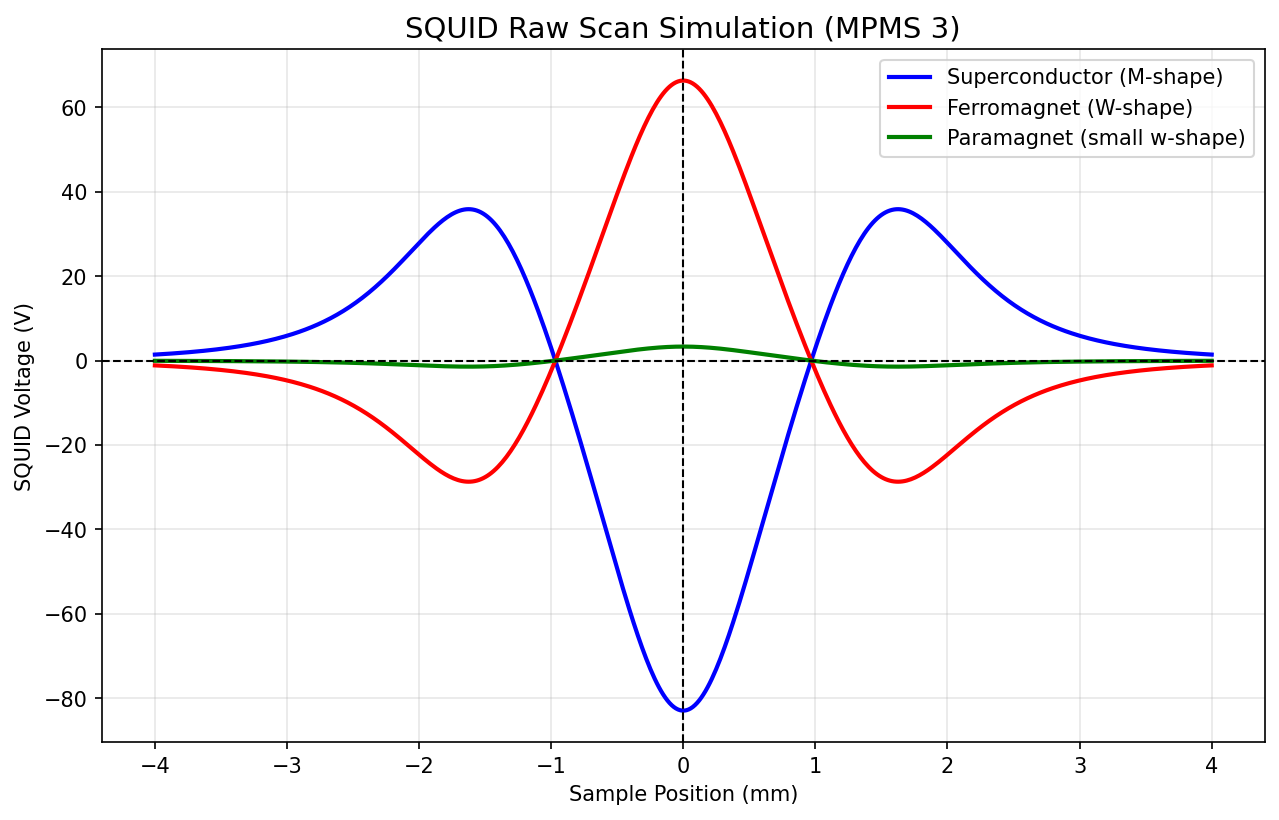}
        \caption{Simulated SQUID scans for superconductor, ferromagnet, and paramagnet, when coil is in -1, +2, -1 set up.}
        \label{fig:SQUIDScanSimulate}
    \end{figure}

    These different magnetic responses produce characteristic voltage curves as the sample translates through the gradiometer pickup coils (Fig.~\ref{fig:SQUIDScanSimulate}). For the counterwound coil configuration ($-1, +2, -1 $), a diamagnetic sample that reduces the local flux density produces a curve with a central negative peak. A ferromagnetic sample that increases the local flux density produces a curve with a central positive peak of larger amplitude. A paramagnetic sample produces a curve similar in shape to the ferromagnet but with much smaller amplitude, reflecting its weak magnetic response. The amplitude of the voltage curve is proportional to the sample's magnetic moment, and the software fits the curve to extract the magnetic moment.
    
    \subsubsection{Instrumentation}
    
    Temperature in the MPMS is controlled via helium gas flow through a copper heat exchanger with resistive heating elements, regulated by PID feedback. Magnetic fields are generated by a persistent-mode superconducting solenoid.
    
    \subsubsection{Measurement Protocol}
    
    The voltage-to-magnetic moment conversion is standardized using a palladium (Pd) sample. Since the Bohr magneton per atom at $T = 0$, density of Pd, mass of sample, and number of atoms $N = m N_A / A_w$ are known, the theoretical magnetic moment $m_{\text{theory}} = 0.6 \mu_B N$ can be calculated. The standard sample is compared to theory, and a calibration factor is applied. The Pd sample can achieve accuracy within 0.1\% for a point-dipole geometry~\cite{quantumdesigninc.MPMS3Users2016}. With this calibration, any future voltage measurement can be converted to a magnetic moment. Magnetic moment $m$ has units of A\,m$^{2}$, where 1~emu $= 10^{-3}$~A\,m$^{2}$. If the volume of the thin film is known, the total magnetization $M$ can be calculated given $m = \int M \, dV$. Often, the volume is unknown, so the normalized magnetic moment is plotted.
    
    The sample must be centered along the length of the superconducting coils so the magnetic moment can be measured appropriately. Zero-field-cooled (ZFC) measurements are the standard measurement technique for reducing trapped flux and maximizing magnetic moment. The trapped field in the superconducting magnet is minimized by oscillating the magnetic field from positive to negative values as it approaches zero, called degaussing. The residual field can be further reduced by measuring the sample's magnetic moment as a function of the magnetic field. While the sample is below $T_c$, the magnetic field is set to a small negative value ($-1$~Oe), and ramped to a positive value ($+1$~Oe) in small steps. The point at which the line crosses the x-axis indicates the actual zero magnetic field value experienced by the sample. This is typically between $-1$ and $+1$~Oe if degaussing is done correctly. This zero value is the new zero-field value and must be used as a correction for the rest of the measurement. If trapped flux is not removed from the sample, the sample exhibits a smaller magnetic moment and may transition faster depending on the impurity concentration.
    
    For thin-film samples, demagnetization effects can be minimized by applying the field parallel to the thin-film surface. Specific measurement protocols are detailed in Appendix~\ref{appsec:SQUIDProtocol}.
    
    \subsection{Superconductor Response}
    
    \subsubsection{Bulk Samples}
    
    When a superconductor perfectly expels a magnetic field from the bulk ($B = 0$), as in the Meissner state, the magnetization $M \approx -H_0$. This gives the maximum negative magnetic moment, $m = -1$ when normalized. Even when the superconductor is not perfectly in the Meissner state, the sample exhibits a diamagnetic response. The screening currents, as discussed in Sec.~\ref{sec:superconductor}, are induced perpendicular to the external magnetic field to expel the magnetic field from the bulk, setting $B = 0$. If the sample thickness is much larger than the penetration depth, $d \gg \lambda$, the superconductor perfectly screens the external magnetic field, and the magnetic moment is maximized. As temperature increases, $\lambda$ gets larger, but as long as $\lambda \ll d$, perfect diamagnetic response continues.
    
\begin{figure}[H]
        \includegraphics[width=\textwidth]{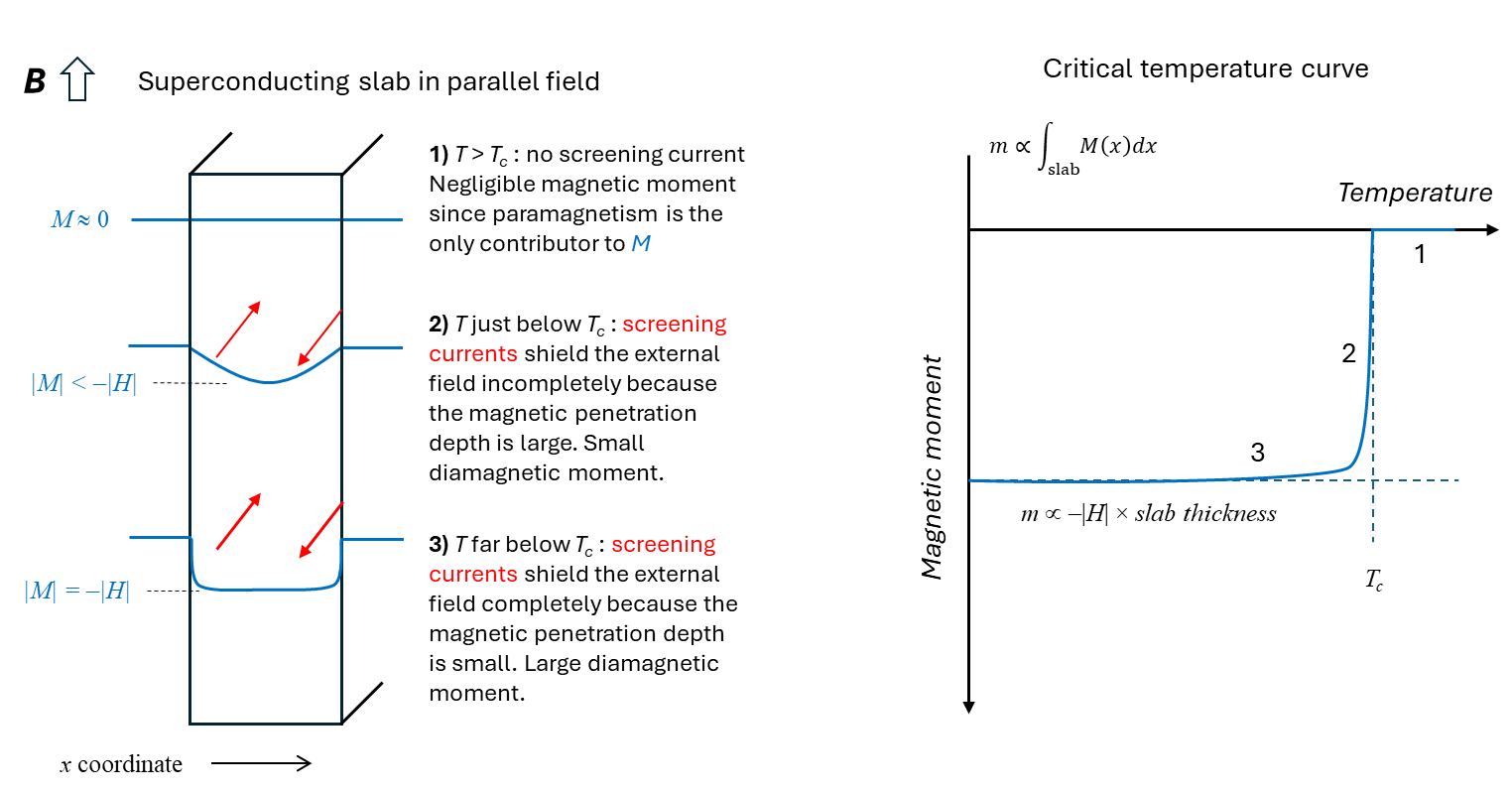}
        \caption{Superconducting slab in parallel field with example critical temperature curve}
        \label{fig:MomentvsTempLance}
    \end{figure}

    As the temperature increases, the areas of the sample with poor superconducting properties transition from diamagnetic (large negative $M$) to paramagnetic (small positive $M$), when $T>T_c$. This shift from superconducting to non-superconducting occurs very rapidly in uniform samples and very gradually in non-uniform films. This transition continues until the entire sample becomes diamagnetic, and all areas have reached their respective critical temperatures. A schematic explaining the change in magnetic moment with temperature and the resulting critical temperature curve is given in Fig.~\ref{fig:MomentvsTempLance}. If the sample is divided by cracks or has polluted grain boundaries, the magnetic field can penetrate intergranularly, thereby reducing the magnetic moment. 

    \subsubsection{Thin Films}
    
\begin{figure}[tb]
        \includegraphics[width=\textwidth]{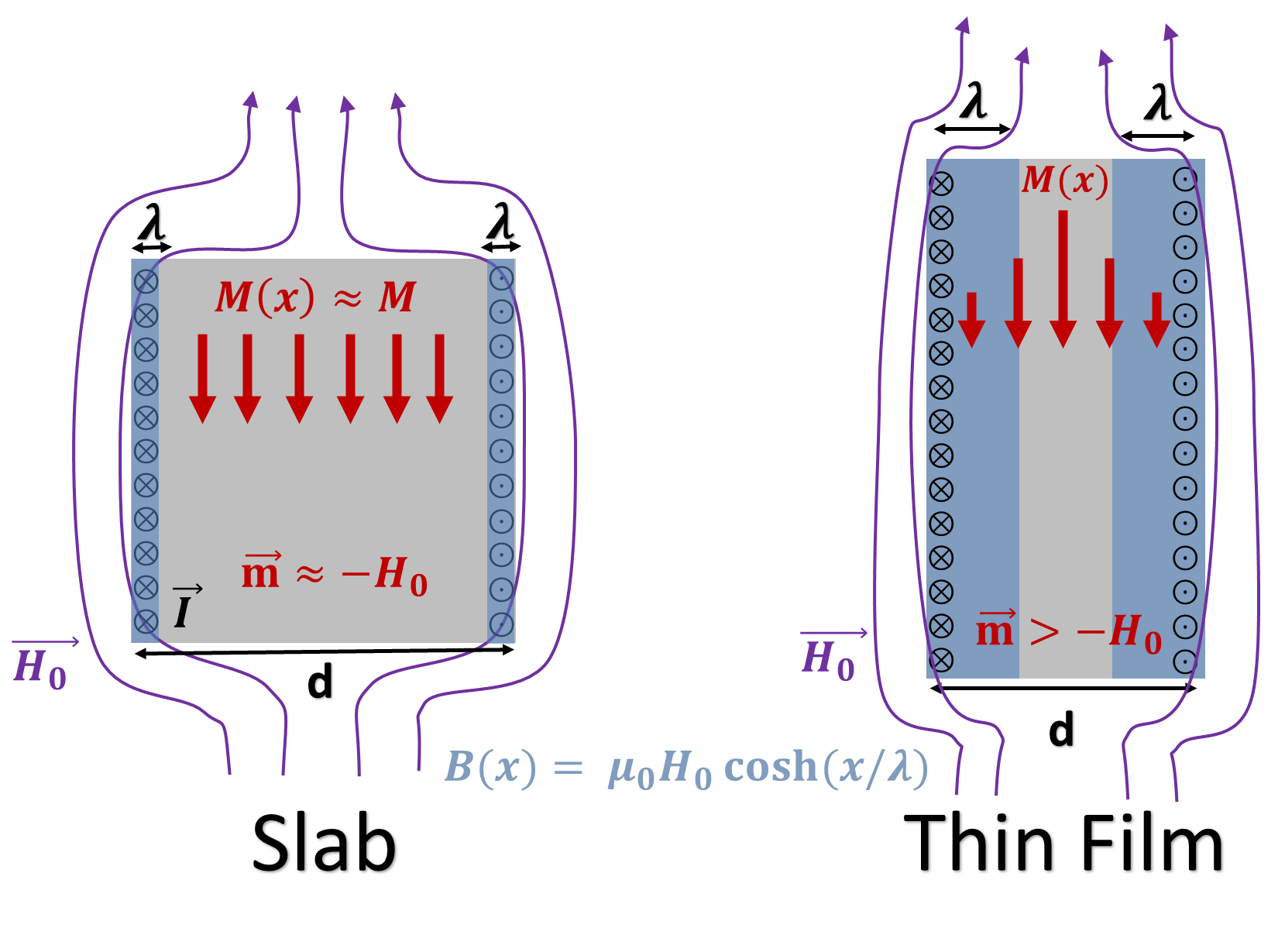}
        \caption{Magnetization of large samples where $d \gg \lambda$ and thin films where $d \sim \lambda$.}
        \label{fig:MagnetizationofDifferentGeometries}
    \end{figure}

    If the superconducting material has a thickness on the same order as the penetration depth, $d \sim \lambda$, the magnetic field cannot be perfectly screened. The material is still diamagnetic but has a smaller magnetic moment $|M| < H_0$ (seen in Fig.~\ref{fig:MagnetizationofDifferentGeometries}). As the temperature increases, more and more of the material is ineffectively screened even if the sample is perfectly uniform, resulting in a decreasing magnetic moment. This appears as a transition width even in uniform films.
    
    For example, consider Nb$_3$Sn with $\lambda \approx 100$~nm. In a 1~mm-thick bulk sample, the field penetrates only $\sim$200~nm total (100~nm from each surface parallel to external field), representing 0.02\% of the volume, the interior is fully screened and the sample exhibits near-ideal diamagnetism. In a 300~nm thin film, however, the field penetrates approximately 2/3 of the thickness, significantly reducing the diamagnetic moment. The reduced moment arises not only from the smaller volume but also from incomplete screening throughout the film.

\begin{figure}[H]
        \centering
        \includegraphics[width=\textwidth]{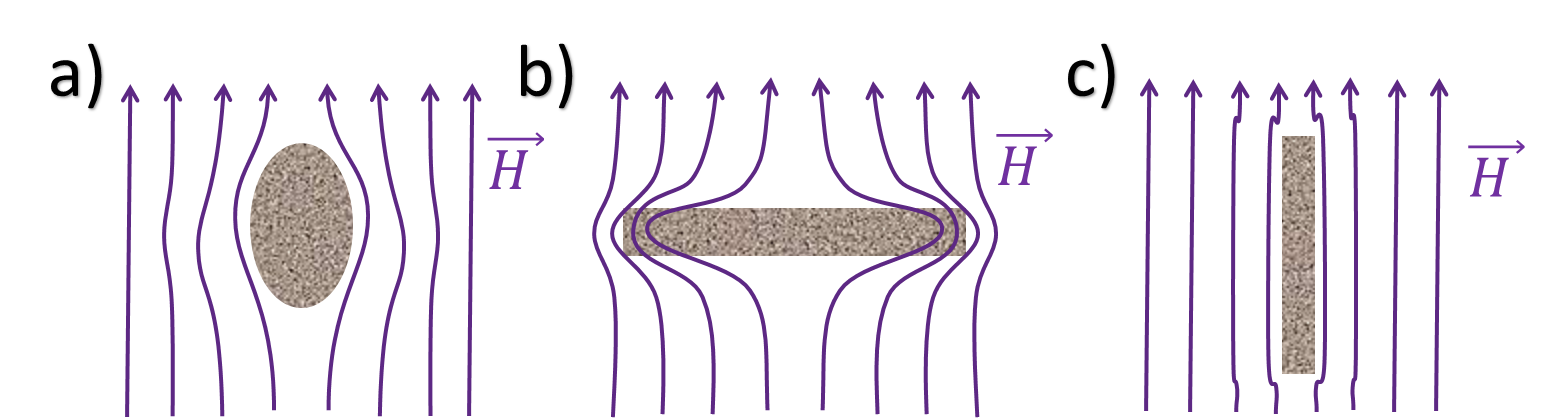}
        \caption[Demagnetization affecting magnetic field penetration at the edges for (a) spheroid and (b,c) thin film geometries.]{Demagnetization affecting magnetic field penetration at the edges for (a) spheroid and (b,c) thin film geometries. Perpendicular field orientation (b) maximizes edge effects, while parallel field orientation (c) minimizes edge effects in thin films. Parallel field orientation was used for all measurements.}
        \label{fig:Demag}
    \end{figure}

    Sample geometry also affects the apparent critical field. Flux lines bend around a superconductor and concentrate at edges and corners, increasing the local field relative to the applied field. This demagnetization effect causes flux to penetrate edges at lower applied fields than would penetrate a flat parallel surface. A prolate spheroid with its long axis parallel to the field minimizes this effect, yielding nearly uniform flux penetration (Fig.~\ref{fig:Demag}a). A thin film oriented perpendicular to the field maximizes edge effects, causing early flux penetration at the edges (Fig.~\ref{fig:Demag}b). Orienting the film parallel to the field significantly reduces demagnetization (Fig.~\ref{fig:Demag}c); this geometry was used for all measurements in this work. Nevertheless, demagnetization complicates accurate determination of the lower critical field $H_{c1}$.
    
    \subsubsection{Interpretation of $T_c$ Curves}
    
\begin{figure}[H]
        \centering
        \includegraphics[width=\textwidth]{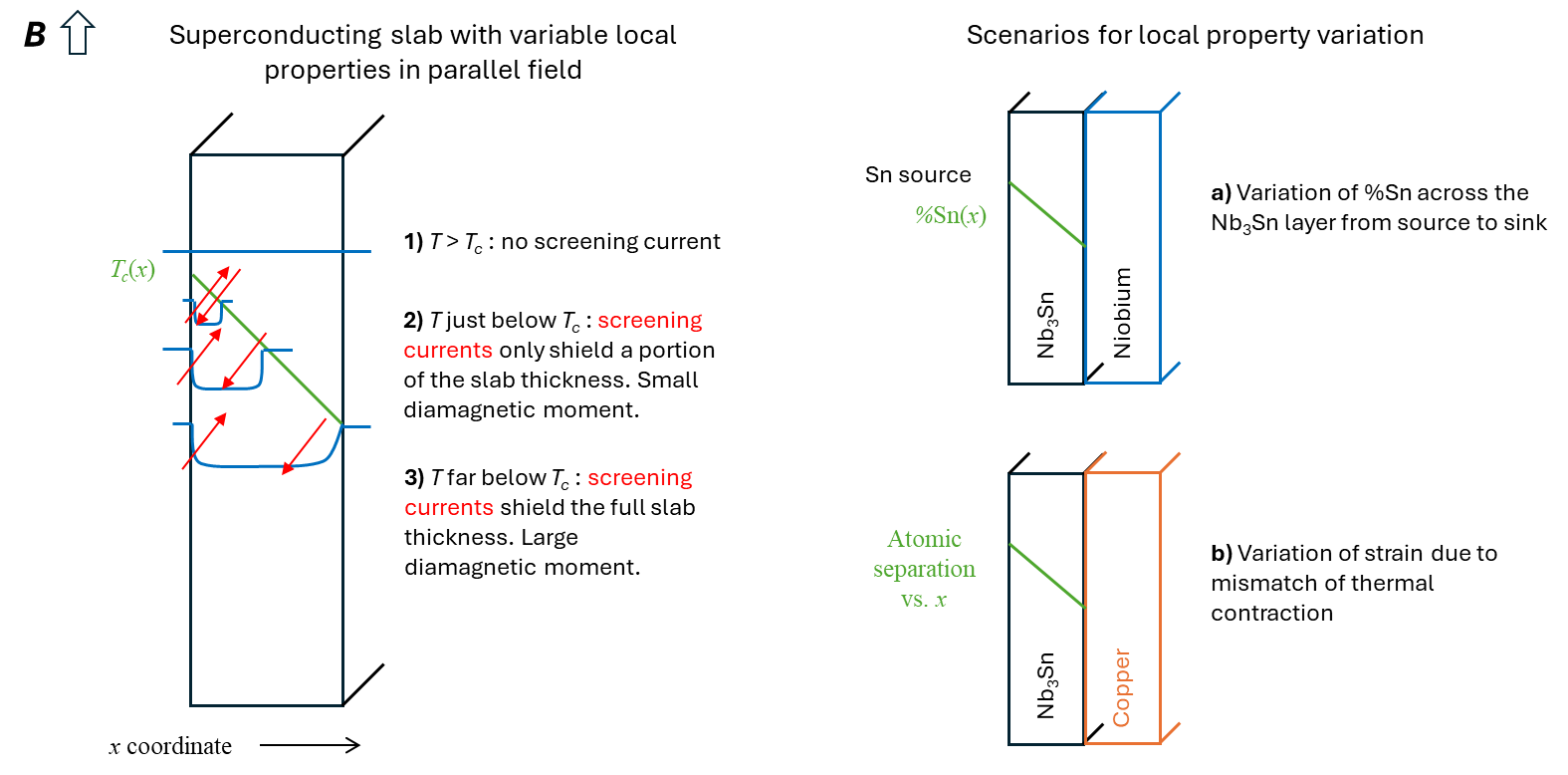}
        \caption{Schematic of superconducting slab with variable Sn or strain properties}
        \label{fig:SnandStrainCriticalTempLance}
    \end{figure}

    The $T_c$ transition provides diagnostic information about film quality. As the sample is heated from low temperature (2~K), regions with lower critical temperatures transition to the normal state, while higher-quality regions transition to normal state at temperatures close to maximum $T_c$. The critical temperature of Nb$_3$Sn depends on local variations in strain and Sn content, with zero strain and stoichiometric composition giving the highest $T_c$ (see Sec.~\ref{sec:Tceffectch5}). Film properties vary along the thickness. The transition width of the critical temperature curve reflects this variation, while the $T_c$ onset indicates the strain and Sn content in the highest-quality regions. High-$T_c$ regions can magnetically shield low-$T_c$ regions, complicating interpretation (see Sec.~\ref{sec:SapphireExp}). This effect is illustrated schematically in Fig.~\ref{fig:SnandStrainCriticalTempLance}. As temperature increases, less of the sample becomes diamagnetically screened, uncovering the magnetic response from low-Sn and high-strain regions.

\begin{figure}[H]
        \includegraphics[width=\textwidth]{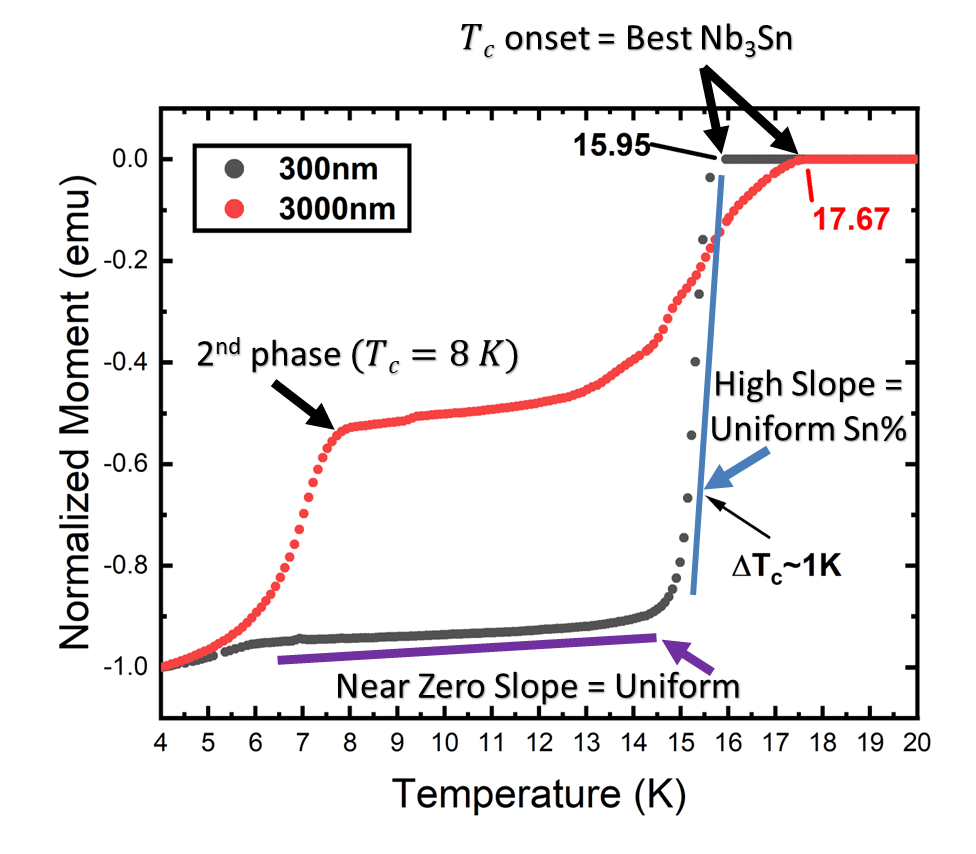}
        \caption[Critical temperature curves for two Nb$_3$Sn samples on Cu substrates.]{Critical temperature curves for two Nb$_3$Sn samples on Cu substrates. The black curve exhibits a sharp transition ($\Delta T_c \approx 0.6$~K) with $T_c$ onset at $\approx 16$~K. The red curve shows a broader transition with a higher $T_c$ onset. Secondary transitions at 6~K (black) and 8~K (red) indicate additional phases.}
        \label{fig:TcExplanation}
    \end{figure}

    Figure~\ref{fig:TcExplanation} shows $T_c$ curves for two Nb$_3$Sn films on Cu substrates. The black curve exhibits a sharp transition ($\Delta T_c \approx 0.6$~K), indicating high compositional uniformity, while the red curve shows a broader transition due to a wider range of Sn concentration. The higher $T_c$ onset in the red curve is attributed to strain relaxation in the thicker film. Secondary transitions at 6~K and 8~K indicate additional phases, likely Sn-rich intermetallics or unreacted Nb. Because of the interrelated effects of strain and composition on the critical temperature, we rely on theory and other experimental methods to further characterize the sample. However, this measurement provides rapid screening of superconducting film quality before proceeding to more time-intensive characterization.
    
\section{RF Characterization of Pillbox Cavities}
    
    \subsection{Measurement Principles and Instrument Configuration}
    To characterize the RF properties of copper and determine its surface resistance, the material is fabricated into a cylindrical "pillbox" cavity. This common geometry is often used because it can be solved analytically easily, enabling comparison among simulation, theory, and experiment.
    
    The fundamental instrument for these measurements is the Vector Network Analyzer (VNA). The VNA serves as a phase-coherent source and receiver, facilitating S-parameter measurements. In a two-port configuration:
    \begin{itemize}
        \item \textbf{Transmission ($S_{21}$):} Measures power delivered from Port 1 to Port 2 through the cavity.
        \item \textbf{Reflection ($S_{11}$):} Measures power reflected from Port 1 back to Port 1.
    \end{itemize}
    
\begin{figure}[H]
        \centering
        \includegraphics[width=0.9\linewidth]{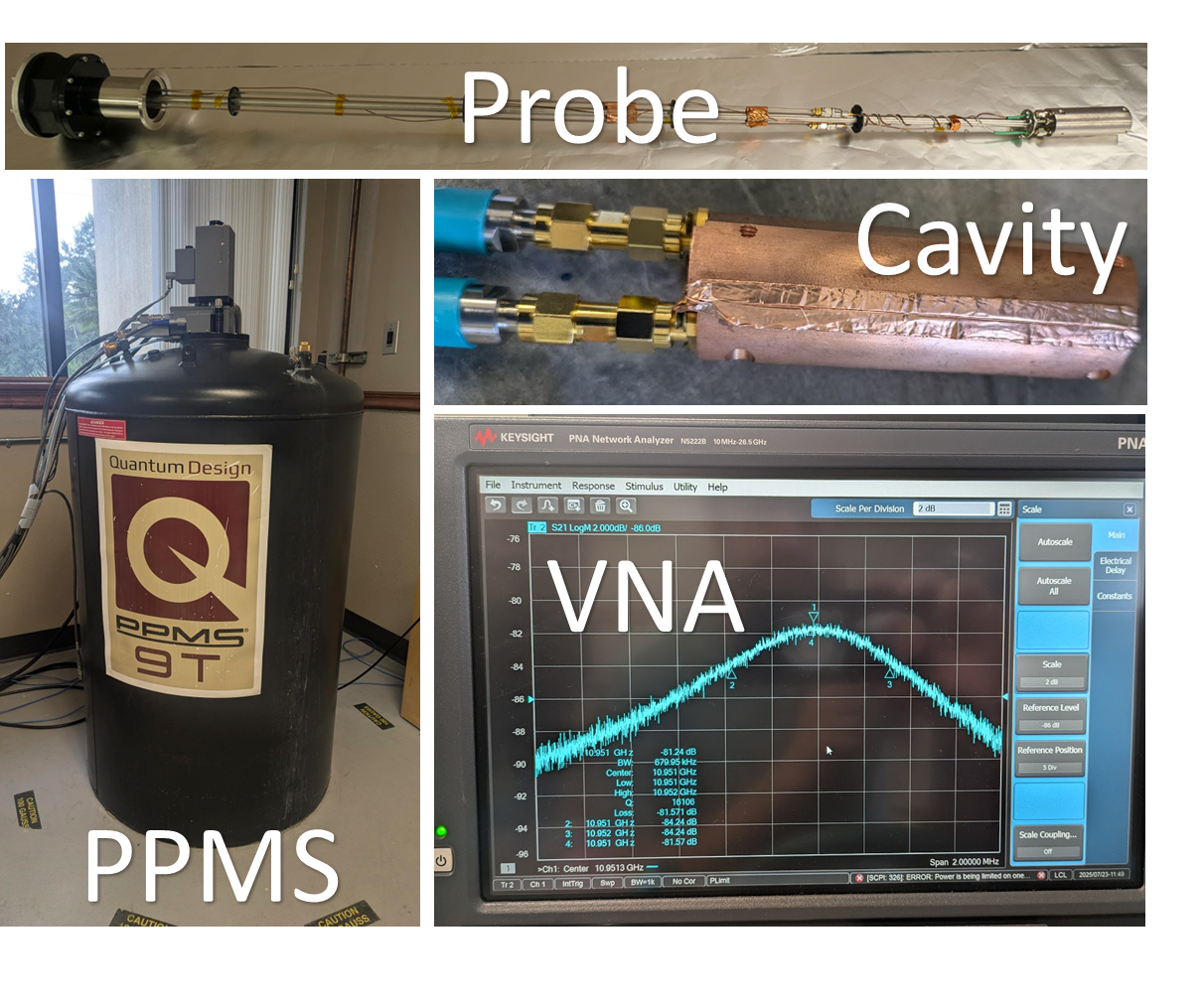}
        \caption{Experimental apparatuses needed for Q measurement: probe~\cite{Braine:2024nzi}, PPMS magnet, cavity, and VNA.}
        \label{fig:ProbePPMSVNACavity}
    \end{figure}

    These S-parameters are frequency dependent. When a circuit component (cavity) has a resonant frequency equal to the stimulating AC frequency, the transmitted signal is resonantly enhanced. This signal peak occurs because the power is resonantly enhanced by the cavity walls when the input AC frequency is in resonance with the cavity mode. The frequency locations of the resonance modes for the cavity geometry are already known using simulation (Sec.~\ref{sec:cavitydesign}). The VNA is set to these frequencies to check for resonant behavior. The measurement circuit is impedance-matched to 50~$\Omega$. While the coaxial shields are grounded via the VNA chassis, the cavity enclosure must be connected to a common ground to suppress external electromagnetic interference. The PPMS, probe design, VNA, and cavity used are shown in Fig.~\ref{fig:ProbePPMSVNACavity}.

    When transitioning from room temperature to cryogenic environments, the conductivity of copper increases, leading to a significant rise in $Q_0$. Lowering the temperature also reduces skin depth, allowing current to navigate through even smaller gaps. However, at 10~GHz, the surface resistance reaches a limit defined by the Anomalous Skin Effect, where the electron mean free path exceeds the skin depth. Furthermore, thermal contraction of the copper reduces the cavity dimensions, shifting $f_0$ to higher frequencies. These shifts can be tracked in real-time.
    
    \subsection{Determining the Quality Factor of an Ideal Cavity}
    
    The Quality Factor ($Q$) represents the ratio of energy stored in the cavity to the energy dissipated per cycle. Three $Q$ values can be distinguished: the loaded ($Q_L$), the external ($Q_{ext}$), and the intrinsic ($Q_0$).
    
    \subsubsection{Loaded Quality factor $Q_L$}
    $Q_L$ is extracted from the $S_{21}$ transmission peak by measuring the resonance frequency ($f_0$) and the 3~dB full-width at half-maximum ($\Delta f_{3dB}$):
    \begin{equation}
        Q_L = \frac{f_0}{\Delta f_{3dB}}
        \label{eq:QVNA_fixed}
    \end{equation}
    \noindent A narrower peak signifies a higher $Q$, indicating lower surface resistance in the copper walls.
    
    \subsubsection{Intrinsic $Q_0$}\label{sec:Qantenna}
    The antennas used to probe the cavity introduce their own losses and must be accounted for to isolate the material-dependent intrinsic quality factor ($Q_0$) using the coupling coefficients $\beta_1$ and $\beta_2$:
    
    \begin{equation}
        Q_0 = Q_L (1 + \beta_1 + \beta_2)
        \label{eq:Q0_calculation}
    \end{equation}
    
    \noindent The coupling coefficients can be calculated from:
    
    \begin{equation}
        \beta = \frac{1 + \text{sign}(\angle\Gamma(f_0) - \pi)|\Gamma(f_0)|}{1 - \text{sign}(\angle\Gamma(f_0) - \pi)|\Gamma(f_0)|}~.
        \label{eq:beta}
    \end{equation}
    
    \noindent where $|\Gamma|$ the reflection coefficient and its phase $\angle\Gamma$ is another name for the S-parameter. The sign function indicates whether the antenna is over-coupled or under-coupled to the cavity mode, as determined by the phase of the reflection coefficient relative to $\pi$ radians. Practically, $\beta$ is a measure of the depth of the dip from the reflection measurements on resonance. 
    
    In this work, "weak coupling" is used, in which antenna penetration is minimized until the reflection dips are $<1$~dB. The antenna penetration depth can be adjusted by trimming or pulling the antennas out of the cavity. Strong coupling (deep antenna penetration) heavily loads the cavity, presenting a low impedance that reflects significant power and degrades $Q$. Weak coupling (shallow penetration) minimally perturbs the cavity's resonance while maintaining sufficient signal for measurement. When the reflection dip is less than 1~dB, the antenna interacts minimally with the cavity, and the reduction in the antenna's quality factor is also minimal (<5\% loss in $Q$ per antenna). For example, if $Q_L$ is measured to be 100,000, and reflection dips $S_{11}$ and $S_{22}$ are about 1~dB deep on resonance, $Q_0$ will be about 110,000. For initial prototyping, a 10\% reduction in $Q$ is considered negligible, so, for simplicity, we set $Q_L = Q_0$.
    
        \begin{figure}[H]
            \centering
            \includegraphics[width=.75\textwidth]{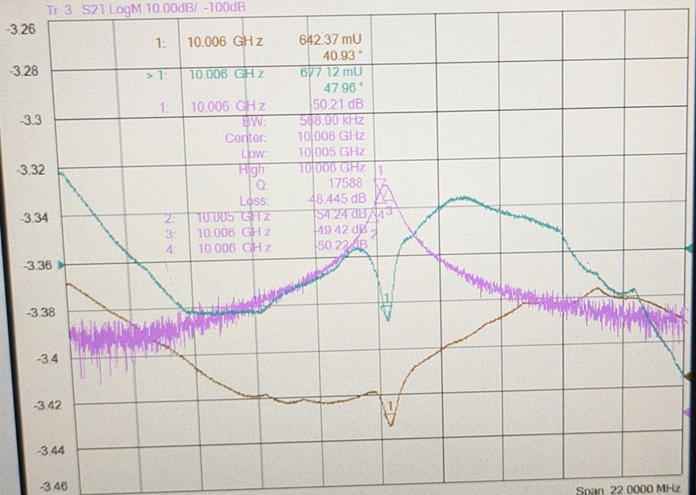}
            \caption[Example vector network analyzer scan for a specific mode at 10 GHz.]{Example vector network analyzer scan for a specific mode at 10 GHz. The purple curve shows the transmission measurement $S_{21}$ or $S_{12}$, and the quality factor can be read from the 3 dB full-width at half-maximum of the peak. The blue and yellow curves show the reflection measurements $S_{11}$ and $S_{22}$, and the coupling from the antenna to the cavity can be inferred from the depth of the dip. A dip of less than 1 dB from background to trough is adequate.}
            \label{fig:CouplingExample}
        \end{figure}
        
        It should be noted that different modes will couple differently to the antenna. For example, specific modes may isolate the current away from the antenna, leading to reduced coupling.

    \subsection{Quality Factor Correction for Hybrid Material Cavities with RF Leakage}\label{sec:QHybrid}

    When characterizing a superconducting coating deposited on only a portion of a copper cavity, the measured quality factor reflects contributions from both materials. Additionally, multi-piece cavity assemblies exhibit RF leakage at mechanical seams. To extract the intrinsic surface resistance of the superconducting material, corrections must be applied for both effects.

    \subsubsection{General Framework}
    
    For a resonant cavity, the quality factor relates surface resistance to geometric properties:
    \begin{equation}
        Q_0 = \frac{G}{R_s}
        \label{eq:Q_basic}
    \end{equation}
    where $G$ is the geometric factor (units of $\Omega$) and $R_s$ is the surface resistance (units of $\Omega$). The geometric factor encapsulates how the electromagnetic field distribution weights different surfaces:
    \begin{equation}
        G_i = \frac{\int_V |\mu \vec{H}|^2 \, dV}{\int_{S_i} |\vec{H}_\text{tan}|^2 \, dS}
        \label{eq:G_definition}
    \end{equation}
    where the numerator represents the magnetic energy stored in the cavity volume and the denominator is the integral of the tangential magnetic field squared over the surface $i$.
    
    For a cavity with multiple surfaces, losses add in parallel:
    \begin{equation}
        \frac{1}{Q_\text{total}} = \sum_{i} \frac{R_{s,i}}{G_i}
        \label{eq:Q_multisurf}
    \end{equation}
    Surfaces sharing the same material can be grouped, with effective geometric factors calculated as:
    \begin{equation}
        \frac{1}{G_\text{group}} = \sum_{i \in \text{group}} \frac{1}{G_i}
        \label{eq:G_grouped}
    \end{equation}
    
    \subsubsection{Multi-Material Correction}
    
    Consider a hybrid cavity with copper on most surfaces and Nb$_3$Sn on a single wall. The measured quality factor combines losses from both materials:
    \begin{equation}
        \frac{1}{Q_\text{measured}} = \frac{R_{s,\text{Cu}}}{G_\text{Cu}} + \frac{R_{s,\text{Nb}_3\text{Sn}}}{G_{\text{Nb}_3\text{Sn}}}
        \label{eq:Q_hybrid}
    \end{equation}
    where $G_\text{Cu}$ and $G_{\text{Nb}_3\text{Sn}}$ are the effective geometric factors for the copper and Nb$_3$Sn surfaces, respectively.
    
    For the hexagonal pillbox cavity used in this work, finite element simulation (COMSOL Multiphysics) yields the following geometric factors for the TM$_{010}$ mode:
    \begin{itemize}
        \item Single Nb$_3$Sn-coated wall: $G_{\text{Nb}_3\text{Sn}} = 2647~\Omega$
        \item Remaining copper surfaces (five walls and two endcaps): $G_\text{Cu} = 411~\Omega$
        \item Total geometric factor for uniform material: $G_\text{total} = 356~\Omega$
    \end{itemize}
    
    The copper surface resistance is determined from a separate all-copper reference cavity measurement:
    \begin{equation}
        R_{s,\text{Cu}} = \frac{G_\text{total}}{Q_\text{Cu,measured}}
        \label{eq:Rs_Cu}
    \end{equation}
    
    With $R_{s,\text{Cu}}$ known, the Nb$_3$Sn surface resistance can be extracted:
    \begin{equation}
        R_{s,\text{Nb}_3\text{Sn}} = G_{\text{Nb}_3\text{Sn}} \left(\frac{1}{Q_\text{measured}} - \frac{R_{s,\text{Cu}}}{G_\text{Cu}}\right)
        \label{eq:Rs_Nb3Sn}
    \end{equation}
    
    The quality factor for a hypothetical fully-coated Nb$_3$Sn cavity is then:
    \begin{equation}
        Q_{\text{Nb}_3\text{Sn,full}} = \frac{G_\text{total}}{R_{s,\text{Nb}_3\text{Sn}}}
        \label{eq:Q_full}
    \end{equation}
    
    \subsubsection{RF Leakage Correction}\label{sec:QRFleakage}

    At 10~GHz ($\lambda \approx 3$~cm), the skin depth ($\delta$) of copper is approximately 0.6~$\mu$m at 300~K (0.2~$\mu$m at 4~K). Any physical gap in the cavity assembly, even as small as 0.1~mm (approximately one sheet of paper), disrupts the surface current paths and allows radiative leakage. Because this leaked energy is indistinguishable from internal ohmic dissipation, the measured $Q$ is artificially suppressed. Effective characterization requires ensuring that all seams are much smaller than $\lambda/50$ ($\approx 0.6$~mm) or sealed with conductive gaskets.

    $\alpha$ is a correction factor that quantifies this RF leakage:
    \begin{equation}
        \alpha = \frac{Q_\text{ideal}}{Q_\text{measured}}
        \label{eq:alpha_def}
    \end{equation}
    where $Q_\text{ideal}$ is obtained from simulation assuming perfect electrical contact and $Q_\text{measured}$ is the experimental value including seam losses.
    
    This factor is determined using the copper reference cavity:
    \begin{equation}
        \alpha = \frac{Q_\text{Cu,ideal}}{Q_\text{Cu,measured}}
        \label{eq:alpha_Cu}
    \end{equation}
    
    Assuming the same fractional leakage applies to all measurements taken with the same assembly procedure, the corrected Nb$_3$Sn quality factor is:
    \begin{equation}
        Q_{\text{Nb}_3\text{Sn,corrected}} = \alpha \times Q_{\text{Nb}_3\text{Sn,full}}
        \label{eq:Q_corrected}
    \end{equation}
    
    Together, the multi-material and RF leakage corrections enable the extraction of the intrinsic quality factor of the superconducting coating, representing the expected performance of a hypothetical cavity fully coated with Nb$_3$Sn and free of seam losses. The complete correction procedure is detailed in Appendix~\ref{app:Qcorrection}.

\chapter{\texorpdfstring{Nb$_3$Sn}{Nb3Sn} for High-Field Applications}\label{sec:whynb3sn}

As discussed earlier, axion detection experiments could achieve faster scanning of the exploration space by using superconducting RF cavities operating in multi-tesla magnetic fields. This chapter examines the merits of  Nb$_3$Sn for this application.

\section{\texorpdfstring{Nb$_3$Sn}{Nb3Sn} vs Other Superconducting Materials}

Nb$_3$Sn is the most technologically mature high-field superconductor other than NbTi alloy. Among strongly type~II superconductors (NbTi, Nb$_3$Sn, REBCO, BSCCO), Nb$_3$Sn offers the optimal balance of properties for RF cavity applications~\cite{gurevichUseNotUse2011}. Its high critical temperature and upper critical field ($T_c = 18.3$~K and $H_{c2} = 30$~T) exceed NbTi's capabilities, while its larger coherence length ($\xi_0 \sim 3$~nm) compared to high-temperature superconductors such as REBCO ($\xi_0\sim 1{-}2$~nm), reduce grain boundary resistance~\cite{gurevichUseNotUse2011}. When the coherence length ($\xi_0$) becomes smaller than typical grain boundary widths ($\sim3{-}5$~nm), supercurrent cannot cross grain boundaries coherently, causing severe weak-link behavior and increased dissipation~\cite{alexgurevich33ThermalRF2006}. Nb$_3$Sn's coherence length ($\xi_0 \approx$ GB width) makes grain boundary quality important, but it is potentially manageable with proper processing. In contrast, REBCO's much smaller $\xi_0 \ll$ GB width means grain boundaries act as intrinsic weak links regardless of processing. High-temperature superconductors also exhibit strong crystalline anisotropy, leading to properties varying dramatically with crystal orientation. This may lead to degraded properties for unfavorable orientations of the magnetic field. Table~\ref{tab:SC_comparison} summarizes key superconductor properties of relevant superconducting materials.

\begin{table}[ht]
    \centering
    \caption{Comparison of superconductor properties relevant for high-field RF applications}
    \begin{tabular}{lcccc}
    \hline
    Material & Type & $T_c$ (K) & $H_{c2}$ (0 K, T) & $\xi_0$ (nm) \\
    \hline
    Nb & LTS & 9.2 & 0.4 & 40 \\
    NbTi & LTS & 9.1 & 11.3 & 5.2 \\
    Nb$_3$Sn & LTS & 18.3 & 30 & 3 \\
    REBCO$^*$ & HTS & 93 & $>$100 & $\sim$1.5$^{**}$ \\
    BSCCO$^*$ & HTS & 110 & $>$100 & $\sim$2$^{**}$ \\
    \hline
    \multicolumn{5}{l}{\small $^*$REBCO = REBa$_2$Cu$_3$O$_{7-x}$ (RE = rare earth element)} \\
    \multicolumn{5}{l}{\small \hspace{0.5em} BSCCO = Bi$_2$Sr$_2$Ca$_2$Cu$_3$O$_{14+x}$} \\
    \multicolumn{5}{l}{\small $^{**}$In-plane coherence length; c-axis values are 3-10× smaller}
     \end{tabular}
    \label{tab:SC_comparison}
\end{table}

Nb$_3$Sn benefits from decades of optimization by the high-field magnet community, with thousands of established magnets delivered to customers~\cite{cooleyBusinessModelsAssure2023}. The accelerator community is also pursuing Nb$_3$Sn cavities to surpass Nb's fundamental accelerating gradient limit of 50~MV/m. However, as an intermetallic compound rather than a ductile metal or alloy, Nb$_3$Sn requires fundamentally different fabrication approaches than those used for most cavity resonators. Traditional metal-forming and welding techniques cannot be applied to intermetallic structures because these materials are brittle and prone to cracking. A wide range of opportunities for fabricating axion detector cavities is afforded by thin-film deposition onto suitable substrates. Similar opportunities exist for other high-field intermetallic superconductors.

\section{Fundamental Material Properties of \texorpdfstring{Nb$_3$Sn}{Nb3Sn}}
 
    \subsection{Crystal Structure (A15)}

\begin{figure}[H]   
        \centering
        \includegraphics[width=.5\textwidth]{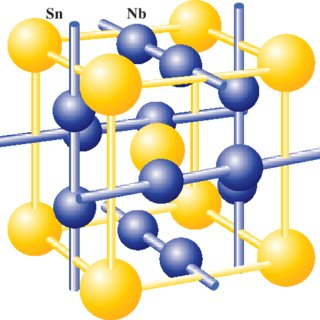}
        \caption{Arrangement of Nb and Sn atoms in Nb$_3$Sn crystal structure~\cite{godekeReviewPropertiesNb3Sn2006}.}
        \label{fig:CrystalStructure}
    \end{figure}

    The stoichiometry of Nb$_3$Sn is 3:1 (Nb:Sn), corresponding to 75 at.\% Nb and 25 at.\% Sn. It is part of a class of materials called A15 compounds. All of these materials have a chemical composition of A${_3}$B, where A is a transition metal, and B is any element. Many of these materials have surprisingly good superconducting properties, with high $T_c$ and $H_{c2}$. The Nb atoms couple up in twos, forming linear chains along the alternating faces of a cubic unit cell. The Sn atoms occupy a body-centered cubic (BCC) position, as shown in Fig.~\ref{fig:CrystalStructure}. These aligned Nb chains facilitate electron transport between unit cells. This cubic unit cell leads to isotropic superconducting properties, unlike HTS materials. 
    
    
    \subsection{Nb-Sn Phase Diagram}
  
\begin{figure}[H]   
        \centering
        \includegraphics[width=.7\textwidth]{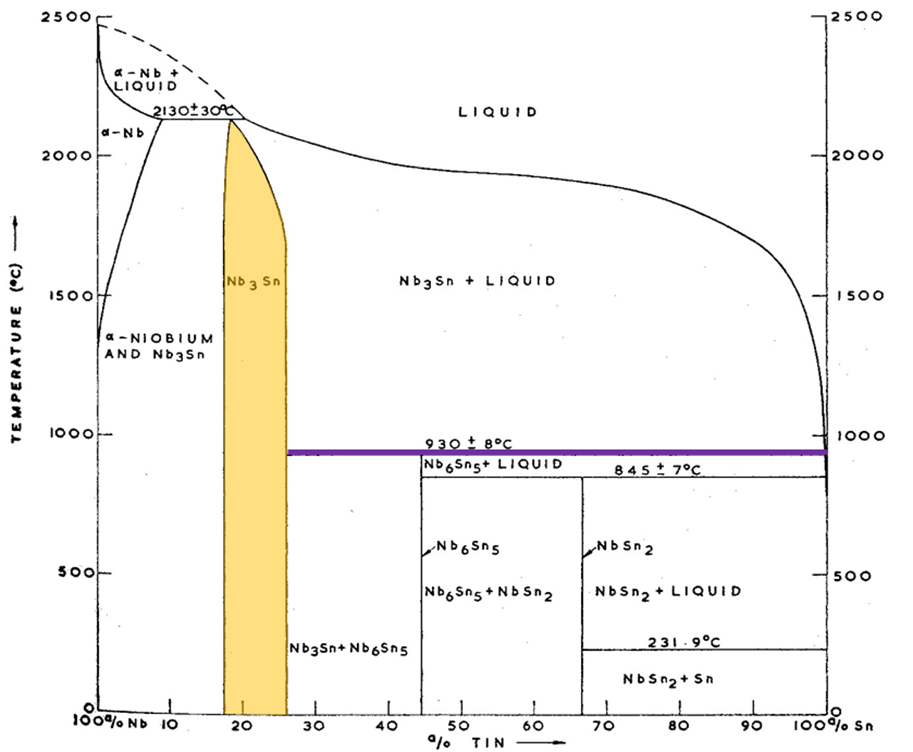}
        \caption{Nb-Sn Phase diagram, where the Nb$_3$Sn compound is highlighted in yellow, and above the purple line shows where Nb$_6$Sn$_5$ and NbSn$_2$ break down~\cite{charlesworthExperimentalWorkNiobiumtin1970}.}
        \label{fig:NbSnphasediagram}
    \end{figure}

    The Nb$_3$Sn compound can form with a varying degree of Sn\%, from 16 to 26~at.\% Sn, where 25~at.\% Sn is the most desirable for superconducting properties~\cite{devantayPhysicalStructuralProperties1981}. There are other Nb-Sn compounds, Nb$_6$Sn$_5$ and NbSn$_2$, with poor superconducting properties ($T_c = 2.8$ K, and $T_c = 2.68$ K respectively). These compounds can form when reacting Nb and Sn at temperatures below 930~$^\circ$C (Fig.~\ref{fig:NbSnphasediagram})~\cite{charlesworthExperimentalWorkNiobiumtin1970}.

    \FloatBarrier
    
    \subsection{Nb-Cu-Sn Reactions}\label{sec:NbCuSn}

\begin{figure}[htb]         
        \centering
        \includegraphics[width=.6\textwidth]{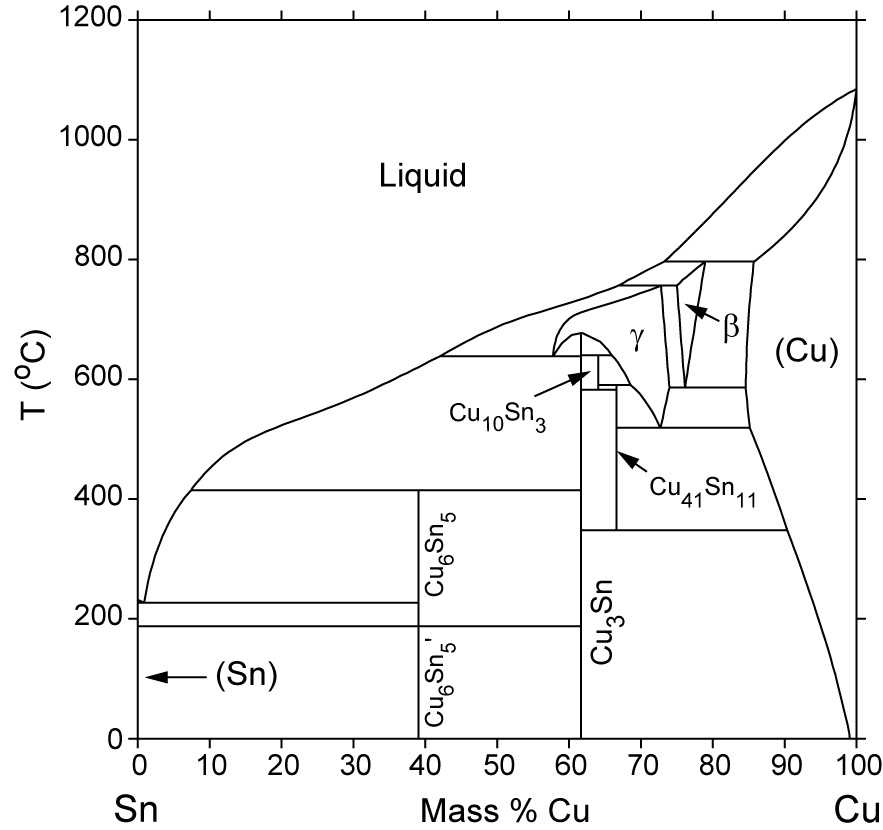}
        \caption{The binary Cu-Sn phase diagram~\cite{shimThermodynamicAssessmentCuSn1996}.}
        \label{fig:CuSnPhasediagram}
    \end{figure}

\begin{figure}[htb]          
        \centering
        \includegraphics[width=.6\textwidth]{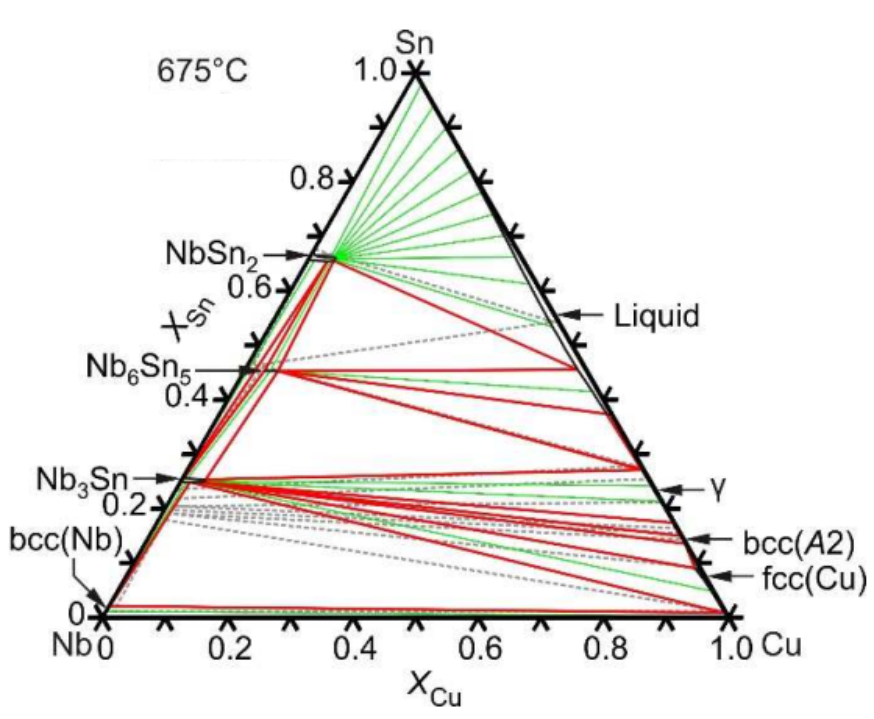}
        \caption{Ternary Cu-Nb-Sn phase diagram at 675 $^{\circ}$C isothermal~\cite{neijmeijerTernarySystemNb1987, lachmannThermodynamicRemodellingCu2022}.}
        \label{fig:CuSnNbTernaryPhaseDiagram}
    \end{figure}

    Adding Cu to a Nb-Sn reaction enables Nb$_3$Sn formation at lower temperatures (650--750~°C instead of $>$930~°C). Understanding this ternary reaction requires examining both the Cu-Sn binary system (Fig.~\ref{fig:CuSnPhasediagram}) and the Nb-Cu-Sn ternary diagram (Fig.~\ref{fig:CuSnNbTernaryPhaseDiagram}). Because Cu and Sn have much lower melting temperatures than Nb, Cu-Sn phases form first during heating before reacting with Nb.
    
    The ternary phase diagram reveals which compositions lead to Nb$_3$Sn. Tie lines indicate phases that coexist in thermodynamic equilibrium at a given temperature. At 675~°C, Nb$_3$Sn is in equilibrium with Cu-Sn phases containing $<$1 to 26~at.\% Sn (corresponding to the $\alpha$, $\gamma$, and $\epsilon$ Cu-Sn phases) and with Nb$_6$Sn$_5$. This defines the reaction pathway: Cu-Sn alloys within this composition range can react directly with Nb to form Nb$_3$Sn. Starting from Cu-Sn compositions $>$26~at.\% Sn requires intermediate transformations—either to lower-Sn Cu-Sn phases or to Nb$_6$Sn$_5$—before Nb$_3$Sn can form. This pseudo-binary approach (reacting pre-formed Cu-Sn with Nb) is the basis for commercial Nb$_3$Sn wire manufacturing~\cite{ambrosioNb3SnHighField2015}.
    
    \FloatBarrier
    
    \subsection{Kinetics of \texorpdfstring{Nb$_3$Sn}{Nb3Sn} Formation}

    Nb$_3$Sn forms by a solid-state diffusion reaction. Growth of the Nb$_3$Sn phase at an interface between Cu-Sn and Nb depends on nucleation, followed by continued supply of Sn as Nb$_3$Sn grains grow. However, as the diffusion reaction proceeds, Sn becomes depleted at the reaction front. As the Nb$_3$Sn layer grows, Sn must diffuse from deeper in the Cu-Sn layer to sustain the reaction. Maintaining the 25~at.\% Sn stoichiometric concentration across a Nb$_3$Sn layer, therefore, requires high-Sn activity and a small diffusion path.
        
    If the diffusion path length from the Sn reservoir to the reaction front is short, grain boundary diffusion rapidly supplies Sn, and the resulting Nb$_3$Sn film maintains approximately constant Sn content throughout its thickness. Conversely, long diffusion paths create Sn concentration gradients, with lower Sn content in regions far from the reservoir. Sn activity is also important, where a mild Cu-Sn alloy, such as $\alpha$, sets up a lower concentration difference to drive diffusion than a Cu-Sn compound with higher Sn content, like $\epsilon$. 
    
\begin{figure}[H]         
        \centering
        \includegraphics[width=.7\textwidth]{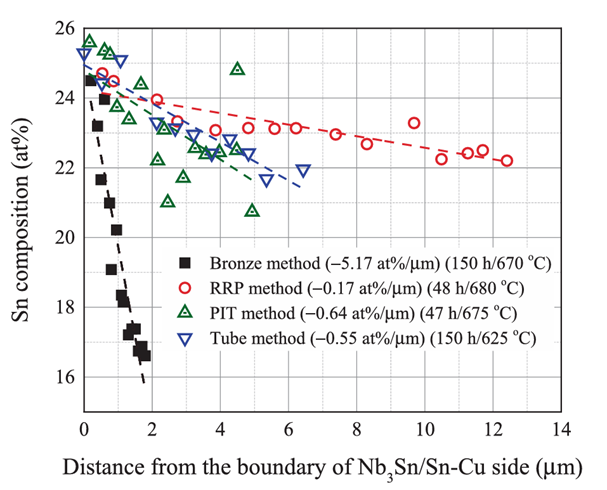}
        \caption{Sn composition vs. distance from Nb$_3$Sn/Cu-Sn interface for different wire architectures~\cite{bannoLowtemperatureSuperconductorsNb3Sn2023}.}
        \label{fig:SnGradient}
    \end{figure}

    The size of the Sn reservoir also determines the extent of reaction completeness. If the Cu-Sn source layer becomes depleted before all Nb has reacted, the film will exhibit spatial non-uniformity with low-Sn content in regions distant from the original reservoir. The Nb$_3$Sn wire magnet community addressed this challenge by minimizing diffusion distances between Sn sources and Nb. The Restacked-Rod-Process (RRP) wire architecture (red data points in Figure~\ref{fig:SnGradient}) demonstrates how reduced diffusion paths minimize Sn compositional gradients.
    
    \section{Properties}\label{sec:Tceffectch5}

    The RF properties of Nb$_3$Sn thin films depend strongly on the film microstructure, which determines the critical temperature $T_c$. The maximum critical temperature can be achieved with a stoichiometric Nb$_3$Sn film ($\sim$25~at.\% Sn) under zero strain. Because both strain and Sn content affect the critical temperature, it is challenging to deconvolve their individual contributions to measured $T_c$ values. Microanalysis of the Sn content in the films helps separate these effects.

    Impurities in the film can also affect the RF properties and cannot be fully characterized with critical temperature measurements alone. Impurities can be found in the bulk or near grain boundaries of the film. If the reaction has sufficient time to equilibrate, impurities are expelled from the bulk into the grain boundaries. Understanding the film morphology at nm-scale resolution is essential for correlating final $Q$ measurements with thin-film properties.

    \subsection{Sn Content Effect}

\begin{figure}[tb]         
        \centering
        \includegraphics[width=.8\textwidth]{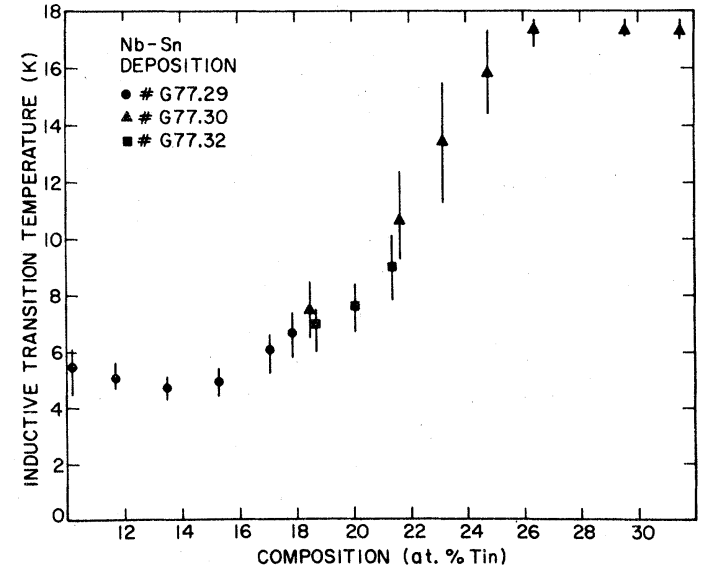}
        \caption[Plot of critical temperature $T_c$ as a function of at. Sn\% in Nb$_3$Sn.]{Plot of critical temperature $T_c$ as a function of at. Sn\% in Nb$_3$Sn. The maximum $T_c$ is associated with at. Sn 25\% to 26\%~\cite{mooreEnergyGaps151979}.}
        \label{fig:TcvsSnMoore}
    \end{figure}

\begin{figure}[tb]          
        \centering
        \includegraphics[width=.8\textwidth]{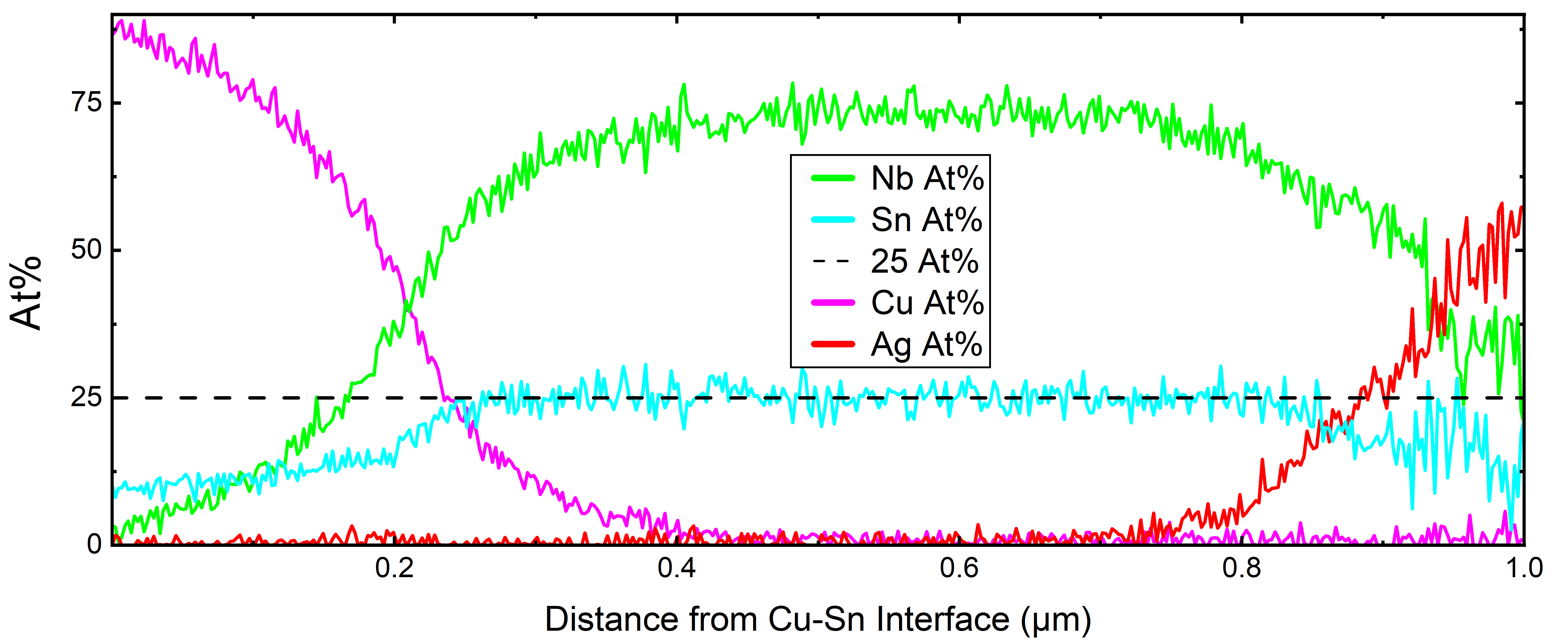}
        \caption{EDS linescan of Nb$_3$Sn film with 25\% Sn in the Nb$_3$Sn layer.}
        \label{fig:25SnNb3SnAS011423}
    \end{figure}

    Only Nb$_3$Sn films with $25{-}26$~at.\% Sn achieve the optimal critical temperature of $T_c = 18.3$~K (Fig.~\ref{fig:TcvsSnMoore}). An example film with $\sim$25~at.\% Sn in the Nb$_3$Sn layer is confirmed with an EDS line scan seen in Fig.~\ref{fig:25SnNb3SnAS011423}.
    
    \FloatBarrier



    \subsection{Strain Sensitivity}
    
\begin{figure}[htb]               
        \centering
        \includegraphics[width=.6\textwidth]{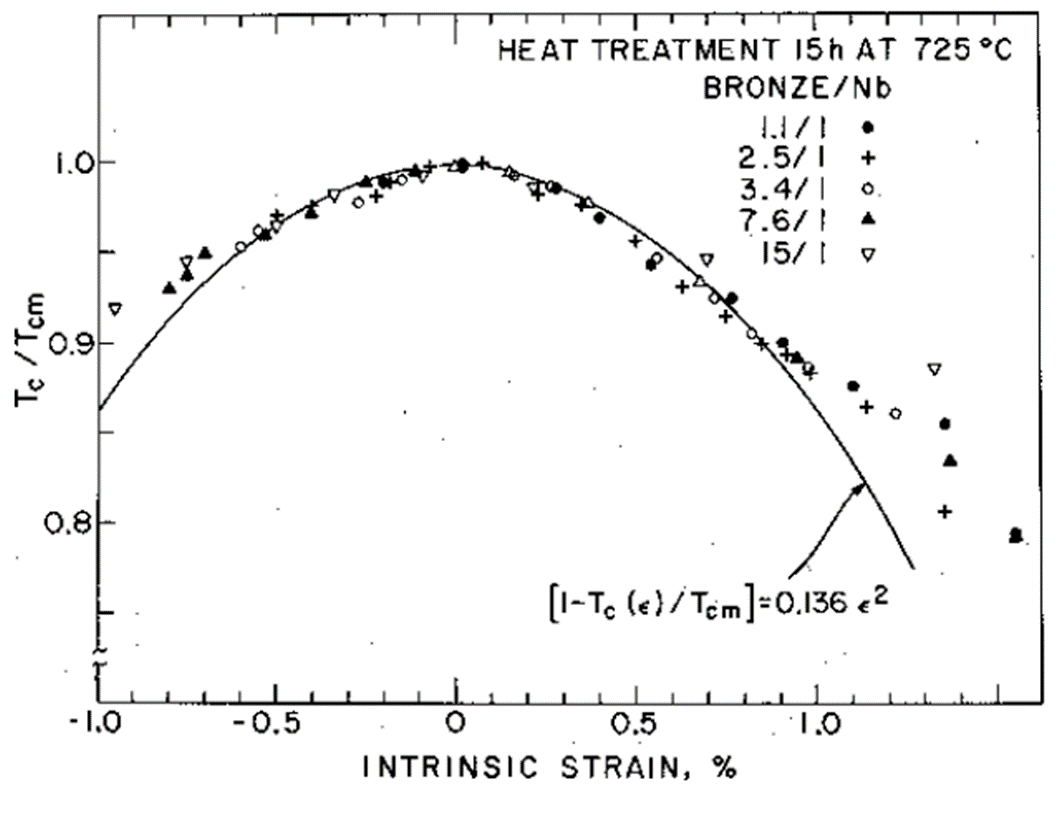}
        \caption{Nb$_3$Sn's strain-dependent critical temperature for bronze processed Nb$_3$Sn wires under tension and compression~\cite{ekinStrainScalingLaw1980}.}
        \label{fig:TcvsStrain}
    \end{figure}

\begin{figure}[htb]       
        \centering
        \includegraphics[width=.6\textwidth]{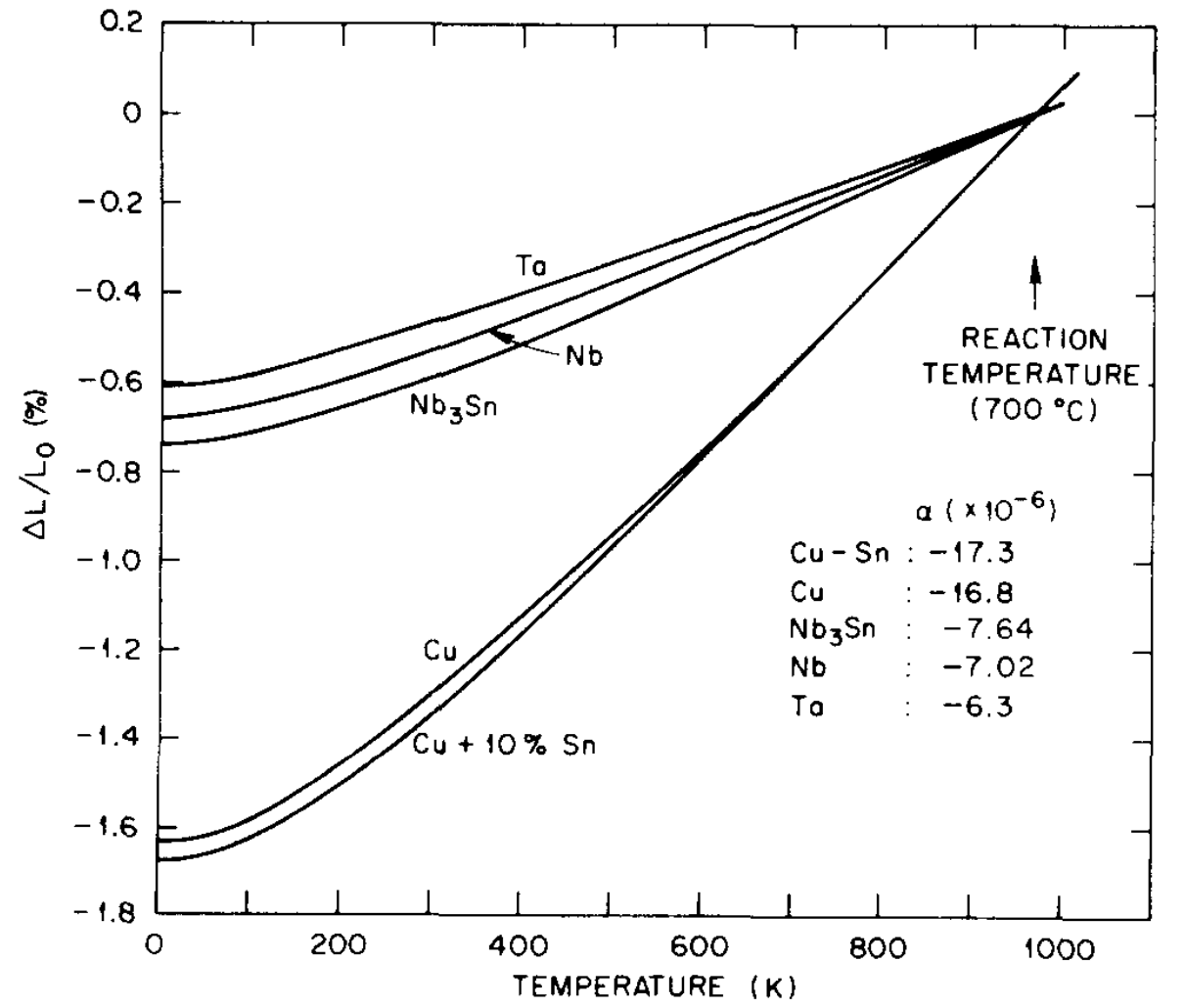}
        \caption{Coefficient of Thermal Expansion (CTE) or Cu, Nb, and Nb$_3$Sn starting with zero strain at 700~$^\circ$C~\cite{eastonPredictionStressState1980}.}
        \label{fig:CTECuNbNb3Sn}
    \end{figure}

    Nb$_3$Sn's critical temperature is also strongly strain-dependent (Fig.~\ref{fig:TcvsStrain}). With films $100$~nm to 1~\unit{\um} thickness deposited on copper, the coefficient of thermal expansion (CTE) mismatch is the dominant strain contribution after cooling the material below the critical temperature~\cite{ekinStrainScalingLaw1980}. For example, 1~\unit{\um} Nb$_3$Sn films formed on Nb substrates at a reaction temperature of $\sim700~^\circ$C and then cooled to 4.2~K, have very little strain $\approx 0.1\%$, while films made on Cu substrates have $\approx 0.9\%$ strain. A comparison between the thermal expansion of Nb, Nb$_3$Sn, Cu, and other materials is given in Fig.~\ref{fig:CTECuNbNb3Sn}. With sufficient thickness ($\sim 30$~\unit{\um}~\cite{fonnesuRecipeOptimizationSRF2025}), strain relaxation can occur in a thin film even with significant CTE mismatch between film and substrate. High strain can also lead to cracks in the brittle Nb$_3$Sn~\cite{Vallone:2023jue, Cheggour:2018pge}. Even though films may have the ideal Sn\% as shown in Fig.~\ref{fig:25SnNb3SnAS011423}, they may have reduced critical temperature as illustrated in the substrate experiment in Sec.~\ref{sec:substratestrainandSn}.
     
    \FloatBarrier

\section{Microstructure}

    \subsection{Common Features of the \texorpdfstring{Nb$_3$Sn}{Nb3Sn} Microstructure}

\begin{figure}[tbh]      
        \centering
        \includegraphics[width=.9\textwidth]{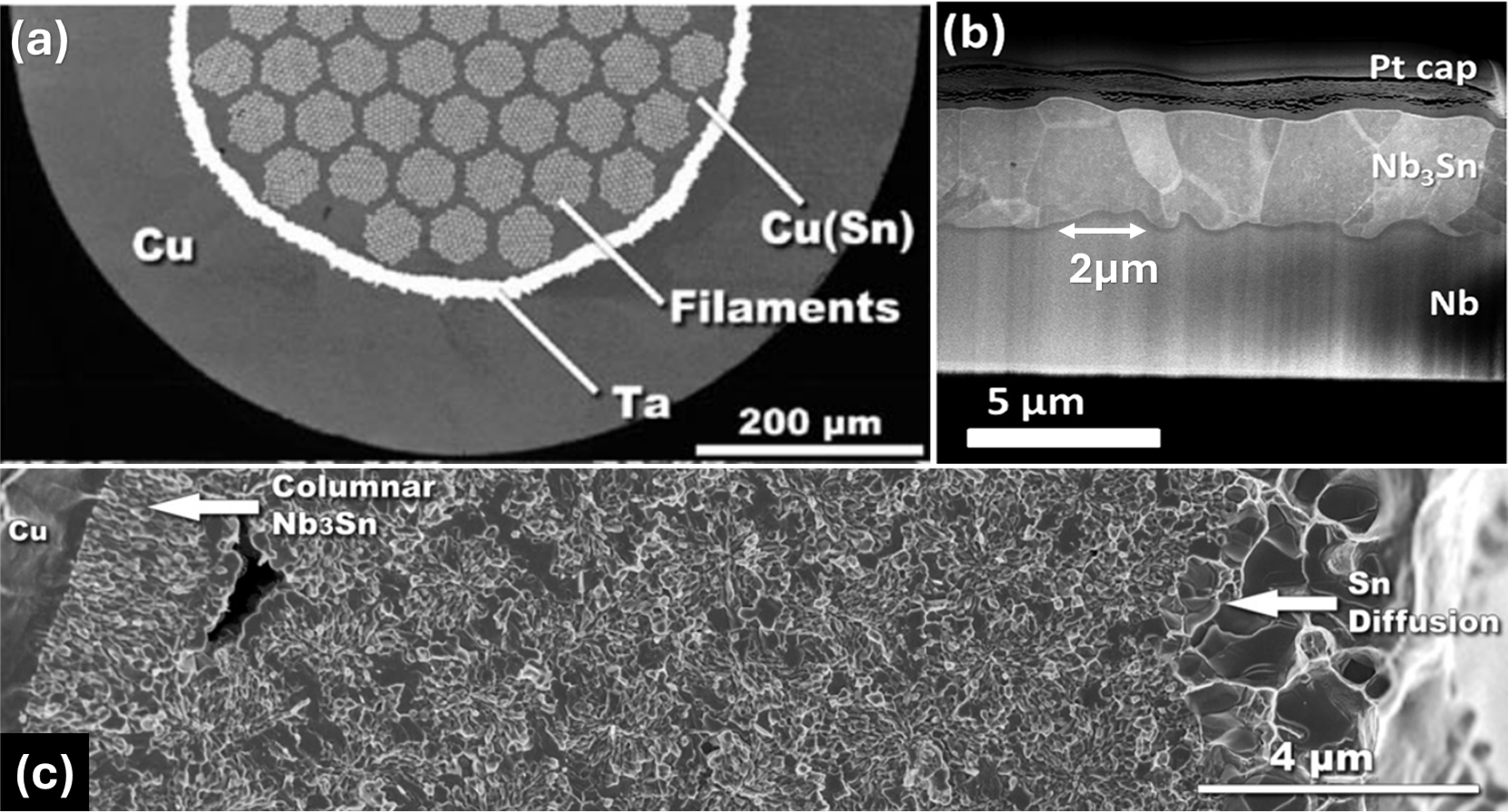}
        \caption[(a) Wire cross section with Nb$_3$Sn filaments~\cite{jewellInfluenceNb3SnStrand2003}, and (c) closer look at microstructure with small grains near the Sn source and large columnar grains farther from the Sn source~\cite{leeMicrostructuralFactorsImportant2008} commonly found in Nb$_3$Sn wires.]{(a) Wire cross section with Nb$_3$Sn filaments~\cite{jewellInfluenceNb3SnStrand2003}, and (c) closer look at microstructure with small grains near the Sn source and large columnar grains farther from the Sn source~\cite{leeMicrostructuralFactorsImportant2008} commonly found in Nb$_3$Sn wires. (b) Cross-section microscopy showing thin film Nb$_3$Sn grains made with the Sn vapor route~\cite{leeGrainboundaryStructureSegregation2020}. }
        \label{fig:Nb3SnMicrostructure}
    \end{figure}

    Nb$_3$Sn forms through solid-state grain bo undary diffusion. The resulting microstructure has been studied comprehensively in the magnet community. Example wire cross-section and microstructure are shown in (a) and (c) of~Fig.\ref{fig:Nb3SnMicrostructure}. The Nb$_3$Sn near the Sn source forms as small equiaxed grains that are high in Sn content. Nb$_3$Sn, further from the Sn source, forms as long columnar grains because the Sn is supplied at a lower rate, having to go through all the grains from the Sn source to the last Nb$_3$Sn layer. Example microstructure from a  Nb$_3$Sn thin film using the Sn vapor route can be seen in (b) of Fig.~\ref{fig:Nb3SnMicrostructure}.

    \FloatBarrier

    \subsection{Grain Boundaries}

\begin{figure}[tb]
        \centering
        \includegraphics[width=\textwidth]{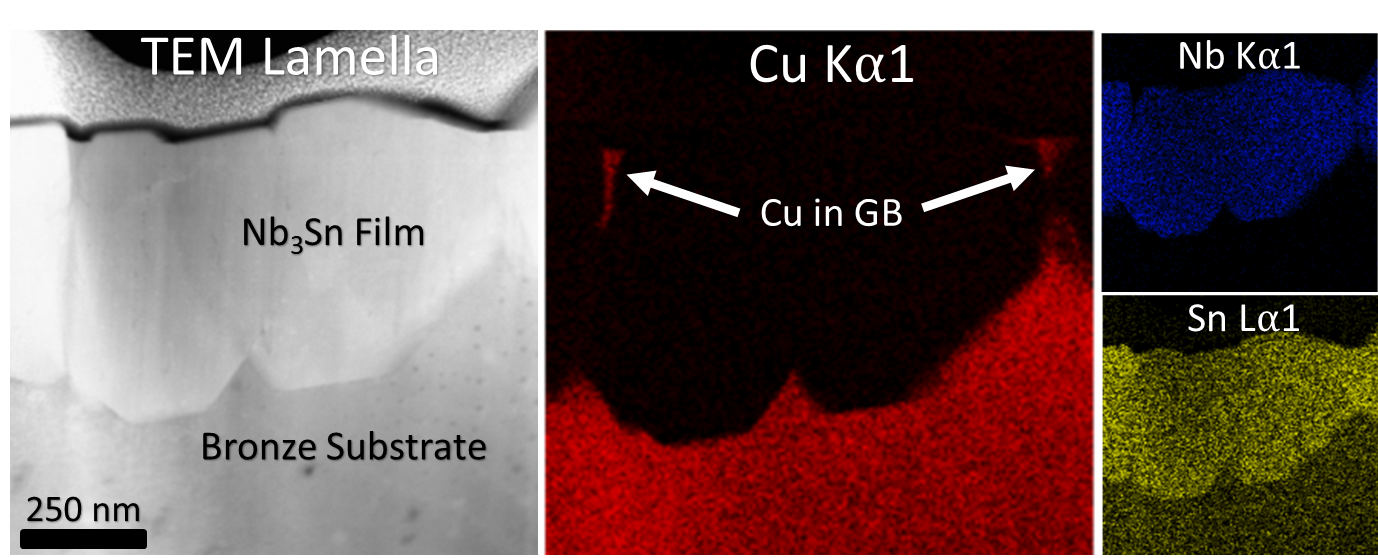}
        \caption{TEM image from a hot-bronze film on a bronze substrate, where Cu was found at the edge of some of the grain boundaries.}
        \label{fig:TEMHotBronze}
    \end{figure}

    Grain boundaries affect Nb$_3$Sn performance differently depending on the application. In magnets, grain boundaries provide beneficial flux pinning at high fields. In RF cavities operating near zero field, grain boundaries contribute to residual surface resistance. Because grain boundaries are narrow ($\sim$5~nm), transmission electron microscopy (TEM) is required to resolve their composition. Figure~\ref{fig:TEMHotBronze} shows a hot-bronze film on a bronze substrate with Cu impurities segregated at the grain boundaries.

    \subsubsection{Accelerator Applications: Minimizing Grain Boundary Losses}
    
\begin{figure}[tb]
    \centering
        \includegraphics[width=\textwidth]{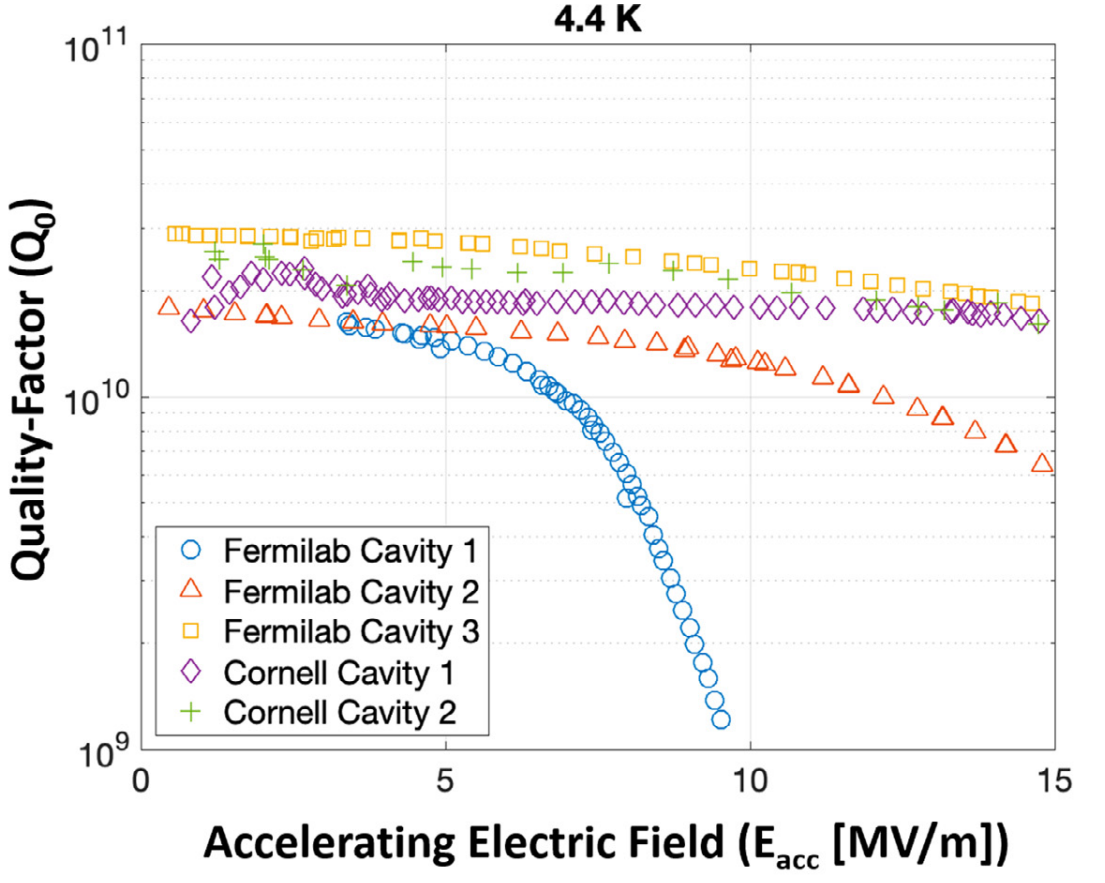}
        \caption{Q factor measurements showing improved performance in samples with Sn removed from grain boundaries via high-temperature annealing~\cite{leeGrainboundaryStructureSegregation2020}.}
        \label{fig:QfactorSngrain}
    \end{figure}

    Solid-state diffusion routes can trap impurities such as Cu and excess Sn in Nb$_3$Sn grain boundaries. When the grain boundary width approaches the coherence length ($\sim$3~nm), these impurities significantly obstruct supercurrent flow. In RF cavities, this increases the residual resistance $R_0$. \citeauthor{leeGrainboundaryStructureSegregation2020}~\cite{leeGrainboundaryStructureSegregation2020} showed that annealing at 1100~°C for 3~hours drives Sn out of grain boundaries, increasing the quality factor from $2\times10^{10}$ to $3\times10^{10}$ without degrading other properties (Fig.~\ref{fig:QfactorSngrain}). Ab initio calculations have confirmed that grain boundary impurities degrade superconducting properties in the Cu-Nb-Sn system~\cite{kelleyInitioTheoryImpact2020}.
    
    \subsubsection{High-Field Applications: Enhancing Flux Pinning}

    For high-field applications, grain boundaries serve as beneficial flux pinning centers. Pinning force scales inversely with grain size ($F_p \propto 1/d$)~\cite{scanlanFluxPinningCenters1975}, so higher grain boundary density reduces Campbell resistance $R_C$ in applied fields. The pinning mechanism arises because grain boundaries locally suppress the superconducting order parameter, creating sites that interact strongly with vortex cores~\cite{gurevichNonlocalJosephsonElectrodynamics1992, gurevichTimeScalesFlux1993, gurevichAnisotropicFluxPinning1994}. These models predict that vortices depin at currents less than 10\% of the current circulating within vortex cores, so the current-blocking effect of grain boundaries discussed in the previous section is negligible for pinning applications. The trade-off is that increased grain boundary density may raise the residual resistance $R_0$ at zero field.
    
    
    \subsubsection{Implications for Axion Detection}
    
    For axion detector cavities operating in 9~T fields, grain boundaries present a trade-off: they provide beneficial flux pinning (reducing $R_C$) but may increase residual resistance ($R_0$). The optimal microstructure depends on which loss mechanism dominates at the operating frequency, temperature, and field. Key open questions include whether films with smaller grains improve high-field performance through enhanced pinning, and whether Cu impurities in grain boundaries degrade performance in applied fields as they do at zero field.
    




    

\section{Fabrication Methods for Films and Coatings}\label{sec:coatings}
    
    Due to the low toughness and thermal conductivity of Nb$_3$Sn, it is neither practical nor favorable to make a cavity from bulk Nb$_3$Sn. Nb$_3$Sn, therefore, must be fabricated as a thin film, and ideally on a thick, high thermal conductivity metal. There are many different techniques to deposit Nb$_3$Sn as a thin film. The industry-standard accelerator method, the Sn vapor method~\cite{peinigerNb3SnSUPERCONDUCTINGACCELERATORS1989,posenMeasurementHighQuality2022}, is the most developed Nb$_3$Sn thin film process, but is limited to Nb substrates. Other methods, such as the ternary route explored in this work, enable thin film fabrication on Cu substrates, which also present challenges.

    For Nb$_3$Sn fabrication to be compatible with Cu substrates, the reaction temperature must be well below the melting temperature (1084 $^\circ$C) of Cu. High-temperature processes promote annealing and recrystallization, which cause the substrate to lose strength~\cite{okudaFabricationTestingLband1991}. This temperature constraint divides Nb$_3$Sn axion detector cavity fabrication methods into two categories: high-temperature routes limited to Nb substrates ($>950~^\circ$C), and low-temperature routes compatible with Cu substrates ($<750~^\circ$C). 

    \subsection{Substrate Requirements}

    Cu is an ideal substrate for maintaining an isothermal cavity environment due to its high thermal conductivity, which prevents thermal gradients during operation. However, fabricating Nb$_3$Sn on Cu presents a critical challenge: the Cu substrate acts as a Sn sink, depleting Sn from the Cu-Sn/Nb$_3$Sn reaction zone. Since $T_c$ depends strongly on Sn content (Fig.~\ref{fig:TcvsSnMoore}), this depletion degrades superconducting performance. A diffusion barrier of refractory metals (Ta, Mo, Nb, W) between the Cu substrate and the Cu-Sn source layer can block Sn migration while maintaining thermal contact.
    
    Substrate surface roughness and cleanliness critically affect thin film quality, since deposited films replicate the underlying surface morphology. The challenge intensifies when underlying layers undergo structural changes during heat treatment, including recrystallization and grain growth. For example, if a Cu-Sn source layer sits between the Cu substrate and the Nb$_3$Sn film, the Cu-Sn layer will undergo phase transformations and volume changes during the reaction (Sec.~\ref{sec:NbCuSn}). These structural rearrangements can induce strain, surface roughness, or, in severe cases, cracking and delamination of the overlying Nb$_3$Sn film. The choice of substrate and its effect on the thin film must be carefully considered when selecting a method and recipe for thin film fabrication.
    
    \subsection{High-Temperature Routes (Nb Substrates Only)}
    
    \subsubsection{Sn Vapor Method} \label{sec:Snvapordetail}

    The Sn vapor method is the most mature route for Nb$_3$Sn thin film fabrication~\cite{peinigerNb3SnSUPERCONDUCTINGACCELERATORS1989,posenMeasurementHighQuality2022}. Bulk Nb cavities are exposed to Sn vapor at elevated temperatures, forming Nb$_3$Sn through solid-state diffusion. Thermodynamically, the binary Nb-Sn reaction requires temperatures above 950~$^\circ$C to destabilize competing phases (Nb$_6$Sn$_5$ and NbSn$_2$) and stabilize stoichiometric Nb$_3$Sn~\cite{xuxingchenReviewProspectsNb3Sn2017}. However, kinetic limitations necessitate a reaction temperature near 1100~$^\circ$C to drive the reaction to completion. This method has achieved the highest quality factors ($Q_0 \sim 10^{11}$) at zero field among Nb$_3$Sn fabrication techniques for accelerator applications~\cite{Posen:2018zjb}.

    The Sn Vapor method has good surface coverage, compared to line-of-sight deposition methods. This enables a relatively uniform coating on the inside of complex geometries. However, since the wettability of Sn on Nb is poor~\cite{heinSuperconductingIntermetallicCompounds1973}, tin chloride (SnCl$_2$) gas is used to improve wetting and help seed Nb with Sn atoms. This initiates the nucleation of initial Nb$_3$Sn growth, which is further driven by a post-reaction at 1100 $^\circ$C. The resulting Nb$_3$Sn has sharp superconducting transitions ($\Delta T_c < 1$~K) with critical temperatures near the optimal value of $T_c\approx18$~K~\cite{Posen:2018zjb}.
    
    The Sn vapor method is continuing to improve, advancing toward higher accelerating gradients~\cite{posenNb3SnSuperconductingRadiofrequency2022, Posen:2018zjb, Posen:2020kei}. Recent work by Viklund et al. demonstrated that cavities can be mechanically polished to remove defective Nb$_3$Sn layers and recoated to achieve performance equal to or exceeding the original coating~\cite{Viklund:2023uxj, Viklund:2023nlq, Viklund:2024lew}. Still, this process has shortcomings, including poor uniformity and the presence of Sn domains near the center of the Sn seed~\cite{Viklund:2023uxj}.

    Jefferson Lab and Fermilab have developed hybrid approaches that retain the good superconducting properties of the Sn vapor method while incorporating Cu for thermal management. They use Cu cooling clamps, and electroform Cu in a shell around the completed Nb$_3$Sn/Nb cavity. These modifications sufficiently improve thermal conductivity to enable cryocooler operation (rather than liquid helium), thereby reducing operating costs. However, thermal performance remains limited due to the significant distance (mm-scale) between the Nb$_3$Sn film and the outer Cu shell~\cite{ciovatiDesignCwLowenergy2018, posenNb3SnSuperconductingRadiofrequency2022}. The Cu clamping approach may be a feasible way to increase thermal conductivity for axion detector cavities. However, the room is very limited in the bore of a magnet, and the Cu clamp occupies valuable bore space that could be better used to increase the axion-photon cross section (more cavity volume) than to increase thermal conductivity.
        
    \subsubsection{Nb-Sn Multilayer Sputtering}
    
    In another high-temperature method, Nb and Sn are separately deposited in alternating layers using sputtering and a Nb substrate, followed by a reaction~\cite{vandenbergNewPhaseFormation1985}. The short diffusion distances between the Nb and Sn layers allow a slightly lower reaction temperature of 950 $^\circ$C, rather than the 1100~$^\circ$C required by the Sn vapor method. A full 2.6 GHz accelerator cavity has been coated with a cylindrical magnetron sputtering system~\cite{Shakel:2023cdz}, and reached $Q = 1.1 \times10^9$. This method has not achieved the performance levels of the Sn vapor method.

    \subsection{Low-Temperature Routes (Cu Substrate Compatible)}

    \subsubsection{Cu Compatible Nb-Sn Routes}\label{sec:StoichiometricDetail}
    
    The diffusion path between Nb and Sn can be reduced even further using a stoichiometric Nb$_3$Sn target or co-sputtering Nb and Sn simultaneously. This allows for Nb$_3$Sn fabrication on Cu substrates with a binary Nb-Sn route at temperatures around $700$~°C. Co-sputtering methods suffer from non-uniformity, off-stoichiometry, and Sn islands on the film's surface~\cite{Tan:2020hmh}. Additionally, films on Cu substrates cannot be post-reacted at the high temperatures ($>950$°C) typically used to optimize Nb$_3$Sn stoichiometry and microstructure after deposition. The as-deposited film is essentially final, requiring precise control of deposition parameters since deficiencies cannot be corrected through subsequent heat treatment.
    
    \citeauthor{Ilyina-Brunner:2019iay} at CERN uses Nb$_3$Sn stoichiometric targets to deposit on Cu, and they found large Cu inclusions on the surface of the Nb$_3$Sn~\cite{Ilyina-Brunner:2019iay}. These Cu inclusions are detrimental to RF properties, so subsequent experiments used a thick diffusion barrier between the Cu substrate and Nb$_3$Sn film to prevent Cu diffusion to the surface.
    
    
    Recent progress~\cite{fonnesuRecipeOptimizationSRF2025, piraProgressEuropeanThin2023} using optimized recipes with a stoichiometric target has shown a world record low surface resistance $R_s\approx 7$~\unit{n\ohm} using a Nb substrate quadrupole resonator (QPR) sample. \citeauthor{Ilyina-Brunner:2019iay} have near-term plans of coating a full cavity with their recipe, and have results showing that their process has higher pinning properties than those of the Sn vapor route~\cite{fonnesuRecipeOptimizationSRF2025}. This is the first result showing that Nb$_3$Sn can be fabricated with a non-Sn vapor method and display competitive properties in specific applications.
 
    Reza et al. at Daresbury noted~\cite{valizadehSynthesisNbAlternative2022, turnerFacilityCharacterisationPlanar2022, turnerInvestigationMagneticField2026} that the divergence of the sputtered plume of material from a planar magnetron was sufficient to coat the walls of a cavity resonator by simply moving the magnetron along the cavity axis, where the sputter target plane is perpendicular to the axis. This avoids complications of cylindrical magnetrons that sputter radially. An advantage of the planar magnetron is more efficient cooling, which allowed the Daresbury team to heat the cavity wall without overheating the sputter source. They have been able to make good-quality films with this process and aim to test full cavities soon. With these low-temperature binary Nb-Sn reactions, the energy and rate of the deposition are crucial for achieving optimal stoichiometry.
    
    \subsubsection{Cu-Nb-Sn Route: Electrochemical Cu-Sn Process}
    
    The ternary route leverages technology developed for Nb$_3$Sn high-field magnets. Using a pseudo-binary reaction between a Cu-Sn compound and Nb avoids the unfavorable superconducting compounds and lowers the practical Nb$_3$Sn reaction temperature. This is often termed the bronze route~\cite{suenagaChemicalCompositionsGrain1983, tachikawaHighFieldNb3SnSuperconductors1997} in reference to the alpha Cu-Sn phase called bronze.

    The bronze route has been successfully applied to Nb cavities by \citeauthor{luDevelopmentPerformanceFirst2022} in China~\cite{luDevelopmentPerformanceFirst2022}. A bronze layer was electroplated onto the inner surface of a Nb cavity and subsequently reacted, achieving a quality factor of $1.2 \times 10^9$. This Chinese group also coated Cu substrate samples using the same electrochemical process, in which they first sputtered a thick 5~$\mu$m Nb layer onto the Cu substrate, electroplated Cu-Sn, heat-treated in vacuum at 700~$^\circ$C, and then removed the Cu-Sn layer by electropolishing~\cite{Lu:2023wyr}. The group hopes to extend the process to a full-sized Cu cavity. The bronze route can also be used to fabricate multiple electroplated layers of alternating Cu and Sn~\cite{franzSYNTHESISNbSnCOATINGS2017}. The resulting films using this method have characteristics similar to those of the above electroplated Cu-Sn films.

    The advantage of the electrochemical processing method is that it can uniformly coat complex geometries. However, one potential pitfall of this method is the large impurity concentration created when using aqueous electrochemical solutions, which introduce oxygen and organic contaminants into the Cu-Sn fil, resulting in broader superconducting transitions ($\Delta T_c = 1.5-4$~K)~\cite{luDevelopmentPerformanceFirst2022, luElectrochemicalThermalSynthesis2021, luImpactCuSn2025}.
    
    \subsubsection{Cu-Nb-Sn: Thermal Evaporation Cu-Sn Process (This Work)}\label{sec:ourworkdescribe}
    
    This dissertation explored the deposition of Cu-Sn layers by thermal evaporation. This work adds to studies conducted at the Applied Superconductivity Center (ASC) in the early 2000s~\cite{cooleyShiftFluxpinningForce2001}, in which Nb was deposited on bronze samples. The pinning force was analyzed and found to be higher in films with $20-50$~nm grains (hot-bronze films) than in films with $50-250$~nm grains (post-reacted films).

    The pseudo-binary reaction path is very similar to the electrochemical process described above, except that thermal evaporation is used to deposit the Cu-Sn. Evaporation allows control of the Sn concentration in the range $10-40$~at.\% Sn, and since it takes place in a high-vacuum environment, impurities from aqueous solutions are avoided. Narrow superconducting transitions ($\Delta T_c \sim 1 $~K) have been obtained. Obtaining good Nb$_3$Sn films critically depends on the cleanliness of the environment during deposition and reaction~\cite{withanageRapidNbSn2021}. The sputter chamber for Nb deposition and the thermal evaporator for Cu-Sn deposition both use vacuum systems designed to limit oxygen and carbon from entering the film. This work employed two line-of-sight physical vapor deposition techniques: thermal evaporation and magnetron sputtering.


    Two variants using this thermal evaporation approach were developed. The first is a post-reaction solid-state diffusion technique. This method requires a Cu-Sn film in contact with a Nb layer, where Sn diffuses out of the Cu-Sn film into the Nb, forming Nb$_3$Sn at the Cu-Sn/Nb interface. The speed of this reaction is limited by the grain boundary diffusion rate of Sn through the Nb$_3$Sn grains. The second variant is a vapor reaction method termed the hot-bronze technique. A Cu-Sn film is heated, and Nb atoms arrive and immediately react with a high concentration of Sn vapor right above the Cu-Sn film~\cite{withanageRapidNbSn2021}. During Cu-Sn heating, the bronze target surface turns silver, confirming its Sn-rich composition. Hot-bronze films can be further improved with an additional post-reaction heat treatment. The Sn in the remaining Cu-Sn layer continues to feed the Nb$_3$Sn during this anneal, further saturating the Nb$_3$Sn grains and increasing $T_c$. The resulting film microstructure is fundamentally different between the two variants: the hot-bronze variant shows large columnar grains spanning the film thickness, whereas the post-reaction method shows standard equiaxed grains near the Sn source and columnar grains farther from the Sn source.

    \subsection{Summary of Fabrication Techniques}
    
    All the fabrication methods are listed in Table~\ref{fig:Nb3SnFabrication2}. 
    
    
    
    \begin{table}[ht]      
    \centering
        \caption{Comparison of Nb$_3$Sn fabrication methods showing reaction temperature, equipment requirements, substrate compatibility, and key performance characteristics~\cite{posenNb3SnSuperconductingRadiofrequency2017,luElectrochemicalThermalSynthesis2021, sayeedFabricationSuperconductingNb3Sn2023, schaferKineticallyInducedLowtemperature2020, Ilyina-Brunner:2019iay, withanageRapidNbSn2021}.}
    \includegraphics[width=.9\textwidth]{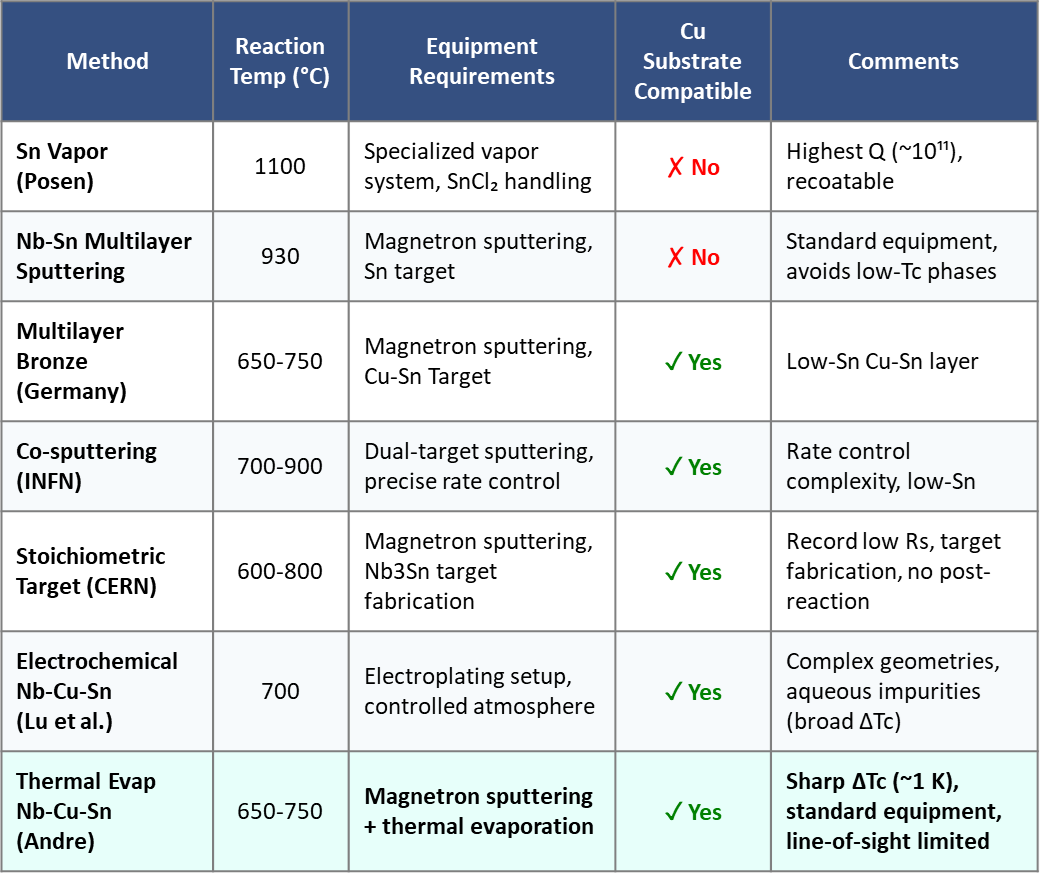}
    \label{fig:Nb3SnFabrication2}
    \end{table}
    \FloatBarrier

    \subsection{Future Directions}

    The Nb$_3$Sn thin-film cavity community has adopted many different approaches that trade off benefits and constraints. Future work might consider combining the strengths of various approaches to overcome weaknesses. For instance, the fabrication process could start with an electrochemical Cu-Sn seed layer to achieve uniform coverage and increased Sn wettability, and then deposit a higher-Sn, impurity-free Cu-Sn film via thermal evaporation. Collaborative development may help accelerate Nb$_3$Sn fabrication towards a reproducible high-$Q$, high-field superconducting RF resonator. 
\chapter{Achieving High Quality \texorpdfstring{Nb$_3$Sn}{Nb3Sn} Films on Cu}\label{sec:nb3snresults}

This chapter describes how the thermal evaporation approach discussed in the previous chapter was studied and then optimized using small coupon samples. The primary objective was to synthesize Nb$_3$Sn thin films on copper substrates with (1) uniform microstructure and Sn content throughout the film thickness, and (2) maximize $T_c$. 

\begin{figure}
    \centering
    \includegraphics[width=.9\textwidth]{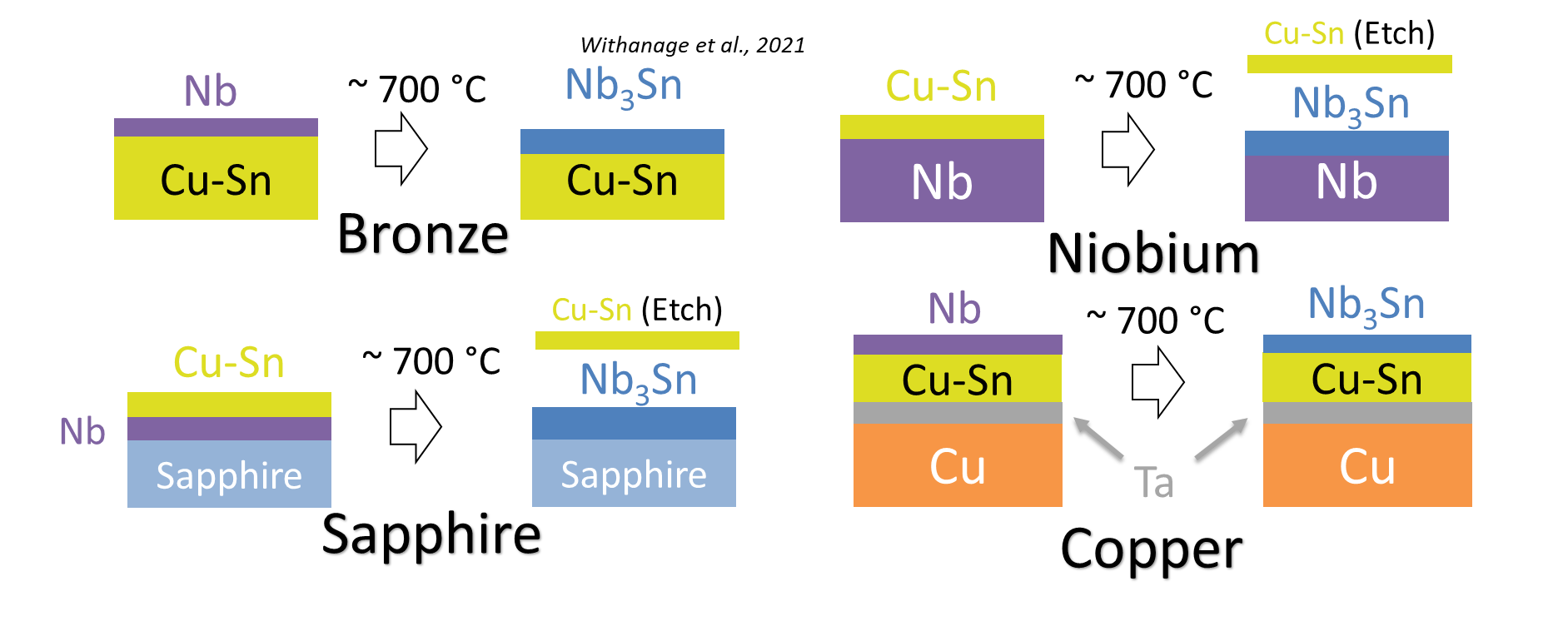}
    \caption{Schematic showing the different multilayers needed for Nb$_3$Sn reaction on bronze, Nb, sapphire, and Cu substrates.}
    \label{fig:SubstrateSchematic}
\end{figure}

Cu is the ideal substrate due to its high thermal conductivity. However, it suffers from a significant CTE mismatch that induces compressive strain during cooldown, as well as rapid Sn diffusion into the Cu substrate, depleting Sn at the reaction front. The Nb$_3$Sn formation was systematically optimized using different substrate materials (bronze, Nb, sapphire, Cu Fig.~\ref{fig:SubstrateSchematic}), diffusion barrier composition (Ta, Nb, Mo), Cu-Sn layer thickness, initial Sn concentration in the Cu-Sn layer, Nb seed layer thickness, and heat treatment profiles. Studies on bronze, Nb, and sapphire substrates isolated the effects of strain and Sn depletion before transferring this knowledge to form complex Cu-substrate multilayers. Challenges with thermal evaporation were discovered, including Sn gradients and film roughness. Some of these challenges remain as opportunities for future refinement. 


This chapter also reports multiple proof-of-principle studies to downselect three recipes with the most potential. Two recipes using the hot-bronze technique (see Sec:~\ref{sec:ourworkdescribe}) were separately achieved: (1) $\Delta T_c< 1$~K transitions proving compositional uniformity, and (2) $T_c = 17.7$~K demonstrating that strain can be managed through film thickness. Recipe (3) achieved an exceptionally uniform film morphology using an Nb-first post-reaction approach and was therefore selected as the RF cavity coating recipe in the next chapter.



\section{Constraints on the Experiments}

Constraints on the temperature and time of heat treatments were enforced when reacting using the thin-film sputtering chamber. While this chamber provides excellent vacuum conditions ($10^{-9}$~torr at room temperature, degrading to $\sim10^{-7}$~torr at $750~^\circ$C due to outgassing), superior to the available tube furnace ($10^{-5}$ to $10^{-6}$~torr), heat treatments were restricted to 6 hours and a maximum temperature of $750~^\circ$C to prevent thermal damage to the chamber components. Samples may have achieved better phase purity and homogeneity with longer annealing times or higher reaction temperatures, which represents an avenue for future improvement. Additionally, the homogeneity of the Cu-Sn film could have been improved if the thermal evaporator had been equipped with substrate rotation and/or heating capabilities.

\section{Experimental Methods}

Experimental methods in thin film analysis and characterization is provided in Appendix~\ref{app:Methods}, useful for future graduate students in this field. Given here however, is a description of the specific methods used to fabricate and characterize the films made in this work.

    \subsection{Thin Film Fabrication}
    
    \subsubsection{Substrate Preparation}

    Single-crystal substrates (Si(100) and c-plane sapphire) were used as-received after standard ultrasonic cleaning in acetone and IPA. High-Sn bronze (13~wt.\% Sn), high-purity niobium, and oxygen-free high thermal conductivity (OFHC) copper were cut into 7.5x7.5 mm square samples from a 1 mm thick sheet using a guillotine blade. These substrates were mechanically polished to a mirror finish using progressive carbide paper ($400-1200$ grit), diamond slurry ($5-1$ \unit{\um}), and vibratory colloidal silica (0.05~\unit{\um}) polishing. Samples were ultrasonically cleaned in acetone and isopropyl alcohol (IPA) and stored in a desiccator until deposition.
    
    \subsubsection{Diffusion Barrier Deposition}
    
    The Cu substrates required diffusion layers before the Cu-Sn layer deposition. Without a diffusion barrier, too much Sn diffuses into the substrate, leading to poor-quality Nb$_3$Sn. A thick diffusion barrier can also allow for strain relaxation in the film. Diffusion-barrier materials with high melting points and high strength were used: $100 - 500$~nm-thick Ta, W, or Nb diffusion barriers. A 10 mTorr argon working gas was used for sputtering Ta and W, and 8 mTorr for Nb. All sputtering was done using a 5-target system manufactured by Kurt J. Lesker. The typical background pressure was $\sim 5 \times 10^{-9}$ Torr, and cryopumping enabled efficient removal of water and hydrogen. 
    
    The cleaned substrate was placed into the magnetron sputtering system, heated to 200~$^{\circ}$C for 10 minutes, and the diffusion barrier was sputtered onto the surface. After waiting for samples to cool, they were transferred into an Edwards Auto 306 Thermal evaporator with a base pressure of $\sim8 \times10^{-7}$ Torr using a liquid nitrogen-enhanced pump. The thermal evaporator was used to deposit Cu-Sn alloys. 
    

    \subsubsection{High Sn\% Cu-Sn Alloys}

    Cu (99.9\% purity) and Sn (99.8\% purity) spherical powders (Alfa Aesar) were mixed to create four Cu-Sn compositions: 7, 14, 21, and 26~at.\% Sn. When these compositions were heated and vaporized, the resulting deposited film was single-phase or multiphase, depending on the initial composition. Cu-Sn films were made at thicknesses of 2, 5, and 10 \unit{\um}. Initially, a tungsten boat was used, but it resulted in very Sn-poor Cu-Sn films. A switch was made to alumina-coated tungsten baskets from R.D. Mathis for the remaining films. 
    
    \noindent The required Cu-Sn configuration depends on the substrate:
    
    \begin{itemize}
        \item \textbf{Nb substrates:} Cu-Sn deposited directly on Nb (no diffusion barrier needed)
        \item \textbf{Bronze substrates:} No additional Cu-Sn layer required---the substrate itself serves as the Sn reservoir  
        \item \textbf{Cu substrates:} Cu-Sn deposited on a diffusion barrier (Ta, Nb, or Mo) to prevent Sn loss into bulk Cu
    \end{itemize}
    
    \noindent This leads to two processing routes on Cu substrates:
    
    \begin{itemize}
        \item \textbf{Cu-Sn-first route:} Cu/Barrier/Cu-Sn/Nb $\rightarrow$ enabling a hot-bronze reaction and/or a post-reaction optimization after deposition
        \item \textbf{Nb-first route:} Cu/Barrier/Nb/Cu-Sn $\rightarrow$ only allows for a post-reaction (the thermal evaporation system was not equipped with a heater to facilitate in-situ reaction experiments)
    \end{itemize}
  
\begin{figure}[ht]
        \centering                
        \includegraphics[width=.8\textwidth]{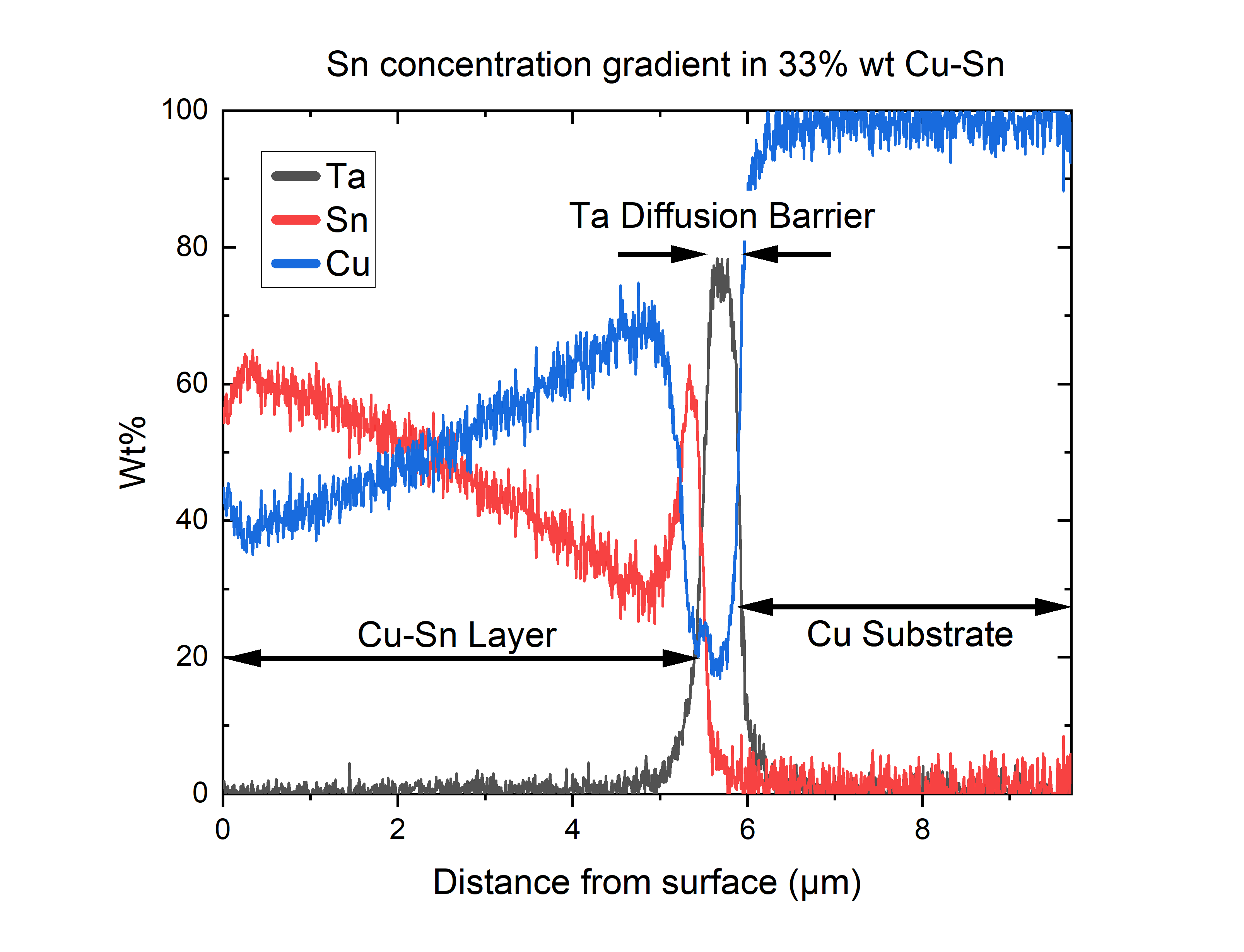}
        \caption[Thermally evaporated Cu-Sn film with concentration gradient.]{Thermally evaporated Cu-Sn film with concentration gradient. This gradient was found in the as-deposited film before reaction with Nb.}
        \label{fig:CuSnGradient}
    \end{figure}

    Thermal evaporation enables rapid deposition of thick Cu-Sn films ($\sim$10 \unit{\um}/hour) with high purity and compositional control. When depositing materials with different vapor pressures and melting temperatures, such as Cu and Sn from the same thermal evaporation boat, composition gradients can develop across the film thickness during deposition (seen in Fig.~\ref{fig:CuSnGradient}). The substrate becomes hot due to radiative heating, causing diffusion during deposition that further alters the spatial distribution of Sn throughout the film. Since Sn redistributes during the subsequent Nb$_3$Sn formation reaction, these as-deposited composition gradients were not a primary concern for this application. 

    Poor wetting of the Cu-Sn layer on smooth substrates can lead to film dewetting and delamination. Two approaches were found to mitigate peeling: substrates with surface roughness $>10$~nm showed reduced delamination, and depositing Cu-Sn films in 5 steps at 20 \AA/s with 30-minute intervals between steps allowed for film cooling and strain relaxation, further reducing peeling. Both techniques were used in this study to mitigate Cu-Sn film peeling.
    
    
    
    %

    \FloatBarrier

    \subsubsection{Nb deposition and \texorpdfstring{Nb$_3$Sn}{Nb3Sn} reaction}

    The next step was deposition of Nb and Nb$_3$Sn reaction. The multilayer structure underwent a pre-heating step at 110~\unit{\celsius} in the sputtering chamber load lock to remove adsorbed water after transfer from the thermal evaporator to the sputter chamber. A Nb film was deposited using 8~mTorr working gas, 250~W power, and 0~V substrate bias. Three distinct recipes were employed:

    \begin{enumerate}
    
        \item Hot-Bronze: Nb deposited on heated Cu-Sn (650-750 °C) for the duration of the deposition process.
        \item Post-Reaction: Nb deposited at 200 °C, followed by in-situ heating (650-750 °C) for ~3 hours. 
        \item Hot-Bronze with Post-Reaction: Extended in-situ heating (650-750 °C) after the hot-bronze method. 
    
    \end{enumerate}
    
    \begin{figure}[ht]      
        \centering
        \includegraphics[width=.7\textwidth]{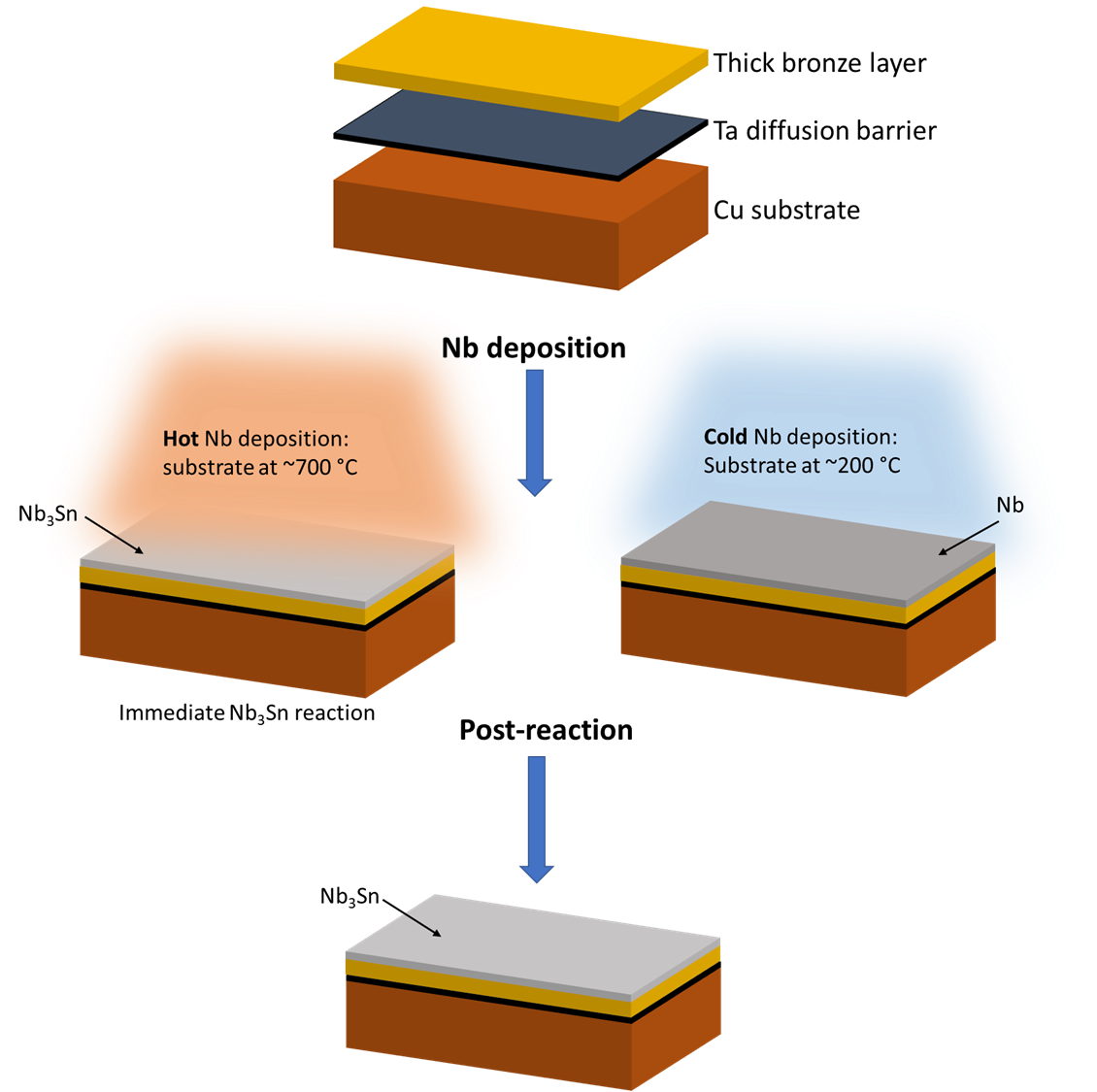}
        \caption[A schematic representation of the Cu substrate with a diffusion layer and bronze beneath a Nb$_3$Sn film.]{A schematic representation of the Cu substrate with a diffusion layer and bronze beneath a Nb$_3$Sn film. This film is formed by hot- or cold-deposition of Nb, followed by a post-reaction. Notably, if Nb is deposited hot, it results in distinct microstructural differences and a higher $T_c$ onset due to the higher mobility of atoms at the elevated temperature. (Adapted from \citeauthor{withanageRapidNbSn2021})}
        \label{fig:CuMultilayer}
    \end{figure}

    \noindent The temperature recorded is inferred from a thermocouple nearby the sample, and therefore is not the input temperature from the furnace heating element.
    
    \subsection{Characterization Techniques}

        \subsubsection{Critical Temperature \texorpdfstring{$T_c$}{Tc}}

        Measurements of magnetic moment [emu] vs.\ temperature [K] were collected using a superconducting quantum interference device (SQUID) magnetometer (Quantum Design MPMS); Appendix~\ref{app:Methods} discusses the theory of operation and mounting technique. Samples with cross-sections no larger than $7.5 \times 7.5$~mm and a thickness of 2~mm were zero-field cooled below $T_c$, and any trapped flux was removed using the manufacturer's oscillatory degauss procedure. A 1~mT field was then applied parallel to the film surface to minimize demagnetization effects. The magnetic moment was measured in $\sim$0.1~K steps as the sample warmed to 20~K.
        
        The $T_c$ onset is defined as the last measured point before the normalized magnetic moment reaches zero, representing the highest-quality superconducting region in the film. The transition width $\Delta T_c$ is determined from the 90\%--10\% span of the superconducting transition, and the baseline is established by extrapolating the normal-state response. A narrow $\Delta T_c$ indicates uniform Sn stoichiometry throughout the film.
        
        However, $T_c$ measurements alone do not fully predict RF performance: they lack spatial resolution, are influenced by magnetic screening, and do not distinguish where high-quality Nb$_3$Sn resides within the film thickness. Microstructural characterization is therefore required to complement these results.
        
        \subsubsection{Microanalysis}

        
        
        An FEI Helios G4 UC field-emission scanning electron microscope (SEM) with a focused ion beam (FIB) was used to study the surface and cross-sectional morphology in secondary-electron (SE) mode. The elemental composition with a spatial resolution of $\sim1$~\unit{\um} was analyzed using energy-dispersive spectroscopy (EDS) with an Oxford Instruments X-MaxN SDD x-ray detector and standardless analysis in the same SEM/FIB chamber. Roughness was measured using a DSX optical microscope, with a height resolution of less than 1~\unit{\um}. The piezoelectric stepper motors in the optical microscope allow for height resolution significantly better than its lateral resolution (limited by the wavelength of light).
        
        The film cross-section was also prepared (when a larger area was needed than FIB could provide) by slicing the coupon samples with a precision diamond saw perpendicular to the film and gluing the halves. A thermally evaporated silver film was deposited on the Nb$_3$Sn film surface to protect it before slicing, thereby reducing saw-induced damage. This sandwiched thin film was placed in a conductive powder mount and mechanically polished to a mirror finish for SEM analysis.

\section{Systematic Studies}

    In this section, the results of reactions between Nb and Cu-Sn under different conditions are provided. Specifically, these experiments helped determine the impact of strain from other substrates and neighboring layers near the Nb$_3$Sn film, as well as the effect of Sn concentration and mobility on the resulting Nb$_3$Sn properties. These experiments led to an optimized recipe for the RF cavity experiment, allowing further thin film characterization with quality factor measurements.

    \subsection{Sapphire Strain Experiment}\label{sec:SapphireExp}

    Initial analysis on bronze substrates experiments~\cite{withanageRapidNbSn2021}, found a depression of critical temperature by $\sim3~$K for the hot-bronze films, even when the Sn content was near maximum (26\%). Using XRD and COMSOL, the strain was determined to be dominated by CTE, with some additional strain from the growth mode. Since strain significantly affects superconducting properties in Nb$_3$Sn, a new experiment was designed to measure the $T_c$ of a film under three different strain conditions. Strain in the film resulted from thermal contraction mismatch between the sapphire substrate, Nb$_3$Sn film, and Cu-Sn layer. To isolate the contributions from each component, three sample states were characterized: (1) as-deposited film on sapphire, (2) film peeled from sapphire, and (3) film peeled and etched to remove the Cu-Sn layer.

    A 2~\unit{\um} Cu-13~at.\% Sn film was thermally evaporated in steps with 30-minute cooling intervals between each 1~\unit{\um} increment at a base pressure of $\sim3\times10^{-7}$ Torr onto a c-plane sapphire substrate. The Nb was then sputtered at 0.37~nm/s using 250~W power, 8~mTorr argon working gas, and a substrate temperature of 680~$^\circ$C with no pre-heat treatment.

\begin{figure}
        \centering
        \includegraphics[width=.9\textwidth]{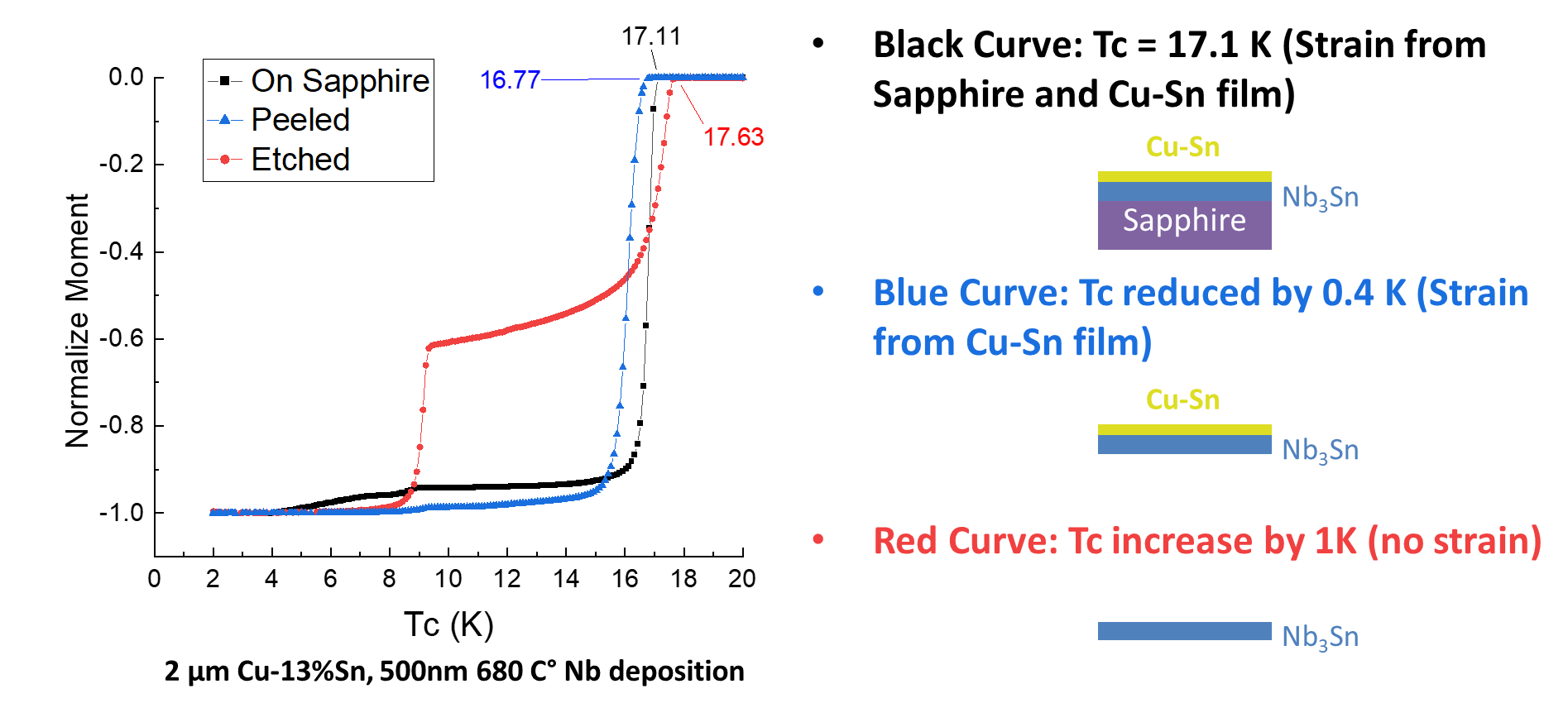}
        \caption{$T_c$ curves for a hot bronze film on sapphire, that has been peeled, and etched and measured at each stage.}
        \label{fig:SapphireStrain}
    \end{figure}

    This presented an opportunity to compare different strain states due to thermal contraction mismatch from different layers, since the thermal contraction of sapphire is close to that of Nb$_3$Sn~\cite{eastonPredictionStressState1980}, and Cu-Sn CTE is significantly higher than either (Fig.~\ref{fig:CTECuNbNb3Sn}). $T_c$ before and after peeling from the sapphire substrate, leads to a critical temperature onset decrease from 17.11~K (black curve) to 16.77~K (blue curve Fig.~\ref{fig:SapphireStrain}). We attribute this to an increased strain impact from the nearby Cu-Sn layer, with the sapphire substrate mitigating the compressive strain in the Cu-Sn film. After peeling, the thin Cu-Sn/Nb$_3$Sn bi-layer was etched to study the impact the 2~\unit{\um} Cu-Sn film had on the Nb$_3$Sn layer, and found an increase of $T_c$ onset by almost 1~K (red curve Fig.~\ref{fig:SapphireStrain}). A Nb transition also appeared after etching, suggesting that the Cu-Sn layer removed some Nb$_3$Sn material that magnetically screened the Nb from the SQUID measurement. This study reveals that the residual Cu-Sn film beneath the Nb$_3$Sn layer induces sufficient strain to depress $T_\text{c,onset}$ by $\sim$1~K. Additionally, etching uncovered unreacted Nb and Sn-poor Nb$_3$Sn phases that were magnetically screened in initial SQUID magnetization measurements, which indicates hidden inhomogeneities in the film composition.

    \FloatBarrier

    \subsection{Substrate Experiments}\label{sec:substratestrainandSn}

    By understanding the impact the substrate has on $T_c$, the effects of low Sn concentration and strain on superconducting performance can be separated. All samples used 2~\unit{\um} Cu-13~at.\% Sn films thermally evaporated in steps with 30-minute holds, followed by 500~nm Nb sputtered at 0.37~nm/s (250~W, 8~mTorr) at 680~$^\circ$C substrate temperature, with no additional pre- or post-heat treatment. This recipe matches the sapphire substrate conditions from Section~\ref{sec:SapphireExp}. These film conditions were chosen because they represent a minimum: low Sn volume in the Sn source and a relatively low reaction temperature. When Sn conditions are poor, strain and Sn effects are more readily observed.

\begin{figure}[tb]
        \centering
        \includegraphics[width=\textwidth]{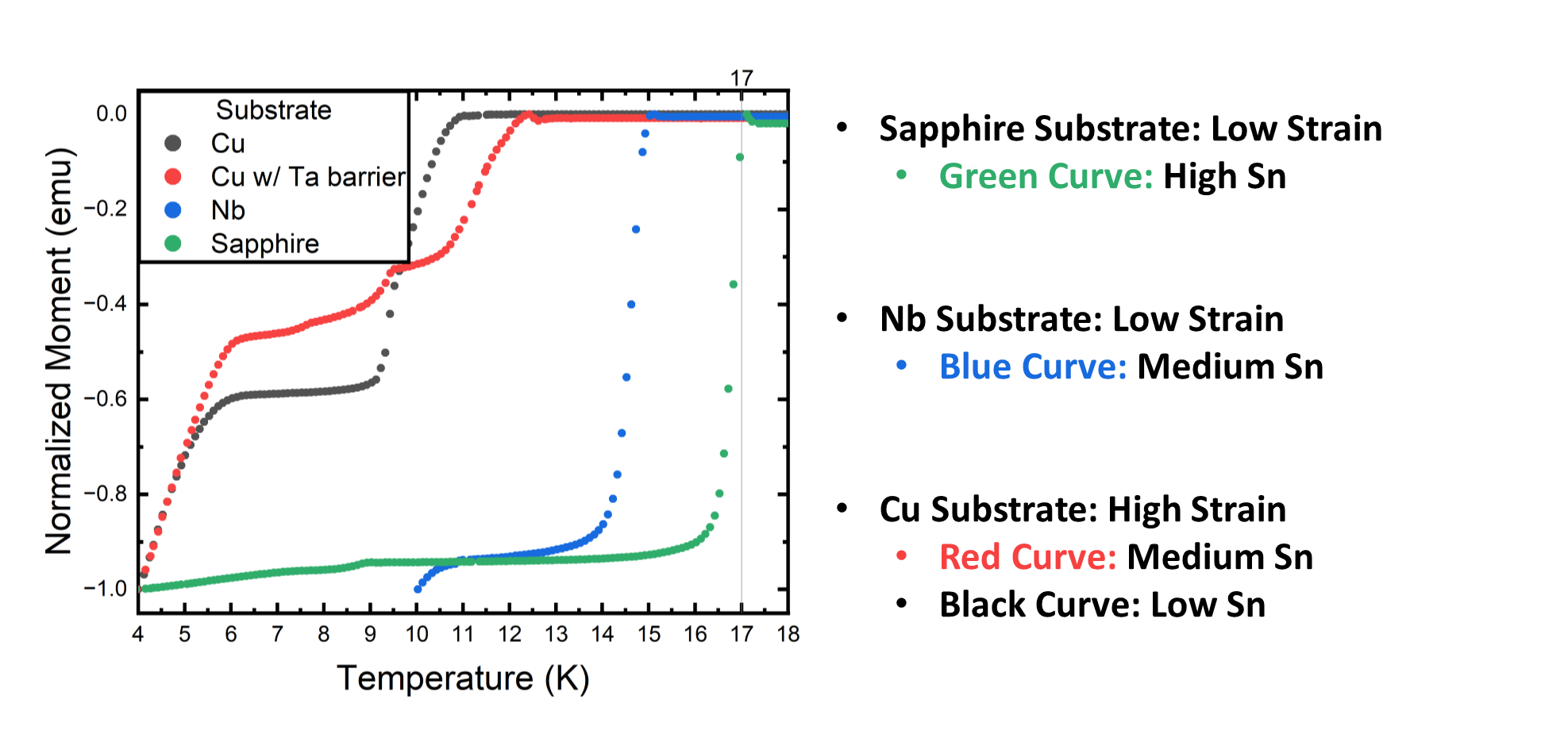}
        \caption{$T_c$ curves for substrates: Cu, Cu w/ Ta diffusion barrier, Nb, and sapphire.}
        \label{fig:SubstrateExpTc}
    \end{figure}

    Starting with the sapphire substrate (green curve in Fig.~\ref{fig:SubstrateExpTc}), a high critical temperature onset around 17~K with a very sharp transition was observed. The high $T_\text{c,onset}$ indicates high Sn concentration and low strain, with the sharp transition indicating film homogeneity. 

    In contrast, the Nb substrate sample (blue curve) exhibited a similar transition profile but was shifted downward to 15~K. Since Nb has a CTE similar to Nb$_3$Sn (Fig.~\ref{fig:CTECuNbNb3Sn}), the decreased $T_c$ can be attributed to reduced Sn concentration rather than strain effects. For the Nb substrate geometry, two Nb sources are present during the reaction: the Nb substrate itself and the deposited Nb film. This creates competing reactions for Nb$_3$Sn formation at both the substrate interface and within the deposited layer. With a finite Sn supply from the thin 2~\unit{\um} Cu-Sn film and excess Nb from both sources, the resulting Nb$_3$Sn is Sn-deficient, explaining the reduced critical temperature compared to the sapphire substrate sample.
    
    Moving to the Cu substrate with a Ta diffusion barrier (red curve in Fig.~\ref{fig:SubstrateExpTc}) sample, a further reduction in $T_c$ onset to 12 K and a broad transition was seen. This was due to significant strain from the Cu substrate and a medium Sn content. The Ta diffusion barrier blocks some Sn from migrating into the Cu substrate; however, because the original Sn content is low, this results in a tin deficiency in the Nb$_3$Sn layer. The adverse effects of the Cu substrate are even further pronounced when no Ta diffusion barrier is present, and no Sn is blocked from entering the Cu substrate, as is the case with the black curve in Fig.~\ref{fig:SubstrateExpTc}. The $T_c$ onset further decreases to 10 K, almost at the Nb transition, indicating very low Sn concentration and a considerable strain from the Cu substrate. 

    \FloatBarrier

    \subsection{Nb Substrate - High Sn Cu-Sn}\label{sec:Nbsub}

\begin{figure}[htb]
        \centering
        \includegraphics[width=.9\textwidth]{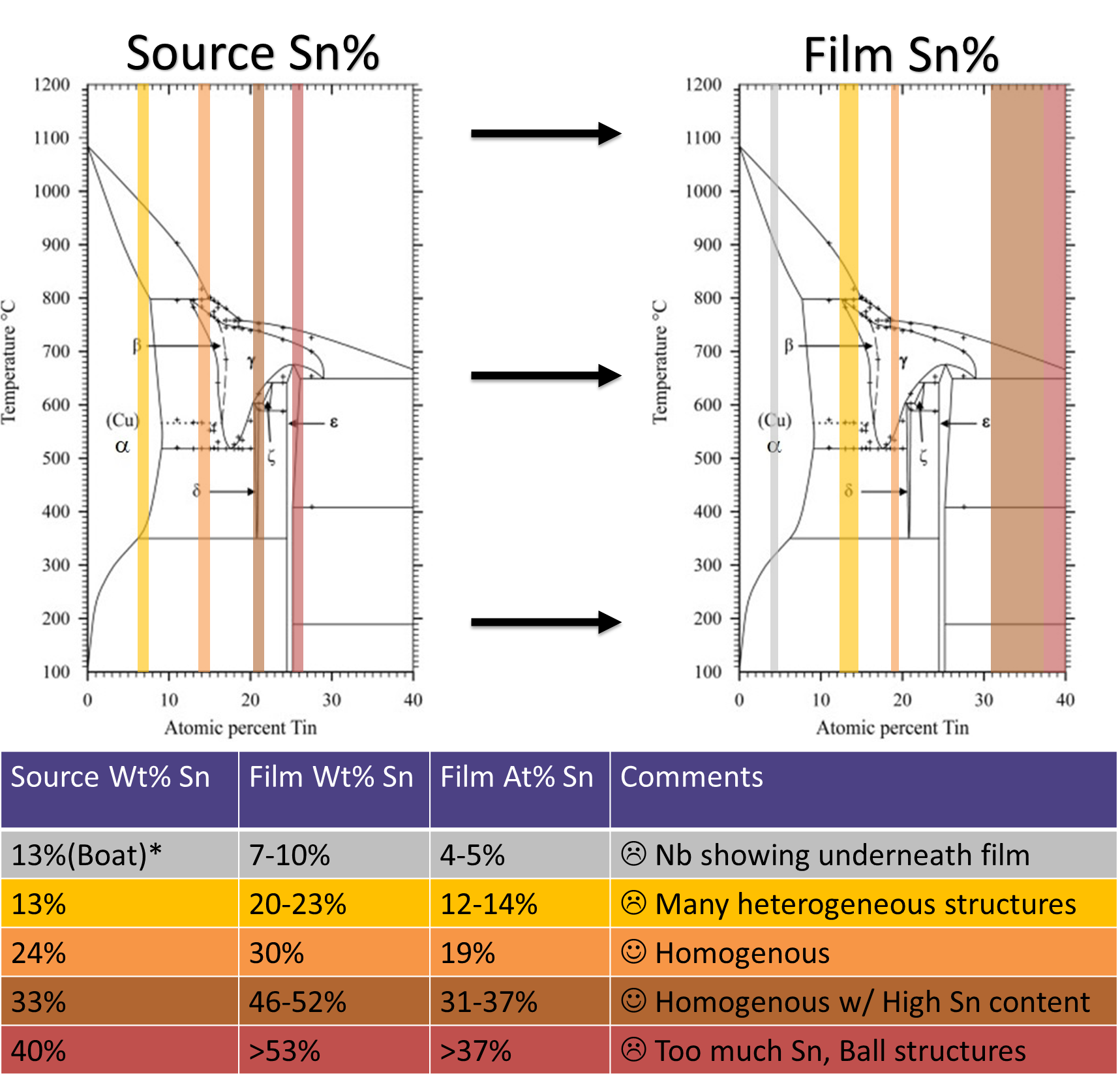}
        \caption{Cu-Sn phase diagram showing initial source Sn concentration and final Cu-Sn layer Sn concentration, with a comment on resulting microstructure.}
        \label{fig:SourcevsFilmCuSn}
    \end{figure}

\begin{figure}[H]
        \includegraphics[width=\textwidth]{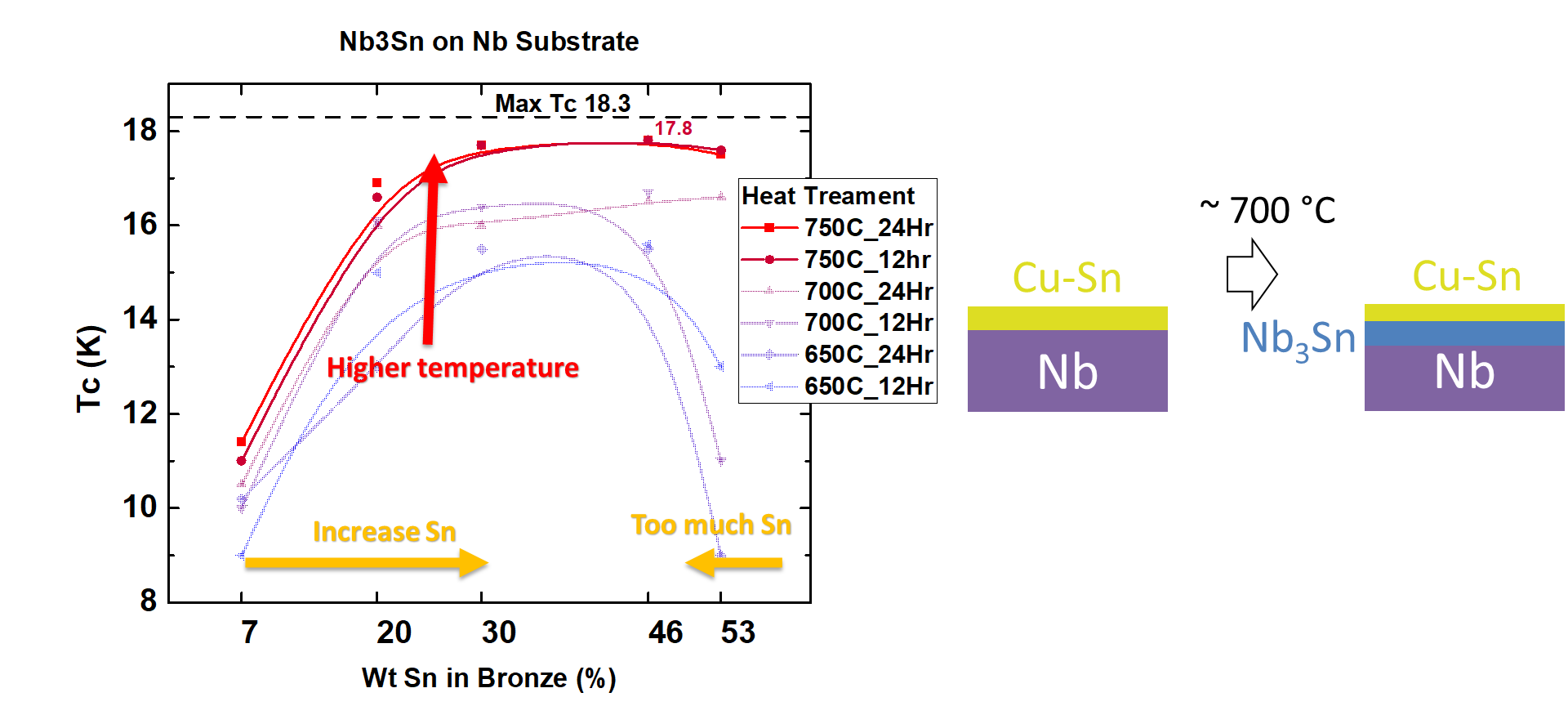}
        \caption{$Tc$ curves for Nb$_3$Sn films Nb substrates made with varying Sn content, Cu-Sn films, and heat-treatment profiles.}
        \label{fig:NbSubTcandSChem}
    \end{figure}

    The following Nb substrate experiment varied the Sn content in the Cu-Sn films (Fig.~\ref{fig:SourcevsFilmCuSn}) and heat-treatment profiles to determine an optimal recipe determined by the critical temperature (Fig.~\ref{fig:NbSubTcandSChem}). All of these films used a post-reaction method to convert Nb into Nb$_3$Sn. 

    \FloatBarrier

\begin{figure}[htb]
        \centering
        \includegraphics[width=\textwidth]{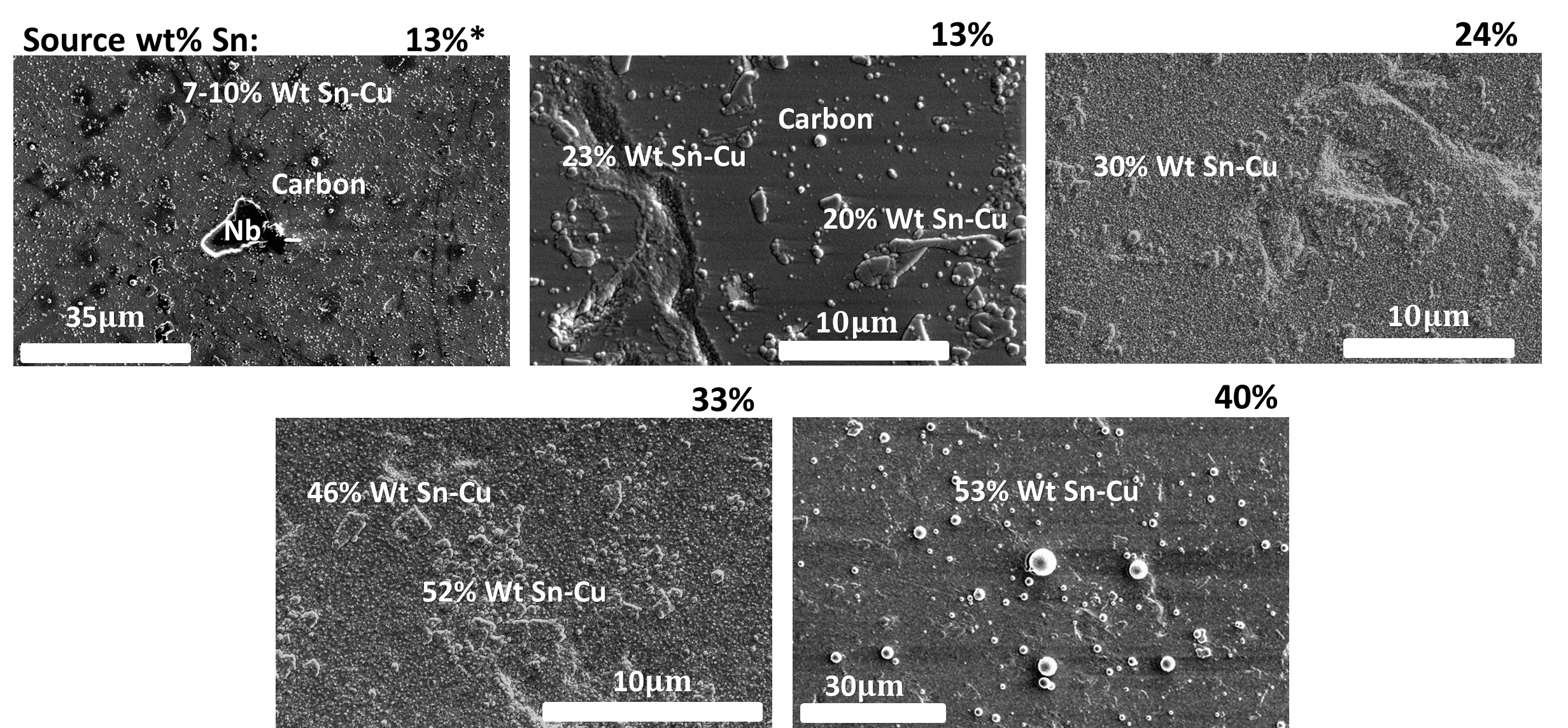}
        \caption{4 \unit{\um} Cu-Sn films with varying Sn content on Nb substrates.}
        \label{fig:BronzeOnNbSnVariation}
    \end{figure}

    Cu-Sn films with five different compositions (4, 13, 19, 33, and 37~at.\% Sn as measured by EDS after deposition and top surfaces images in Fig.~\ref{fig:BronzeOnNbSnVariation}) were thermally evaporated onto Nb substrates to a thickness of 4~\unit{\um}, deposited in four 1~\unit{\um} steps with 30-second holds between steps to allow for film cooling and strain relaxation. The thermal evaporator base pressure was $\sim3\times10^{-7}$ Torr before deposition. These samples were then transferred to a positive-pressure argon furnace equipped with a PPM oxygen filter and ultra-high-purity (UHP) argon gas at a flow rate of 0.2 L/min. The furnace was pumped down to $4\times10^{-5}$ Torr before introducing argon flow. The furnace was ramped at 10~$^\circ$C/min to the reaction temperature. Heat treatments were performed at 650, 700, and 750~$^\circ$C for either 12 or 24 hours, testing all five Sn compositions at each temperature and time combination (30 samples total). The Nb$_3$Sn formation occurred entirely within this tube furnace; no sputtering chamber processes were used for this experiment.

    Higher temperature consistently increased the critical temperature, given our reaction temperatures of $650 - 750$ $^\circ$C (Fig.~\ref{fig:NbSubTcandSChem}). This is consistent with observations in literature, which showed increased Sn mobility at elevated temperatures leading to an increased Sn content in the Nb$_3$Sn~\cite{chadmatthewfischerINVESTIGATIONRELATIONSHIPSSUPERCONDUCTING2002, hawesMeasurementsMicrostructuralMicrochemical2006, nausOptimizationInternaltinNiobiumtin2002, godekeReviewPropertiesNb3Sn2006, pongCuDiffusionNb3Sn2013}. Little change was observed in the superconducting properties between 12- and 24-hour reactions, except that the 24-hour reaction yielded a thicker Nb$_3$Sn layer. The separate tube furnace with positive-pressure argon was used for these experiments, enabling longer heat-treatment times.

    The Sn concentration plotted on the x-axis of Fig.~\ref{fig:NbSubTcandSChem} represents the composition of the deposited Cu-Sn film (measured by EDS after deposition), which differed from the starting powder composition in the evaporation boat. This is due to the difference in Sn and Cu vapor pressure, as explained in Appendix~\ref{app:Methods}. The lowest Sn concentration film (7~wt.\% Sn) was produced using the 13~wt.\% Sn starting powder, and the uncoated tungsten boat. The remaining Cu-Sn concentrations were fabricated using the alumina-coated tungsten boats.
    
    
    An optimal Sn concentration range for the final deposited Cu-Sn layer was found to be between 30 and 46~wt.\% Sn. Below 30~wt.\% Sn, the Cu-Sn reservoir could not supply sufficient Sn during reaction, producing Sn-depleted Nb$_3$Sn with reduced $T_c$. Above 46~wt.\% Sn, the Cu-Sn film exhibited very rough surface morphology, creating spatial inhomogeneities in Sn availability. This non-uniform reaction produced regions with varying Nb$_3$Sn stoichiometry, broadening the transition and depressing the $T_c$ onset.
    
    Note that these $T_c$ measurements were performed without etching away the Cu-Sn layer after reaction. As demonstrated in the sapphire substrate experiments (Section~\ref{sec:SapphireExp}), residual Cu-Sn can affect measured $T_c$ values due to strain effects, though the compositional trends observed here remain valid.

    \FloatBarrier

    \subsection{High-Sn Routes Combined with Cu Substrates}

    The final proof-of-principle experiment aimed to understand how the Sn concentration in the final Nb$_3$Sn film was affected by the Cu substrate. This experiment used three different Cu-Sn starting powder concentrations (13, 24, and 33~wt.\% Sn) thermally evaporated to 5~\unit{\um} thickness in 1~\unit{\um} steps with 30-second holds between steps. A 500~nm Ta diffusion barrier was first sputtered onto the Cu substrate at 200~$^\circ$C. The Cu-Sn layers were deposited via thermal evaporation at a base pressure of $\sim4\times10^{-9}$ Torr. After a 110~$^\circ$C, 10-minute load lock preheat, 300~nm of Nb was sputtered at 0.37~nm/s using 250~W power and 8~mTorr argon working gas (802~s total deposition time). Three reaction profiles were tested: (1) \textbf{hot-bronze}: Nb deposited at 715~$^\circ$C with 30-minute substrate preheat, no post-reaction; (2) \textbf{post-reaction}: Nb deposited at 200~$^\circ$C, followed by 715~$^\circ$C heat treatment for 4 hours; and (3) \textbf{hot-bronze + post-reaction}: Nb deposited at 715~$^\circ$C with 30-minute preheat, followed by an additional 715~$^\circ$C heat treatment for 3 hours after deposition was completed.
    
\begin{figure}[H]
        \centering
        \includegraphics[width=\textwidth]{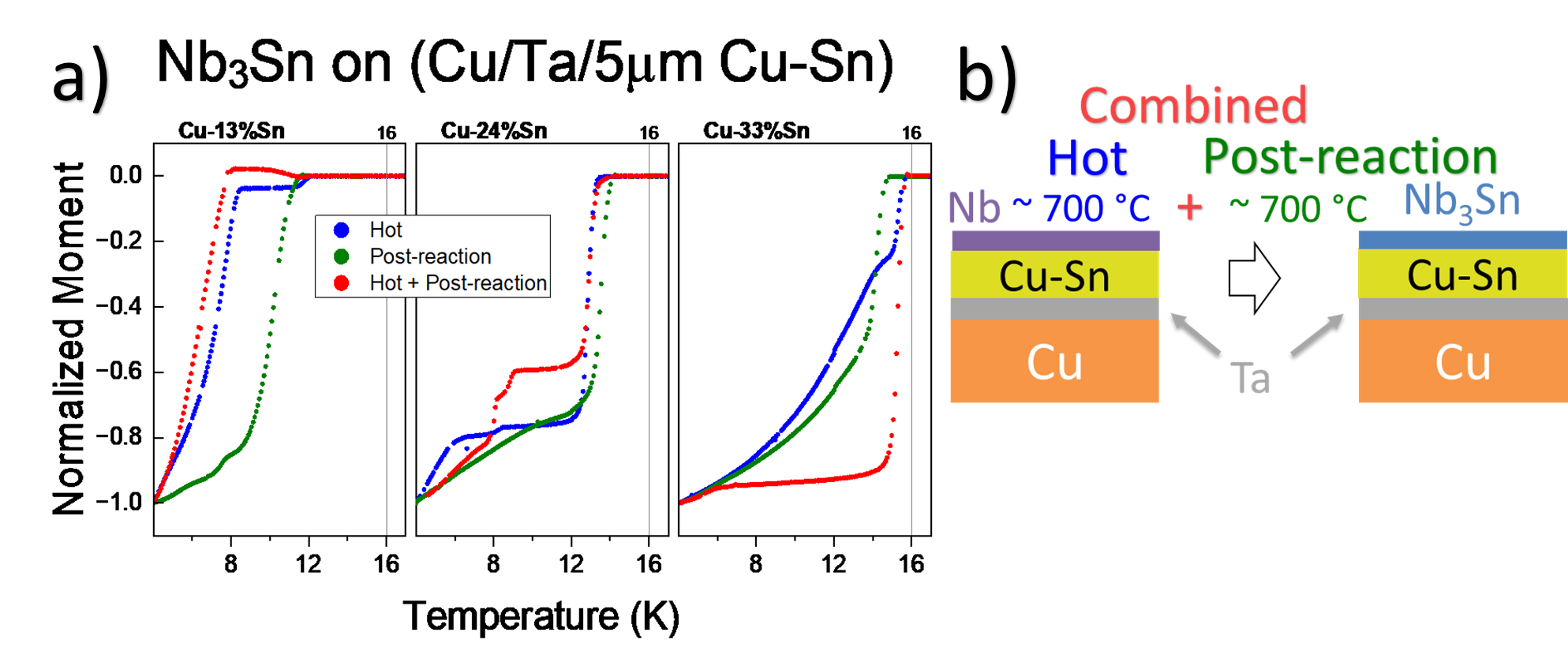}
        \caption{(a) $T_c$ curves for Nb$_3$Sn on Cu substrates with varying reactions and Sn concentrations, and (b) a schematic showing the multilayer structure.}
        \label{fig:Nb3SnCuSubSnconcentration}
    \end{figure}

    A schematic of the multilayer and all $T_c$ curves is given in Figure~\ref{fig:Nb3SnCuSubSnconcentration}. The prior work on Nb substrates identified 33 wt.\% Sn as optimal for achieving high $T_c$. This experiment again confirmed that 33 wt.\% Sn in the Cu-Sn layer was optimal, verifying for Cu substrates as well, across the full composition range: 13 wt.\% Sn produced $T_c \sim 11$~K, 24 wt.\% Sn yielded $T_c \sim 13$~K, and 33 wt.\% Sn reached $T_c \sim 16$~K. The hot-bronze + post-reaction process achieved the highest critical temperatures and sharpest transition. The reduction in $T_c$ by $\sim 2$~K from the ideal 18~K, is mainly due to the CTE mismatch between the Nb$_3$Sn and the Cu substrate.

    
    \section{Implications of the Proof-of-Principle Experiments for High-Quality \texorpdfstring{Nb$_3$Sn}{Nb3Sn} Films}

    \subsection{Hot-Bronze + Post-Reaction}\label{sec:2StepProcess}

    The hot-bronze deposition method, where Nb is sputtered at 715~°C onto Cu-Sn, offers rapid Nb-Sn reaction kinetics, approximately 12× faster than with post-reaction alone, and results in a higher initial Sn content~\cite{withanageRapidNbSn2021}. In this work, to enable hot-bronze films on Cu substrates, a Ta/Cu-Sn multilayer was deposited on a Cu substrate. The high substrate temperature provides thermal energy for immediate reaction as Nb atoms arrive at the Cu-Sn surface. As the Nb$_3$Sn film thickness increases with longer deposition time, Sn concentration at the surface appears to deplete. To address Sn depletion, a post-reaction heat treatment of 3 hours at 715~°C helped saturate Sn throughout the Nb$_3$Sn film by grain-boundary diffusion. During this step, Sn redistributes into Sn-deficient regions of the Nb$_3$Sn film, making it more homogeneous. 
    
    Post-reaction alone can also produce $\sim1$~\unit{\um} thick Nb$_3$Sn films. However, based on the results for 3-hour reactions in Fig.~\ref{fig:Nb3SnCuSubSnconcentration}, there is evidently a sluggish start to the layer growth that is avoided by the hot-bronze deposition.


    Neither approach alone achieves optimal results for this configuration on Cu (an alternative for Cu-Sn on Nb is discussed later). The combination of an initial hot-bronze deposition followed by further annealing after the deposition is stopped enables both a high critical temperature onset and a sharp superconducting transition. This demonstrates that high-quality Nb$_3$Sn films can be fabricated on Cu substrates when using the Cu-Sn route.

    
    \subsection{Nb First or Cu-Sn first}

    \begin{table}[h]
        \centering
        \caption{Microstructural trade-offs for Cu-Sn layer position}
        \label{tab:CuSn_position_comparison}
        \begin{tabular}{lcc}
        \hline
        Property & Cu-Sn First & Nb First \\
        \hline
        Hot-bronze capable? & \ding{51} & \ding{55} \\
        Nb$_3$Sn thickness uniformity & Poor (follows rough Cu-Sn) & Excellent \\
        Ta/Nb$_3$Sn interface quality & Rough (via Cu-Sn) & Sharp (direct contact) \\
        Tc (onset) & 16K & 15K \\
        $\Delta$Tc & $<$1K & $\sim$2K \\
        Requires etching? & \ding{55} & \ding{51} \\
        \hline
        \end{tabular}
    \end{table}

    So far, all methods explored place the Cu-Sn layer and the Sn source below the Nb. This requires the reaction to convert the entire Nb layer, leaving Nb$_3$Sn on the RF-facing side of the cavity. However, in this situation, the RF-facing material would be farthest from the Sn source, and could thus be depleted in Sn. While the results in the prior sections show that the hot-bronze plus post-reaction approach is viable for $\sim500$~nm thick Nb$_3$Sn layers, it would still be preferable to ensure that the Sn-rich Nb$_3$Sn region is at the RF surface. As outlined in section~\ref{sec:ourworkdescribe}, an alternative approach is to deposit Cu-Sn on Nb or a suitably prepared Cu/Nb base material, react the layers after deposition to form Nb$_3$Sn, and then remove Cu-Sn by etching or electropolishing. 

    The Nb$_3$Sn film can be synthesized with the Sn source (Cu-Sn layer) positioned either underneath or on top of the Nb film. These two geometries each have distinct advantages and limitations.

\begin{figure}
        \centering
        \includegraphics[width=.6\textwidth]{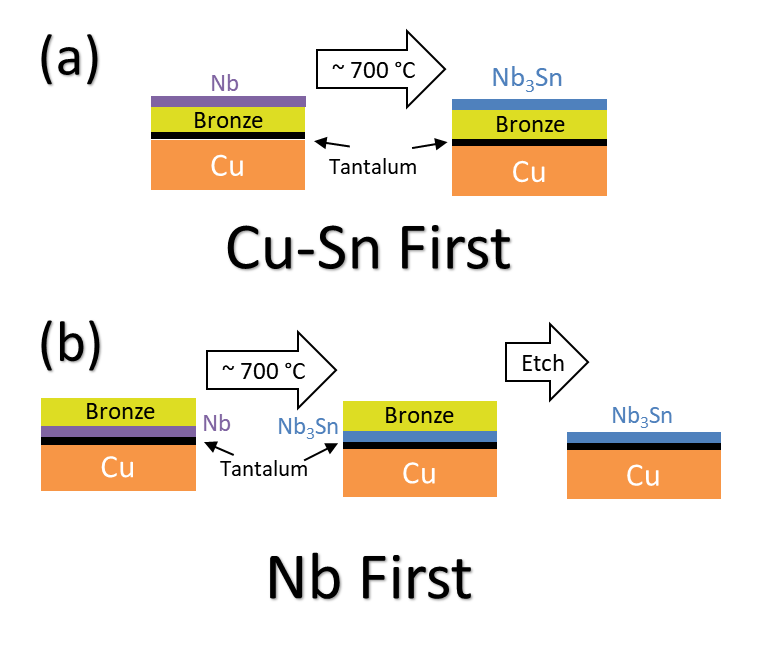}
        \caption{Schematic showing (a) hot-bronze viable route Nb$_3$Sn on bronze, and uniform microstructure sample (b) bronze on Nb$_3$Sn requiring etching the Cu-Sn away after reaction.}
        \label{fig:BronzeonNbandNbonbronzediagram}
    \end{figure}

    The hot-bronze reaction requires Nb atoms to arrive on a hot ($\sim715$~°C) Cu-Sn surface. This imposes a strict deposition order: Cu-Sn must be on the substrate and in the deposition chamber before Nb sputtering begins:

    \begin{itemize}
        \item \textbf{Cu-Sn first} (Fig.~\ref{fig:BronzeonNbandNbonbronzediagram}a): 
        Hot-bronze capable
        \item \textbf{Nb first} (Fig.~\ref{fig:BronzeonNbandNbonbronzediagram}b): 
        Hot-bronze impossible (Cu-Sn arrives after Nb already deposited)
    \end{itemize}

   A plausible Nb-first route on a copper substrate would begin with a well-polished Cu surface, followed by a diffusion barrier (Ta) and, optionally, interface layers to improve adhesion and morphology. Fortunately, it was found that smooth Ta films adhered well when deposited directly on polished Cu. Subsequent deposition of Nb resulted in exceptionally uniform morphology. A sharp interface between Ta and Nb was observed with little to no oxidation (Fig.~\ref{fig:BronzeOnNb}). Large defects were not observed, unlike the porous layers sometimes observed for Cu-Sn layers ( seen in a) Fig.~\ref{fig:FullNb3Sn10micron2}). After sputtering Ta/Nb, Cu 33~wt.\% Sn was evaporated onto the Nb layer and reacted at $715~^\circ$C for 3 hours. After the reaction, the Cu-Sn was removed by etching in ammonium persulfate (APS). The resulting Nb$_3$Sn film displayed a $T_c$ of 15~K and $\Delta T_c$ of 2~K (blue curve in Fig.~\ref{fig:Recipe123Tc}). These properties are consistent with other Nb$_3$Sn films on Cu substrates.

\begin{figure}[htb]
        \centering
        \includegraphics[width=.7\textwidth]{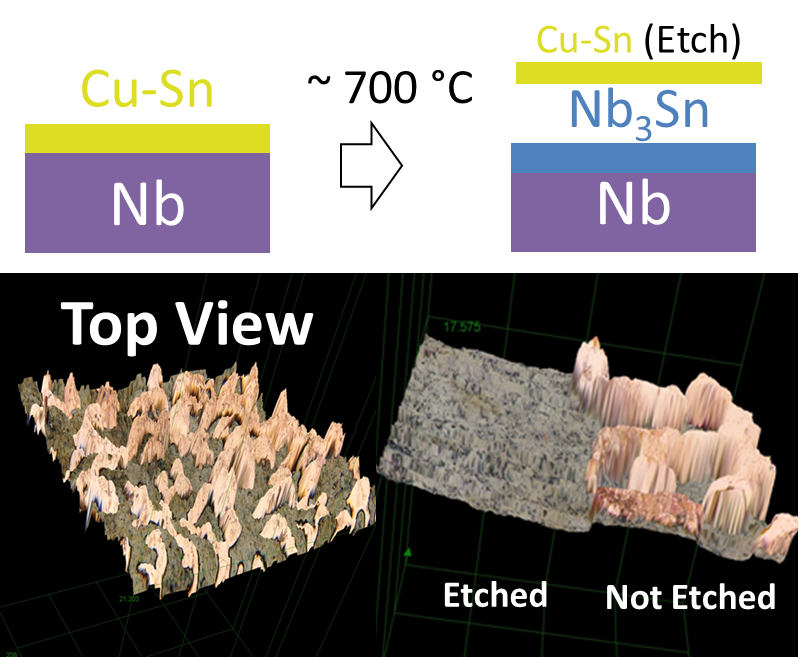}
        \caption{Schematic of the Nb$_3$Sn film on a Nb substrate, with a 3-D optical imaging showing the surface morphology before and after etching.}
        \label{fig:Nbsubetching}
    \end{figure}

    A schematic of the multilayer is shown before and after etch, and includes a top-down 3-D optical image of the film (Fig.~\ref{fig:Nbsubetching}). The Cu-Sn film was selectively removed in one area by taping one side, since the etchant could not penetrate the tape. The same etched and non-etched film cross-section was analyzed (Fig.~\ref{fig:NbsubetchSEM}). The etched and non-etched sides show continuous Nb$_3$Sn throughout the film, verifying the success of the etching process.

    \section{Three Optimized Recipes for Cu Substrates}
   
\begin{figure}[htb]
        \centering
        \includegraphics[width=.7\textwidth]{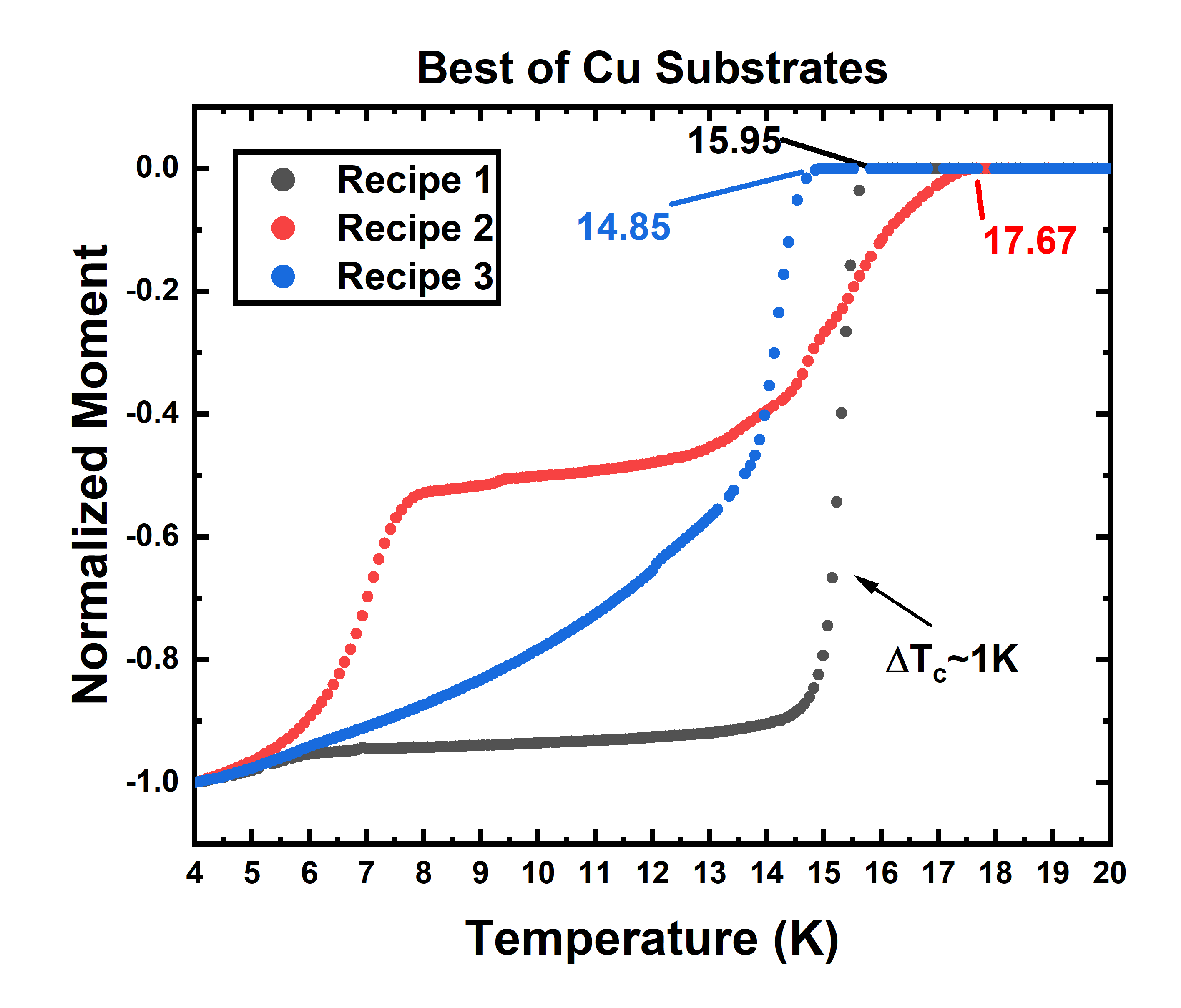}
        \caption[$T_c$ curves for best Nb$_3$Sn samples on Cu substrates.]{$T_c$ curves for best Nb$_3$Sn samples on Cu substrates. Black curve: Recipe 1 (thin Cu-Sn first) with sharp transition, $\Delta T_c \approx 1$~K. Red curve: Recipe 2 (thick Cu-Sn first) with the highest $T_c = 17.7$~K onset, but low-Sn Nb$_3$Sn near the surface. Blue curve: Recipe 3 (Nb first) with uniform morphology, but needing longer reaction.}
        \label{fig:Recipe123Tc}
    \end{figure}

    From the systematic studies on bronze~\cite{withanageRapidNbSn2021}, Nb, sapphire, and Cu substrates, three recipes were chosen optimized for $T_c$ (Fig.\ref{fig:Recipe123Tc}) and RF cavity application. 

    \subsection{Recipe 1: Thin Cu-Sn First}

\begin{figure}[htb]
        \centering
        \includegraphics[width=.5\textwidth]{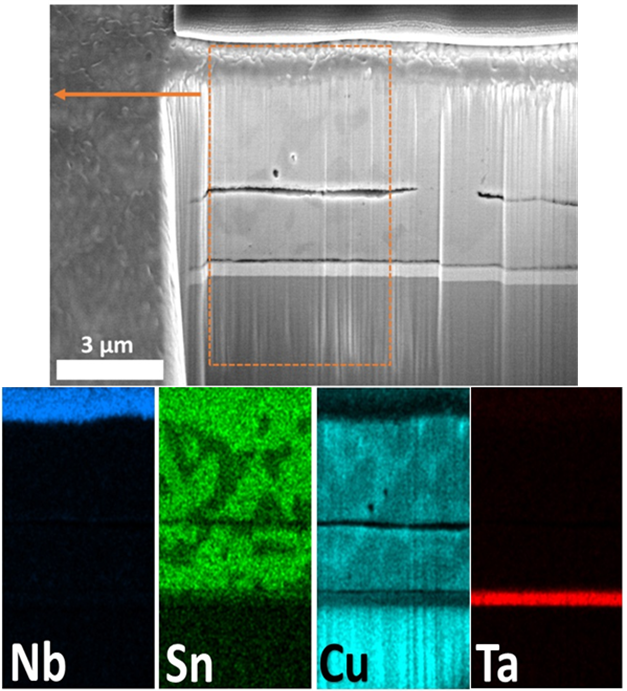}
        \caption[Recipe 1, thin Cu-Sn first: Cross-sectional microscopy imaging structure of thin Nb$_3$Sn film.]{Recipe 1, thin Cu-Sn first: Cross-sectional microscopy imaging structure of thin Nb$_3$Sn film. This sample was made using a Cu substrate, a 500nm Ta diffusion barrier, a 5 \unit{\um} Cu-33\% at. Sn film, and then depositing 300 nm of Nb at 700 $^\circ$C for 30 minutes and post-reacting for 3 hours. Delamination is observed in the CuSn layer, consistent with Ta diffusion-barrier samples.}
        \label{fig:300nmNb3sn16K}
    \end{figure}

    The superconducting transition (black curve in~\ref{fig:Recipe123Tc}) and microstructure (Figure~\ref{fig:300nmNb3sn16K}) of the most stoichiometrically uniform Nb$_3$Sn film on Cu are reported. This sample was made using a 500~nm Ta diffusion barrier, a 5 \unit{\um} Cu- 33~at.\% Sn film, then deposit 300~nm of Nb at 700 $^\circ$C, and post-react for 3 hours. This sample exhibited a $T_c$ onset of 16~K and a sharp transition of $\Delta T_c\sim1$ K. It is of note that a small amount of the film had a $T_c < 6$~K, which is not used for calculating $\Delta T_c$. The 0.9\% compressive strain estimated from the Cu substrate CTE mismatch (Fig.~\ref{fig:CTECuNbNb3Sn}) depresses $T_c$ by approximately 2~K (Fig.~\ref{fig:TcvsStrain}). A small $\Delta T_c$ indicates that the film is highly uniform in stoichiometry, owing to its thin width and ample Sn supply. Delamination and cracking were observed in the Cu-Sn layer, as seen with other Ta barrier films. This sample employed the hot-bronze + post-reaction two-step process described in Section~\ref{sec:2StepProcess}.

    This sample represents an improvement in transition sharpness compared to other Cu-compatible methods, which typically exhibited $\Delta T_c\sim 1.5-4$ K due to compositional inhomogeneities~\cite{fonnesuRecipeOptimizationSRF2025, luImpactCuSn2025}. The data for this Cu-Sn first specimen on Cu substrates is especially encouraging for two reasons: (1) the Ta diffusion barrier is evidently effective in this geometry at preventing Sn diffusion into the Cu substrate, resulting in a complete conversion of the Nb into Sn-rich Nb$_3$Sn at $700~^\circ$C reaction temperature; (2) Sn variations in the Nb$_3$Sn appear to be suppressed despite obvious multiple phases forming in Cu-Sn, which suggests high tin mobility, possibly like the situation in high-quality Nb$_3$Sn wires.

        \FloatBarrier
    
    \subsection{Recipe 2: Thick Cu-Sn First}

\begin{figure}[htb]
        \centering
        \includegraphics[width=.5\textwidth]{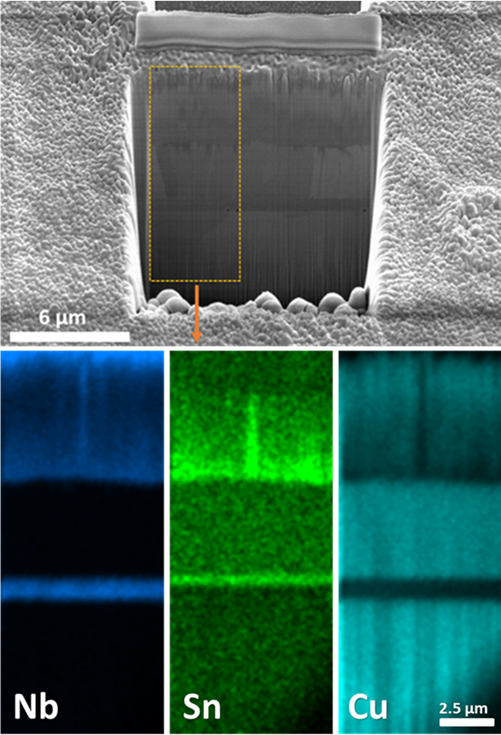}
        \caption[Recipe 2, thick Cu-Sn first: first Cross-sectional microscopy imaging multilayer structure of a thick Nb$_3$Sn film.]{Recipe 2, thick Cu-Sn first: first Cross-sectional microscopy imaging multilayer structure of a thick Nb$_3$Sn film. This sample was made on a Cu substrate, with a 500nm Nb diffusion barrier and a 5 \unit{\um} Cu-33\% at. Sn film, and then depositing 3000 nm of Nb at 700 $^\circ$C for 3 hours. No delamination or cracking is observed in the Cu-Sn layer, consistent with other Nb diffusion-barrier samples.}
        \label{fig:Thicknb3sn17K}
    \end{figure}

    This film was made using a 500 nm Nb diffusion barrier, a 5~\unit{\um} Cu-33\% at. Sn film, and a 3000 nm film of Nb deposited at 700 $^\circ$C with no post-reaction. This film took 3 hours to deposit, so it was decided that a post-reaction would only produce an inferior film due to the limited Sn supply compared to the volume of Nb. The film's critical temperature (red curve in Fig.~\ref{fig:Recipe123Tc}) and microstructure (Fig.~\ref{fig:Thicknb3sn17K}) are reported, which exhibited an unusually high $T_c$ onset of 17.7~K for a Cu substrate Nb$_3$Sn sample. It is believed that the high $T_c$ onset is a result of the thicker film having more room to relax from the CTE substrate strain. However, a broad transition occurs because the thick film has an inadequate Sn supply throughout the reaction. 
    
    This sample is nearing the upper limit of $T_c \sim 18.3$ K, proving this method is capable of creating a near-perfect $T_c$ onset on Cu substrates. This sample also demonstrates that the critical temperature depends on thickness due to strain induced by the substrate. This thickness dependence aligns with recent results from \citeauthor{Ilyina-Brunner:2019iay}~\cite{Ilyina-Brunner:2019iay}, where they found that a 30 \unit{\um} Nb diffusion barrier was thick enough to mitigate strain and chemical effects due to the Cu substrate.

    \FloatBarrier

    \subsection{Recipe 3: Nb First}\label{sec:nbfirst}

\begin{figure}[htb]
        \centering
        \includegraphics[width=.5\textwidth]{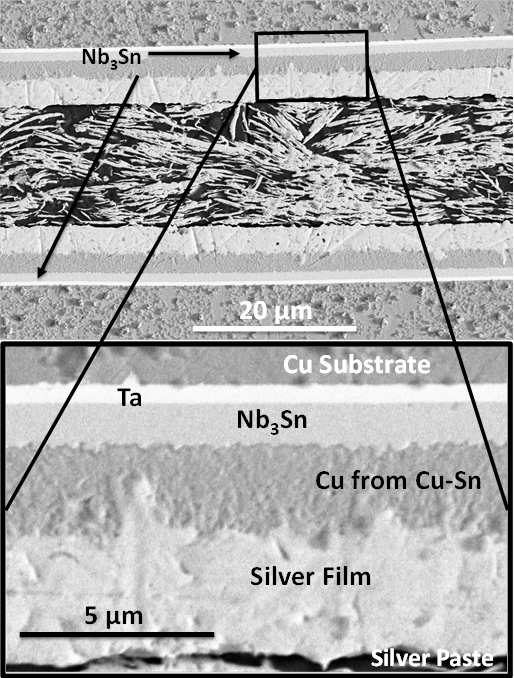}
        \caption[Recipe 3, Nb First: SEM Image of Nb$_3$Sn film made by depositing Nb onto a Cu substrate/Ta diffusion bilayer.]{Recipe 3, Nb First: SEM Image of Nb$_3$Sn film made by depositing Nb onto a Cu substrate/Ta diffusion bilayer. Then Cu-Sn was evaporated onto the Nb and post-reacted. This film had particularly uniform morphology due to the clean interface between Ta and Nb.}
        \label{fig:BronzeOnNb}
    \end{figure}

    The microstructure (Fig.~\ref{fig:BronzeOnNb}) and the critical temperature transition (blue curve in Fig.~\ref{fig:Recipe123Tc}) are reported for a film prepared using the Nb-first approach, which yields the most uniform morphology. This recipe was not fully reacted due to time constraints during the reaction, so the film exhibited a broadened transition compared to recipe 1 (black curve in Fig.~\ref{fig:Recipe123Tc}), but achieved superior lateral uniformity, critical for RF applications (Fig.~\ref{fig:BronzeOnNb}). The $T_c$ transition suggests there is some inhomogeneity and unreacted Nb. However, the unreacted Nb might have less impact because it lies below the RF surface. This sample was made with a 500 nm Ta barrier, a 500~nm Nb seed layer, a 5~\unit{\um} Cu-33~at.\% Sn layer, a post-reaction at 715~°C for 3 hours, and finally, the Cu-Sn layer was removed via APS etching. 

\chapter{Prototype Axion Detectors}

\begin{figure}[htb] 
    \centering
    \includegraphics[width=\textwidth]{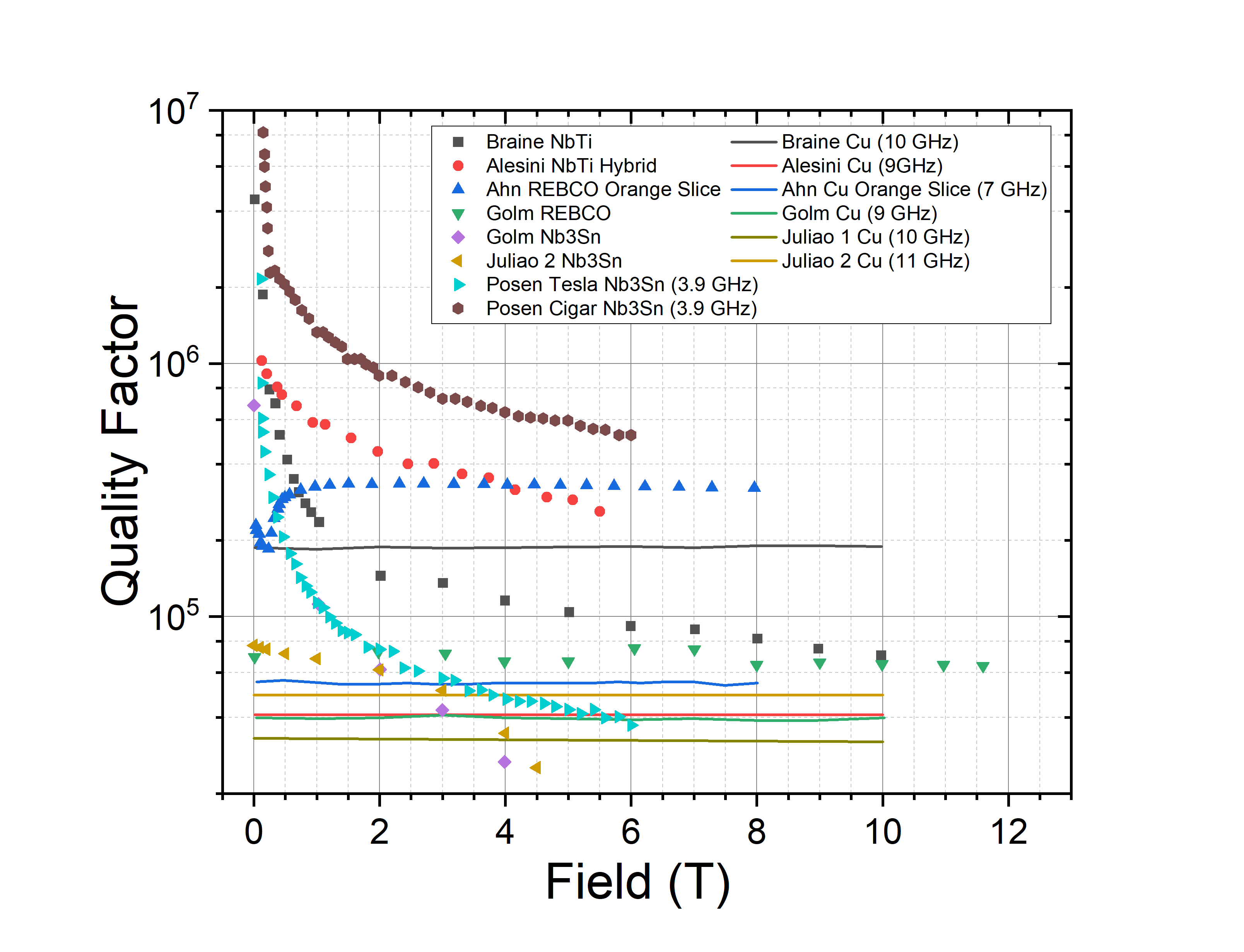}
    \caption[Quality factor vs magnetic field measurements for superconducting axion detector prototypes.]{Quality factor vs magnetic field measurements for superconducting axion detector prototypes. All cavities, except the Posen cavities, had a comparison Cu cavity with the same geometry. The cavities have different frequencies, so data points cannot be compared directly~\cite{ahnSuperconductingCavityHigh2020, Braine:2024nzi, posenNb3SnSuperconductingRadiofrequency2022, Golm:2021ooj, Alesini:2019ajt}.}
    \label{fig:axionprototypes}
\end{figure}

Chapters 1, 2, and 3 explained why an axion haloscope is placed in a large magnetic field. It was also explained why a superconducting cavity (compared to the present copper cavities) increases the scan rate, allowing faster probing across the broad range of possible axion masses. It is argued in Chapter 5 that the available superconducting ductile metal cavities (Bulk Nb) and alloys (NbTi) are not suitable for axion detectors because their maximum upper critical field lies at or below the fields being employed in existing and planned axion searches. This chapter surveys superconductor axion detector prototypes developed for high field and high $Q$, as summarized in the $Q$ vs $B$ plot in Fig.~\ref{fig:axionprototypes}.

    \section{Nb-Ti Cavities}

    NbTi's upper critical field is too low for this material to be applicable for planned high-field applications. Prior experiments conducted at lower magnetic fields~\cite {ADMX:2023rgo} and development efforts with easy-to-fabricate Nb-Ti help us understand the dependence of $Q$ on the magnetic field for a given superconducting surface orientation with respect to the magnetic field.

    \subsection{ADMX Nb-Ti Cavity}
    
    ADMX currently uses pure copper cavities, but recent work has fabricated a bulk NbTi superconducting cavity, which has been shown to increase sensitivity by $10\times$ compared to Cu at zero field~\cite{Braine:2022qal}. 
    
    \begin{figure}[htb]        
        \centering
        \includegraphics[width=.5\textwidth]{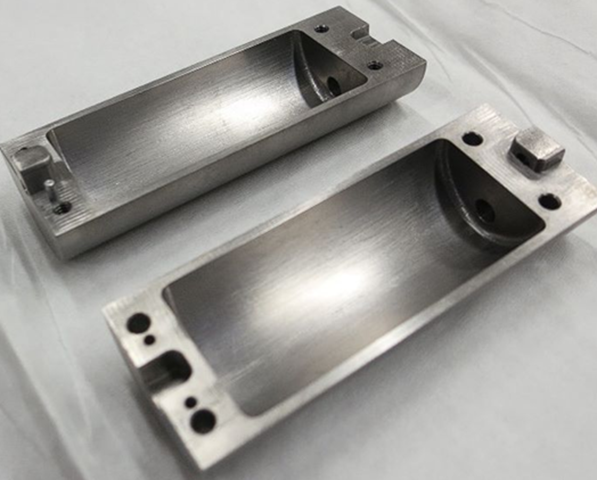}
        \caption{NbTi Bulk Cavity~\cite{Braine:2022qal}.}
        \label{fig:TomNbTi}
    \end{figure}
    
\begin{figure}[htb]
        \centering
        \includegraphics[width=.5\linewidth]{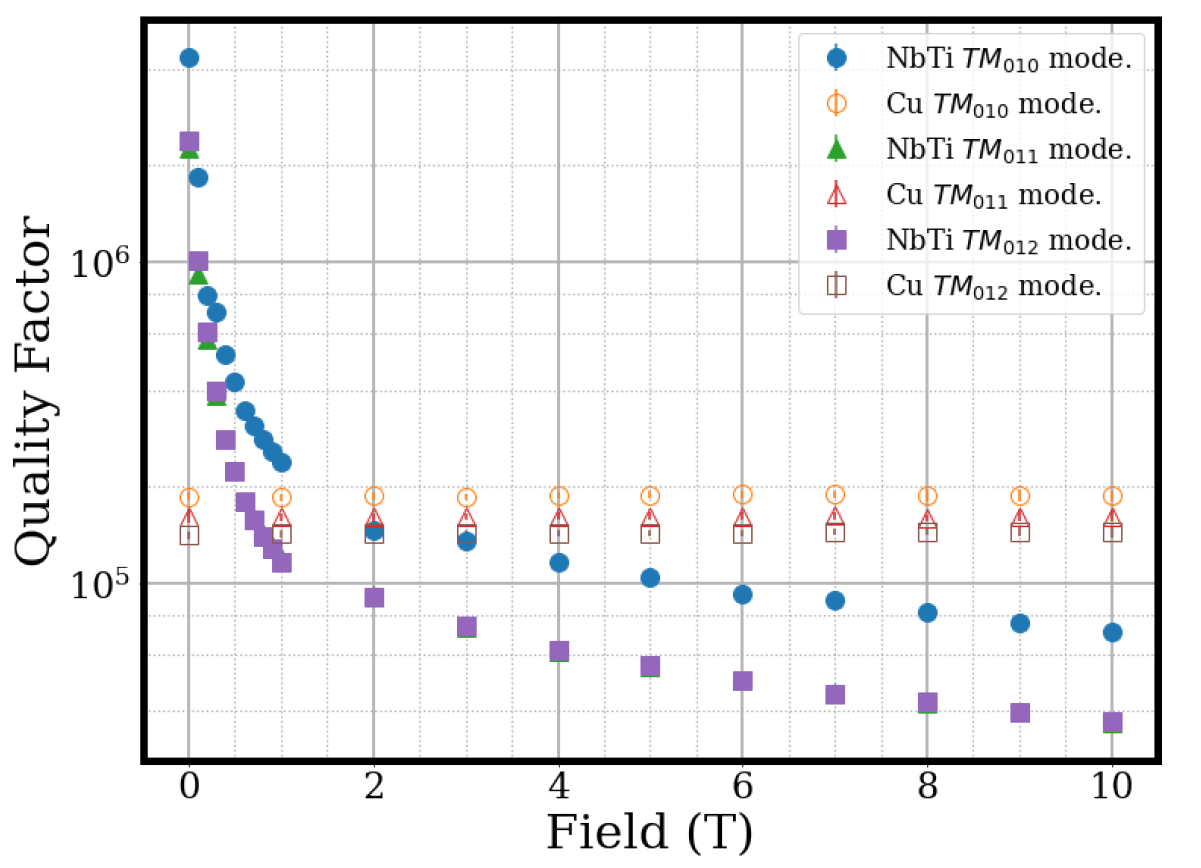}
        \caption{NbTi versus copper cavity Quality factor vs Magnetic field~\cite{Braine:2022qal}.}
        \label{fig:NbTiQvsT}
    \end{figure}

\begin{figure}[htb]
        \centering
        \includegraphics[width=0.5\linewidth]{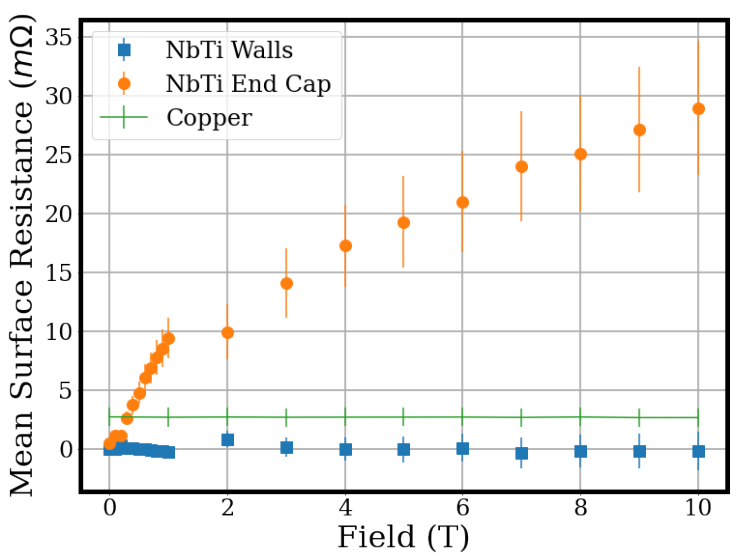}
        \caption{NbTi surface resistance for both the walls and end caps calculated using a mode decomposition technique, compared to the copper cavity resistance~\cite{Braine:2022qal}.}
        \label{fig:NbTiendcapsl}
    \end{figure}

    \citeauthor{Braine:2022qal}~\cite{Braine:2024nzi} modeled and machined a Cu and a bulk NbTi prototype cavity that fits inside a laboratory-size magnet (9~T, 11~GHz, 26~mm diameter). The Cu cavity achieved a quality factor of 180,000 at 2~K, with behavior remaining constant up to 9~T (suggesting that magnetoresistance is not significant). In contrast, the Nb-Ti cavity had a quality factor of 1 million but dropped to 75,000 at 9~T, as seen in~\ref{fig:NbTiQvsT}. A mode decomposition technique introduced by~\cite{Braine:2024nzi}, using different cavity modes and their varying geometric factors, he was able to determine that the NbTi endcaps were more resistive than Cu after just 0.3~T. However, the NbTi wall resistance was at a constant, very low resistance all the way out to 9~T (seen in Fig.~\ref{fig:NbTiendcapsl})~\cite{Braine:2024nzi}. The dependence of surface resistance on magnetic field orientation is explained in Sec.~\ref{sec:campbell}.
    
    \FloatBarrier

    \subsection{European Nb-Ti Cavity}
    
    A European group coated a 7 and 9~GHz Cu cavity with a NbTi film~\cite{Alesini:2019ajt, marconatoNbTiThinFilmSRF2024}. The cavity geometry was optimized for minimal magnetic field components perpendicular to the walls, using tapered normal conducting endcaps. This design masked the endcaps during deposition, allowing for a hybrid design with NbTi-coated walls and Cu endcaps. With this optimized design, the quality factor reached $10^6$ at zero field and $10^5$ up to 9~T for the best 7~GHz cavity. 
    
    \begin{figure}[H]        
        \centering
        \includegraphics[width=.5\textwidth]{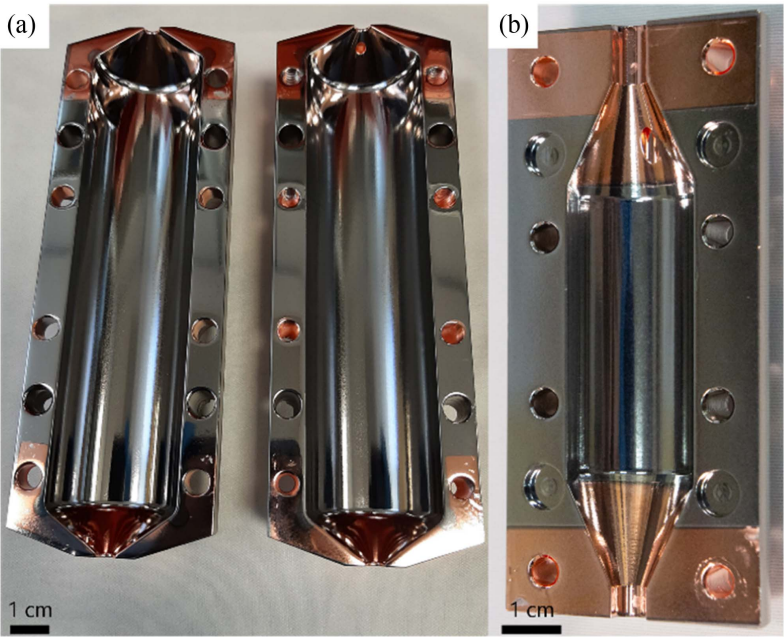}
        \caption{(a) 7-GHz and (b) 9-GHz cavities halves made by the European collaboration~\cite{Alesini:2019ajt, marconatoNbTiThinFilmSRF2024}.}
        \label{fig:AlesaniNbTi}
    \end{figure}
    \FloatBarrier
    
\section{REBCO Cavities}

REBCO cavities are a strong candidate for high-field applications, as they are already optimized in tape form by third-party high-temperature superconductor (HTS) manufacturers. Necessary research related to the fabrication of these REBCO cavities focuses on extracting the REBa$_2$Cu$_3$O$_7$ (RE = rare earth element) superconductor from the materials encapsulating it in the manufactured conductor and soldering the extracted material to the resonating cavity, with the REBCO surface face up. There will be gaps between superconducting tapes, so the REBCO tapes should have the maximum possible width to reduce the number of discontinuities.

\begin{figure}[H]        
    \centering
    \includegraphics[width=\textwidth]{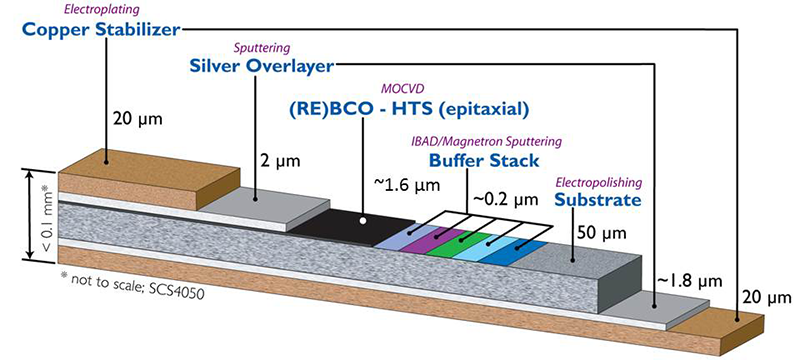}
    \caption{A schematic of the REBCO HTS tape’s architecture~\cite{superpowerOurTechnologySuperPower}.}
    \label{fig:REBCO}
\end{figure}

\subsection{South Korean REBCO Cavity}\label{sec:KoreanRebco}

\begin{figure}[H]        
    \centering
    \includegraphics[width=.5\textwidth]{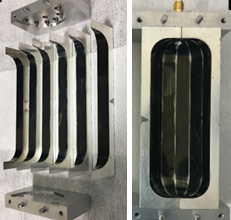}
    \caption{REBCO orange slice cavity~\cite{ahnSuperconductingCavityHigh2020}.}
    \label{fig:AhnREBCO}
\end{figure}

The Institute for Basic Science (IBS) Center for Axion and Precision Physics Research (CAPP) our of Korea, made a REBCO cavity with a stainless steel substrate (Fig.~\ref{fig:AhnREBCO}) for axion detection~\cite{ahnSuperconductingCavityHigh2020}, and achieved the highest quality factor in high magnetic field ($3\times10^5$ at 9~T). They delaminated the REBCO from either the buffer-layer side (between the REBCO superconductor and the Hastelloy tape) or the silver side, and soldered the plated Cu on the outside of the conductor directly to the cavity. REBCO's upper critical field exceeds 100~T at low temperature~\cite{bai40SuperconductingMagnet2020}, which is a possible reason why $Q$ is practically independent of field on the plot in~\ref{fig:axionprototypes}. The REBCO tapes had gaps that aligned with the external magnetic field (solenoid magnet). This led to currents in the TM$_{010}$ mode that do not cross the gaps between tapes.

\FloatBarrier

\subsection{European CERN REBCO Cavity}

\begin{figure}[H]        
    \centering
    \includegraphics[width=.5\textwidth]{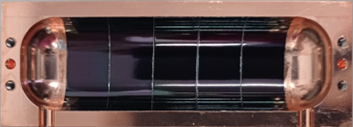}
    \caption{A REBCO cavity made with the tapes perpendicular to cavity axis~\cite{Golm:2021ooj}.}
    \label{fig:GolmREBCO}
\end{figure}

\citeauthor{Golm:2021ooj}~\cite{Golm:2021ooj} made a REBCO cavity that showed promising results, with minimal magnetic field degradation out to 9~T (Fig.~\ref{fig:GolmREBCO}). This REBCO cavity has tapes wound circumferentially, perpendicular to the cavity axis. A dipole magnet generated the external magnetic field, so the field remained parallel to the REBCO seams. This prevents the TM$_{010}$ mode currents from crossing the gaps between tapes.

\FloatBarrier

\section{\texorpdfstring{Nb$_3$Sn}{Nb3Sn} Cavities}

Section~\ref{sec:whynb3sn} explains the rationale for interest in Nb$_3$Sn, which is the material explored in this dissertation.
                        
    \subsection{Sn Vapor \& Optimized Geometry}
    
\begin{figure}[H]         
        \centering
        \begin{subfigure}{0.45\textwidth}
        \centering
        \includegraphics[height = 4cm, angle = 90]{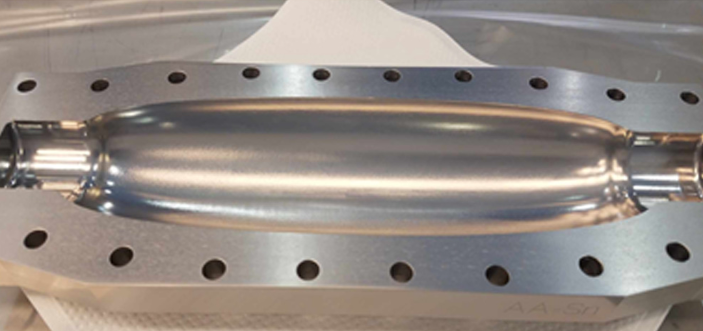}
        \caption{}
        \label{fig:Cigar}
        \end{subfigure}
        \begin{subfigure}{0.45\textwidth}
        \centering
        \includegraphics[width = .7\textwidth]{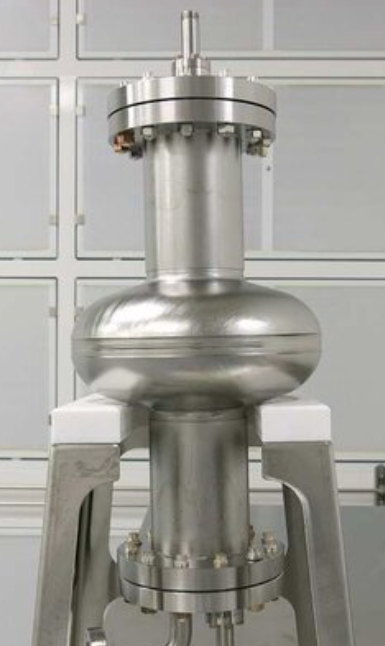}
        \caption{}
        \label{fig:Teslacavity}
        \end{subfigure}
        \caption{Two different cavity geometries: a) optimized for high field use called the cigar-shaped cavity at 3.9 GHz~\cite{Posen:2022tbs} and b) a donut-shaped cavity designed for high-velocity particle acceleration at 1.3 GHz~\cite{jiangFeasibilityStudySRF2013}.}
        \label{fig:combinedcigartesla}
    \end{figure}

\begin{figure}[H]        
        \centering
        \includegraphics[width=.7\textwidth]{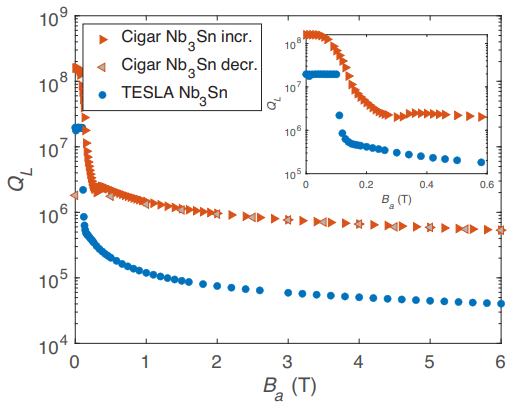}
        \caption[Loaded quality factor versus applied magnetic field for a small 3.9 GHz TESLA cavity shape and the high field designed Cigar cavity.]{Loaded quality factor versus applied magnetic field for a small 3.9 GHz TESLA cavity shape and the high field designed Cigar cavity. Measurements were made in both increasing and decreasing fields for the cigar-shaped cavity~\cite{Posen:2022tbs}.}
        \label{fig:SamCigarQvsB}
    \end{figure}

    \citeauthor{Posen:2022tbs}~\cite{Posen:2022tbs} has taken the Sn vapor method (Sec.~\ref{sec:Snvapordetail}) developed for accelerators and applied it to a specialized cigar-shaped cavity. This shape is designed to reduce magnetic field components perpendicular to the superconducting walls when the external field is aligned with the cavity's long axis. When the DC magnetic field is perpendicular to the superconducting surface, the flux lattice reacts maximally to the Lorentz force from the AC currents, which leads to vortex motion and higher losses (Section~\ref{sec:campbell}). RF testing out to 6~T, comparing the donut-shaped cavity (Fig.~\ref{fig:SamCigarQvsB}) produced by the same process, suggested that the change to a cigar-shaped cavity showed an order-of-magnitude increase in $Q(B)$ (Fig.~\ref{fig:combinedcigartesla}), effective in reducing vortex-induced effects.

    
    \FloatBarrier
        
    \subsection{Stoichiometric \texorpdfstring{Nb$_3$Sn}{Nb3Sn} Target}
        
\begin{figure}[H]  
        \centering
        \includegraphics[width=.5\textwidth]{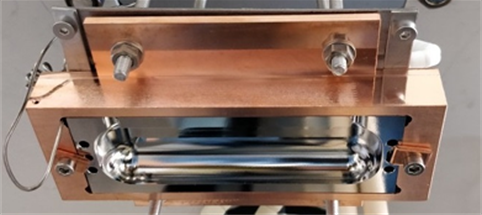}
        \caption{The Nb$_3$Sn cavity made by sputtering from a stoichiometric Nb$_3$Sn target~\cite{Golm:2021ooj}.}
        \label{fig:GolmNb3sn}
    \end{figure}

    Using a similar recipe to the coupon samples from the stoichiometric Nb$_3$Sn target group (discussed in Sec.~\ref{sec:coatings}), \citeauthor{Golm:2021ooj}~\cite{Golm:2021ooj} set out to make a superconducting 9 GHz axion cavity using a Nb$_3$Sn thin film on bulk Cu (Fig.~\ref{fig:GolmNb3sn}) using a similar recipe described in Sec.~\ref{sec:StoichiometricDetail}. Specifically, this experiment used a high-impulse magnetron sputtered (HIPIMS) Ta diffusion barrier (for densification) directly on the Cu substrate, and a stoichiometric Nb$_3$Sn target sputtered at elevated temperature. 
            
    \FloatBarrier


\chapter{Superconducting RF Results}\label{sec:CavityRF}

In this dissertation, the critical temperature and the width of the magnetic transition to the superconducting state have been used to assess the quality of Nb$_3$Sn films and to infer information about composition homogeneity and strain state. However, magnetometry techniques do not provide information about RF behavior, making quality factor measurements in a resonator setup a more conclusive evaluation of superconducting films for RF cavity applications. Quality factor measurements in a magnetic field are especially useful for providing feedback to film synthesis that is directly relevant to axion detector applications, whether that field is entirely composed of the AC RF magnetic field or a combination of a large background DC field and a small resonator AC field.

Two strategies were used to measure the RF properties of superconducting samples:
\begin{enumerate}
    \item \textbf{Small-sample method}: Place a small superconducting sample in a large resonant cavity of known material, and extract the quality factor of the superconducting sample
    \item \textbf{Full-coating method}: Coat the entire resonating surface with the superconducting material, and measure the total quality factor
\end{enumerate}

\noindent For the small-sample method (1), the measured $Q$ depends on both the sample and the host cavity through a weighted average:

\begin{equation}
    R_{s,\text{total}} = \frac{\sum_i A_i R_{s,i}}{\sum_i A_i}
\end{equation}

\noindent where $A_i$ is the surface area and $R_{s,i}$ is the surface resistance of region $i$. The measured quality factor is then $Q_\text{measured} = G/R_{s,\text{total}}$, where $G$ is the geometric factor. If the host cavity has an area $A_\text{host} \gg A_\text{sample}$ and $R_{s,\text{host}} \gg R_{s,\text{sample}}$, the host resistance dominates the measurement regardless of the sample's superior properties. This error can be reduced by using a high-$Q$ superconducting host cavity where the sample's contribution is more significant.

In both methods, antenna coupling losses can be accounted for (Sec.~\ref{sec:Qantenna}). However, the full-coating method (2) requires separate calculation of RF leakage at seams (Sec.~\ref{sec:QRFleakage}), which introduces additional uncertainty in the extracted intrinsic quality factor. When multiple surfaces interact with the RF field, as occurs in hybrid geometries using method~(2) and always in method~(1), the intrinsic $Q$ of the surface of interest can be extracted using the calculations described in Sec.~\ref{sec:QHybrid}. Results from both methods are described below, though this work primarily employed the full-coating method.

\section{RF Results for \texorpdfstring{Nb$_3$Sn}{Nb3Sn} on Bronze Coupons}

Multiple accelerator groups offered to collaborate during this dissertation project to test the quality factor of our films using method~(1). No group was able to offer $Q_0$ as a function of DC field $B_0$, however, Nb$_3$Sn films on bronze made during this work were tested for quality factor vs. temperature at zero field at SLAC National Accelerator Laboratory using a mushroom cavity. 

 \subsection{Mushroom Cavity Testing}\label{sec:mushroomCavity}

\begin{figure}[htb]
            \centering
            \includegraphics[width=0.7\linewidth]{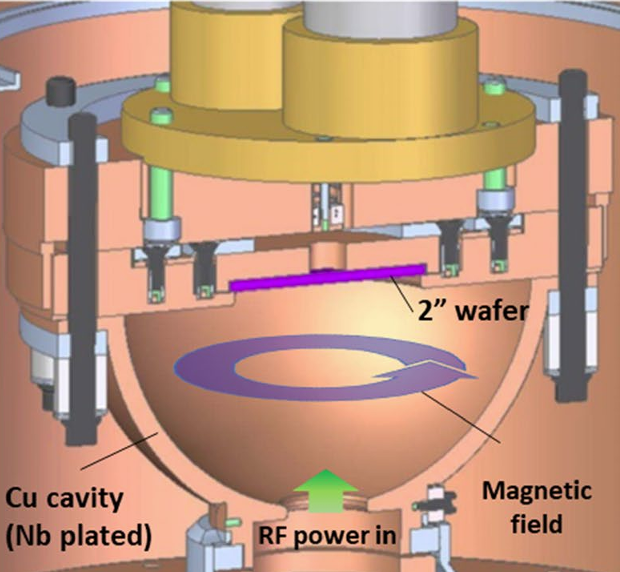}
            \caption{SLAC mushroom cavity schematic, where the 2-inch wafer is the superconducting sample of interest, and the Cu cavity below is the host structure~\cite{sundahlDevelopmentCharacterizationNb3Sn2021}.}
            \label{fig:MushroomCavity}
        \end{figure}

    Mushroom cavities (Fig.~\ref{fig:MushroomCavity})~\cite{tantawiSuperconductingMaterialsTesting2007, thomsonSRF2015Whistler2015} are RF test structures used to measure the surface resistance of superconducting materials on flat disk samples. The design uses a 2-inch planar sample pressed against a Nb or Cu host cavity, with RF fields flowing in the radial direction concentrated parallel to the sample surface. The mushroom cavity experiment exploits higher-order frequency modes that localize surface currents to small regions of the cavity, allowing the use of a small superconducting sample confined within the cavity while avoiding high-resistance boundaries. Thermometers on the back side of the sample record temperature, with cooling provided by a cold head attached to the cavity. This setup allows measurement of $Q$ vs $T$ on small test samples without making a full cavity. SLAC developed mushroom cavity measurements for testing superconducting thin films, and similar setups exist at other accelerator labs for material screening \cite{oikawaFabricationNbMushroom2019}. 

    Mushroom-cavity testing separates material properties from full-cavity complications such as seam resistance, geometric errors, or coupling losses. Testing multiple small samples from one batch uses less material and time than coating full cavities. However, mushroom measurements have significant limitations: they cannot perform magnetic-field dependence measurements because they are too large to fit inside magnets, and the Nb host cavities would also break down in high magnetic fields. This means mushroom cavity tests cannot assess actual axion detector conditions near 9~T. 
    
    The Cu-Sn first approach discussed in Sec.~\ref{sec:nb3snresults} had its origin in depositing Nb on bronze substrates~\cite{withanageRapidNbSn2021}. An opportunity to measure these films in RF became available. Although it took longer than expected due to the pandemic and other delays, this work was completed because the results are directly relevant to the coatings being developed and the questions being investigated in this dissertation. Mushroom cavity experiments were never able to explore everything required, and coating the entire cavity (option~2) was always envisioned.

\begin{figure}[htb]
            \centering
            \includegraphics[width=0.9\linewidth]{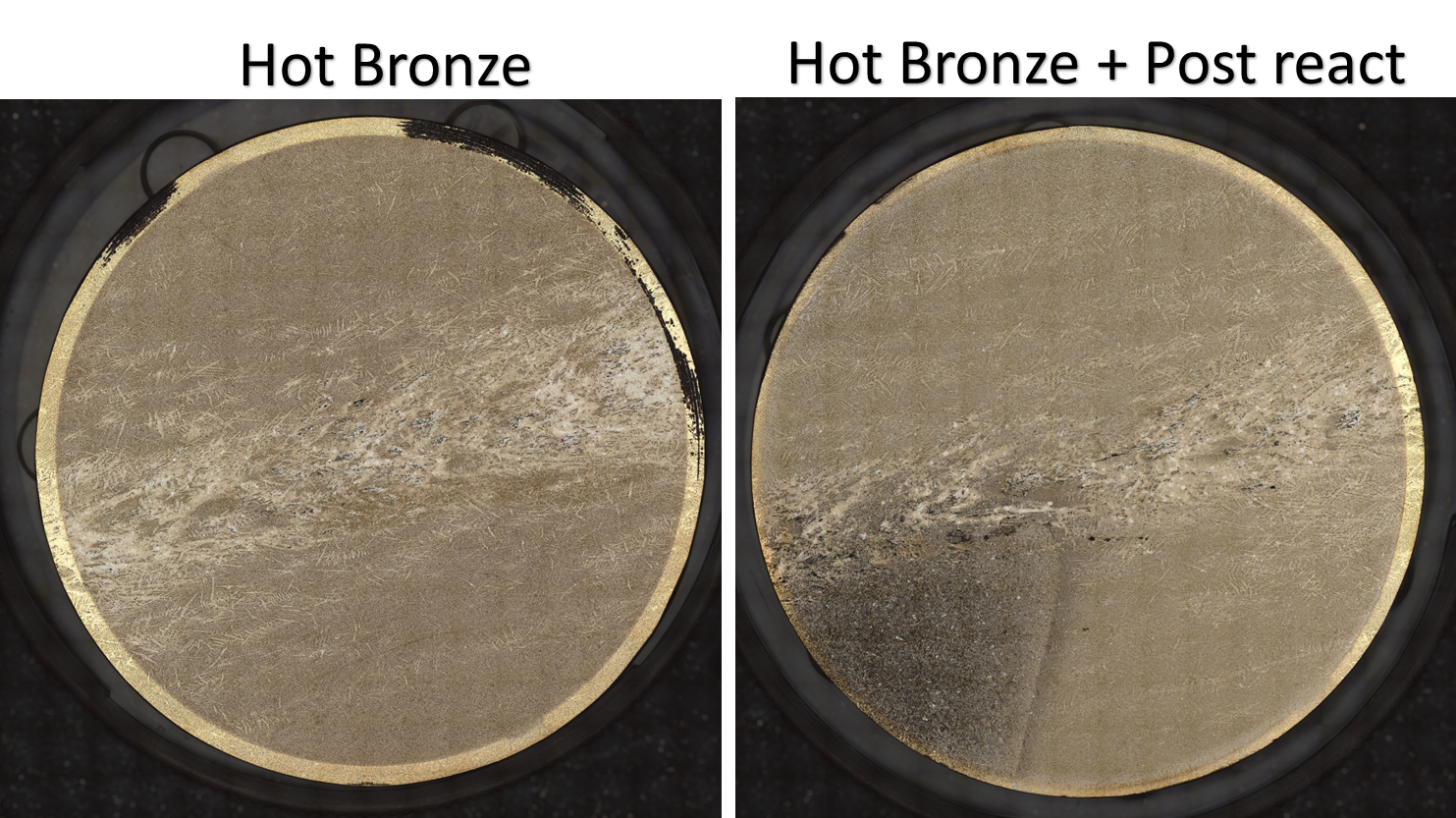}
            \caption[The 2-inch bronze substrate disks, after Nb deposition and reaction, showing both the hot-bronze reaction, and the hot-bronze + a post-reaction.]{The 2-inch bronze substrate disks, after Nb deposition and reaction, showing both the hot-bronze reaction, and the hot-bronze + a post-reaction. The texture visible in the coatings is indicative of the underlying texture of the bronze substrates. These samples underwent the same heat-treatment as in bronze substrate samples found in \citeauthor{withanageRapidNbSn2021}~\cite{withanageRapidNbSn2021}.}
            \label{fig:2inchBronzeSub}
        \end{figure}

\begin{figure}[htb]
            \centering
            \includegraphics[width=0.5\linewidth]{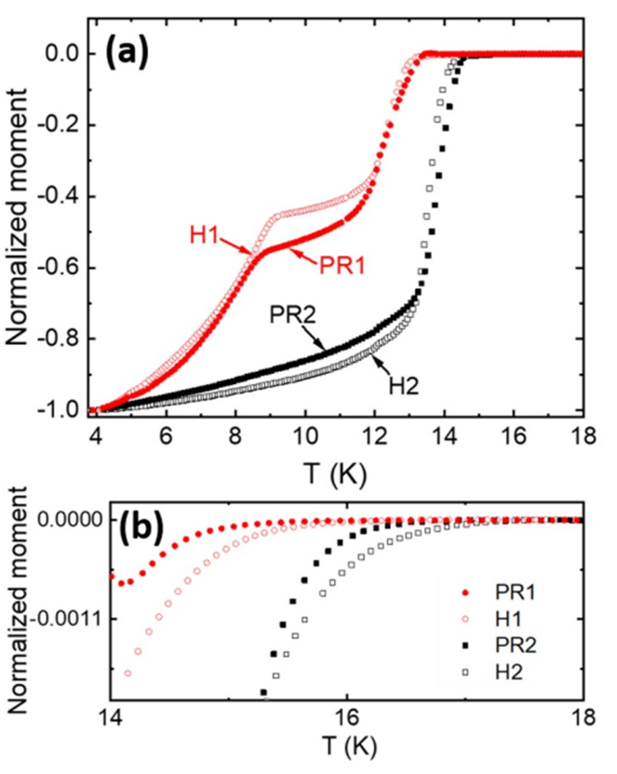}
            \caption{Samples marked with H2 and PR2 are coupon samples that underwent the same recipe as the 2-inch disk samples from the mushroom cavity measurement: hot-bronze, and hot-bronze + post-reaction, respectively~\cite{withanageRapidNbSn2021}.}
            \label{fig:WenuraBronzeTc}
        \end{figure}

\begin{figure}[H]
            \centering
            \includegraphics[width=0.7\linewidth]{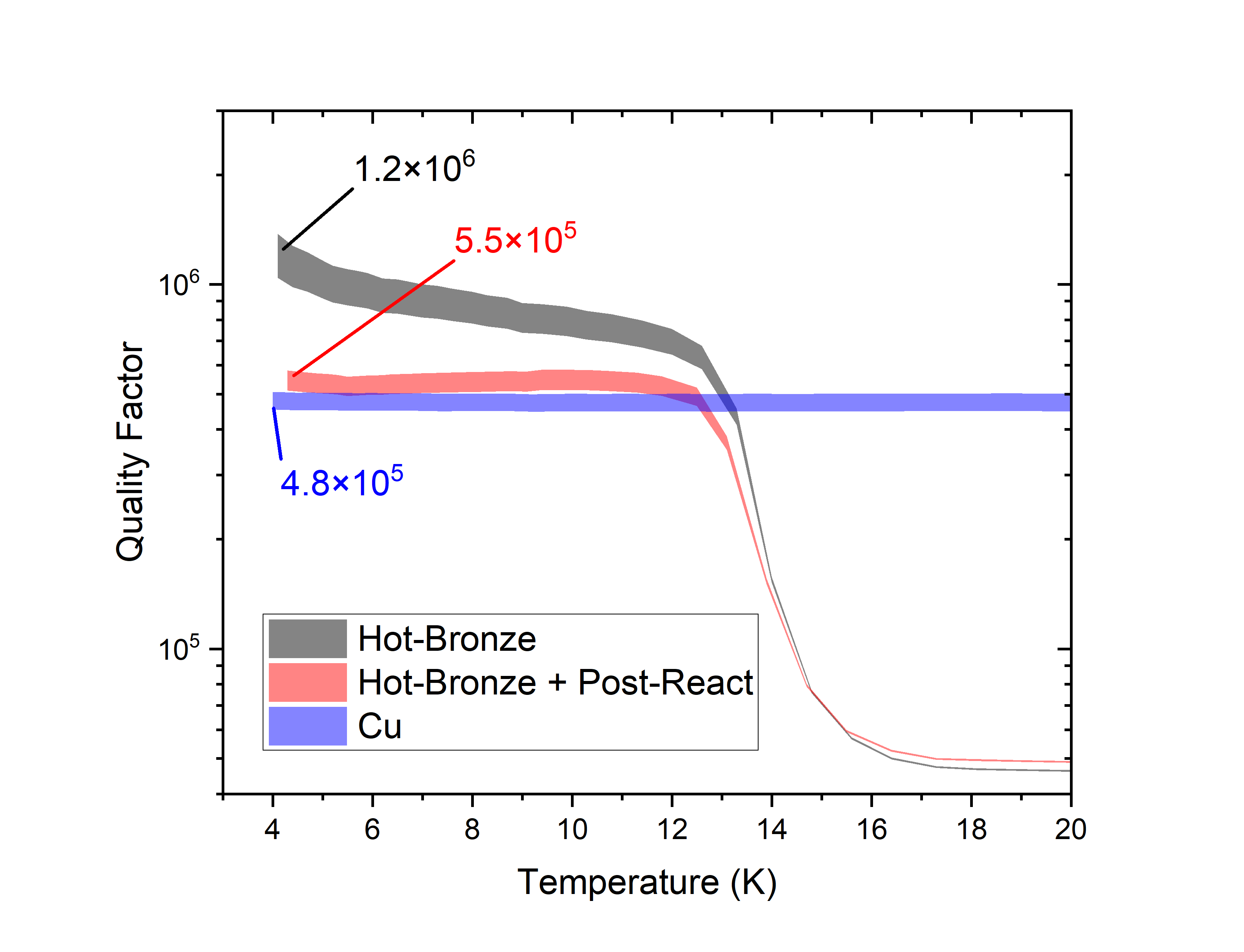}
            \caption[Quality factor as a function of temperature for 2-inch bronze disk samples measured using the mushroom cavity at SLAC.]{Quality factor as a function of temperature for 2-inch bronze disk samples measured using the mushroom cavity at SLAC. The hot-bronze sample (deposited at elevated temperature) achieved $Q_0 = 1.2\times10^6$, while the hot-bronze sample with an additional post-deposition heat treatment achieved $Q_0 = 5.5\times10^5$. Both Nb$_3$Sn samples demonstrate higher quality factors than the bare Cu reference sample, confirming superconducting performance. Shaded regions indicate measurement uncertainty.}
            \label{fig:MushroomQvsT}
        \end{figure}

    Two-inch bronze substrate Nb$_3$Sn samples were sent to SLAC for $Q$ measurements in a mushroom cavity. These samples were made with the same recipe found in~\cite{withanageRapidNbSn2021}, which is the basis for the Cu-Sn first approach in this dissertation, i.e., the hot-bronze and the hot-bronze + post-reaction method (seen in Fig.~\ref{fig:2inchBronzeSub}). The critical temperature curves of the bronze coupon samples from \citeauthor{withanageRapidNbSn2021}~\cite{withanageRapidNbSn2021}, are shown in Fig.~\ref{fig:WenuraBronzeTc}. These samples had a superconducting transition around $14-15$~K. The quality factor results from the SLAC mushroom cavity are shown in Fig.~\ref{fig:MushroomQvsT}. The experiment resulted in a value of $Q=1.2\times10^6$ and $Q=5.5\times10^5$ for the hot-bronze and the hot-bronze + post-reacted samples, respectively. This mushroom cavity result was measured using a Cu host cavity, which increased the error. SLAC plans to repeat these measurements with a Nb mushroom cavity, which will reduce the error in high-$Q$ films. 

    The hot-bronze + post-reacted sample has a darker region in the lower left, different from the rest of the sample (Fig.~\ref{fig:2inchBronzeSub}). This may be due to post-reaction effects that cause significant grain growth in the underlying bronze substrate. This can explain the drop in $Q$ for the post-reacted sample, even though this sample should have higher and more uniform Sn content than the hot-bronze sample based on $T_c$ curves (Fig.~\ref{fig:WenuraBronzeTc}). The patterning at the center of both samples is an artifact of rolling during the preparation of the bronze substrates. While mushroom cavity measurements provided initial $Q$ validation, $Q$ vs $B$ measurements were needed for accurate axion-detection relevant results. This led to the investigation and development of RF testing infrastructure in-house, to test $Q$ vs $B$.

    \FloatBarrier

    \section{Superconducting RF Cavity}




    A previous ADMX PhD student, Thomas Braine from the University of Washington, designed a cavity and probe that fits into small laboratory magnets (probe seen in Fig.~\ref{fig:ProbePPMSVNACavity})~\cite{Braine:2024nzi}. This dissertation partly adapts Braine's approach to allow measurements of $Q$ vs $B$ in the superconducting state using magnet systems available at the Applied Superconductivity Center. Braine's results show that a superconducting surface perpendicular to an external magnetic field incurs greater losses than when it is parallel to the field. This suggests a hybrid approach in which the top and bottom endcaps, which are perpendicular to the magnetic field, are normal-conducting Cu, and the walls parallel to the magnetic field are superconducting.

    The quality factor is the sum of all resonant surface losses in the cavity and depends on both surface resistance and geometry, as outlined in Chapter~\ref{sec:RFCharacterization}. Typical geometric factors for the first fundamental mode of 1.3~GHz Tesla cavities and 10~GHz cylindrical cavities are 270~$\Omega$~\cite{padamseeRFSuperconductivity2009} and 350~$\Omega$, respectively. The geometric factor for a given cavity and mode was calculated with a simulation.

    \subsection{Axion Detector Cavity Methodology} \label{sec:RFCharacterization}

    To fully characterize the properties of a material for an axion detector application, quality factor measurements in external magnetic fields at low temperatures are required. This necessitates cooling the superconducting cavity to a low temperature, increasing the external magnetic field around the cavity, and measuring the quality factor. Cooling to cryogenic temperatures can be achieved with a helium cryostat or a cryocooler. A magnetic field can be created with a solenoid magnet. The magnet bore size (usable inner diameter of the magnet) presents the most significant constraint on the size and, therefore, frequency of the cavity. Measuring the quality factor requires stimulating the RF cavity surface of interest with AC power and measuring the ratio of reflected to absorbed power by the cavity walls. 
    
    It is important in axion detection to maximally couple the cavity's AC electric field and the external DC magnetic field, which depends on their dot product $\mathbf{E}_{\text{AC}} \cdot \mathbf{B}_0$. This requires a cavity mode where the electric field is parallel to the applied magnetic field. For a cylindrical cavity oriented vertically inside a solenoid magnet, the TM$_{010}$ mode produces an axial electric field aligned along the direction of the solenoid's axial magnetic field. The TM$_{010}$ is also the fundamental resonance, utilizing the full cavity volume without nodal planes that would reduce the effective interaction region. Therefore, when testing the cavity quality factor, a cylindrical cavity must be aligned with its axis along the magnetic field direction and the $Q$ of the TM$_{010}$ mode measured, thereby replicating the axion-detection electromagnetic field orientations.

    \subsubsection{Cavity Design}\label{sec:cavitydesign}

    The initial cavity design begins with an enclosed surface that provides the boundary conditions necessary for the formation of standing electromagnetic waves. A ``pillbox" cavity (cylinder) is commonly the starting geometry when describing a resonant cavity. This cylinder can be modeled using any computer-aided design (CAD) software and imported into a finite element simulation software. The cavity walls are modeled as boundary conditions with a specified conductivity, and an eigenfrequency study of this model will reveal the electromagnetic resonance behavior. This usually includes outputting the different oscillating modes with their respective geometric and quality factors. 

    Once the initial design and electromagnetic response of the cavity box is modeled, antenna ports (holes that lead from the exterior to the inner surface) must be added. This allows for input and output signals to be read from the cavity. Cavities are often made in two pieces due to machining and deposition constraints. It's difficult to machine a hollow volume with limited ports and deposit a superconductor on the inner surface. TESLA cavities used for accelerators have primarily been made in half-shells and electron-beam-welded together, but recent advances using hydroforming and other techniques have enabled the fabrication of full cavities from bulk, and Nb$_3$Sn recipes with cylindrical magnetrons or vapor are compatible with coating the inside of an enclosed cavity. However, it is much easier to inspect the surface when prototyping, so the design used here can be disassembled into two pieces, enabling machining, coating, and inspection with limited instrumentation. 

    \subsubsection{Physical Property Measurement System (PPMS)}

    The RF cavity testing stage presents particular experimental challenges. Measuring $Q$ at $9$~T and millikelvin temperatures requires careful consideration of magnet geometry. Magnet bore size limits the cavity dimensions and, therefore, the resonant frequency. For prototype designs using the widely available Physical Property Measurement System (PPMS) magnet, the magnet bore limits the cavity outer diameter to $<26$~mm. Accounting for wall thickness, this corresponds to resonant frequencies $\gtrsim 10$~GHz for cylindrical TM$_{010}$ cavities. This size constraint necessitates specialized probe design and cryogenic RF techniques. While larger-bore magnets ($40-100$~mm diameter) would permit lower-frequency cavities closer to axion-detection targets ($\sim4$~GHz), such systems require substantially more resources: higher capital costs, increased helium consumption, and limited availability. Constraining the design to PPMS-compatible dimensions enables rapid prototyping and accessibility across multiple laboratories.


    The PPMS (Quantum Design) provides high magnetic fields up to 9~T and cooling to 2~K. For quality factor testing, a specialized RF probe must connect the cavity inside the cold, high-field environment to a room-temperature vector network analyzer (VNA) for frequency response measurements. The probe and cavity must be constructed from non-magnetic materials (typically plastic (G10), copper, brass, or beryllium-copper), and all components must remain mechanically stable at cryogenic temperatures. The probe design, including RF feedthrough, is detailed in Ref.~\cite{Braine:2024nzi} and shown in Fig.~\ref{fig:ProbePPMSVNACavity}.

    Thermal load and thermal leakage to the environment must be considered, since the PPMS has limited cooling power. The PPMS is designed for small samples, so the cavity's large thermal mass increases helium consumption and cool-down times. However, if the probe sufficiently isolates the cavity from the external environment, the PPMS should be able to cool down the cavity to a stable 2~K using pot-fill mode. A Cernox temperature sensor is attached to the top of the cavity to ensure that the bottom and top are at thermal equilibrium. A Cernox can be attached to a cable that runs to the top of the probe, allowing for temperature measurements at the top of the cavity. The Cernox must be calibrated before cavity measurement to correctly correlate voltage to temperature.

    \subsubsection{Quality Factor vs Temperature}
            
    The cavity is cooled to base temperature (2~K) in zero applied field, and $Q$ is measured at increasing temperature steps (typically 0.5-1 K increments) past $T_c$. At low temperatures, the residual resistance due to trapped flux and material defects determines $Q$. As temperature increases, thermally-activated quasiparticles contribute additional losses through BCS surface resistance $R_\text{BCS} \propto \exp(-\Delta/k_BT)$, causing $Q$ to decrease exponentially. The temperature at which $Q$ begins to drop rapidly indicates the onset of a significant thermal quasiparticle population and helps verify the film's superconducting transition temperature.
    
    \subsubsection{Quality Factor vs Magnetic Field}
    
    Field-dependent measurements characterize the cavity's tolerance to external magnetic fields, which are critical for axion-detection applications. The cavity is maintained at a constant low temperature (2~K) while the applied magnetic field is increased from 0 to 9~T in steps of $0.25-1$~T. Below the lower critical field H$_{c1}$, the superconductor completely expels the magnetic field from its interior (Meissner state) and $Q$ remains relatively constant. Above H$_{c1}$, magnetic flux penetrates as quantized vortices, and their motion under RF creates additional dissipation. The field dependence of $Q$ reveals the effectiveness of flux pinning, the Campbell resistance $R_C$.

    \subsubsection{Dilution Fridge Millikelvin Measurements}
    
    Dilution refrigerators use a mixture of $^3$He and $^4$He isotopes to reach temperatures as low as 10-100 mK. At these temperatures, the BCS surface resistance $R_\text{BCS} \propto \exp(-\Delta/k_BT)$ becomes exponentially suppressed. Millikelvin measurements isolate the temperature, independent contributions of surface resistance from trapped magnetic flux, grain boundaries, surface oxides, and other material defects, revealing the residual resistance R$_{0}$.
    
    Dilution refrigerator measurements are challenging due to complex setups, amplifiers, low input power required to limit stage heating, longer cooldown times, and limited availability, particularly for systems with high-field magnet capability. Given these constraints, the magnetic-field dependence was assessed using the PPMS, and the dilution fridge separately assessed ultra-low-temperature performance. Combining PPMS field data with dilution fridge temperature data allows performance estimation in axion detector operating conditions (mK temperatures, 9~T fields).

    \subsection{Recipe Selection for Cavity Fabrication}\label{sec:recipe3}

    Recipe 3 (Nb-first geometry) was selected for cavity coating (see Sec.~\ref{sec:nbfirst}) based on microstructural uniformity. SEM cross-sections demonstrated excellent Nb$_3$Sn thickness uniformity (Fig.~\ref{fig:BronzeOnNb}) and a sharp Ta/Nb$_3$Sn interface resulting from sequential in-vacuum deposition. This route sacrificed chemical uniformity (broader transitions and a lower $T_c$ onset compared to the Cu-Sn first routes, Recipe 1 \& 2) in favor of microstructural uniformity. However, Recipe 3 was chosen because RF surface resistance depends critically on lateral uniformity across the cavity surface.
    
    \subsection{RF Cavity Design Constraints}
        
\begin{figure}[htb]
        \centering
        \includegraphics[width=0.75\linewidth]{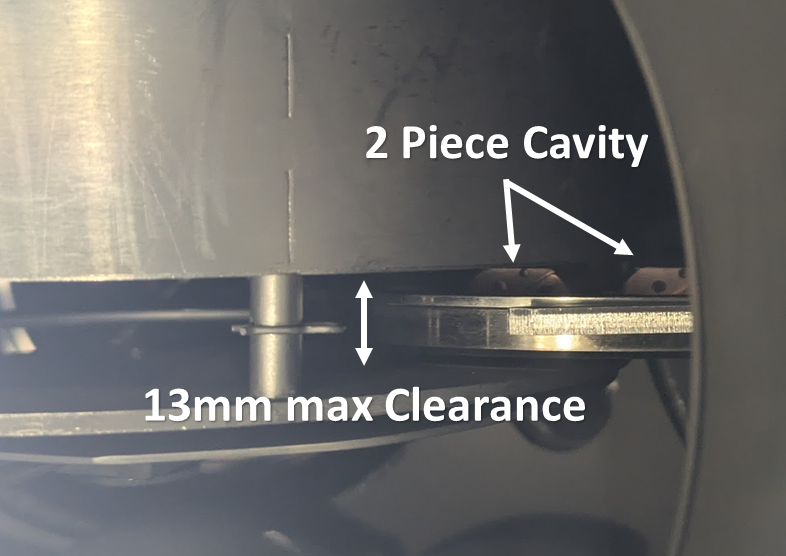}
        \caption{Sputtering chamber max clearance, for substrate thickness.}
        \label{fig:ClearenceSputter}
    \end{figure}

    \noindent The design of RF cavities within the constraints discussed above should satisfy six goals:
    \begin{enumerate}
        \item Build a cavity that facilitates an optimization process to increase $Q$ at a temperature of 2~K and an external magnetic field of 9~T.
        \item The cavity pieces must fit inside the low-profile evaporator and sputtering chamber, i.e., no more than 13~mm thick (Fig.~\ref{fig:ClearenceSputter}).
        \item When assembled, the cavity outer diameter must be less than 26~mm to fit inside the bore of the magnet. This constrains the cavity frequency to be above 10 GHz. Comparing cavity properties with the literature requires accounting for the frequency difference.
        \item Cavity surfaces must be flat planes to make best use of the line-of-sight deposition sources.
        \item The design should allow for hybrid designs that interchange superconducting and normal-conducting materials for the surfaces of the walls and the endcaps.
        \item The resonator must fit within the magnet's 55 mm homogeneous field region.
    \end{enumerate}

    Braine's design, the starting point for the cavity design, had a curved surface that was not tailored to the deposition systems at the Applied Superconductivity Center (ASC). Inspiration came from the Korean group's REBCO "orange wedge" cavity discussed in Sec.~\ref{sec:KoreanRebco} and Fig.~\ref{fig:AhnREBCO}. By changing Braine's upright cylinder shape to an upright polygon, it became possible to conceive of a cavity that could be fabricated using methods directly derived from the Nb$_3$Sn thin-film approaches developed in this dissertation. Furthermore, both research groups noted opportunities to use electromagnetic models to account for behavior at end caps and seams. Thus, the two approaches were merged in this work to create a compact footprint for testing in actual RF conditions using equipment available to ASC.


    \subsection{Rational for Hexagonal Prisms}\label{sec:Simulation}

\begin{figure}[htb]
        \centering
        \includegraphics[width=0.8\linewidth]{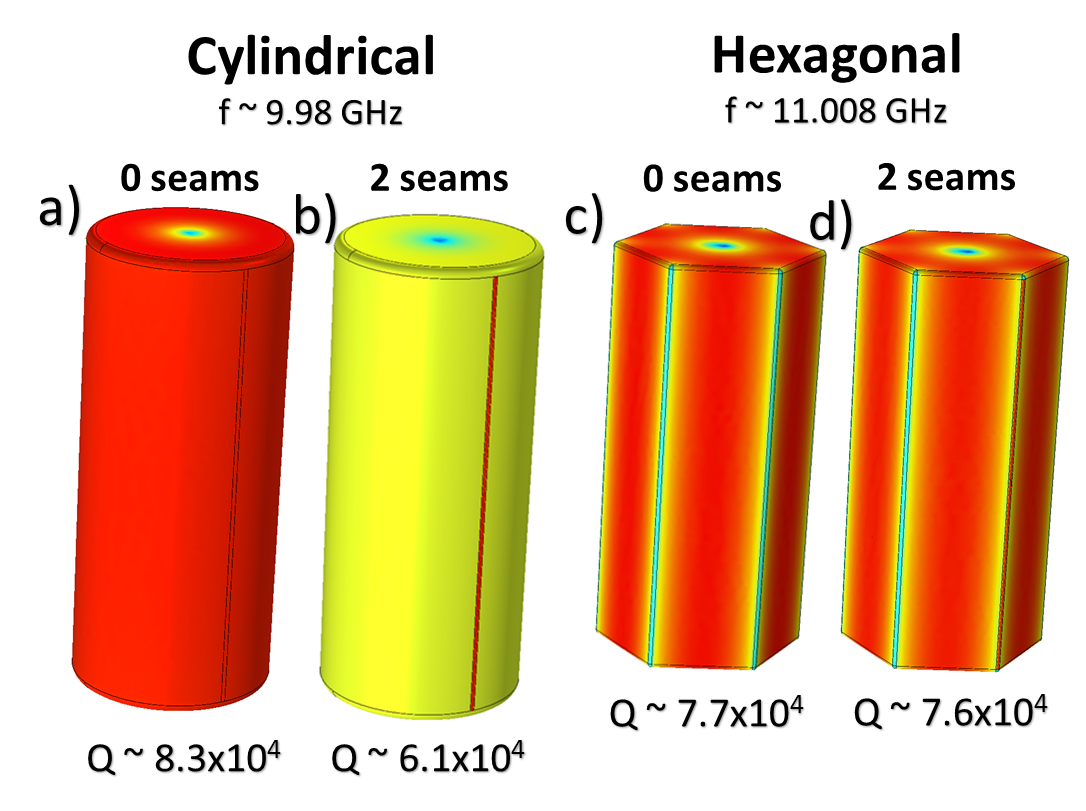}
        \caption[COMSOL simulation showing the change of frequency and quality factor for the cylinder (a, b) and hexagon (c, d) geometries.]{COMSOL simulation showing the change of frequency and quality factor for the cylinder (a, b) and hexagon (c, d) geometries. The color scale shows the surface current, with red indicating maximum and green indicating minimum. Input surface resistance is 5 \unit{\mohm} for the walls and 1 \unit{\ohm} for the gap. Comparing b and d, we note that using the hexagonal cavity reduces the surface resistance lost when a seam is at the edge of the hexagonal cavity.}
        \label{fig:GapnoGapCylinderHexagon}
    \end{figure}

    Finite element simulations including resistive seam interfaces (Fig.~\ref{fig:GapnoGapCylinderHexagon}) revealed a key advantage for hexagonal geometry. Without seams, the hexagon has ~9\% lower $Q$ than a cylinder due to geometry alone. However, introducing two resistive seams on the wall (simulating an interface in a two-piece cavity), reduces cylinder $Q$ by ~27\% but hexagon $Q$ by only ~1\%. The hexagonal geometry naturally concentrates current away from the corners (current is concentrated at the center of the faces). If the seams are located on the corners, dissipation is reduced. In contrast, the cylinder's axial symmetry maintains uniform current distribution along the wall, leading to more current losses at the seams.

    \subsection{RF Quality Factor Measurement Protocol}

    
    All cavity quality factor measurements were performed using a Keysight PNA vector network analyzer (VNA) with a frequency range up to 20~GHz. The VNA output power was set to 10~dBm, with an IF bandwidth of 1~kHz and 200 sweep points. For quality factor extraction, the frequency range was swept over three times the full-width at half-maximum (FWHM) centered on the resonance frequency. 
    
    Cavities were coupled to the VNA using weak coupling via transmission ($S_{21}$) measurements. Weak coupling was verified by reflection measurements showing a dip of less than 1~dB, ensuring that the loaded quality factor $Q_L$ differed from the unloaded quality factor $Q_0$ by less than 10\%. All quality factors reported in this chapter are loaded values ($Q_L$); the intrinsic cavity quality factor $Q_0$ is approximately 10\% higher.
    
    Temperature-dependent measurements in the Physical Property Measurement System (PPMS) allowed 1-minute equilibration time after the PPMS temperature stabilized before recording data. Magnetic field-dependent measurements used the fastest available ramp rate in the PPMS, with 1-minute hold time at each field value. Measurements were performed in 0.5~T steps from 0 to 9~T with the cavity axis aligned parallel to the magnetic field direction.
            
    \subsection{8-Piece Hexagonal Cavity}

    \subsubsection{8-Piece Cu Cavity Fabrication}

    \begin{figure}[ht]
        \centering
        \includegraphics[width=0.75\linewidth]{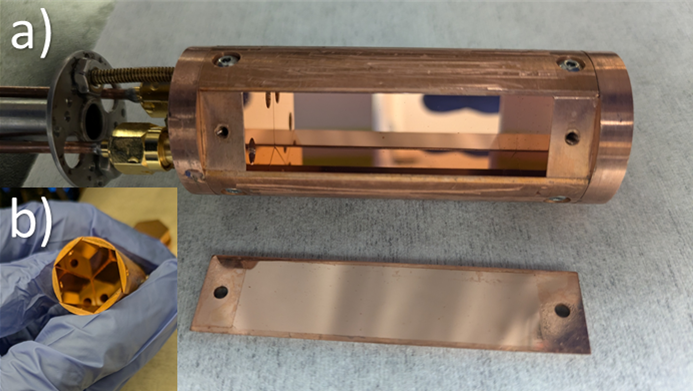}
        \caption{a) Side view and b) cross section of the 8-piece Cu cavity (6 walls, 2 endcaps).}
        \label{fig:1stcavity}
    \end{figure}


    The hexagonal cavity consisted of six separate walls and two endcaps machined from annealed oxygen-free high-conductivity (OFHC) copper at the National High Magnetic Field Laboratory machine shop. The cavity measured 75~mm in height with a hexagonal cross-section; each flat face was 13~mm wide. Wall thickness tapered from approximately 2~mm at the center to $<$0.5~mm near the vertices.

    Individual pieces were mechanically polished using the same procedure as coupon samples: progressive carbide paper (400--1200 grit), diamond slurry (5--1~\unit{\um}), and vibratory colloidal silica (0.05~\unit{\um}). The cavity was assembled using 12 hand-tightened stainless steel 2-56 screws. Antenna coupling was achieved through two gold-plated SMA female connectors mounted on the top endcap.

    
    This segmented geometry enabled polishing and deposition on flat surfaces, reducing the risk of coating non-uniformity. However, the design required a trade-off: thinner walls reduced interior volume (increasing frequency-dependent losses), while the minimal thickness near the vertices compromised mechanical stability under thermal stresses during cooldown, induced currents during magnet ramping, and assembly misalignment.

\begin{figure}[ht]
        \centering
        \includegraphics[width=0.75\linewidth]{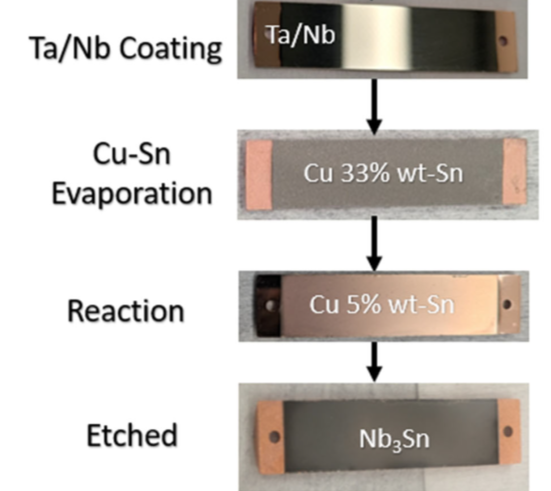}
        \caption{Cavity wall from the 8-piece design that went through a full downselected Nb-first recipe.}
        \label{fig:8pieceRecipe}
    \end{figure}

    Nonetheless, the 8-piece Cu prototype allowed RF measurement, which encouraged coating one of the six walls with Nb$_3$Sn using the Nb-first approach described in Sec.~\ref{sec:nbfirst}. This process is depicted in Fig.~\ref{fig:8pieceRecipe}. 

    \FloatBarrier
    
    Because of the mechanical stability limitations encountered, it was decided not to attempt an 8-piece cavity with either 6 superconducting walls and 2 Cu endcaps or all 8 pieces coated with superconductor. Future work that overcomes challenges associated with mechanical stiffness and alignment could further explore this design with increased wall thickness. Increasing the magnet bore size for RF measurements would facilitate adding volume for thicker cavity pieces.

    \subsubsection{8-Piece Cavity Coating}

\begin{figure}[htb]
        \centering
        \includegraphics[width=\linewidth]{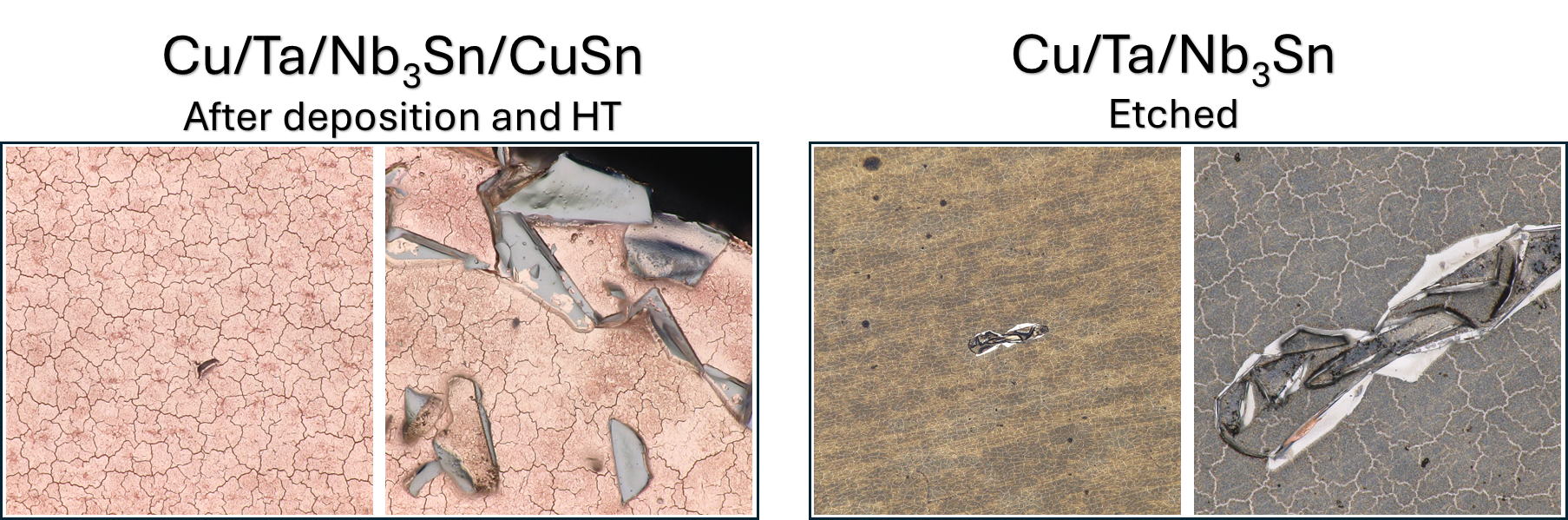}
        \caption{Delamination near the edge after heat-treatment for 8-piece cavity wall before and after etching.}
        \label{fig:8pieceDelamination}
    \end{figure}

    One cavity wall was coated with Nb$_3$Sn using Recipe 3 (Nb-first approach, Section~\ref{sec:nbfirst}). Due to the non-standard geometry, a custom sample holder was fabricated to hold the hexagonal wall piece at the top edge, where the wall would interface with the endcaps, leaving the interior surface exposed for coating. The wall underwent the standard Recipe 3 process: a 500~nm Ta diffusion barrier, a 500~nm Nb seed layer, and a 5~\unit{\um} Cu-33~at.\% Sn thermally evaporated in steps, followed by post-reaction at 715~$^\circ$C for 3 hours, and finally Cu-Sn removal via ammonium persulfate etching. Visual inspection after coating revealed some film delamination, particularly near the edges where the holder contacted the substrate and at regions with steep geometry relative to the deposition flux direction (Fig.~\ref{fig:8pieceDelamination}).\

    \subsubsection{8-piece Cavity Results}

\begin{figure}[htb]
        \centering
        \includegraphics[width=0.7\linewidth]{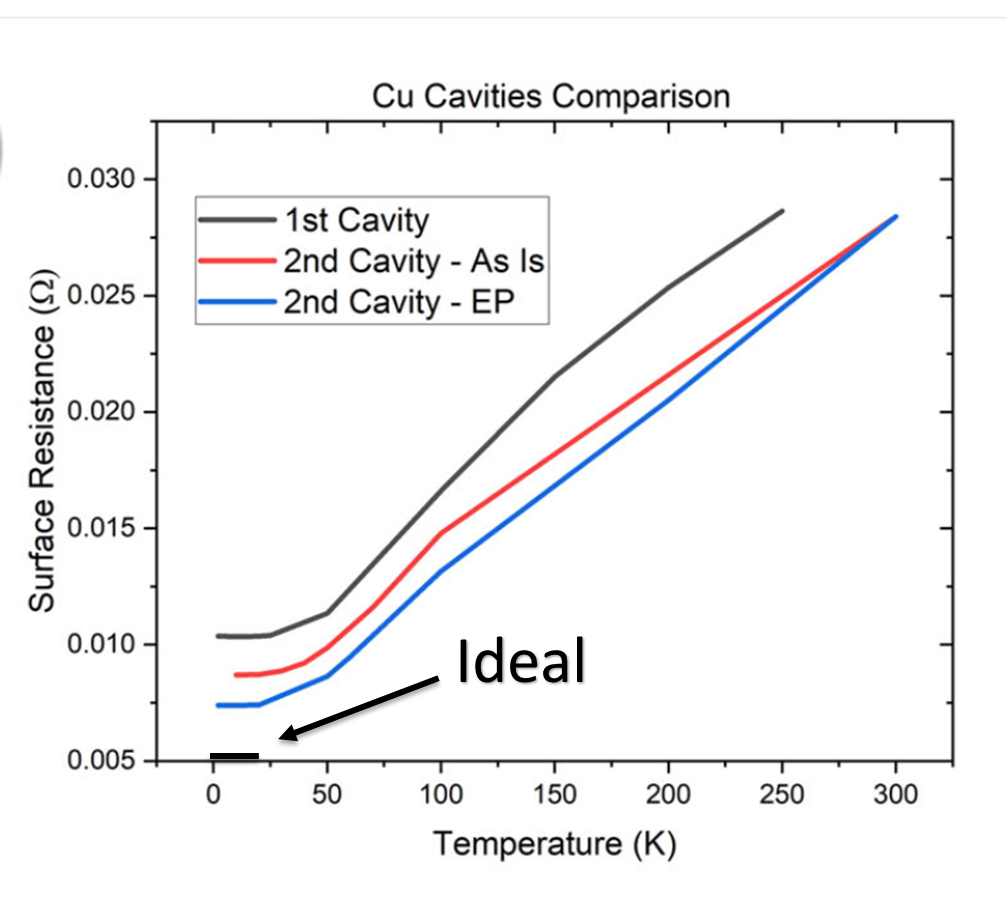}
        \caption[Surface Resistance vs Temperature for the two Cu cavity designs, and a comparison between no polish, and electro-polished for the 2 piece cavity.]{Surface Resistance vs Temperature for the two Cu cavity designs, and a comparison between no polish, and electro-polished for the 2 piece cavity. The ideal 5~\unit{m\ohm} line is numerically calculated by simulating the cavity geometry and imputing the conductivity of the walls of high-purity (RRR=300) Cu at 11~GHz.}
        \label{fig:CuCavitiesRsvsT}
    \end{figure}

    The assembled bare Cu cavity with antenna couplers achieved $Q = 9{,}000$ at room temperature and $Q = 33{,}000$ ($R_s = 11$~\unit{m\ohm}) at 4~K (1st cavity in Fig.~\ref{fig:CuCavitiesRsvsT}). These values fall below our simulated expectations for high-purity Cu ($RRR > 300$) with no RF leakage, which predict $Q = 13{,}500$ at room temperature and $Q = 77{,}000$ ($R_s = 5$~\unit{m\ohm}) at 4~K (as seen in simulation Fig.~\ref{fig:GapnoGapCylinderHexagon}). Temperature dependence on surface resistance is given in Fig.~\ref{fig:CuCavitiesRsvsT} with a comparison to the ideal at low temperature.

\begin{figure}[htb]
        \centering
        \includegraphics[width=0.5\linewidth]{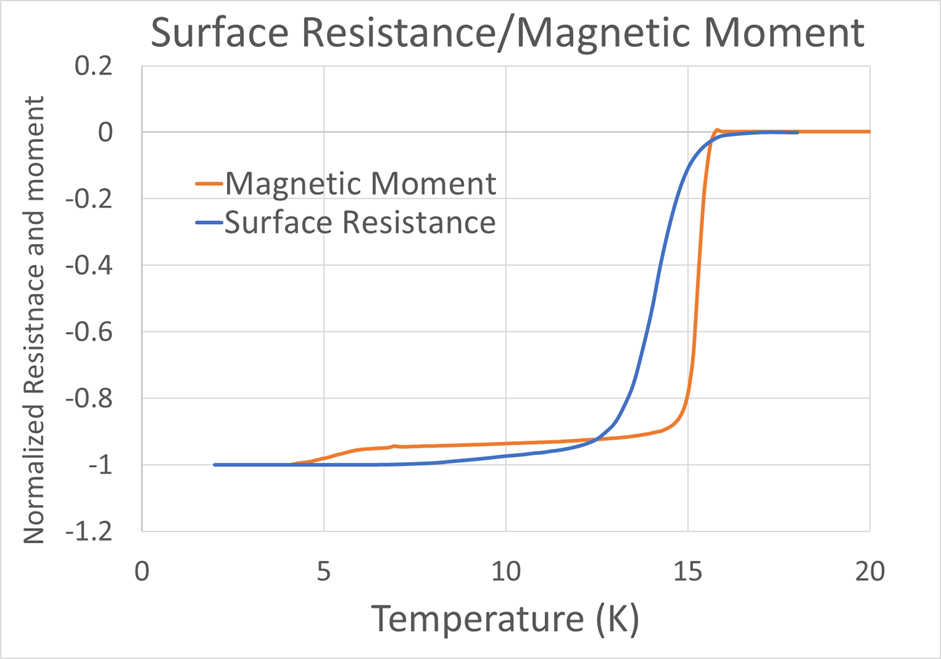}
        \caption[Comparison between normalized cavity measurement and coupon sample SQUID measurement.]{Comparison between normalized cavity measurement and coupon sample SQUID measurement. This confirms the cavity's sensitivity to the superconducting material.}
        \label{fig:RSandMoment8Piece}
    \end{figure}

    The measured quality factor vs. temperature shows a clear superconducting transition, yielding a higher quality factor than that of bare Cu. Comparing the normalized surface resistance curve of the cavity to the critical temperature curve from our coupon sample (Fig.~\ref{fig:RSandMoment8Piece}) with the same recipe, the curves are in reasonable agreement, showing a $T_c$ around 15~K. This confirms that transferring our recipe from a coupon sample to a larger-scale cavity wall creates quantitatively similar Nb$_3$Sn. 

\begin{figure}
        \centering
        \includegraphics[width=0.9\linewidth]{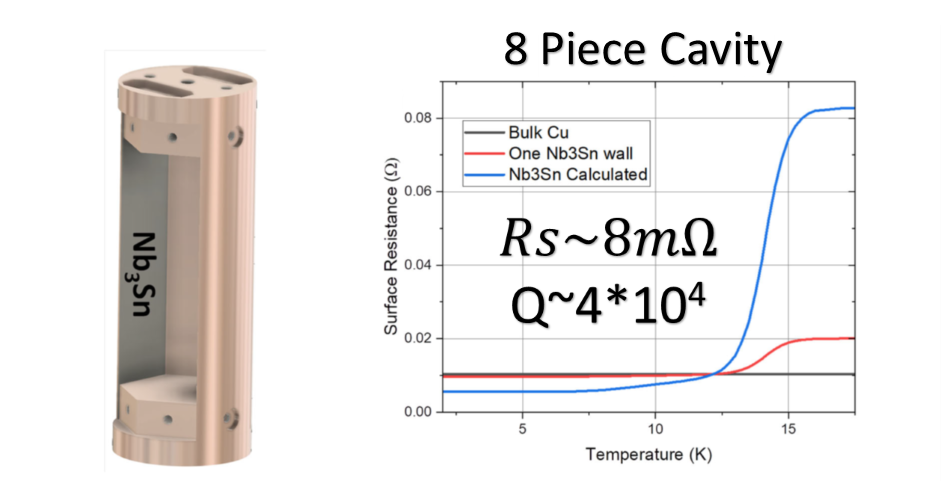}
        \caption{Model of 8 piece cavity with one Nb$_3$Sn wall and the $R_s$ vs $T$ curve for bulk Cu, cavity with one superconducting wall, and isolated superconducting wall.}
        \label{fig:8pieceCavityRCorrected}
    \end{figure}

    The surface resistance of the one superconducting wall was extracted by comparing the pure Cu cavity to the Cu cavity with one superconducting wall. This calculated surface resistance is overlaid with the pure Cu and single wall cavity in Fig.~\ref{fig:8pieceCavityRCorrected}. The surface resistance of individual surfaces is calculated using the math found in Appendix \ref{app:Qcorrection}. The superconducting transition is apparent at 15~K, and the cavity losses become dominated by the Cu walls at around 13~K. This leads to a calculated surface resistance of $R_s \sim \qty{8}{m\ohm}$ for the Nb$_3$Sn wall compared to $R_s \sim \qty{11}{m\ohm}$ for Cu. 
    

\begin{figure}[htb]
        \centering
        \includegraphics[width=0.75\linewidth]{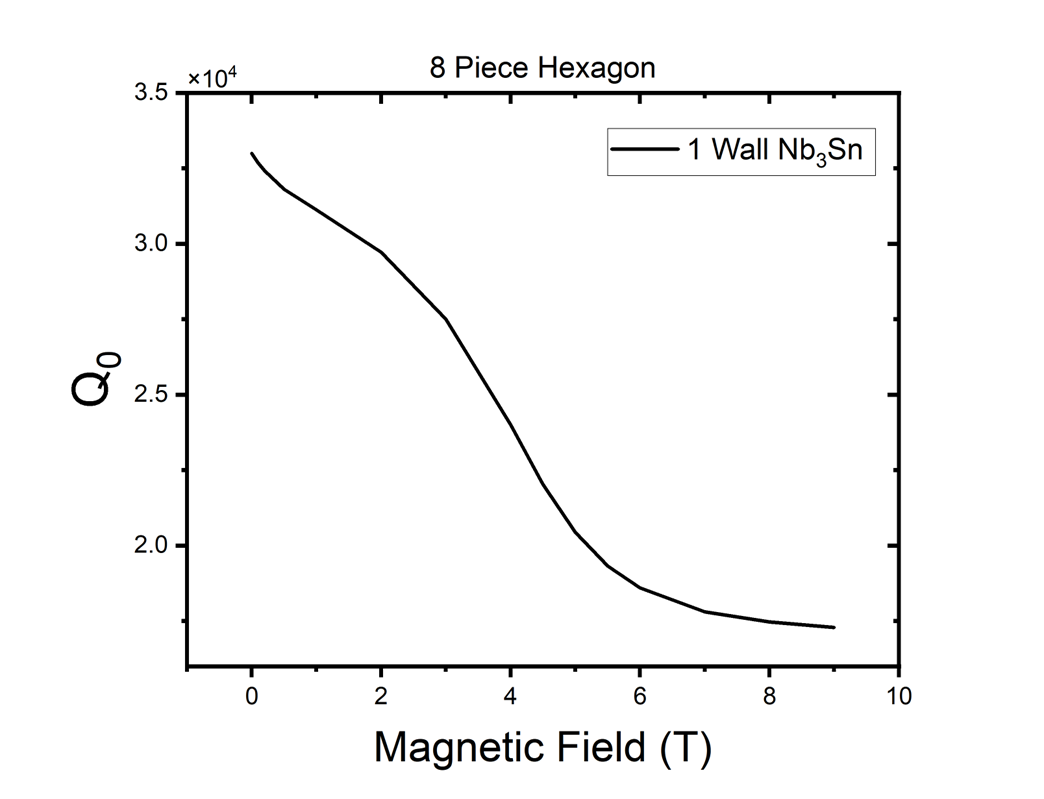}
        \caption[Quality factor vs Magnetic field for the single wall 8-Piece cavity.]{Quality factor vs Magnetic field for the single wall 8-Piece cavity. At 8 tesla, the superconducting wall becomes normal conducting.}
        \label{fig:8pieceQvsB}
    \end{figure}

    The quality factor vs. magnetic field for this cavity, shown in Fig.~\ref{fig:8pieceQvsB} at 2~K, exhibits a linear dependence on magnetic field. The superconducting wall was parallel to the magnetic field, so from geometrical considerations, low resistance should have been achieved with this hybrid geometry. At 8~T the superconducting state is lost, which is much lower than the predicted $>H_{c2}=20$~T commonly found in wires. At very low field $B_0<1$~T, the superconductor was more resistive than Cu. 
    
    
    
    This experiment proved promising as the first prototype. This design allowed a direct comparison of results from DC measurements on coupon samples with those from RF measurements on the larger cavity walls, demonstrating a new capability for ASC. Another important outcome was the first indication that the RF properties of the Nb$_3$Sn thin films produced by the methods researched in this dissertation can indeed lead to improvement of the quality factor over Cu-only alternatives. However, the results from the cavity were unsatisfactory due to both RF leakage in the design and the substantial degradation of $Q$ with the magnetic field. The main limitation of this design was the large number of high-resistance Cu-to-Cu interfaces (18). 
    
    \subsection{2-Piece Hexagonal Cavity: DOE Office of Science Graduate Student Fellowship project}

    The encouraging RF results from the mushroom cavity and 8-piece hexagonal cavity experiments motivated the writing of a proposal for a Department of Energy Office of Science Graduate Student Research (SCGSR) fellowship. The proposal was partnered with Dr. Gianpaolo Carosi at Lawrence Livermore National Laboratory (LLNL) and the research team that previously worked with the ADMX experiment and mentored Dr. Braine's thesis. The LLNL facility has equipment that permits testing of RF cavities for axion detectors down to $< 100$~mK temperature and in magnetic fields up to 9~T, similar to the capabilities at the ASC. The goal of the fellowship was to augment this dissertation by extending measurement capabilities to the millikelvin range and leveraging expertise in cavity design.
    
    \subsubsection{2-Piece Cavity Design and Fabrication}    
    
    The SCGSR proposal considered reducing the number of cavity interfaces by constructing a hexagonal cavity out of 2 pieces of solid Cu. This geometry was almost identical to Braine's cavity, except that the inner surface was hexagonal rather than cylindrical. The hexagon shape was retained to (a) fit in the sputtering chamber, (b) replicate the inner shape of the 8-piece cavity, (c) facilitate reasonably easy surface preparation and inspection, and (d) make use of the electromagnetic models that were developed to understand the RF properties of the 8-piece cavity. The 2-piece cavity has much thicker walls than the 8-piece cavity, thereby reducing uncertainties associated with RF leakage caused by low stiffness and misaligned edges.

    The 2-piece cavity design trades the improvements discussed above for more difficult surface polishing and less certainty in thin-film deposition. Four cavities were ordered from Xometry and machined from bulk OFHC 101 Cu. Each cavity consisted of two hexagonal prism halves with a total height of 72~mm, a hexagonal flat face length of 12.25~mm, and a minimum wall thickness of 2~mm. 
    
    After receiving the cavities, the inner hexagonal surfaces were hand-polished. Due to the acute inner corners, standard polishing procedures were adapted by wrapping sandpaper around an eraser to reach tight curves. Each grit level (400--1200) was polished for approximately 1 minute per surface. Before pursuing more sophisticated methods such as electropolishing, it was decided that the 1200-grit hand polish was sufficient for initial testing. 
    
    The two cavity halves were assembled using four hand-tightened stainless steel 4-40 screws and three alignment pegs to ensure close positioning of the mating faces. Despite the alignment pegs, a small gap remained at the center of the seam, where paper could be slipped through, indicating imperfect mechanical contact. The cavity was then fitted with antenna couplers and tested for quality factor in the Physical Property Measurement System (PPMS).


    \begin{figure}[htb]
        \centering
        \includegraphics[width=0.75\linewidth]{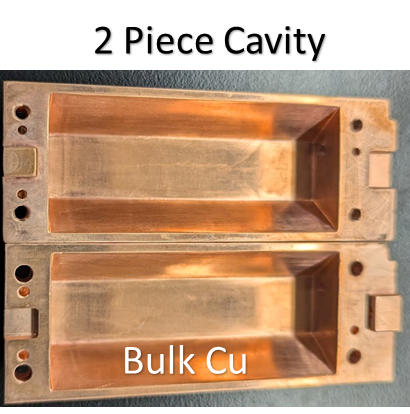}
        \caption{2-piece Cu cavity.}
        \label{fig:2piececavityCu}
    \end{figure}

    \begin{figure}[htb]
        \centering
        \includegraphics[width=0.75\linewidth]{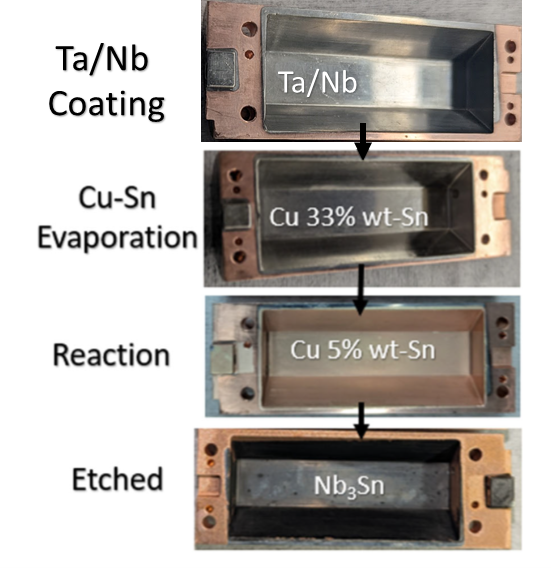}
        \caption{Downselected Nb-first recipe coated on the 2-piece cavity design.}
        \label{fig:2piececavityrecipe}
    \end{figure}

    \subsubsection{2-piece Cavity Coating}

    All coating was performed at Florida State University using a modified version of Recipe 3 (Nb-first approach). The primary modification was the use of high-power impulse magnetron sputtering (HiPIMS) for the Ta diffusion barrier instead of standard DC magnetron sputtering. HiPIMS parameters used were 800~V peak voltage and a 50~\unit{\um} duty cycle, producing a Ta layer approximately 300~nm thick. The remainder of the recipe followed the standard Nb-first process: a Nb seed layer sputtered onto the substrate, with the substrate rotated and heated, followed by Cu-33~at.\% Sn thermal evaporation deposited in steps with 30-second holds between increments. The thermal evaporator lacked both rotation and substrate heating capabilities, raising concerns about the uniformity of the Cu-Sn film on the concave hexagonal surface. After Cu-Sn deposition, the cavity underwent post-reaction at 715~$^\circ$C for 3 hours in the sputtering chamber, followed by Cu-Sn removal via ammonium persulfate etching.


    Due to the cavity's concave geometry, the first non-flat substrate attempted in our fabrication process lacked a mask protecting the endcaps during deposition. A decision not to use a mask was made due to concerns about the added height and risk of contacting the heating element during sample transfer. Consequently, both the hexagonal walls and endcaps received Nb$_3$Sn coating.
    
    Before coating, the cavity was processed at Lawrence Livermore National Laboratory (LLNL) using the facility-standard stainless steel etching protocol, consisting of sequential treatments with soap, HCl, a dichromate solution, and an antioxidant. Following the initial RF measurements of the coated cavity at ASC, the cavity was annealed at 250~$^\circ$C for 6 hours in a positive-pressure argon tube furnace (evacuated to $4\times10^{-5}$ Torr before argon backfill, 0.2 liters per minute flow rate) to improve the Cu residual resistivity ratio. After substrate preparation and coating, the cavity was tested at ASC from 2~K to 20~K and then shipped to LLNL for dilution-refrigerator testing down to 50~mK. All RF measurements at LLNL follow a protocol established at ASC (see Sec.\ref{sec:RFCharacterization}).

    \subsubsection{2-piece Cavity Results}
    
    The first experiments with the 2-piece hexagonal Cu cavity were aimed at understanding the extent to which material grade and machining artifacts affected its resonant properties relative to those of the 8-piece hexagonal cavity prototype. The 2-piece cavity displayed $Q=12,000$ at room temperature and $Q=35,000$ at 4~K. This was an improvement to the previous design, with lower RF leakage (higher $Q$ at room temperature), but the Cu surface purity remained a concern (same $Q$ at low temperature).
    
\begin{figure}[htb]
        \centering
        \includegraphics[width=0.9\linewidth]{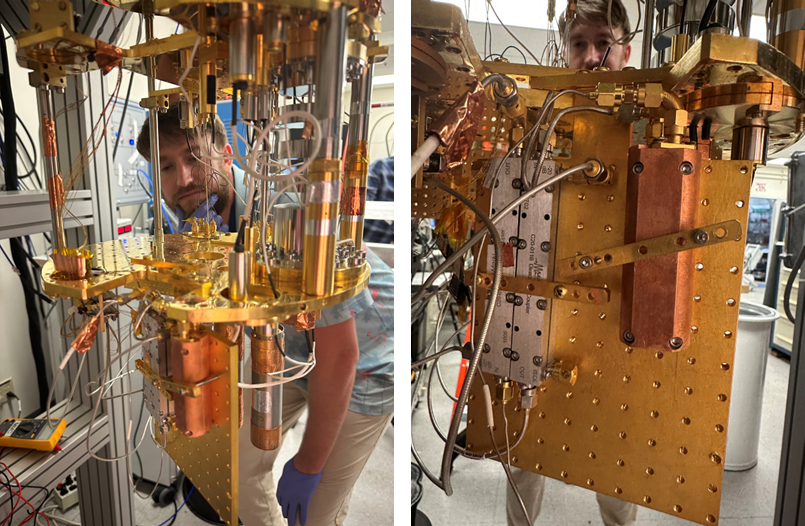}
        \caption{Dilution fridge setup for $Q$ measurements of the 2-piece cavity in mK temperatures.}
        \label{fig:LLNLDilutionQtest}
    \end{figure}

\begin{figure}[htb]
        \centering
        \includegraphics[width=0.9\linewidth]{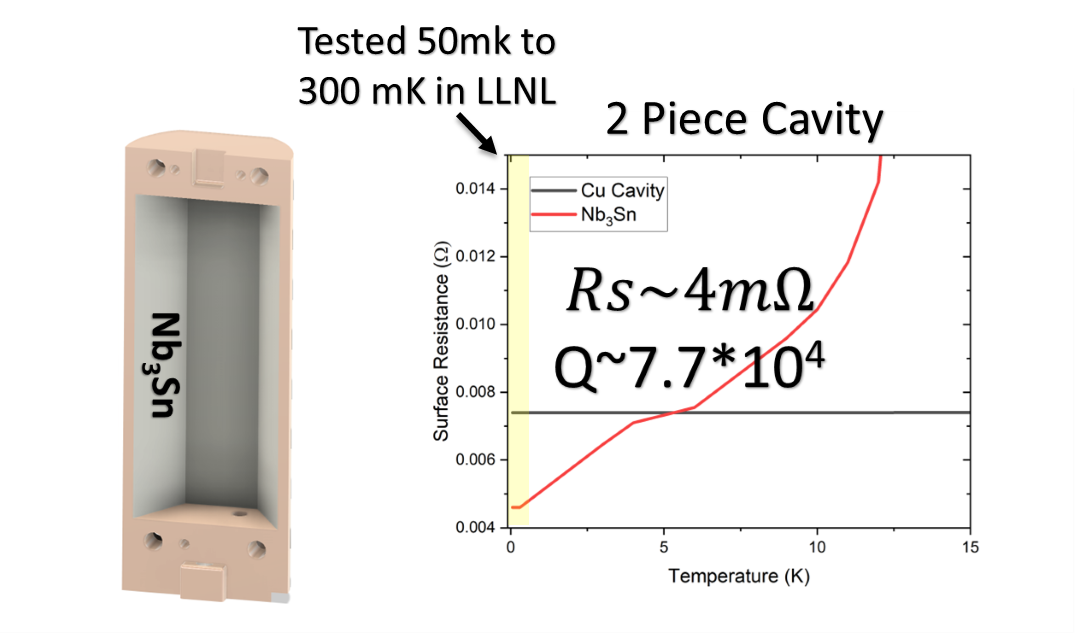}
        \caption{Model and $R_s$ vs T for the 2-piece Nb$_3$Sn cavity.}
        \label{fig:2pieceCavityResults}
    \end{figure}

    The etching process at LLNL increased $Q$ to $49,000$, and annealing at 250~$^\circ$C for 6 hours in a vacuum furnace further raised $Q$ to $55,000$. With the new etched and annealed Cu cavity achieving higher $Q$'s, the cavity's inner surface was fully coated with the downselected recipe 3 (see Sec.~\ref{sec:nbfirst}). We measured the cavity quality factor down to 2~K in the PPMS and to 50~mK in a $_3$He dilution refrigerator at Livermore (Fig.~\ref{fig:LLNLDilutionQtest}). The loaded quality factor measured from 50~mK to 300~mK was constant at $Q_L=77,000$ ($R_s=5~$\unit{m\ohm}) and at 2~K was $Q_L=59,000$ ($R_s=6~$\unit{m\ohm}) as seen in Fig.~\ref{fig:2pieceCavityResults}. As discussed in Section~\ref{sec:Qantenna}, weak coupling conditions ensure that the unloaded quality factor $Q_0$ is approximately 10\% higher than the reported loaded values, i.e., $Q_0 \approx 85{,}000$ at 50~mK and $Q_0 \approx 65{,}000$ at 2~K.

    \subsubsection{Corrections to 2-Piece Cavity Quality Factor}

\begin{figure}[htb]
        \centering
        \includegraphics[width=0.75\linewidth]{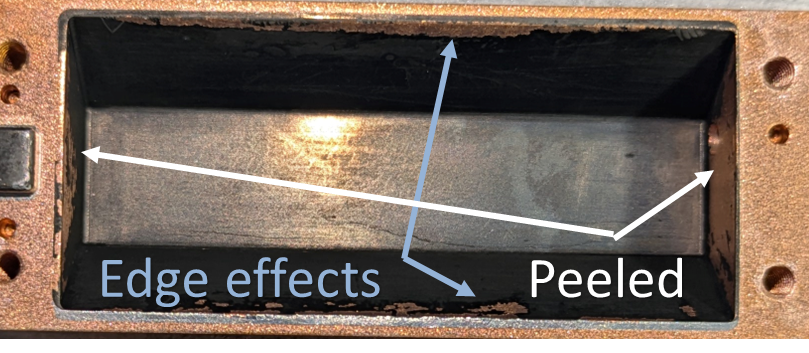}
        \caption{Coating defects at edge regions and endcaps of the two-piece cavity, showing film delamination and non-uniform coverage.}
        \label{fig:EdgeEffect2piece}
    \end{figure}

    Visual inspection revealed significant coating defects: the film on both endcaps peeled away during Cu-Sn etching (Fig.~\ref{fig:EdgeEffect2piece}), indicating that the line-of-sight deposition did not effectively coat surfaces perpendicular to the deposition flux. Additionally, the walls showed clear non-uniformities, including dark spots, scratches, and rough patches, which are artifacts of surface preparation and deposition on steeply inclined surfaces. 
    
    Thus, the measured $Q = 77{,}000$ includes losses from both the uncoated Cu endcaps (Nb$_3$Sn peeled away) and RF leakage at the seams. To estimate the performance of a fully coated Nb$_3$Sn cavity with no seam losses, two corrections were applied:
    
    \textbf{Correction 1: Accounting for Cu Endcaps.} Visual inspection confirmed that approximately one-third of the cavity surface remained uncoated Cu (both endcaps and nearby wall regions). For a hybrid cavity with geometric factor $G_\text{total} = 350~\Omega$, the measured quality factor reflects losses from both materials:
    
    \begin{equation}
    \frac{1}{Q_\text{measured}} = \frac{R_{s,\text{Cu}}}{G_\text{Cu}} + \frac{R_{s,\text{Nb$_3$Sn}}}{G_\text{Nb$_3$Sn}}
    \end{equation}
    
    \noindent where $G_\text{Cu} = 1050~\Omega$ (1/3 surface coverage) and $G_\text{Nb$_3$Sn} = 525~\Omega$ (2/3 coverage). Using the measured Cu cavity surface resistance $R_{s,\text{Cu}} = 6.36~\text{m}\Omega$ at 50~mK and solving for the Nb$_3$Sn contribution yields $R_{s,\text{Nb$_3$Sn}} = 3.64~\text{m}\Omega$. A fully coated Nb$_3$Sn cavity would therefore achieve:
    
    \begin{equation}
    Q_\text{Nb$_3$Sn,full} = \frac{G_\text{total}}{R_{s,\text{Nb$_3$Sn}}} = \frac{350~\Omega}{3.64~\text{m}\Omega} = 9.6 \times 10^4
    \end{equation}
    
    \textbf{Correction 2: Accounting for RF Leakage.} The measured Cu cavity showed RF leakage at seams, reducing $Q_\text{Cu,measured} = 55{,}000$ below the simulated ideal value $Q_\text{Cu,ideal} = 76{,}000$ for seamless geometry. Assuming the same fractional leakage affects the Nb$_3$Sn measurement, the correction factor is:
    
    \begin{equation}
    \alpha = \frac{Q_\text{Cu,ideal}}{Q_\text{Cu,measured}} = \frac{76{,}000}{55{,}000} = 1.38
    \end{equation}
    
    Applying this correction:
    
    \begin{equation}
    Q_\text{Nb$_3$Sn,corrected} = \alpha \times Q_\text{Nb$_3$Sn,full} = 1.38 \times (9.6 \times 10^4) = 1.3 \times 10^5
    \end{equation}
    
    \noindent at $B = 0$ and $T = 50$~mK. This corrected value represents the expected performance for an ideal fully coated Nb$_3$Sn cavity with no seam losses. It should be noted that all reported quality factors in this analysis are loaded values ($Q_L$); the intrinsic unloaded quality factor $Q_0$ would be approximately 10\% higher, giving $Q_{0,\text{corrected}} \approx 1.4 \times 10^5$ for the ideal Nb$_3$Sn cavity. The detailed correction procedure is provided in Appendix~\ref{app:Qcorrection}. This is solid evidence that pathways exist for Nb$_3$Sn cavities on Cu substrates to provide improved performance for future dark-matter searches.

\chapter{Conclusions and Future Work}

\section{Conclusion}

This dissertation developed optimized Nb$_3$Sn coatings on Cu substrates using the thermally evaporated Cu-Sn route. The following notable results were achieved:

\begin{enumerate}
    \item Cu-Sn First: $T_c = 16$~K with transition width $\Delta T_c < 1$~K, among the sharpest reported for Cu-compatible fabrication routes
    \item Nb First: Uniform morphology with $T_c = 15$~K and transition width $\Delta T_c = 2$~K
    \item  Evidence suggesting that strain due to thermal contraction mismatch can be mitigated, e.g., by thick layers
\end{enumerate}

This dissertation also developed and demonstrated processes for optimizing Nb$_3$Sn coatings on copper substrates, including polishing, diffusion barriers, enrichment of the tin source, avoidance of unwanted diffusion reactions, and reaction parameters within the constraints imposed by the thin-film chamber (discussed below). 

The particular advantage explored by this dissertation is the use of thermal evaporation to adjust the Sn content of a deposited Cu-Sn layer flexibly. This allowed exploration of Cu-Sn compositions richer in Sn than standard bronze, which is the solid solution alloy of up to 15 wt.\% Sn in Cu. Sn compositions up to 46 at.\% were explored. The higher Sn content reproduced advantages seen in the manufacture of internal-tin Nb$_3$Sn superconducting wires, where higher tin content supported faster diffusion and higher tin activity in the growing Nb$_3$Sn film, leading to more of the film having composition at or close to the stoichiometric 25\% Sn. Since the superconducting properties are best at the stoichiometric Nb:Sn ratio, the ability to deposit Sn-rich Cu-Sn layers was found to be crucial to optimizing the superconducting properties of films.

While measurements of Nb$_3$Sn cavities in RF conditions are well established for high-gradient accelerator designs, hardly any measurements exist for Nb$_3$Sn films or cavities made on Cu or Cu-alloy substrates. This dissertation presents among the first measurements of RF performance of large-area Nb$_3$Sn films made on Cu substrates as a function of temperature and magnetic field. The results from both mushroom cavity and in-house RF hexagonal cavity experiments indicated that a cavity with Nb$_3$Sn-coated walls could attain a quality factor significantly higher than that of a cavity made from Cu. These measurements validated the excitement in the ADMX community over the potential use of Nb$_3$Sn superconducting cavities for dark-matter detectors.

Hexagonal cavity prototypes were designed, modeled, fabricated, coated with Nb$_3$Sn, and tested. The hexagonal prism geometry, adapted from previous work~\cite{Ahn:2021fgb, Braine:2024nzi}, accommodates RF testing in standard laboratory magnets and is compatible with planar thin-film deposition. RF modeling revealed that if hexagonal rather than cylindrical cavity geometries are used and the seams are located on the edges of the hexagon, the degradation of $Q$ from high-resistance seams can be reduced (see Fig.~\ref{fig:GapnoGapCylinderHexagon}).



\section{Topics For Future Work}

This dissertation was completed under time and temperature constraints for the heat treatments available for the Nb$_3$Sn reaction. Reactions were carried out in the ultra-high vacuum environment of the thin-film sputtering chamber to avoid contamination. The negative impact of overheating the deposition chamber and burning out the chamber heater constrained the reaction temperature to $\sim750~^\circ$C and the reaction time to 3 hours. Therefore, the superconducting properties displayed by several films could have been improved by longer and/or hotter reactions. The use of a high-vacuum tube furnace was found to introduce contamination, suggesting that an ultra-high-vacuum tube furnace would be necessary for future work.

The sputter chamber used in this dissertation imposed geometric constraints on the profiles and curvatures of the pieces used to form the cavities. Future studies could benefit from access to deposition equipment capable of coating more complex surfaces, especially curved ones. Several approaches used by other research groups were reviewed in this dissertation, and a future apparatus to produce Nb$_3$Sn ADMX cavities could incorporate many of the best features of these approaches, including those presented here.


\section{Challenges}

The wall non-uniformities are believed to result from poor cavity surface preparation prior to coating, and the thermal evaporators' lack of substrate rotation. The film near the interface between the two half-cavity pieces began to peel, suggesting that the substrate holder overlapped too much into the inner cavity surface (Fig.~\ref{fig:8pieceDelamination}). Future depositions will decrease the substrate holder mask size so the film can coat past the inner walls, reducing edge effects. With these improvements, this design can be improved with the current infrastructure. This design enables surface resistance and high-Q characterization in magnetic fields using standard laboratory equipment, allowing thin-film groups with limited resources to assess the quality of their films for high-field RF applications.

\addtocontents{toc}{\setcounter{tocdepth}{0}}
\appendix

\chapter{Experimental Methods}\label{app:Methods}

The purpose of this section is to provide future graduate students with the knowledge to repeat and extend the work done here. 

\noindent The fabrication of the superconductor film began with the following procedure for small samples:

\begin{itemize}
    \item Substrate preparation
    \item Magnetron sputtering
    \item Thermal evaporation
    \item Heat treatment
    \item Etching 
\end{itemize}

\noindent To qualify the thin film we just made, we analyze the coupon samples using:

\begin{itemize}
    \item Critical temperature curves ($T_c$)
    \item Scanning Electron Microscopy (SEM)
    \item Energy Dispersive Spectroscopy (EDS)
\end{itemize}

\noindent Next, we utilized COMSOL to design and simulate the radio-frequency (RF) cavity geometries. The RF cavity walls were coated using an optimized recipe and then placed in a PPMS to test RF properties at cryogenic temperatures and in an external magnetic field:

\begin{itemize}
    \item Physical Property Measurement System (PPMS)
    \item Quality factor vs temperature \& magnetic field measurements
\end{itemize}

\section{Thin Film Fabrication}

The following thin-film fabrication techniques and discussion are expanded upon in \cite{seshanHandbookThinFilm2002}.
    
    \subsection{Substrate Preparation}\label{appsec:substrateprep}
    
    Surface preparation is critical for thin film adhesion and quality. Metallic substrates (Nb, Cu) typically require in-house mechanical polishing, while commercial single-crystal silicon (Si) and sapphire substrates are available pre-polished. The samples are cut to a specified area ($7.5\times7.5$~mm), and mounted on a 2-inch metal puck using hot wax. These samples are now ready for polishing.
    
    Mechanical polishing progressively reduces surface roughness through abrasive media. A roughness of approximately 10~nm should be targeted for several reasons:
    \begin{itemize}
        \item \textbf{RF surface resistance}: Surface roughness increases the effective surface area exposed to RF fields, directly increasing losses
        \item \textbf{Film uniformity}: Smooth substrates promote uniform nucleation and grain growth during deposition
    \end{itemize}
    
    The standard polishing process followed in this work includes the following steps:
        
    \begin{enumerate}
        \item \textbf{Carbide paper}: 400 → 600 → 800 → 1200 grit
        \begin{itemize}
            \item Wet polishing with deionized (DI) water to flush debris
            \item Manual polishing on rotating disk
            \item Pressure varies by material (Cu requires lighter pressure than Nb)
            \item Continue each grit until a uniform scratch pattern is achieved
        \end{itemize}
        
        \item \textbf{Diamond slurry}: 5 → 3 → 1 \unit{um}
        \begin{itemize}
            \item Automated polisher with cloth disk
            \item 2 lbs pressure, 2 min per step
            \item DI water rinse between steps
        \end{itemize}
        
        \item \textbf{Vibratory polishing}: 0.05 \unit{um} colloidal silica
        \begin{itemize}
            \item Vibromat polisher
            \item 6-24 hours depending on material
            \item Check surface after 30 min to ensure sample is polishing, if not clean vibromat and add new solution
        \end{itemize}
    \end{enumerate}
    
    \textbf{Post-polish cleaning:}
    \begin{itemize}
        \item Remove samples from mounting wax by heating
        \item Ultrasonic clean: acetone (3×15 min), isopropyl alcohol (IPA) (1×5 min)
        \item Dry with filtered nitrogen gas (N${_2}$) and stored in a desiccator
    \end{itemize}
    
    \textbf{Key considerations:}
    \begin{itemize}
        \item Too much pressure and media will embed into the sample, and too little pressure leads to no material removal
        \item Soft metals (Cu) require lighter pressure than hard metals (Nb)
        \item Wet polishing prevents particle re-deposition
        \item Mounting wax must be completely removed in acetone before the IPA step
    \end{itemize}
    
    

    \subsection{Thermal Evaporation}\label{appsec:ThermalEvapProtocol}
    
    \textbf{Daily Startup:}
    \begin{itemize}
        \item Ensure water supply and N$_2$ gas are open to chamber
        \item Turn on roughing pump and high-vac valve supply
        \item Run oil pump for at least 30 minutes before pumping to high-vac
        \item Note: Water must remain on while oil pump is running and for 30 min after shutdown
    \end{itemize}
    
    \textbf{Source Preparation:}
    \begin{itemize}
        \item Select source material; must fit inside crucible
        \item For alloys: weigh correct concentration of each element, mix powders
        \item Load material into boat
        \item For films $>$5~\unit{\um}: load multiple crucibles (each boat deposits $\sim$5~\unit{\um})
    \end{itemize}
    
    \textbf{Chamber Setup:}
    \begin{itemize}
        \item Load substrate onto holder, desired surface facing down toward source
        \item Align substrate and crucible for perpendicular flux
        \item Ensure positioning spring is in place
        \item Clean plastic window with acetone
        \item If using two boats: verify you can rotate to correct position while blind
        \item Move substrate shutter to closed position
        \item Check QCM crystal life $>$60\%; replace if needed
        \item Tune deposition rate monitor for your material
        \item Clean rubber gasket and metal seal of debris
        \item Lower glass top and ensure proper seal
    \end{itemize}
    
    \textbf{Pump-Down:}
    \begin{itemize}
        \item Close all valves (HV valve and Vent valve)
        \item Open roughing valve; wait 5 min or until $<$1E-2~Torr
        \item Close roughing valve
        \item Open high-vac valve and chamber valve
        \item Turn on fine-tune pressure monitor when pressure falls below 1E-4~Torr
        \item Wait 30 min; pressure should reach $\sim$1E-6~Torr
        \item Add liquid nitrogen to dewar reservoir
        \item Wait 30 min; pressure should reach $>$1E-7~Torr
    \end{itemize}
    
    \textbf{Deposition:}
    \begin{itemize}
        \item Increase current at 1 power/min in low-current mode
        \item At $\sim$7 power: source melts and condenses into a molten ball
        \item At $\sim$9 power: deposition rate increases (visible on QCM)
        \item At $\sim$10 power: stable deposition rate achieved
        \item Wait for stable rate of $\sim$20~\AA/s
        \item Zero QCM monitor
        \item Open substrate shutter to begin coating
        \item Deposit 1~\unit{\um}, then close shutter and turn off current
        \item Wait 30 min between layers
        \item For subsequent layers: can ramp faster to $\sim$6 power, wait for stable 20~\AA/s
        \item Repeat until desired thickness reached
    \end{itemize}
    
    \textbf{Shutdown:}
    \begin{itemize}
        \item After final layer, wait at least 30 min before venting (hot sample, metal vapor)
        \item Turn off HV valve, chamber valve, and fine-tune pressure gauge
        \item Open vent valve
        \item Remove sample; take care with spring on substrate holder
        \item Lower glass top and pump with roughing pump
        \item Turn off oil pump; turn off water 30 min later
    \end{itemize}
    
    \textbf{Troubleshooting:}
    \begin{itemize}
        \item Current won't turn on: check bottom doors are closed (door trip switch)
        \item Also try pressing strip button near current dial
    \end{itemize}

    \subsection{Magnetron Sputtering}\label{appsec:SputteringProtocol}
    
    \textbf{Chamber Preparation (after air exposure):}
    \begin{itemize}
        \item Target pressure: low $10^{-9}$~Torr
        \item Pump down chamber
        \item Bake out: 200~°C for 24~hrs (central heater), 100~°C (external heaters)
        \item Keep cryo gate valve open during bakeout
        \item Use RGA to check residual gas species:
        \begin{itemize}
            \item H$_2$, O$_2$: remove by bakeout
            \item Carbon: indicates organic contamination
        \end{itemize}
    \end{itemize}
    
    \textbf{If Pressure Won't Reach Target:}
    \begin{itemize}
        \item Check bolts around top flange (especially if recently opened)
        \item May need to open chamber and clean
    \end{itemize}
    
    \textbf{Opening Chamber:}
    \begin{itemize}
        \item Remove 36 bolts around top plate
        \item Top flange has knife edge that cuts into copper gasket
        \item Use single-use copper gasket if heating; rubber gasket OK for temporary/no heating
        \item Can replace targets, vacuum chamber interior with clean vacuum
    \end{itemize}
    
    \textbf{Closing Chamber:}
    \begin{itemize}
        \item Clean copper gasket and groove with lint-free cloth, IPA, N$_2$ gas
        \item Lower top plate slowly; avoid pinching wires
        \item Ensure screws slide into holes; check even gap around circumference
        \item Hand-tighten all bolts
        \item Torque in star pattern (1-18-9-27, then 2-19-10-28, etc.):
        \begin{itemize}
            \item First pass: 10~lb/ft
            \item Second pass: 20~lb/ft
            \item Third pass: 30~lb/ft
        \end{itemize}
        \item Start pump-down; if error, check bolt tightness and gasket for debris
        \item Pump overnight; check for $<$5E-8~Torr in morning
        \item Bake out to reach low $10^{-9}$~Torr
    \end{itemize}
    
    \textbf{Loading Substrates:}
    \begin{itemize}
        \item Place substrate and holder in load-lock
        \item Pump load-lock to $>$1E-5~Torr
        \item Run load-lock heat recipe: 110~°C for 10~min
        \item Wait 15~min before loading to main chamber
        \item Run sample load recipe
        \item If software bugs: manually set substrate/empty variable
    \end{itemize}
    
    \textbf{Deposition:}
    \begin{itemize}
        \item Use automatic recipes for single-target DC sputtering (bias/non-bias)
        \item For HiPIMS, multi-target, or RF: create custom recipe or run manually
        \item Record all parameters in main Excel spreadsheet
        \item For new targets/parameters: determine deposition rate via ex-situ characterization
    \end{itemize}
    
    \textbf{Available Modes:}
    \begin{itemize}
        \item \textbf{DC sputtering}: requires working gas pressure and target wattage; do not exceed material-dependent W/m$^2$ (Cu-backed 2-inch targets: max 60~W)
        \item \textbf{Substrate bias}: adjusts incoming atom energy
        \item \textbf{RF sputtering}: for insulators or substrate/target cleaning (20~W, $<$2~min; excess power causes pitting)
        \item \textbf{HiPIMS}: pulsed deposition; set duty cycle, peak voltage, kick; produces denser films but requires tuning
    \end{itemize}

    \subsection{Heat Treatment}\label{appsec:HeatTreatmentProtocol}
    
    \textbf{Option 1: Sputtering Chamber}
    \begin{itemize}
        \item Can reach $10^{-9}$~Torr (cleanest option)
        \item Throttle cryo gate valve before reaching 150~°C
        \item At 700~°C: pressure rises to $\sim10^{-7}$~Torr due to outgassing
        \item Ion gauge turns off during heating; use RGA to monitor pressure
        \item Substrate temperature $\neq$ heater setpoint; use calibration table
        \item Maximum: 6~hrs at 750~°C
    \end{itemize}
    
    \textbf{Option 2: Tube Furnace}
    \begin{itemize}
        \item Can only reach $10^{-5}$~Torr; use Ar flow to reduce impurities
        \item Load sample and pump to $10^{-5}$~Torr for 30~min
        \item Turn on PPB argon filter; allow 30~min to heat up
        \item Turn off pump; flow Ar into chamber
        \item Once positive pressure reached, open outlet valve for continuous flow
        \item Ar flow rate: $\sim$0.2~L/min
        \item Set furnace to desired temperature
        \item Use calibration table for temperature correction
        \item Can run $>$6~hrs (longer than sputtering chamber)
    \end{itemize}

    \subsection{Chemical Etching}\label{appsec:EtchingProtocol}
    
    \textbf{Safety:}
    \begin{itemize}
        \item Work in fume hood only
        \item Complete ISM training for all chemicals used
        \item Ammonium persulfate ((NH$_4$)$_2$S$_2$O$_8$): store in container that can breathe
    \end{itemize}
    
    \textbf{Cu-Sn Etching Procedure:}
    \begin{itemize}
        \item Prepare ammonium persulfate solution
        \item Heat to 70~°C on hot plate
        \item Use magnetic stirrer to agitate continuously (reduces bubble pitting)
        \item To protect areas from etching: cover with Kapton tape before immersion
        \item Immerse sample; Cu-Sn fully etches in $\sim$10~min
        \item Remove sample, rinse, dry
    \end{itemize}
    
\section{Small Sample Characterization}


    \subsection{SQUID Magnetometry Protocol}\label{appsec:SQUIDProtocol}
    
    \textbf{Sample preparation:}
    \begin{itemize}
        \item Cut samples to $7.5\times7.5$~mm, maximum 2~mm thickness (must fit inside of straw)
        \item Mount inside of straw by cutting perpendicular slit 7 cm from end of straw and pushing sample through slit, with supercondcuting surface parallel to straw axis
        \item Ensure sample is centered and secure with Kapton tape
    \end{itemize}
    
    \textbf{Measurement procedure:}
    \begin{enumerate}
        \item Load straw with sample onto end of probe at room temperature
        \item Evacuate sample chamber, load sample down into magnet (ensure sample is all the way down by pressing top of probe), and purge sample chamber
        \item Initialize DC centering (puts sample into center solenoid)
        \item Cool to temperature right below transition (typically 12~K) in zero field (ZFC protocol)
        \item Apply 0.1~mT (1~Oe) field parallel to film surface
        \item Center sample with DC centering option
        \item Heat up past transition (20 K)
        \item Cool down to 4~K and measure residual field
        \item Find residual field and input residual field
        \item Heat back up past transition (20 K), and cool down with input residual field
        \item Apply 1~mT (10~Oe) field parallel to film surface, while accounting for residual field (if residual field is +1~Oe, apply 11~Oe, so sample actually feels 10~Oe)
        \item Wait 2 minutes for thermal equilibrium
        \item Increase temperature in steps (typically 0.5--1~K increments)
        \item Record magnetic moment at each temperature point
        \item Continue through $T_c$ until moment returns to zero (sample fully normal)
    \end{enumerate}
    
    \textbf{Data analysis:}
    \begin{itemize}
        \item Plot moment vs temperature
        \item Define $T_c(\text{onset})$ at first diamagnetic deviation from zero magnetic moment
        \item Define value at 90\% of maximum diamagnetic signal
        \item Define value at 10\% of maximum diamagnetic signal
        \item Transition width $\Delta T_c = T_c(90\%) - T_c(10\%)$
        \item Sharp transitions ($\Delta T_c < 0.5$~K) indicate high uniformity
    \end{itemize}
    
    \textbf{Troubleshooting:}
    \begin{itemize}
        \item If moment is noisy: apply a degauss, increase averaging, check sample mounting stability
        \item If transition is very broad: check for phase separation or compositional gradients
        \item If no transition observed: verify centering is correctly followed, the sample may not be superconducting
    \end{itemize}
    
    \subsection{Scanning Electron Microscopy (SEM)}\label{appsec:SEMProtocol}
    
    This protocol covers basic operation of the Zeiss SEM for imaging conductive samples.
    
    \textbf{Sample Mounting:}
    \begin{enumerate}
        \item Mount sample on aluminum stub using carbon tape or silver paint
        \item For cross-sections, ensure polished face is perpendicular to stub surface
        \item Load stub into holder and record sample position
    \end{enumerate}
    
    \textbf{Startup and Venting:}
    \begin{enumerate}
        \item Log into the computer and open SmartSEM software
        \item Vent the chamber (Vacuum $\rightarrow$ Vent)
        \item Once vented, open chamber door and load sample holder onto stage
        \item Close door and pump down (Vacuum $\rightarrow$ Pump)
        \item Wait for vacuum to reach operating pressure ($<10^{-5}$~mbar)
    \end{enumerate}
    
    \textbf{Imaging:}
    \begin{enumerate}
        \item Turn on EHT (beam voltage): start with 10--15~kV for most samples
        \item Select detector: SE2 for topography, BSD for compositional contrast
        \item Use stage controls (joystick or software) to navigate to sample
        \item Focus at low magnification first, then increase magnification
        \item Adjust working distance (WD) to ~4~mm for standard imaging
        \item Use focus and stigmation (X and Y) to sharpen image
        \item Adjust brightness and contrast as needed
        \item To save image: Freeze frame, then File $\rightarrow$ Save Image
    \end{enumerate}
    
    \textbf{Shutdown:}
    \begin{enumerate}
        \item Turn off EHT
        \item Retract stage to load position
        \item Vent chamber and remove samples
        \item Pump chamber back down before leaving
        \item Log out of software and logbook
    \end{enumerate}

    \subsection{Energy Dispersive Spectroscopy (EDS)}\label{appsec:EDSProtocol}
    
    This protocol covers Oxford AZtec EDS operation for elemental analysis.
    
    \textbf{Setup:}
    \begin{enumerate}
        \item Open AZtec software (separate from SmartSEM)
        \item Ensure SEM is at working distance of 8.5~mm (optimal for EDS geometry)
        \item Set EHT to 15--20~kV (must exceed the absorption edge of elements of interest)
        \item Insert EDS detector if retracted
    \end{enumerate}
    
    \textbf{Point Analysis:}
    \begin{enumerate}
        \item In AZtec, select Point \& ID mode
        \item Click location on live SEM image
        \item Acquire spectrum for 30--120~s (longer = better statistics)
        \item Software auto-identifies peaks; verify assignments manually
        \item Quantification results appear in atomic \% and weight \%
    \end{enumerate}
    
    \textbf{Line Scan:}
    \begin{enumerate}
        \item Select Linescan mode
        \item Draw line across region of interest (e.g., film cross-section)
        \item Set number of points (64--256 typical) and dwell time per point
        \item Run acquisition; results show element concentration vs.\ position
    \end{enumerate}
    
    \textbf{Elemental Mapping:}
    \begin{enumerate}
        \item Select Map mode
        \item Define rectangular region of interest
        \item Select elements to map (e.g., Nb, Sn, Cu, Ta)
        \item Set resolution (256$\times$256 for quick survey, 512$\times$512 or higher for publication)
        \item Acquire for 30--60+ minutes depending on required quality
        \item Maps display X-ray intensity for each element vs.\ position
    \end{enumerate}
    
    \textbf{Tips:}
    \begin{itemize}
        \item Higher beam current increases X-ray signal but may damage sensitive samples
        \item For accurate quantification, sample surface should be flat and perpendicular to beam
        \item Process number (dead time) should be 30--50\%; adjust beam current if outside this range
        \item Save project file to retain all data and settings
    \end{itemize}
    
    \subsection{FIB Lamella Preparation}\label{appsec:FIBProtocol}
        
    \textbf{Equipment:} FEI Helios or equivalent dual-beam FIB-SEM
    
    \textbf{Procedure:}
    \begin{enumerate}
        \item Identify region of interest using SEM imaging
        \item Deposit protective Pt layer (2~\textmu m thick) over region of interest (ROI) to prevent ion damage
        \item Mill coarse trenches (30 kV, high current) on both sides of ROI
        \item Thin lamella to $\sim$1--2~\textmu m using lower currents
        \item For SEM/EDS: Mount lamella perpendicular to beam on stub
        \item For TEM: Further thin to $<$100~nm using 5 kV, 50 pA final polish
    \end{enumerate}
    
    \textbf{Typical parameters:}
    \begin{itemize}
        \item Coarse mill: 30 kV, 20 nA
        \item Thinning: 30 kV, 0.5--2 nA
        \item Final polish (TEM): 5 kV, 50 pA
    \end{itemize}

\section{Full Cavity RF Characterization}

\subsection{PPMS Cavity Measurement Setup}\label{appsec:PPMSProtocol}

This protocol covers RF cavity quality factor measurements using the Physical Property Measurement System (PPMS) and vector network analyzer (VNA).

\textbf{Sample Mounting:}
\begin{enumerate}
    \item Attach RF antennas to cavity ports; ensure antennas are at equal insertion depth
    \item Connect cavity to RF probe using SMA cables
    \item Mount probe assembly in PPMS sample chamber
    \item Verify all connections are secure before cooldown
\end{enumerate}

\textbf{VNA Setup:}
\begin{enumerate}
    \item Connect RF probe cables to VNA ports 1 and 2
    \item Set frequency span to include expected resonance (check COMSOL model $\pm$50~MHz)
    \item Set IF bandwidth to 100~Hz for initial survey, reduce to 10~Hz for final measurements
    \item Calibrate VNA at room temperature before cooldown
\end{enumerate}

\textbf{Cooldown:}
\begin{enumerate}
    \item Set PPMS to cool to 2~K at default rate
    \item Monitor resonance frequency during cooldown---it will shift as cavity contracts
    \item Allow 30+ minutes thermal stabilization at base temperature before measurements
\end{enumerate}

\subsection{Quality Factor Measurement}\label{appsec:QProtocol}

\textbf{Antenna Coupling Check:}
\begin{enumerate}
    \item Measure $S_{11}$ (reflection from port 1) and $S_{22}$ (reflection from port 2)
    \item Find the dip at the resonant frequency
    \item Dip depth should be $<$1~dB from background for weak coupling
    \item If dip is deeper, antennas are overcoupled---retract slightly and remeasure
\end{enumerate}

\textbf{Transmission Measurement:}
\begin{enumerate}
    \item Measure $S_{21}$ (transmission from port 1 to port 2)
    \item Locate the resonance peak
    \item Record peak frequency $f_0$ and peak amplitude
    \item Measure the 3~dB bandwidth: frequencies where signal drops 3~dB below peak
    \item Calculate $Q_L = f_0 / \Delta f_{3\text{dB}}$
\end{enumerate}

\textbf{Calculating $Q_0$:}
\begin{enumerate}
    \item Calculate coupling coefficients $\beta_1$ and $\beta_2$ from $S_{11}$ and $S_{22}$ dip depths
    \item Apply correction: $Q_0 = Q_L(1 + \beta_1 + \beta_2)$
    \item For weak coupling ($<$1~dB dips), $Q_0 \approx 1.1 \times Q_L$
\end{enumerate}

\subsection{Quality Factor Correction Procedure} \label{app:Qcorrection}

This appendix details the procedure for extracting the intrinsic quality factor of a superconducting coating from measurements in a hybrid material cavity with RF leakage at seams.

\subsubsection{Step-by-Step Procedure}

\begin{enumerate}
    \item \textbf{Simulate geometric factors.} Using finite element electromagnetic simulation (e.g., COMSOL Multiphysics), model the cavity geometry with perfectly conducting walls and solve for the resonant mode of interest. Extract the magnetic energy $U_m = \int_V \frac{1}{2}\mu|\vec{H}|^2 \, dV$ and, for each surface $i$, integrate the tangential field $I_i = \int_{S_i} |\vec{H}_\text{tan}|^2 \, dS$. Calculate surface-specific geometric factors as $G_i = 2U_m / I_i$.
    
    \item \textbf{Group surfaces by material.} Calculate effective geometric factors for each material group using:
    \begin{equation}
        \frac{1}{G_\text{group}} = \sum_{i \in \text{group}} \frac{1}{G_i}
    \end{equation}
    Also, calculate $G_\text{total}$ for uniform material throughout the cavity.
    
    \item \textbf{Measure the copper reference cavity.} Measure $Q_\text{Cu,measured}$ for an all-copper cavity of identical geometry. Calculate the copper surface resistance:
    \begin{equation}
        R_{s,\text{Cu}} = \frac{G_\text{total}}{Q_\text{Cu,measured}}
    \end{equation}
    
    \item \textbf{Measure the hybrid cavity.} Measure $Q_\text{measured}$ for the cavity containing the Nb$_3$Sn-coated sample.
    
    \item \textbf{Extract the Nb$_3$Sn surface resistance.} Using the known copper contribution:
    \begin{equation}
        R_{s,\text{Nb}_3\text{Sn}} = G_{\text{Nb}_3\text{Sn}} \left(\frac{1}{Q_\text{measured}} - \frac{R_{s,\text{Cu}}}{G_\text{Cu}}\right)
    \end{equation}
    
    \item \textbf{Calculate full-coverage quality factor.} Determine the quality factor for a hypothetical cavity fully coated with Nb$_3$Sn:
    \begin{equation}
        Q_{\text{Nb}_3\text{Sn,full}} = \frac{G_\text{total}}{R_{s,\text{Nb}_3\text{Sn}}}
    \end{equation}
    
    \item \textbf{Determine leakage correction factor.} Compare simulated ideal performance to measured performance for the copper reference:
    \begin{equation}
        \alpha = \frac{Q_\text{Cu,ideal}}{Q_\text{Cu,measured}}
    \end{equation}
    
    \item \textbf{Apply final correction.} The corrected quality factor, representing intrinsic Nb$_3$Sn performance:
    \begin{equation}
        Q_{\text{Nb}_3\text{Sn,corrected}} = \alpha \times Q_{\text{Nb}_3\text{Sn,full}}
    \end{equation}
\end{enumerate}

\subsubsection{Assumptions and Limitations}

This methodology assumes:
\begin{itemize}
    \item Geometric factors are accurately determined from simulation
    \item Surface coverage and material boundaries are well-defined
    \item Reference material surface resistance is accurately characterized
    \item Leakage factor remains consistent across measurements using the same assembly procedure
    \item Current distributions are similar between materials at the operating frequency
\end{itemize}

Uncertainties in these assumptions propagate to the final corrected quality factor and should be estimated for rigorous error analysis.

\subsection{Quality Factor vs Temperature}\label{appsec:QvsTProtocol}

\begin{enumerate}
    \item Cool cavity to 2~K in zero applied field
    \item Wait 5 minutes for thermal equilibrium
    \item Measure $Q$ using transmission method above
    \item Increase temperature by 0.5--1~K
    \item Wait 5 minutes, then measure $Q$
    \item Repeat until $Q$ drops significantly (above $T_c$, cavity goes normal)
    \item Plot $Q$ vs $T$; sharp drop indicates superconducting transition
\end{enumerate}

\subsection{Quality Factor vs Magnetic Field}\label{appsec:QvsBProtocol}

\begin{enumerate}
    \item Cool cavity to 2~K in zero field
    \item Measure $Q$ at 0~T
    \item Increase field in 0.25--1~T steps
    \item Wait 5 minutes for field stabilization at each step
    \item Measure $Q$ at each field value
    \item Continue to maximum field (9~T)
    \item Plot $Q$ vs $B$; note field where $Q$ begins to degrade
\end{enumerate}

\textbf{Tips:}
\begin{itemize}
    \item Always check coupling before starting a measurement sweep
    \item Resonance frequency shifts with temperature and field---re-center VNA span as needed
    \item Save full VNA traces, not just extracted $Q$ values
    \item Record PPMS temperature and field readings with each VNA measurement
\end{itemize}

\renewcommand*{\bibname}{References}

\printbibliography

\begin{biosketch}
\vspace{12pt}
\begin{center}
\noindent\textbf{\Large Andre Robert Juliao}\\
\vspace{6pt}
Tallahassee, FL $\mid$ U.S. Citizen
\end{center}
\vspace{12pt}

\noindent\textit{Experimental physicist specializing in superconducting thin films for high-field applications. Expertise in Nb$_3$Sn synthesis, RF cavity development, and cryogenic characterization. 6+ years advancing DOE-funded axion dark matter detection technology.}

\vspace{12pt}

\noindent\textbf{Education}
\vspace{6pt}

\noindent\textbf{Florida State University}, Tallahassee, FL
\begin{itemize}
    \item Ph.D. in Physics (Experimental Condensed Matter), Expected December 2025
    \item M.S. in Physics, 2021 (GPA: 3.80)
    \item B.S. in Physics with Mathematics Minor, 2018 (Major GPA: 3.50)
\end{itemize}
\vspace{12pt}

\noindent\textbf{Awards and Fellowships}
\vspace{6pt}
\begin{itemize}
    \item \textbf{DOE Office of Science Graduate Student Research (SCGSR) Fellowship}, Solicitation 1 2024
\end{itemize}
\vspace{12pt}

\noindent\textbf{Research Experience}
\vspace{6pt}

\noindent\textbf{Full-Time Operations Research Assistant} \hfill September 2018--August 2019\\
\textit{Applied Superconductivity Center, National High Magnetic Field Laboratory}
\begin{itemize}
    \item Managed weekly filling of superconducting magnets with liquid helium and nitrogen
    \item Conducted critical current measurements of high-temperature superconductor BSCCO-2212
\end{itemize}
\vspace{12pt}

\noindent\textbf{Graduate Research Assistant} \hfill August 2019--Present\\
\textit{Applied Superconductivity Center, National High Magnetic Field Laboratory}
\begin{itemize}
    \item Led operations of \$1M magnetron sputtering system with HiPIMS capability
    \item Pioneered on-site RF cavity testing, eliminating external characterization costs
    \item Fabricated and characterized over 100 samples, developing three novel Nb$_3$Sn deposition recipes that achieved $T_c > 17$~K, performance comparable to world-leading films
    \item Served as Principal Experimentalist for DOE-HEP funded axion dark matter cavity development
\end{itemize}
\vspace{12pt}

\noindent\textbf{Publications}
\vspace{6pt}

\noindent\textit{Peer-Reviewed Articles}
\begin{itemize}
    \item Withanage, W.K., \textbf{Juliao, A.R.}, Cooley, L.D. (2021). ``Rapid Nb$_3$Sn film growth by sputtering Nb on hot bronze.'' \textit{Superconductor Science and Technology}, 34, 065010. (10 citations)
    \item Kossler, W.J., Klein, S., Morrow, D., \textbf{Juliao, A.R.} (2016). ``Modern Cavendish experiment for high school students.'' \textit{American Journal of Physics}, 84, 448.
\end{itemize}

\noindent\textit{Manuscripts in Preparation}
\begin{itemize}
    \item ``Nb$_3$Sn Healing Joints for Superconducting Applications'' -- \textit{To be submitted to Superconductor Science and Technology December 2025}
    \item ``High-Field RF Performance of Nb$_3$Sn Films on Copper Substrates via Cu-Sn Route'' -- \textit{In preparation}
\end{itemize}
\vspace{12pt}

\noindent\textbf{Conference Presentations}
\vspace{6pt}

\noindent\textit{Oral Presentations}
\begin{itemize}
    \item ``Nb$_3$Sn for Dark Matter Detection,'' LLNL Quantum Group Seminar, Livermore, CA (August 2025)
    \item ``Nb$_3$Sn on Cu in High Field,'' Thin Film SRF Workshop, CEA Paris, France (2024)
    \item ``Nb$_3$Sn Films for Axion Dark Matter Detection,'' ADMX Collaboration Meeting, Chicago, IL (2024)
    \item ``Nb$_3$Sn Films on Cu Base Materials,'' Thin Film SRF Workshop, Jefferson Lab, VA (2022)
\end{itemize}

\noindent\textit{Poster Presentations}
\begin{itemize}
    \item ``Nb$_3$Sn Thin Films for Dark Matter Detection,'' SRF Conference, Tokyo, Japan (2025)
    \item ``Nb$_3$Sn Thin Films for Dark Matter Detection,'' AXION Workshop, Berkeley, CA (May 2025)
    \item ``Bronze Route Nb$_3$Sn Development,'' Applied Superconductivity Conference, Salt Lake City, UT (2024)
    \item ``Nb$_3$Sn Thin Films on Different Substrates,'' Applied Superconductivity Conference, Honolulu, HI (2022)
\end{itemize}

\noindent\textbf{Technical Skills}
\vspace{6pt}
\begin{itemize}
    \item \textbf{Fabrication:} Magnetron sputtering, HiPIMS, thin film deposition optimization
    \item \textbf{Characterization:} SEM (Zeiss, Thermo Fisher), XRD (Rigaku), AFM, FIB, PPMS (Quantum Design), RF cavity testing (VNA)
    \item \textbf{Computational:} COMSOL Multiphysics, Python (NumPy, SciPy, matplotlib), LaTeX
    \item \textbf{Specialized Expertise:} Cu-Sn route Nb$_3$Sn on copper substrates for high-field applications, cryogenic measurements (4K-300K)
\end{itemize}
\vspace{12pt}

\noindent\textbf{Teaching and Mentorship}
\vspace{6pt}
\begin{itemize}
    \item Teaching Assistant for 4 courses: Materials Science, Physics I \& II, Radiation in Materials (2019--2023)
    \item Mentored 4 student researchers (2 undergraduate, 2 high school) in thin film fabrication, resulting in 3 poster presentations
    \item Received consistently excellent student evaluations
\end{itemize}
\vspace{12pt}

\noindent\textbf{Collaborations}
\vspace{6pt}
\begin{itemize}
    \item \textbf{Lawrence Livermore National Laboratory:} mK RF characterization using dilution refrigeration in near quantum-limited regime
    \item \textbf{ADMX Collaboration:} Superconducting cavity design and testing for axion dark matter detection
    \item \textbf{Applied Superconductivity Center:} Nb$_3$Sn for magnet applications
\end{itemize}

\end{biosketch}

\end{document}